%% file: triple-point-draft.tex
\documentclass[12pt]{article}
\pdfoutput=1

\usepackage{latexsym, amsmath, amsfonts, amssymb, braket}
\usepackage{tikz-cd}
\usepackage[framemethod=TikZ]{mdframed}
\usetikzlibrary{decorations.pathmorphing, cd, decorations.markings, calc, patterns, shapes.symbols}
\usetikzlibrary{decorations.pathreplacing, calligraphy}
\usetikzlibrary{arrows}
\usetikzlibrary{shapes}
\usetikzlibrary{matrix}
\usetikzlibrary{positioning}
\usetikzlibrary{shapes.multipart}
\usepackage{diagbox}
\usetikzlibrary{positioning}
\usetikzlibrary{calc}
\usepackage{xstring}
\usepackage{mathtools}
\usepackage{hhline}
\usepackage{cancel}

\usepackage{mathrsfs}
\usepackage[american]{babel}
\usepackage{graphicx}
\usepackage{bbm}
\usepackage[nosort]{cite}
\usepackage{amsfonts}
\usepackage{amsmath}
\usepackage{amssymb}
\usepackage{array}
\usepackage{bigints}
\usepackage{booktabs}
\usepackage[nosort]{cite}
\usepackage{comment}
\usepackage{caption}
\usepackage{color}
\usepackage{dsfont}
\usepackage{float}
\usepackage{framed}
\usepackage{graphicx}
\usepackage{indentfirst}
\usepackage{mathrsfs}
\usepackage{multirow}
\usepackage{setspace}
\usepackage{subdepth}
\usepackage{subfig}
\usepackage{titlesec}
\usepackage[dotinlabels]{titletoc}
\usepackage{wrapfig}
\usepackage[all]{xy}
\usepackage{young}
\usepackage{enumitem}
\usepackage[vcentermath]{youngtab}

\numberwithin{equation}{section}

\usepackage[colorlinks=true, citecolor=black!90!black, linkcolor=black!90!black, linktocpage=true, urlcolor=red!70!black]{hyperref}

\usepackage[left=2.5cm,right=2.5cm,top=2.5cm,bottom=3cm]{geometry}
\renewcommand\theparagraph{ \thesubsubsection.\alph{paragraph}}

\titleformat{\paragraph}
  {\normalfont\normalsize\sf}
  {\theparagraph}{1em}{}

\titlespacing*{\paragraph}
  {-5pt}{3.25ex plus 1ex minus .2ex}{1ex}

\def\tilde{\widetilde}
\def\t{\tilde}
\def\ep{\varepsilon}
\def\hat{\widehat}

\def\bar{\overline}
\def\b{\bar}

\def\half{{1 \over 2}}
\def\d{\partial}

\newcommand{\C}{{\mathbb C}}
\newcommand{\R}{{\mathbb R}}
\def\P{\hbox{$\mathbb P$}}

\def\CC{{\mathcal C}}
\def\CD{{\mathcal D}}

\def\CM{{\mathcal M}}
\def\CN{{\mathcal N}}
\def\CO{{\mathcal O}}

\def\CR{{\mathcal R}}

\def\CT{{\mathcal T}}

\def\CV{{\mathcal V}}

\DeclareFontShape{OT1}{cmr}{mx}{n}%
    {<->cmr10}{}
\newcommand{\mytitlefont}{\fontseries{mx}\selectfont}
\DeclareMathAlphabet{\titlemath}{OT1}{cmr}{mx}{n}

\newcommand{\slashed}{{\bf\not}}

\newcommand{\Z}{\mathbb{Z}}

\newcommand{\Dslash}{D\!\!\!\!\slash\,}
\newcommand{\dslash}{\partial \!\!\!\slash\,}

\newcommand{\be}{\begin{equation}} \newcommand{\ee}{\end{equation}}
\newcommand{\bea}{\begin{equation} \begin{aligned}} \newcommand{\eea}{\end{aligned} \end{equation}}

\newcommand{\cC}{\mathcal{C}}
\newcommand{\cD}{\mathcal{D}}

\newcommand{\cM}{\mathcal{M}}

\newcommand{\cO}{\mathcal{O}}

\newcommand{\cT}{\mathcal{T}}

\newcommand{\bC}{\mathbb{C}}

\newcommand{\bR}{\mathbb{R}}

\newcommand{\bZ}{\mathbb{Z}}

\def\repa{\raise4pt\hbox{$\square$}\mkern-14mu\raise-4pt\hbox{$\square$}}
\def\repab{\overline{\raise4pt\hbox{$\square$}\mkern-14mu\raise-4pt\hbox{$\square$}\mkern-1mu}}

\DeclareMathOperator{\Tr}{Tr}
\DeclareMathOperator{\tr}{tr}

\DeclareMathOperator{\sign}{sign}

\def\[#1\]{%
  \begin{equation}\begin{gathered}#1\end{gathered}\end{equation}%
}

\begin{document}

\begin{titlepage}

\begin{center}

~\\[5pt]

{\fontsize{26pt}{0pt} \mytitlefont From QED$_3$ to Self-Dual Multicriticality \\[5pt] in the Fradkin-Shenker Model
   \\[7pt]} 

\vskip25pt

Thomas T.~Dumitrescu,$^1$  Pierluigi Niro$,^{2,3}$  and Ryan Thorngren$^1$

\vskip20pt

${}^{1}$ {\it Mani L.\,Bhaumik Institute for Theoretical Physics, Department of Physics and Astronomy, University of California, Los Angeles, CA 90095, USA} \\
\vskip 2mm

${}^{2}$ {\it SISSA, Via Bonomea 265, 34136 Trieste, Italy} \\
\vskip 2mm

${}^{3}$ {\it INFN, Sezione di Trieste, Via Valerio 2, 34127 Trieste, Italy} \\
\vskip 2mm

\bigskip
\bigskip

\end{center}

\noindent We consider the Fradkin-Shenker~${\mathbb Z}_2$ gauge-Higgs lattice model in 2+1 dimensions, i.e.~the toric code deformed by an in-plane magnetic field. Its phase diagram contains a multicritical CFT with gapless, mutually non-local electric and magnetic particles, exchanged by a~${\mathbb Z}_2^{\mathsf{D}}$ self-duality symmetry. We introduce a staggered generalization of the model in which these particles carry global $U(1)_e$ and $U(1)_m$ charges, respectively, and we propose a continuum QFT description in terms of QED$_3$ with $N_f = 2$ Dirac fermion flavors and a charge-two Higgs field with Yukawa couplings. The conjectured phase diagram harbors a multicritical CFT with $(O(2)_e \times O(2)_m)\rtimes\mathbb{Z}_2^\mathsf{D}$ symmetry, some of which is emergent in the QFT description. We compute the scaling dimensions of some operators using a large-$N_f$ expansion and find agreement with the emergent selection rules. The staggered model admits a deformation to the original Fradkin-Shenker model, which maps to unit-charge monopole operators in Higgs-Yukawa-QED$_3$ that break the $U(1)_e \times U(1)_m$ symmetry. We show explicitly that this deformation reproduces all features of the Fradkin-Shenker phase diagram. Finally, we propose a multicritical duality between Higgs-Yukawa-QED$_3$ and the easy-plane $\mathbb{ CP}^1$ model (i.e.~two-flavor scalar QED$_3$ with a suitable potential), which describes spin-1/2 anti-ferromagnets on a square lattice. This duality implies a first-order line of N\'eel-VBS transitions ending in a deconfined quantum multicritical point, described by the same $O(2)_e \times O(2)_m$ symmetric CFT that arises in the staggered Fradkin-Shenker model, which separates it from a gapped~${\mathbb Z}_2$ spin liquid phase.

\vfill

\end{titlepage}


\hypersetup{linktoc=all}
\pagenumbering{arabic}
\setcounter{page}{1}
\setcounter{footnote}{0}
\setcounter{secnumdepth}{4}
\renewcommand{\thefootnote}{\arabic{footnote}}

{\renewcommand{\baselinestretch}{.88} \parskip=0pt
	\setcounter{tocdepth}{4}

\tableofcontents}


\newpage

\section{Introduction and Summary}

\begin{figure}[t!]
    \centering
    \begin{tikzcd}[column sep=-0.5em]
    & \boxed{\hyperref[sec:hyqed]{\substack{\vspace{0.5mm}\textbf{Higgs-Yukawa-QED$_3$} \vspace{1mm} \\ \text{(HYQED)} \vspace{1mm} \\ S(O(2)_e \times O(2)_m)\rtimes (\bZ_2^{\mathsf{D}} \times \bZ_2^{\mathsf{T}}) \vspace{1mm} \\ \text{Section }\ref{sec:hyqed}}}} \arrow[dr,"\text{emergent mirror}"']
    & & \boxed{\hyperref[sec:easy]{\substack{\textbf{Easy-Plane $\mathbb{CP}^1$} \vspace{0.5mm} \\ \text{(EP$\mathbb{CP}^1$)} \vspace{1mm} \\ (O(2)_e \times O(2)_m)\rtimes  \bZ_2^{\mathsf{T}} \vspace{1mm} \\ \text{Section } \ref{sec:easy}}}} \arrow[dl,"\text{emergent duality}"] \\
    \boxed{\hyperref[secU1FSmodel]{\substack{\textbf{Staggered Fradkin-Shenker Model} \vspace{1mm} \\ \text{(SFS Model)} \vspace{1mm} \\ \text{Section }\ref{secU1FSmodel}}}} \arrow[rr] \arrow[d,"\text{ symmetry breaking deformation}"]
    & & \boxed{\substack{\textbf{SFS CFT} \vspace{1mm} \\ (O(2)_e \times O(2)_m)\rtimes (\bZ_2^{\mathsf{D}} \times \bZ_2^{\mathsf{T}}) \vspace{1mm}}} \arrow[d,"\text{ charge-1 monopoles}"] & \\
    \boxed{\hyperref[secFSmodel]{\substack{\textbf{Fradkin-Shenker Model} \vspace{1mm} \\ \text{(FS Model)} \vspace{1mm} \\ \text{Section }\ref{secFSmodel} }}} \arrow[rr]
    & & \boxed{\substack{\textbf{FS CFT} \vspace{1mm} \\  \bZ_2^{\mathsf{D}} \times \bZ_2^{\mathsf{T}} \vspace{1mm}}} &
    \end{tikzcd}
    \caption{Relation between the $2+1$d models in this paper, with clickable links, and their global symmetries. On the left are two lattice models: the Fradkin-Shenker (FS) model, which refers to the toric code in an in-plane field, and the Staggered Fradkin-Shenker (SFS) model, which is a $U(1)_e \times U(1)_m$-symmetric variant we introduce. When tuned to their self-dual multicritical points, we expect these lattice models to flow to conformal field theories, which we refer to as the FS CFT and the SFS CFT, respectively. On top, we propose two continuum QFTs that flow to the SFS CFT at their multicritical point: Higgs-Yukawa-QED$_3$ (HYQED) and the easy-plane $\mathbb{CP}^1$ (EP$\C\P^1$) model. In both, a $\mathbb{Z}_2$ symmetry (referred to as mirror and duality, respectively) emerges in the IR, which is required to match the global symmetry of the SFS CFT.  From the point of view of the continuum Lagrangians, the symmetry-breaking deformation from the SFS model to the FS model is given by certain unit-charge monopole operators. These deformations preserve a $\bZ_2^{\mathsf{C}_e} \times \bZ_2^{\mathsf{C}_m} \subset O(2)_e \times O(2)_m$ symmetry corresponding to certain lattice rotations; it is not known if they act on the FS CFT in the deep IR, and for this reason we have not included them in the figure (see section~\ref{FSviaMonos} for more detail).}
    \label{fig:overview}
\end{figure}
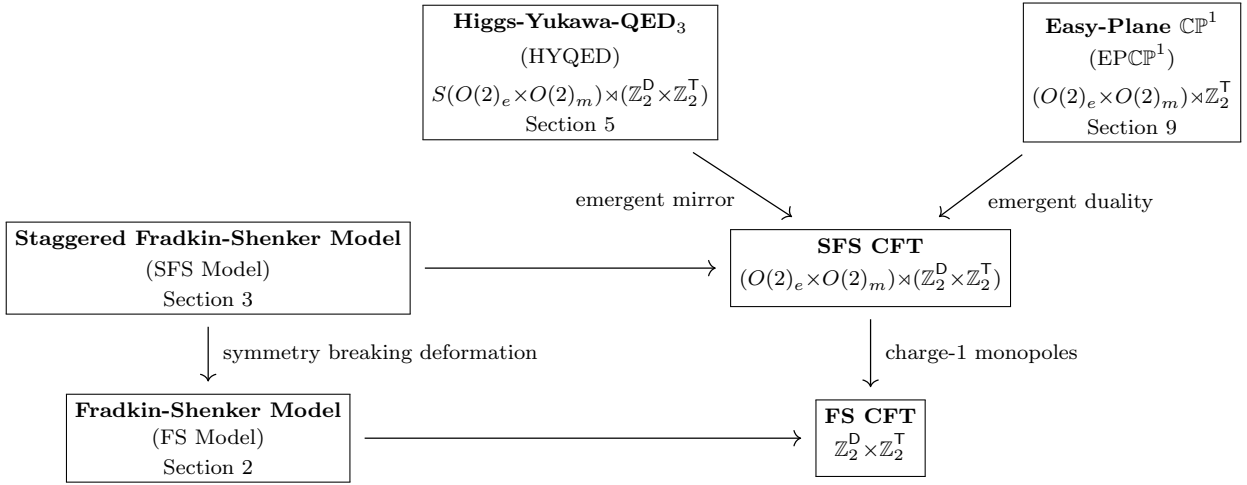

The study of strongly-coupled critical points in three dimensions lies at the crossroads of  quantum many-body physics, statistical mechanics, and quantum field theory (QFT), all of which exhibit some of their richest phenomena in this arena. In this paper we combine lattice and continuum QFT arguments to elucidate the phase diagram of the Fradkin-Shenker lattice model~\cite{Fradkin:1978dv} and its multicritical point~\cite{Tupitsyn:2008ah}.  

\subsection{The Fradkin-Shenker Lattice Model}

Fradkin and Shenker~\cite{Fradkin:1978dv} analyzed~$\Z_2$  lattice gauge theory in three Euclidean dimensions coupled to a~$\Z_2$ Higgs field, famously demonstrating that the Higgs and confining regimes of this model comprise a single connected phase, which we sometimes call the Higgs/confined phase.\footnote{~The $\Z_2$ lattice gauge theory was first studied by Wegner~\cite{Wegner:1971app}, and Higgs-confinement continuity was also demonstrated for other lattice gauge theories around the same time in~\cite{Banks:1979fi}.} In the modern language, this ``Higgs-confinement continuity'' reflects the absence of any one-form or other symmetries in the model distinguishing these non-degenerate, gapped phases~\cite{Gaiotto:2014kfa,Verresen:2022mcr,Dumitrescu:2023hbe}.

It was shown in~\cite{Tupitsyn:2008ah} that the Fradkin-Shenker model (when viewed as a Hamiltonian lattice model in~$2+1$ dimensions) is identical to the toric code~\cite{Kitaev:1997wr} Hamiltonian~$H_\text{TC}$, deformed by an in-plane magnetic field in the~$X$ and~$Z$ directions,
\begin{equation}\label{eq:Hfsintro}
  \qquad  H_\text{FS} = H_\text{TC} - h_e \sum_\ell Z_\ell - h_m \sum_\ell X_\ell~, \qquad h_{e, m} \geq 0~.
\end{equation}
Here~$X_\ell, Z_\ell$ denote the Pauli matrices~$\sigma_x, \sigma_z$ acting on the qubits residing on the links~$\ell$ of the spatial lattice comprising the toric code. See section~\ref{secFSmodel}, which also serves as an extended introduction, for more detail. Throughout this paper, we refer to~\eqref{eq:Hfsintro} as the Fradkin-Shenker (FS) lattice model. (See figure~\ref{fig:overview} for a summary of the various models analyzed in this paper, and how they are related.) The FS model has a unitary~$\Z_2^{\sf D}$ duality symmetry~$\sf D$ that exchanges~$h_e \leftrightarrow h_m$, i.e.~it is a symmetry on the self-dual line~$h_e = h_m$.\footnote{~In addition, the FS lattice model enjoys an anti-unitary~$\Z_2^{\sf T}$ time-reversal symmetry~$\sf T$, as well as a host of lattice symmetries that will be discussed below.}

\begin{figure}[t]
    \centering   
\begin{tikzpicture}[x=4cm,y=4cm] 
  \draw[gray!60, line width=0.3pt] (0,0) rectangle (1.35,1.35);

  \node[below=8pt] at (0.64,0.02) {$h_m$};
  \node at (-0.15,0.64) {$h_e$};
  \draw[->] (0.49,-0.05) -- (0.79,-0.05);
  \draw[->] (-0.05,0.49) -- (-0.05,0.79);

  \node at (0.23,1.06) {};
  \node at (0.3,0.96) {Higgsing};
  \node at (0.23,0.72) {\textcolor{blue}{Ising$^*_e$}};

  \node at (0.81,0.43) {\textcolor{blue}{Ising$^*_m$}};
  \node at (1.0,0.25) {Confinement};
  \node at (1.05,0.18) {};

  \node at (0.23,0.28) {Toric};
  \node at (0.23,0.18) {code};

\draw[<->] (-0.25,0.05) -- (0.05,-0.25);

\node at (-0.15,-0.15) {$\mathsf{D}$};

  \draw[line width=0.8pt,blue]
    (0.0,0.64) -- (0.64,0.64) -- (0.64,0.00);

  \draw[densely dotted, line width=0.8pt]
    (0.64,0.64) -- (1,1);
  \node at (0.91,0.85) {\hskip-12pt $\mathsf{D}$ SSB};  
  \fill[color=teal] (1,1) circle (0.03);
  \fill[color=purple] (0.640,0.640) circle (0.03);
  \node at (0.9,0.640) {\textcolor{purple}{FS CFT}};
  \node at (1.1,1.1) {\textcolor{teal}{Ising}};
\end{tikzpicture}
    \caption{Schematic phase diagram of the Fradkin-Shenker (FS) model~\eqref{eq:Hfsintro}, reproduced from \cite{Wu:2012cj}, as a function of~$h_e,h_m \geq 0$. The~$\Z_2^{\sf D}$ duality symmetry acts as~${\sf D}: h_e \leftrightarrow h_m$ and is preserved along the self-dual diagonal~$h_e = h_m$. In the lower left region there is the toric code phase, described at long distances by an emergent~$\Z_2$ gauge theory, which couples electrically and magnetically to dynamical, massive~$e$ and~$m$ anyons that are exchanged by~$\sf D$. These become massless on the second-order~$\Z_2$-gauged Ising lines (indicated in blue and labeled~Ising$^*_e$ and Ising$^*_m$, respectively), which describe electric and magnetic Higgs transitions into the trivial Higgs/confined phase. The Ising$^*_{e,m}$ lines meet at a multicritical point (indicated in red) on the self-dual diagonal, which is described by the FS CFT. Continuing further along this diagonal into the Higgs/confined phase, we find a first-order (dotted) line on which~$\sf D$ is spontaneously broken, and which ends at a conventional Ising critical point (indicated in green).}
    \label{figfsphasediagram}
\end{figure}
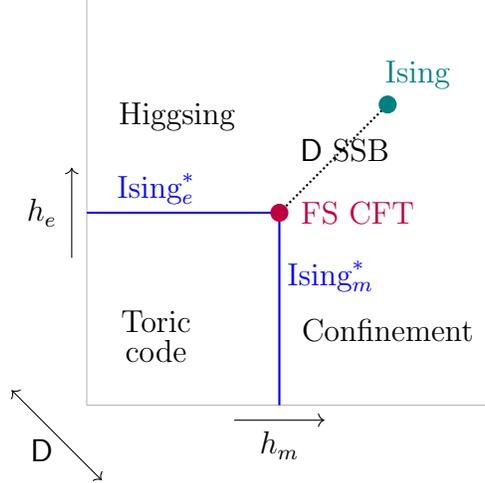

The phase diagram of the Fradkin-Shenker model is depicted in figure~\ref{figfsphasediagram}. The gapped toric code phase at small~$h_{e,m}$ describes an emergent~$\Z_2$ topological gauge theory at long distances, hosting bosonic electric~$e$ and magnetic~$m$ anyons with mutual braiding phase~$-1$, and a composite emergent fermion~$f = em$. In the FS model~\eqref{eq:Hfsintro} these anyons are dynamical, fluctuating particles,\footnote{~This breaks the exact one-form symmetries of the toric code~$H_\text{TC}$.} which are massive in the toric code phase. Note that the microscopic duality symmetry~${\sf D} : e \leftrightarrow m$ swaps the dynamical~$e$ and~$m$ anyons, so that~$f$ is invariant. 

As we increase~$h_e$ (while keeping~$h_m$ small), the dynamical~$e$ anyon becomes lighter, and eventually massless, before affecting a second-order transition to the trivially gapped Higgs phase. (Throughout, the~$m$ anyon remains massive.) As shown in~\cite{Fradkin:1978dv}, this transition is in the gauged Ising universality class~\cite{Wegner:1971app}, which we indicate by the blue Ising$^*_e$ line in figure~\ref{figfsphasediagram}, to emphasize that the~$e$ anyon becomes massless there. Near this transition, the~$e$ anyon is described by a real Ising order parameter~$\phi_e$ that couples electrically to the~$\Z_2$ gauge field of the toric code phase (where~$\phi_e$ is massive). In the Higgs phase~$\phi_e$ acquires a vacuum expectation value (VEV) $\langle \phi_e \rangle \neq 0$, leading to a trivially gapped Higgs vacuum.\footnote{~\label{fn:AC}At the level of TQFTs, the transition from~$\Z_2$ gauge theory to the trivial Higgs phase can be characterized by algebraically condensing the~$e$ anyon, i.e.~by gauging the electric~$(\Z_2^{(1)})_e$ one-form symmetry that is present in the~$\Z_2$ TQFT (see section~\ref{sec:z2gaugereview}). This symmetry is explicitly broken by the dynamical~$\phi_e$ field. (Thus, it cannot be gauged.) Instead, the transition becomes a Higgsing transition for~$\phi_e$, which acquires a VEV in the trivial phase. This is a dynamical version of anyon condensation, which is distinct from the algebraic operation on TQFTs. See~\cite{NatiUQM} for a recent discussion of this point. See section~\ref{subsecsuperconductorintuition} for some further comments on anyon condensation.} Since the~$m$ anyon remains massive, the Ising$^*_e$ theory has an emergent magnetic~$(\Z_2^{(1)})_m$ one-form symmetry, which is spontaneously broken in the toric code phase and unbroken in the Higgs phase. As already discussed above, this symmetry is not present in the full FS model. 

By acting with duality~$\sf D$, we find that increasing~$h_m$ at small~$h_e$ leads to a confinement transition, indicated by~Ising$_m^*$ in figure~\ref{figfsphasediagram}, whose order parameter~$\phi_m$ describes the massless~$m$ anyon and couples magnetically to the~$\Z_2$ gauge field of the toric code. It was shown in~\cite{Fradkin:1978dv} that the Higgs phase at large~$h_e$ and the confining phase at large~$h_m$ are continuously connected, i.e.~they are the same phase. This follows almost immediately from~\eqref{eq:Hfsintro}, as we review in section~\ref{sec:FSdetail}.

The fate of the Ising$^*_{e,m}$ transitions in the interior of the phase diagram has been investigated numerically, starting with~\cite{Tupitsyn:2008ah}, and subsequently in~\cite{Vidal:2008uy, Wu:2012cj, Somoza:2020jkq,Oppenheim:2023uvf}. There it was shown that the second-order Ising$^*_{e,m}$ transitions persist all the way until the self-dual line~$h_e = h_m$, where they meet in a multicritical point. A more recent numerical study~\cite{Somoza:2020jkq} of this multicritical point indicates that it is described by a relativistic conformal field theory (CFT), which we refer to as the Fradkin-Shenker (FS) CFT (indicated by a red dot in figure~\ref{figfsphasediagram}). The scaling dimensions of the operators~$\CO_{ \sf D \, \text{even, odd}}$ that dial the~$\sf D$-even and~$\sf D$-odd tuning parameters~$h_e + h_m$ and~$h_e - h_m$, respectively, were estimated as
\begin{equation}\label{eq:SFCFTDeltas}
    \Delta_{ \sf D \, \text{even}} \simeq 1.51~, \qquad \Delta_{ \sf D \, \text{odd}} \simeq  1.22~.
\end{equation}

Finally, there is a line of first-order phase transitions that emanates from the FS multicritical point into the trivial Higgs/confined phase along the self-dual~$h_e = h_m$ direction. Along this line, the duality symmetry~$\sf D$ is spontaneously broken, leading to two degenerate vacua. (This is indicated by the dotted line in figure~\ref{figfsphasediagram}.) This line ends in a conventional, ungauged, Ising critical point (indicated by a green dot in figure~\ref{figfsphasediagram}), at which the duality symmetry is restored and the theory transitions into the trivial Higgs/confined phase. 

The nature of the Fradkin-Shenker CFT at the multicritical point has remained somewhat mysterious (see however~\cite{Shi:2024pem}, which we will comment on in section~\ref{sec:discussion}). Intuitively, the reason it exists is also what makes it hard to analyze: it sits at the junction of two continuous phase transitions with massless~$e$ and~$m$ anyons, which are mutually non-local because of their non-trivial mutual braiding. In this sense the FS CFT is a (2+1)-dimensional, non-supersymmetric analogue of Argyres-Douglas CFTs~\cite{Argyres:1995jj, Argyres:1995xn}. The latter are superconformal theories 3+1 dimensions, which can be thought of as (limits of) strongly-coupled~$U(1)$ gauge theories with  massless charges, monopoles, and dyons (which are also mutually non-local). Likewise, the FS CFT has massless electric charges $e$, magnetic fluxes $m$, and fermions~$f$, which are bound states of $e$ and $m$. 

In this paper we will propose two dual continuum QFT descriptions of the FS CFT and the Fradkin-Shenker phase diagram in figure~\ref{figfsphasediagram}. Both duals are Abelian gauge theories -- which we refer to as Higgs-Yukawa-QED$_3$ (HYQED) and the Easy-Plane~$\C\P^1$ (EP$\C\P^1$) model -- that are further deformed by the addition of monopole operators (see figure~\ref{fig:overview}). To explain how this description comes about, we first introduce a generalization of the FS model with additional global symmetries.

\subsection{The Staggered Fradkin-Shenker (SFS) Model}

We will make progress towards understanding the Fradkin-Shenker model and its multicritical point by introducing a closely related model with~$U(1)_e \times U(1)_m$ zero-form symmetry in section~\ref{secU1FSmodel}. The corresponding charges~${\sf Q}_e$ and~${\sf Q}_m$ count the number of~$e$ and~$m$ anyons.\footnote{~More precisely, $e$ and~$m$ have fractional charges~$({\sf Q}_e, {\sf Q}_m) = (1/2, 0)$ and~$(0,1/2)$, respectively, which implies a mixed anomaly that is discussed in section~\ref{sec:latanomaly}.} Since these charges are staggered on the vertices and plaquettes of the spatial lattice, with a certain sublattice structure, we refer to this~$U(1)_e \times U(1)_m$ symmetric version of the Fradkin-Shenker model as the Staggered Fradkin-Shenker (SFS) model. (See figure~\ref{fig:overview}, as well as section~\ref{secU1FSmodel}.)  The SFS Hamiltonian is given by a suitable symmetry projection of the Fradkin-Shenker Hamiltonian~\eqref{eq:Hfsintro}. It still involves two tuning parameters~$h_{e, m}$, exchanged by duality~$\sf D$, which now also exchanges~$U(1)_e \leftrightarrow U(1)_m$. 

In section~\ref{secU1FSmodel} we establish the following properties of the SFS model:
\begin{itemize}
    \item It has a mixed anomaly involving~$U(1)_{e, m}$ and time reversal~$\sf T$. Hence, unlike the Fradkin-Shenker model, the SFS model has no trivially gapped phases. 
    
    \item We analyze in detail the action of the lattice symmetries on the charges~${\sf Q}_{e, m}$ and show that they extend~$U(1)_{e,m}$ to~$O(2)_{e,m}$.
    
    \item On the~$h_e$-axis and the~$h_m$-axis, the model can be shown to have a single second-order transition in the~$\Z_2$-gauged~$O(2)$ Wilson-Fisher universality class, which we indicate by~$O(2)_{e}^*$ and $O(2)_{m}^*$, respectively.  
\end{itemize}
We do not have direct access to the interior of the SFS phase diagram. In section~\ref{int:QFT} below we will propose a continuum QFT dual for the SFS model, which will allow us to conjecture its phase diagram in figure~\ref{figU1phasediagram}. In section~\ref{int:fstosfs} we further explain how to recover the full phase diagram of the Fradkin-Shenker model in figure~\ref{figfsphasediagram} by deforming the SFS model and its dual.

\subsection{Continuum QFT Description in terms of Higgs-Yukawa-QED}\label{int:QFT}

We propose in section~\ref{sec:hyqed} that the~$U(1)_e \times U(1)_m$ symmetric SFS model is described at long distances by the following continuum QFT, which we term Higgs-Yukawa-QED (HYQED) (see also figure~\ref{fig:overview}),\footnote{~Here~$D_a$ is the covariant derivative for unit-charge fields.}
\begin{equation}\label{intro:hyqedlag}
\begin{split}
\mathscr{L}_{\rm HYQED} = &- \frac{1}{4e^2} f^{\mu\nu}f_{\mu\nu} - i \bar\psi_i \Dslash_a \psi^i  - |D_{2a} \phi|^2 -  \lambda_4 |\phi|^4 +   \mathscr{L}_\text{Yukawa} + {\mathscr L}_\text{masses}~.
\end{split}
\end{equation}
This theory was also discussed in~\cite{Jian:2017chw, Boyack:2018urb, Dupuis:2021flq}. It can be viewed as a deformation of QED$_3$ with~$N_f = 2$ Dirac fermion flavors~$\psi^{i = 1,2}$ (here~$a_\mu$ is a~$U(1)$ gauge field -- more precisely a spin$_c$ connection -- with field strength~$f_{\mu\nu}$), which we review in section~\ref{section:qed3review}. In particular we recall from~\cite{Chester:2024waw, Dumitrescu:2024jko} why two-flavor QED$_3$ (without additional fields) is believed not to flow to a CFT in the IR, but rather to spontaneously break its global symmetry.

We also add a charge-2 Higgs field~$\phi$ that couples to the fermions via the Yukawa interaction~${\mathscr L}_\text{Yukawa} \sim  \phi^* \psi^1 \psi^2 + (\text{h.c.})$ in~\eqref{intro:hyqedlag}. The symmetries and anomalies of HYQED are discussed in section~\ref{sec:hyqed}. They are matched with those of the SFS lattice model in section~\ref{u1LatToCont}.\footnote{~As we will discuss below, a $\mathbb{Z}_2$ symmetry of the SFS lattice model, which we refer to as mirror symmetry, must emerge in the IR of HYQED.} We find that the duality symmetry~$\sf D$ of the SFS model exchanges~$\psi^1 \leftrightarrow \psi^2$, while leaving~$\phi$ invariant. The~$U(1)_e \times U(1)_m$ symmetries, which are also exchanged by~$\sf D$, are linear combinations of the~$U(1)_f$ flavor symmetry under which~$\psi^1$ and~$\psi^2$ have charges~$q_f = \pm1$, and the~$U(1)_\CM$ monopole symmetry of QED$_3$. The minimal monopole operators~$\CM^1$ and~$\CM^2$ are Lorentz scalars with~$q_\CM = 1$; their~$U(1)_{e,m}$ charges are~$(q_e, q_m) = (1,0)$ and~$(q_e, q_m)=(0,1)$, respectively. 

Together with other symmetry-related considerations discussed in section~\ref{u1LatToCont}, this implies that the tuning parameters~$h_{e,m}$ of the SFS lattice model map to certain mass terms for the fermions~$\psi^i$ and the Higgs field~$\phi$,
\begin{equation}\label{intro:masses}
    {\mathscr L}_\text{masses} = - m_\phi^2 |\phi|^2 + i m_3 \b \psi_i {(\sigma_3)^i}_j \psi^j~, \qquad m_\phi^2, m_3 \in \R~,
\end{equation}
where $\sigma_a$ are the Pauli matrices. Note that the Higgs mass~$m_\phi^2$ is~$\sf D$-even and corresponds to~$h_e + h_m$ on the lattice (up to an additive constant), while the opposite-sign fermion mass~$i m_3 (\b \psi_1 \psi^1 - \b \psi_2 \psi^2)$ is~$\sf D$-odd and corresponds to the lattice parameter~$h_m - h_e$ (see section~\ref{sec:tuningparams}).

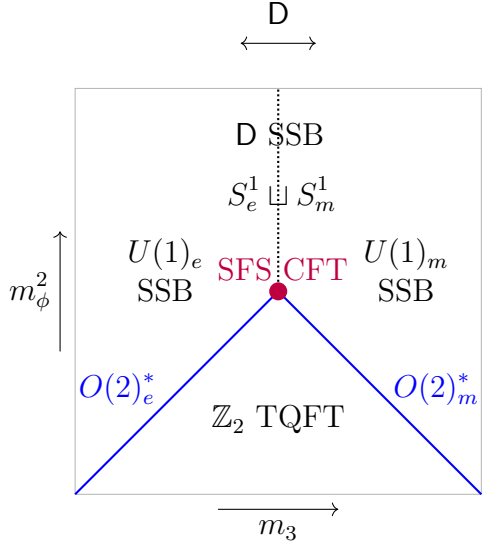
\begin{figure}[t!]
    \centering
\begin{tikzpicture}[x=4cm,y=4cm] 
    \draw[gray!60, line width=0.3pt] (0,0) rectangle (1.35,1.35);
  \node at (0.675,-0.12) {$m_3$};
  \draw[->] (0.475,-0.05) -- (0.875,-0.05);
  
  \node at (-0.15,0.675) {$m^2_\phi$};
  \draw[->] (-0.05,0.475) -- (-0.05,0.875);

  \node at (1.1,0.80) {$U(1)_m$};
  \node at (1.1,0.68) {SSB};

  \node at (1.2,0.35) {\textcolor{blue}{$O(2)^*_m$}};

  \node at (0.3,0.80) {$U(1)_e$};
  \node at (0.3,0.68) {SSB};

  \node at (0.14,0.35) {\textcolor{blue}{$O(2)^*_e$}};

  \node at (0.675,0.25) {$\mathbb{Z}_2$ TQFT};

\node at (0.68,1.2) {$\mathsf D$ SSB};
\node at (0.687,1.0) {$S^1_e \sqcup S^1_m$};

\draw[<->] (0.55,1.5) -- (0.8,1.5);
\node at (0.675,1.6) {$\mathsf{D}$};

  \draw[line width=0.8pt,blue]
    (0.0,0.0) -- (0.675,0.675) -- (1.35,0);
  \draw[densely dotted, line width=0.8pt]
    (0.675,0.675) -- (0.675,1.35);
  \fill[color=purple] (0.675,0.675) circle (0.03);
  \node at (0.69,0.75) {\textcolor{purple}{SFS CFT}};
\end{tikzpicture}
    \caption{Conjectured phase diagram of Higgs-Yukawa-QED as a function of the $\sf D$-even Higgs mass~$m_\phi^2$ and the $\sf D$-odd fermion mass~$m_3$ in~\eqref{intro:masses}. The multicritical point at the origin, described by the SFS CFT, is shown in red. The phase to its right ($m_3 > 0$) spontaneously breaks~$U(1)_m$, leading to an~$S^1_m$ circle of vacua. Its~$\sf D$-reflection at~$m_3 < 0$ spontaneously breaks~$U(1)_e$, with an~$S^1_e$ circle of vacua. These circles coexist on the~$m_\phi^2 > 0$ axis (indicated by the dotted black line), where the vacua are~$S_e^1 \sqcup S_m^1$ and the duality symmetry~${\sf D} : S_e^1 \leftrightarrow S_m^1$ is spontaneously broken. The phase at~$m_\phi^2 < 0$ is gapped, with a~$\Z_2$ TQFT in the IR, whose~$e$ and~$m$ anyons are exchanged by~$\sf D$. The transitions from the TQFT phase to the symmetry-breaking phases (indicated by blue lines) occur when either the~$e$ or the~$m$ anyon of the~$\Z_2$ TQFT becomes massless. The resulting second-order transitions are described by~$\Z_2$-gauged versions of the~$O(2)$ Wilson-Fisher CFT, which we denote by~$O(2)_{e,m}^*$, ending at the multicritical point.}
    \label{figU1phasediagram}
\end{figure}

We explore the phase diagram of Higgs-Yukawa-QED in section~\ref{subseccontinuumphasediagram}. It has various asymptotic regions that can be analyzed reliably. Together with its relationship to the Fradkin-Shenker phase diagram, discussed in section~\ref{int:fstosfs} below, this leads us to conjecture a complete phase diagram for HYQED, which is shown in figure~\ref{figU1phasediagram}:
\begin{itemize}
    \item For sufficiently large, negative~$m_\phi^2 < 0$ the charge-2 Higgs field~$\phi$ acquires a VEV, leading to a gapped~$\Z_2$ TQFT phase, whose dynamical~$e$ and~$m$ anyons arise from~$\pi$-flux vortices. Thanks to the fermions these vortices acquire fractional~$U(1)_e$ and~$U(1)_m$ charge~$1/2$, respectively, which also ensures that they are swapped by duality~$\sf D$. The fermionic anyon~$f = em$ of $\Z_2$ TQFT simply arises from the dynamical fermions~$\psi^i$ of HYQED, which are massive due to the Yukawa coupling.  

\item If we start in the~$\Z_2$ TQFT phases and increase the fermion mass~$m_3$, the~$m$ anyon will get lighter (while the~$e$ anyon and the fermions remain heavy), eventually becoming massless. Using particle-vortex duality~\cite{Peskin:1977kp, Dasgupta:1981zz} we show that this second-order transition is in the~$\Z_2$-gauged~$O(2)$ Wilson-Fisher universality class, which we indicate by the blue~$O(2)_m^*$ line in figure~\ref{figU1phasediagram}. At even larger~$m_3$ the~$U(1)_m$ symmetry is spontaneously broken: the monopole operator~$\langle \CM^2\rangle \neq 0$ condenses, leading to a massless NGB and a circle~$S^1_m$ of vacua. 

\item Thanks to duality~$\sf D$, we find an analogous~$O(2)_e^*$ phase transition at negative~$m_3$, and eventually a~$U(1)_e$ breaking phase with~$\langle \CM^1 \rangle \neq 0$ and an~$S^1_e$ circle of vacua. 

\item We conjecture that the~$\CO(2)_{e,m}^*$ transitions persist until the self-dual~$m_3 = 0$ axis in figure~\ref{figU1phasediagram},\footnote{~We will give some evidence that this happens when the quartic coupling~$\lambda_4$ and the Yukawa coupling in~\eqref{intro:hyqedlag} are sufficiently strong when compared to the gauge coupling~$e^2$. This is a generalization of the type~II regime of Abelian Higgs models/superconductors.} where they meet in a multicritical point (indicated in red) we refer to as the SFS CFT.\footnote{~We work in a scheme in which the multicritical point is at~$m_\phi^2 = 0$.}

\item If we exit the SFS CFT point along the self-dual~$m_3 = 0$ axis in the~$m_\phi^2 > 0 $ direction, we find that the~$S^1_{e,m}$ circles coexist along a first-order line (dotted in figure~\ref{figU1phasediagram}). Since~$\sf D$ swaps the two circles, it is spontaneously broken there. This line does not end, and it matches onto the symmetry-breaking dynamics of QED$_3$ with~$N_f = 2$ Dirac fermions (discussed below~\eqref{intro:hyqedlag}) when the Higgs mas~$m_\phi^2 >0$ is very large and~$\phi$ decouples. We conjecture that the circles persist all the way to the SFS CFT at the multicritical point, where they smoothly shrink to zero.
\end{itemize}

A careful study of the lattice symmetries of the SFS model in section~\ref{subseccryssymmetriesfsmodel} shows that this model has an exact~$\Z_2$ mirror symmetry that is not manifest in  continuum HYQED. (More precisely, mirror symmetry extends the manifest~$S(O(2)_e \times O(2)_m)$ symmetry of HYQED to~$O(2)_e \times O(2)_m$, as indicated figure~\ref{fig:overview}.) We propose that this symmetry emerges in the vicinity of the multicritical point. Since mirror symmetry exchanges the~$U(1)_f$ flavor and~$U(1)_\CM$ monopole symmetries of HYQED, it must be an emergent self-duality of HYQED. In section~\ref{sec:emergemirr} we relate this self-duality to those proposed for pure~$N_f = 2$ QED$_3$ in~\cite{Xu:2015lxa,Karch:2016sxi,Hsin:2016blu}. These proposals are contingent upon the theory flowing to a gapless CFT in the IR -- a fact not borne out for QED$_3$, but significantly more likely for HYQED, as we argue below. 

To estimate the scaling dimensions of the SFS CFT at the HYQED multicritical point, and in particular to test the emergence of mirror symmetry there, we introduce a large-$N$ generalization of HYQED with~$N_f = 2N$ Dirac fermions in section~\ref{sec:largeN}. We calculate the scaling dimensions of several interesting operators at next-to-leading order in~$1/N$. Extrapolating to~$N =1$, we estimate the scaling dimensions of several non-monopole operators,\footnote{~As we explain in section~\ref{sec:largeN}, the equality of the scaling dimensions for~$|\phi|^2$ and~$i\b \psi \sigma_{1,2} \psi$ is not enforced by any symmetry and appears to be an large-$N$ artifact at this order of the $1/N$ expansion.}
\begin{equation}\label{int:deltanonm}
    \Delta[{|\phi|^2}] = \Delta[ i\b \psi \sigma_{1,2} \psi] \simeq 1.46~, \qquad \Delta[i\b \psi \sigma_3 \psi] \simeq 0.65~.
\end{equation}
The large-$N$ scaling dimensions of the lowest-lying monopole operators~$\mathcal{M}_{(1)} \sim \CM^i$, $\CM_{(2), (3), (4)}$ with~$U(1)_\CM$ charges~$q_\CM = 1,2,3,4$ were calculated to the same~$1/N$ order in~\cite{Dupuis:2021flq}; extrapolating to~$N=1$ we find,
\begin{equation}\label{int:deltam}
\Delta[\mathcal{M}_{(1)} \sim \CM^i] \simeq 0.63~,  \quad  \Delta[\mathcal{M}_{(2)}] \simeq 1.53~, \quad \Delta[\mathcal{M}_{(3)}] \simeq 2.64~, \quad \Delta[\mathcal{M}_{(4)}] \simeq 3.92~.
\end{equation}
Our conjectured mirror self-duality of HYQED exchanges~$U(1)_f$ and~$U(1)_\CM$, and thus the fermion bilinear~$i\b \psi (\sigma_{1}- i \sigma_2) \psi$ of~$U(1)_f$ charge~$q_f = 2$ and the~$q_\CM = 2$ monopole~$\CM_{(2)}$. Comparing~\eqref{int:deltanonm} and~\eqref{int:deltam}, we see that their scaling dimensions agree to within about~$5 \%$. We take this as an encouraging indication that the massless HYQED theory does indeed flow to the SFS CFT with emergent mirror symmetry in the IR.

\subsection{Interpolating Between the SFS and Fradkin-Shenker Models}\label{int:fstosfs}

As we explain in section~\ref{secU1FSmodel}, there is a deformation that interpolates between the SFS and Fradkin-Shenker models,
\begin{equation}
    H_\text{FS}(h_e, h_m) = H_\text{SFS}(h_e, h_m) +  \Delta H(h_e, h_m)~.
\end{equation}
Here~$\Delta H$ is deformation of the SFS model by local operators that break its~$U(1)_e$ and~$U(1)_m$ symmetries, while preserving all others.

An important aspect of the matching between the SFS lattice model and continuum HYQED (see section~\ref{u1LatToCont}) is the existence of precisely such a deformation. We find that it is given by a particular linear combination of monopole operators~$\CM^1$ and~$\CM^2$ (and their Hermitian conjugates), which carry~$U(1)_e$ and~$U(1)_m$ charges~$+1$, respectively. We thus propose that we can flow to the Fradkin-Shenker phase diagram, and in particular the FS CFT at its multicritical point, by adding such unit-charge monopole operators to HYQED,
\begin{equation}\label{into:fsmonolag}
    {\mathscr L}_\text{FS} = {\mathscr L}_\text{HYQED} + \left(\text{monopole operators } \CM^1, \CM^2\right)~.
\end{equation}
In section~\ref{FSviaMonos} we analyze the effect of these monopole operators on the HYQED phase diagram in figure~\ref{figU1phasediagram}. Under the assumption that the RG flow is smooth, we precisely recover the full Fradkin-Shenker phase diagram in figure~\ref{figfsphasediagram}. This shows that our conjectured web of dualities and phase diagrams is self-consistent, and in a sense minimal. In particular, the symmetry-breaking dynamics of~$N_f = 2$ QED$_3$, mentioned below~\eqref{intro:hyqedlag}, plays a starring role.

It follows from~\eqref{int:deltam} that the monopole deformation~\eqref{into:fsmonolag} is relevant in the SFS CFT at the HYQED multicritical point, where it triggers the RG flow to the Fradkin-Shenker CFT. However, it is not straightforward to use the Lagrangian~\eqref{into:fsmonolag} to compute the scaling dimensions of operators in the FS CFT, precisely because it contains monopole operators.

\subsection{Dual Description via the Easy-Plane $\C\P^1$ Model}\label{intro:easy}

We have argued that HYQED has an emergent~$\Z_2$ mirror symmetry  -- arising from a self-duality -- near its multicritical point, and that it describes the SFS CFT, whose full unitary global symmetry is~$(O(2)_e \times O(2)_m) \rtimes \Z_2^{\sf D}$ (see figure~\ref{fig:overview}), where duality~$\sf D$ exchanges the two~$O(2)$ factors. It is desirable to search for a dual continuum QFT description in which mirror symmetry is manifest, perhaps at the expense of other manifest symmetries.  

We propose such a dual description in section~\ref{sec:easy}. It is given by the so-called easy plane~$\mathbb{CP}^1$ (EP$\C\P^1$, see figure~\ref{fig:overview}) model, i.e.~scalar QED$_3$ with~$U(1)$ gauge field~$b_\mu$ and field strength~$f_{\mu\nu}$, as well as two flavors of complex scalar fields~$z_{i = 1,2}$,
\begin{equation}\label{int:eqnCP1lagrangian}
\mathscr{L}_{\text{EP}\mathbb{CP}^1} = -\frac{1}{4{\t e}^2} f^{\mu\nu} f_{\mu\nu} - \left|D_b z_i\right|^2 - m_z^2|z_i|^2 - V_{\rm EP}~.
\end{equation}
Here~$V_{\rm EP}$ is the easy-plane scalar potential,
\begin{equation}
\label{int:EP_potential}
V_{\text{EP}} = \lambda_{SO(3)}\left(|z_i|^2\right)^2 + \lambda_{\rm EP} |z_1|^2 |z_2|^2~, \qquad \lambda_{SO(3)}>0 \,, \quad -4 < \lambda_{\rm EP}/\lambda_{SO(3)} < 0~.
\end{equation}
The easy-plane coupling~$\lambda_\text{EP} \neq 0$ breaks the~$SO(3)$ flavor symmetry (under which the scalar doublets~$z_i$ transform projectively) to its~$O(2)_e = U(1)_e \rtimes \Z_2^{{\sf C}_e}$ subgroup. The reflection element~${\sf C}_e$, which exchanges~$z_1 \leftrightarrow z_2$, is precisely the sought-after mirror symmetry; thus it is manifest in the EP$\C\P^1$ description. The monopole symmetry of the model is simply the~$U(1)_m$ symmetry of HYQED and the SFS model. Indeed, the EP$\C\P^1$ model manifests all symmetries of HYQED, except for duality~$\sf D$, which is emergent in our proposal (see below).

Our proposed duality between the EP$\C\P^1$ model~\eqref{int:eqnCP1lagrangian} and HYQED~\eqref{intro:hyqedlag} holds around the multicritical point described by the SFS CFT. Hence there are two relevant tuning parameters:
\begin{itemize}
\item[(1.)] The fermion mass~$m_3$ of HYQED in~\eqref{intro:masses} maps onto the scalar mass~$m_z^2$ in~\eqref{int:eqnCP1lagrangian}. Indeed, dialing~$m_z^2$ in the presence of the easy-plane potential~\eqref{int:EP_potential} reproduces the two~$U(1)_{e,m}$ symmetry-breaking phases, with vacua~$S_{e,m}^1$, of HYQED in figure~\ref{figU1phasediagram}. 

\item[(2.)] The Higgs mass~$m_\phi^2$ of HYQED in~\eqref{intro:masses} maps onto the easy-plane quartic coupling~$\lambda_\text{EP}$, in such a way as to be consistent with the bounds in~\eqref{int:EP_potential}. The multicritical point~$\lambda^*_\text{EP}$ lies somewhere in this range. For larger values~$\lambda_\text{EP} > \lambda^*_\text{EP}$ we find the first-order coexistence line between the two~$S_{e,m}^1$ phases in figure~\ref{figU1phasediagram}; for smaller values~$\lambda_\text{EP} < \lambda^*_\text{EP}$ we find the~$\Z_2$ TQFT phase. 
\end{itemize}
Note that~$m_3$ is~$\sf D$-odd in HYQED, while~$m_z^2$ respects all manifest symmetries of the EP$\C\P^1$ model, which does not have manifest~$\sf D$ symmetry. We propose that~$\sf D$ emerges near the multicritical point. As we explain in section~\ref{sec:easy}, this follows from a self-duality of the EP$\C\P^1$ model proposed in~\cite{Motrunich:2003fz}. Again, this proposal is contingent on the theory flowing to a gapless CFT in the IR -- a fact that was previously assumed to hold for generic~$\lambda_\text{EP}$.\footnote{~However, it was already pointed out in~\cite{Wang:2017txt} that several lattice models believed to be described by the EP$\C\P^1$ model (see below) appear to exhibit first-order transitions as one dials~$m_z^2$.} By contrast, we propose that~$\sf D$ only emerges in the vicinity of the multicritical point~$\lambda_\text{EP} = \lambda^*_\text{EP}$ described by the SFS CFT.

In the same spirit, our proposed multicritical duality between HYQED and the EP$\C\P^1$ model breaths new life into the previous duality proposals~\cite{Karch:2016sxi, Wang:2017txt, Benini:2017aed}, where HYQED was replaced by QED$_3$ with~$N_f = 2$ Dirac fermions, but without the Higgs field~$\phi$. Just as for the self-dualities discussed above, these proposals hinged on both theories flowing to a gapless CFT in the deep IR -- a scenario that now seems unlikely. Note that relative to these earlier proposals,\footnote{~As discussed in~\cite{Wang:2017txt}, the original duality proposals relating~$N_f = 2$ fermionic QED$_3$ and the EP$\C\P^1$ model implied that these theories flow to a CFT with emergent~$O(4)$ unitary symmetry in the IR. Such a CFT is incompatible with conformal bootstrap bounds~\cite{Li:2018lyb,Li:2021emd}.} our multicritical duality web for the SFS CFT only involves the comparatively small unitary symmetry~$(O(2)_e \times O(2)_m) \rtimes \Z_2^{\sf D}$, with only the~$\Z_2$ mirror element~${\sf C}_e$ emerging in HYQED, and only the~$\Z_2$ duality symmetry~$\sf D$ emerging in the EP$\C\P^1$ model (see figure~\ref{fig:overview}). The fact that the full global symmetry is manifest in the SFS lattice model sharpens many aspects of these dualities.  

Since the EP$\C\P^1$ model furnishes a dual description of the SFS CFT, we can repeat the discussion in section~\ref{int:fstosfs} and add suitable (monopole and non-monopole) local operators that break the~$U(1)_e$ and~$U(1)_m$ symmetries, allowing us to flow to the Fradkin-Shenker CFT and its phase diagram.

\subsection{Relation to N\'eel-VBS Deconfined Quantum Multicriticality}\label{intro:NVBS}

Some of the theories and dualities above have previously appeared in the study of deconfined quantum critical points~\cite{Senthil:2003eed,Senthil:2004fuw} (see~\cite{Senthil:2023vqd} for a recent review with references), and in particular the study of spin-$1/2$ anti-ferromagnets on a square lattice that undergo a transition from a~N\'eel phase with spontaneously broken~$SO(3)$ spin-rotation symmetry to a Valence Bond Solid (VBS) phase, where some of the lattice symmetries are spontaneously broken. In situations where the spin-rotation symmetry is reduced from~$SO(3)$ to~$O(2)_e$, this transition is believed to be described by the EP$\C\P^1$ model in~\eqref{int:eqnCP1lagrangian}, with the N\'eel phase corresponding to Higgs phase at large negative~$m_z^2 < 0$, and the VBS phase corresponding to large positive~$m_z^2 > 0$. Crucially, this description requires adding to~\eqref{int:eqnCP1lagrangian} monopole operators whose~$U(1)_m$ charge is a multiple of 4, thus explicitly breaking~$U(1)_m$ to the~$\Z_4^\text{rot.}$ rotation symmetry of the square lattice (as reviewed in~\cite{Senthil:2023vqd}), while preserving~$U(1)_e$. The scenario that this transition might (generically) be continuous was termed deconfined quantum criticality in\cite{Senthil:2003eed,Senthil:2004fuw}, and it would have rendered the dualities proposed in~\cite{Motrunich:2003fz,Xu:2015lxa,Karch:2016sxi,Hsin:2016blu,Wang:2017txt, Benini:2017aed} applicable.

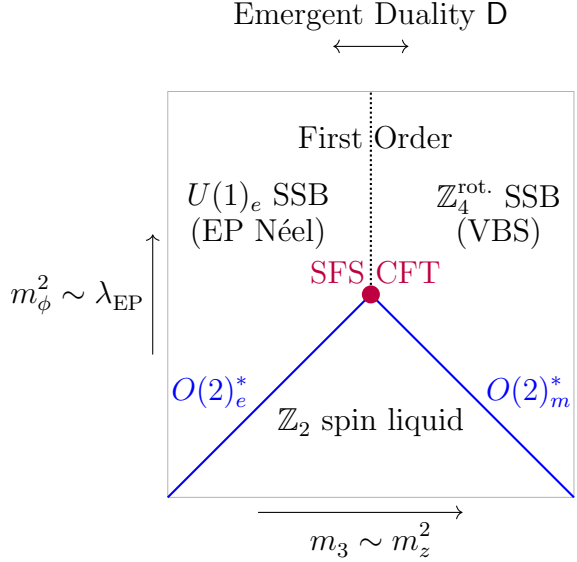
\begin{figure}[t!]
    \centering
\begin{tikzpicture}[x=4cm,y=4cm] 
    \draw[gray!60, line width=0.3pt] (0,0) rectangle (1.35,1.35);
  \node at (0.675,-0.14) {$m_3 \sim m_z^2$};
  \draw[->] (0.3,-0.05) -- (0.98,-0.05);
  
  \node at (-0.3,0.675) {$m^2_\phi \sim \lambda_\text{EP}$};
  \draw[->] (-0.05,0.475) -- (-0.05,0.875);

  \node at (1.1,1) {$\Z_4^\text{rot.}$ SSB};
  \node at (1.1,0.88) {(VBS)};

  \node at (1.2,0.35) {\textcolor{blue}{$O(2)^*_m$}};

  \node at (0.3,1) {$U(1)_e$ SSB};
  \node at (0.3,0.88) {(EP N\'eel)};

  \node at (0.14,0.35) {\textcolor{blue}{$O(2)^*_e$}};

  \node at (0.675,0.25) {$\mathbb{Z}_2$ spin liquid};

\node at (0.69,1.2) {First Order};

\draw[<->] (0.55,1.5) -- (0.8,1.5);
\node at (0.675,1.6) {Emergent Duality~$\mathsf{D}$};

  \draw[line width=0.8pt,blue]
    (0.0,0.0) -- (0.675,0.675) -- (1.35,0);
  \draw[densely dotted, line width=0.8pt]
    (0.675,0.675) -- (0.675,1.35);
  \fill[color=purple] (0.675,0.675) circle (0.03);
  \node at (0.69,0.75) {\textcolor{purple}{SFS CFT}};
\end{tikzpicture}
    \caption{The phase diagram of Higgs-Yukawa-QED$_3$ (see figure~\ref{figU1phasediagram}), deformed by monopole operators of~$U(1)_\CM$ charge~$q_\CM = 4$, reinterpreted in the N\'eel-VBS context via the multicritical duality with the EP$\C\P^1$ model in section~\ref{intro:easy}. The $U(1)_e$ SSB phase corresponds to the easy-plane N\'eel phase, where the~$U(1)_e$ spin-rotation symmetry is spontaneously broken. The~$U(1)_m$ symmetry of HYQED is explicitly broken to the~$\Z_4^\text{rot.}$ lattice rotation group by the monopole operators, leading to a gapped phase with four vacua and~$\Z_4^\text{rot.}$ SSB. The monopole operators are irrelevant on the second-order~$O(2)_{e,m}^*$ lines and at the deconfined quantum multicritical point described by the SFS CFT. The~$\Z_2$ TQFT phase is interpreted as a gapped~$\bZ_2$ spin liquid, where all symmetries are unbroken, and the monopole deformations have no effect in the IR.}
    \label{figNeelVBSinterp}
\end{figure}

Instead, let us reexamine the N\'eel-VBS phase diagram in light of the multicritical duality between the EP$\C\P^1$ model and HYQED summarized in section~\ref{intro:easy} above. The result is shown in figure~\ref{figNeelVBSinterp}. As reviewed above, this involves deforming the EP$\C\P^1$ Lagrangian~\eqref{int:eqnCP1lagrangian} by monopoles that preserve~$U(1)_e$, but break~$U(1)_m \to \Z_4^\text{rot.}$. These map to monopoles of HYQED whose~$U(1)_\CM$ charge in that theory is a multiple of~$4$. Comparing with the estimates in~\eqref{int:deltam}, we see that the lightest such operator~$\CM_{(4)}$ is quite likely irrelevant in the SFS CFT, which thus appears as a deconfined quantum multicritical point in the N\'eel-VBS phase diagram shown in figure~\ref{figNeelVBSinterp}.\footnote{~The possibility that HYQED might appear as a multicritical point in the N\'eel-VBS phase diagram was already mentioned in~\cite{Jian:2017chw}, but the discussion there assumed a continuous N\'eel-VBS transition.}  

In fact, the monopole deformation can be ignored almost everywhere in this phase diagram (including on the~$O(2)_{e,m}^*$ lines), except in the VBS phase, where it is dangerously irrelevant and explicitly breaks~$U(1)_m$ to~$\Z_4^\text{rot.}$, leading to four gapped vacua that are acted on by the spontaneously broken~$\Z_4^\text{rot.}$ lattice symmetry. 

We also see that there should be a gapped~$\Z_2$ TQFT phase, which plays a crucial role in the Fradkin-Shenker and SFS lattice models discussed above. In the context of the N\'eel-VBS phase diagram it is interpreted as a~$\Z_2$ spin liquid phase. Intervening $\bZ_2$ spin liquids are common in systems with competing spin orders, see for instance~\cite{Jiang:2012qea,Gong_2013}. It would be very interesting to identify a natural lattice model that exhibits the easy-plane N\'eel-VBS transition together with this toric code phase.

\bigskip

\noindent {\it Note Added: While this paper was being finalized, we became aware of~\cite{Ji:2026yfj}, where a different continuum description of the FS multicritical point is proposed.}

\section{The Fradkin-Shenker Lattice Model and its  Multicritical Point}\label{secFSmodel}

\subsection{The Toric Code}
We begin by recalling the toric code~\cite{Kitaev:1997wr}. This is a 2+1d Hamiltonian lattice model defined on a square lattice of spin-$1/2$ degrees of freedom, which we think of as occupying the links $\ell$ of the lattice. Its Hamiltonian consists of two types of terms: a \textit{star term} $A_v$ associated to each vertex $v$,
\begin{equation}\label{eqnstarterm}
 \begin{tikzpicture}
    \color{gray}
    \def\length{2}
    
    \draw[-] (-\length/2, 0) -- (\length/2, 0);
    
    \draw[-] (0, -\length/2) -- (0, \length/2);
    
    \filldraw (0, 0) circle (2pt);
    \filldraw (\length/2, 0) circle (2pt);
    \filldraw (-\length/2, 0) circle (2pt);
    \filldraw (0, \length/2) circle (2pt);
    \filldraw (0, -\length/2) circle (2pt);
    \color{blue}
    \node[] at (-\length/4, 0) {$X$}; 
    \node[] at (\length/4, 0) {$X$};  
    \node[] at (0, \length/4) {$X$};  
    \node[] at (0, -\length/4) {$X$}; 
    \color{black}
    \node[left] at (-\length/2,-\length/12) {$A_v = {\displaystyle \prod_{\d \ell \ni v} X_{\ell}} =~$};
\end{tikzpicture}
\end{equation}
where the product is over the four links $\ell$ adjacent to the vertex $v$; and a \textit{plaquette term} $B_p$ associated to each plaquette $p$,
\begin{equation}\label{eqnplaqterm}
    \begin{tikzpicture}
    
    \color{gray}
    \def\length{1}
    
    \draw[-] (-\length/2, -\length/2) -- (\length/2, -\length/2);
    \draw[-] (-\length/2, -\length/2) -- (-\length/2, \length/2);
    \draw[-] (\length/2, \length/2) -- (\length/2, -\length/2);
    \draw[-] (\length/2, \length/2) -- (-\length/2, \length/2);

    \filldraw (\length/2, \length/2) circle (2pt);
    \filldraw (-\length/2, \length/2) circle (2pt);
    \filldraw (-\length/2, -\length/2) circle (2pt);
    \filldraw (\length/2, -\length/2) circle (2pt);
    \color{red}

    \node[] at (-\length/2, 0) {$Z$}; 
    \node[] at (\length/2, 0) {$Z$};  
    \node[] at (0, \length/2) {$Z$};  
    \node[] at (0, -\length/2) {$Z$}; 
    \color{black}
    \node[left] at (-\length/2,-\length/6) {$B_p = {\displaystyle \prod_{\ell \in \d p} Z_{\ell}} =~$};
\end{tikzpicture}
\end{equation}
where the product is over the four links $\ell$ bounding the plaquette $p$. Here and below, we use $X_\ell,Y_\ell,Z_\ell$ to represent the Pauli operators $\sigma_x,\sigma_y,\sigma_z$ on the link $\ell$ of the lattice.

The toric code Hamiltonian is (minus) the sum of these terms over all vertices and plaquettes of the lattice,
\[\label{eqntoriccodeham}H_{\rm TC} = - \sum_v A_v - \sum_p B_p\,.\]
Notice that $A_v$ and $B_p$ are Hermitian operators that square to the identity, $A_v^2=B_p^2=\mathds{1}$, so that their eigenvalues are $\pm 1$. Crucially, the $A_v$ and $B_p$ all commute.\footnote{~This is easy to check using the fact that any $A_v$ and $B_p$ share an even number of links (either 0 or 2), and for a given link~$\ell$, the operators $X_\ell$ and $Z_\ell$ anticommute,~$\{X_\ell, Z_\ell\} = 0$.} The ground states of~$H_{\rm TC}$ are the simultaneous $+1$ eigenstates of all $A_v$ and $B_p$. An important property of~$H_{\rm TC}$ is that it has a four-fold ground-state degeneracy on a lattice with periodic boundary conditions (hence the name toric code). 

We can express the excitations of the toric code in terms of $e$ anyons localized at vertices with $A_v = -1$ and $m$ anyons localized at plaquettes with $B_p = -1$. These are gapped excitations, separated by~$\Delta E=2$ from the ground state. One can also define the $f = em$ anyon, which is given by an $e$ and an $m$ anyon side by side. These three anyon types correspond to the three non-trivial superselection sectors of the model; arbitrary local operators can only create such anyons in pairs. We can consider fusion of superselection sectors by grouping local excitations together and considering the superselection sector of their composite. Together with the trivial superselection section, the three anyon types~$e, m, f$ form the Abelian group $\bZ_2 \times \bZ_2$ under fusion.\footnote{~Using additive notation for~$\Z_2$, we have~$e = (1,0)$, $m = (0,1)$, $f = (1,1)$, and the trivial sector~$(0,0)$.}

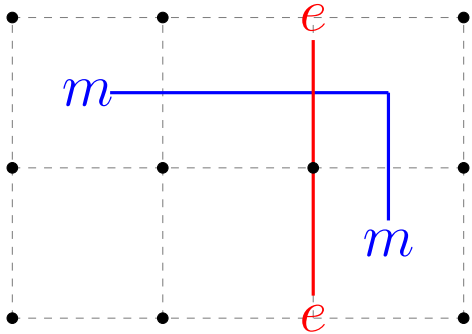
\begin{figure}[t]
    \centering
    \begin{tikzpicture}
    \color{gray}
    \def\length{4}

    \draw[dashed] (0, -\length/2) -- (0, \length/2);
    \draw[dashed] (-\length/2, -\length/2) -- (-\length/2, \length/2);
    \draw[dashed] (\length/2, -\length/2) -- (\length/2, \length/2);
    \draw[dashed] (\length, -\length/2) -- (\length, \length/2);

    \draw[dashed] (-\length/2, 0) -- (\length, 0);
    \draw[dashed] (-\length/2, -\length/2) -- (\length, -\length/2);
    \draw[dashed] (-\length/2, \length/2) -- (\length, \length/2);

    \draw[blue,very thick] (-0.7,1) -- (3,1);
    \draw[blue,very thick] (3,1) -- (3,-0.7);
    \color{blue}
    \node[scale=1.9] at (3,-1) {$m$}; 
    \node[scale=1.9] at (-1,1) {$m$}; 
    \color{black}

    \draw[red,very thick] (2,-1.7) -- (2,1.7);
    \color{red}
    \node[scale=1.9] at (2,-2) {$e$};
    \node[scale=1.9] at (2,2) {$e$};
    \color{black}

    \filldraw (0, 0) circle (2pt);
    \filldraw (\length/2, 0) circle (2pt);
    \filldraw (-\length/2, 0) circle (2pt);
    \filldraw (0, \length/2) circle (2pt);
    \filldraw (0, -\length/2) circle (2pt);
    \filldraw (-\length/2, \length/2) circle (2pt);
    \filldraw (-\length/2, -\length/2) circle (2pt);
    \filldraw (\length, -\length/2) circle (2pt);
    \filldraw (\length, 0) circle (2pt);
    \filldraw (\length, \length/2) circle (2pt);
    
    \color{black}
    
\end{tikzpicture}
    \caption{The $e$ and $m$ string operators can be represented as curves on the lattice and its dual, respectively. Where the red lattice curve crosses a link, we apply $Z_\ell$, and where the blue dual lattice curve intersects a link, we apply $X_\ell$. These anticommute at each point where they intersect, showing the $-1$ braiding of the $e$ and $m$ anyons.}
    \label{figstringoperator}
\end{figure}

One can also consider braiding of anyons. In the toric code this is simply realized via string operators that move or create anyons in pairs. One such string operator is $\prod_{\ell \in \gamma} Z_\ell$, where $\gamma$ is a path on the lattice. This operator creates $e$ anyons at the endpoints of $\gamma$. We call it the $e$ string operator. Likewise, we have an $m$ string operator, $\prod_{\ell \in \gamma'} X_\ell$ where $\gamma'$ is a path on the dual lattice (see figure \ref{figstringoperator}), which creates $m$ excitations on the plaquettes at the endpoints of~$\gamma'$.

Since all $e$ string operators commute with one another, the $e$ anyons are bosons, and the same is true for the~$m$ anyons. By contrast, $e$ and~$m$ have non-trivial mutual statistics: they braid one another with a~$-1$ phase, because intersecting~$e$ and~$m$ string operators anticommute (see figure \ref{figstringoperator}). One consequence of this is that their composite $f = em$ is a fermion. Since $H_{\rm TC}$ is gapped, a very general argument~\cite{Hastings:2005xm} shows that the properties of these anyons are stable to sufficiently small local perturbations of $H_{\rm TC}$, making the toric code a representative Hamiltonian of a whole topologically ordered phase, which we will refer to as the toric code phase.

The properties reviewed above show that the toric code defines an emergent $\bZ_2$ gauge field on the lattice. If we interpret $Z_\ell$ as the Wilson line of this lattice $\bZ_2$ gauge field on the link~$\ell$, then $A_v = 1$ becomes the Gauss law and $B_p = 1$ becomes the flatness (flux-free) constraint.\footnote{~These constraints are imposed energetically below the gap~$\Delta E = 2$.} From this point of view, $e$ anyons are~$\Z_2$ gauge charges, $m$ anyons are~$\Z_2$ fluxes (i.e.~$\pi$-fluxes), and their braiding is due to the usual Aharonov-Bohm effect.

The toric code has a duality symmetry~$\mathsf{D}$ that swaps the~$e$ and~$m$ anyons, which is reminiscent of electric-magnetic duality in (3+1)-dimensional~$U(1)$ gauge theory. This duality~$\sf D$ exchanges~$X_\ell \leftrightarrow Z_\ell$ on each link by applying Hadamard gates, followed by a translation in the $(1/2,1/2)$ direction (here we are writing the square lattice as $\bZ^2$, see section~\ref{subseccryssymmetriesfsmodel} for a more detailed discussion). This symmetry takes $A_v$ to $B_p$, where $p$ is the plaquette to the northeast of $v$, and takes $B_p$ to $A_v$, where $v$ is the vertex to the northeast of $p$. 

\subsection{The Fradkin-Shenker Model}\label{sec:FSdetail}

For reasons explained below, we refer to the toric code Hamiltonian~\eqref{eqntoriccodeham} deformed by~$X$ and $Z$ terms as the Fradkin-Shenker model,
\[\label{eqnfsmodel}H_{\rm FS}(h_e,h_m) = H_{\rm TC} - h_e \sum_\ell Z_\ell - h_m \sum_\ell X_\ell  = - \sum_v A_v - \sum_p B_p - h_e \sum_\ell Z_\ell - h_m \sum_\ell X_\ell~,\]
where $\sum_\ell$ is a sum over the links of the square lattice and~$h_{e,m} \geq 0$ without loss of generality. This model enjoys the $e \leftrightarrow m$ duality $\mathsf{D}$ of the toric code, but now $\mathsf{D}$ exchanges $h_e \leftrightarrow h_m$, i.e.~it is a symmetry only on the self-dual line $h_e=h_m$.

As mentioned above, for small $h_{e,m}$ this model will still be in the toric code phase. However, $Z_\ell$ creates a pair of $e$ anyons at the vertices of the link $\ell$, and so as $h_e \to \infty$ we expect $e$ anyons to condense, giving a Higgs phase.\footnote{~Here we use the word condense somewhat loosely, see section~\ref{subsecsuperconductorintuition} for further comments.} On the other hand, $X_\ell$ creates $m$ anyons on the plaquettes bordering $\ell$, and so $h_m \to \infty$ may be regarded as a magnetic Higgs -- or confined -- phase. One can show this explicitly in the limit where~$h_{e,m} \to \infty$, so that the toric code part of~\eqref{eqnfsmodel} can be dropped and the Hamiltonian simplifies to
\[H^\infty_{\rm FS} =  -\sqrt{h_e^2 + h_m^2} \, \sum_\ell \left(\cos \theta Z_\ell + \sin \theta X_\ell \right)~, \quad \cos \theta = {h_e \over \sqrt{h_e^2 + h_m^2}}~, \quad \sin \theta = {h_m \over \sqrt{h_e^2 + h_m^2}}~.\]
Note that~$\theta \in [0, \pi/2]$. This Hamiltonian is a sum of single-site terms, and its unique ground state is the uniform product state pointing in the direction $\left(\sin \theta,0,\cos \theta\right)$ in the $XZ$ plane of the Bloch sphere. In particular, there is no phase transition as we dial from the Higgs phase at~$\theta = 0$ to the confining phase at~$\theta = \pi/2$, a phenomenon known as Higgs-confinement continuity. This was famously demonstrated in a closely related model by Fradkin and Shenker \cite{Fradkin:1978dv},\footnote{~The fact that Higgsing and confinement are sometimes continuously connected was also independently demonstrated in~\cite{Banks:1979fi}.} and so we also refer to \eqref{eqnfsmodel} as the Fradkin-Shenker (FS) model. The model analyzed by Fradkin and Shenker is actually~$\Z_2$ gauge theory coupled to a~$\Z_2$ Higgs field on a 3d Euclidean Lattice (which had previously been studied in~\cite{Wegner:1971app}). The connection to the Hamiltonian model~\eqref{eqnfsmodel} was established in~\cite{Tupitsyn:2008ah}, by introducing additional, redundant~$\Z_2$ gauge and matter degrees of freedom into~\eqref{eqnfsmodel}.\footnote{~See section~\ref{subsecgaugelatticemodel} for a similar discussion in the context of the staggered Fradkin-Shenker model that we introduce in section~\ref{secU1FSmodel}.}

The FS model is under good theoretical control along the axes~$(h_e, 0)$ and~$(0,h_m)$, along which the FS Hamiltonian $H_{\rm FS}$ can be mapped to a $\bZ_2$-gauged Ising model~\cite{Wegner:1971app, Fradkin:1978dv,Trebst:2006ci},\footnote{~See section~\ref{o2starlat} for an analogous discussion in the staggered Fradkin-Shenker model.} which we indicate by~Ising$^*_e$ or~Ising$^*_m$, and so displays continuous phase transitions as $h_e$ or $h_m$, respectively, is dialed (the two axes are related by duality $\mathsf{D}$). Note that the~$\Z_2$-gauged Ising order parameters~$\phi_e$ and~$\phi_m$ are massive in the toric code phase, massless at the their  respective Ising$^*$ transition, and ultimately acquire vevs~$\langle \phi_{e,m}\rangle \neq 0$ that Higgs the~$\Z_2$ gauge field they couple to, resulting in the trivial Higgs or confined phase, respectively. Note that~$\phi_{e,m}$ cannot couple to the same~$\Z_2$ gauge field, i.e.~one must couple electrically and the other magnetically, because of their mutual braiding.\footnote{~See~\cite{Shi:2024pem} (further discussed in section~\ref{sec:discussion}) for a recent attempt to describe the FS model in the continuum using fields akin to~$\phi_{e,m}$ that couple electrically and magnetically to a~$\Z_2$ gauge field.} It is easy to toggle between the Ising and Ising$^*$ models by gauging a global zero-form or one-form symmetry, respectively. The analogous operation is not available in the Fradkin-Shenker model, which has both electric and magnetic fundamental charges, and hence no one-form symmetries.

The full phase diagram of the FS model (depicted in figure \ref{figfsphasediagram}) has been explored numerically in Monte Carlo simulations, first by \cite{Tupitsyn:2008ah, Vidal:2008uy, Wu:2012cj} and more recently by \cite{Somoza:2020jkq,Oppenheim:2023uvf}. Let us summarize its main features:
\begin{itemize}
    \item The continuous Ising$^*_{e,m}$ transitions along which the~$e$ and~$m$ anyons become gapless persist away from the axes and merge somewhere along the self-dual~$h_e = h_m$ line. We refer to the resulting multicritical point as the FS multicritical point. Numerical evidence~\cite{Somoza:2020jkq} suggests that it is described by a conformal field theory (CFT), which we will refer to as the FS CFT. 
    
    Note that the FS CFT has gapless~$e$ and~$m$ anyons, even though they have non-trivial braiding. In this sense, it is somewhat reminiscent of Argyres-Douglas theories~\cite{Argyres:1995jj, Argyres:1995xn}, i.e.~superconformal fixed points in 3+1 dimensions with simultaneously massless charges and monopoles, which are also mutually non-local.

    \item  There is a line of first-order phase transitions that emanates from the multicritical point into the trivial Higgs-confinement phase along the self-dual~$h_e = h_m$ diagonal. On this first-order line the duality symmetry~$\sf D$ is spontaneously broken, leading to two degenerate vacua. The line ends in an second-order, ungauged Ising transition. 
\end{itemize}
One of our primary goals in this paper is to shed light on this phase diagram by relating it to a version of the Fradkin-Shenker lattice model with more symmetry (the staggered Fradkin-Shenker model, see section~\ref{secU1FSmodel}), as well as to several continuum QFTs.

\subsection{Beyond Landau$^*$ Criticality}\label{subsecbeyondlandau}

One curious result from~\cite{Somoza:2020jkq} is that the duality-even and duality-odd operators tuning away from the Fradkin-Shenker multicritical point, which couple to~$h_e+h_m$ and $h_e-h_m$ respectively, have numerically estimated scaling dimensions~\eqref{eq:SFCFTDeltas} close to those of the $O(2)$ Wilson-Fisher CFT operators $\phi_1^2 + \phi_2^2$ and $\phi_1^2 - \phi_2^2$, respectively, where~$\phi_1, \phi_2$ form the real scalar~$O(2)$ doublet of the model.\footnote{~Recall that in the~$O(2)$ Wilson-Fisher CFT, $\Delta_{\phi_1^2 + \phi_2^2} \simeq 1.51$ and~$\Delta_{\phi_1^2 - \phi_2^2} = \Delta_{2 \phi_1 \phi_2} \simeq 1.24$, see e.g.~\cite{Rychkov:2023wsd}.} This leads to the intriguing possibility that the multicritical point is described by some immediate cousin of the $O(2)$ model \cite{Bonati:2021thy}.

However, a theory for the multicritical point should also predict the correct nearby phase diagram and the action of the global symmetries, in particular of the duality symmetry~$\mathsf{D}$. If we want to use the $O(2)$ model (or indeed any other Ginzburg-Landau theory) to capture a phase diagram with nearby toric code and trivial phases, we need to gauge a $\bZ_2$ symmetry. We could entertain gauging different discrete groups, possibly with Dijkgraaf-Witten twists, but the only way to obtain a toric code is to gauge $\bZ_2$ with trivial twist. A simple proof of this is that the total quantum dimension of any 2+1d Dijkgraaf-Witten theory with gauge group $G$ is $|G|^2$, so to get the total dimension 4 of the toric code, we need to take $G = \bZ_2$. We can then rule out a possible twist, i.e.~the double semion TQFT, because it has the wrong anyon content. 

The resulting theory unavoidably has an exact $\bZ_2$ one-form symmetry, which we may identify with the string operator of one of the bosonic anyons of the toric code, say $m$. On the other hand, we want this theory to have exact duality $\mathsf{D}$, which exchanges $e$ and $m$, so we must also have the exact one-form symmetry corresponding to the $e$ anyon. The only way to achieve this is to completely decouple the toric code from the Ginzburg-Landau theory, i.e.~we gauged a trivial symmetry of the Ginzburg-Landau theory to begin with. This clearly does not have the right phase diagram. Thus, the FS multicritical point is beyond the reach of gauged Ginzburg-Landau models, i.e.~beyond Landau$^*$.

In fact, we do not expect the FS CFT to have any one-form symmetries at all, as was already mentioned above. Away from the axes of the phase diagram in figure \ref{figfsphasediagram} these symmetries are violated in the UV, and along the ${\rm Ising}^*_e$ and ${\rm Ising}^*_m$ lines, there are massless charges and fluxes, respectively. We thus expect massless charges, fluxes, and fermions at the multicritical point.

\subsection{Non-Abelian Anyon Condensation via a Superconductor to Insulator Transition}\label{subsecsuperconductorintuition}

An illuminating point of view on the FS multicritical point begins by gauging~$\mathsf{D}$ on the self-dual line, where it is a symmetry. It is not an on-site symmetry in the Fradkin-Shenker model, so it may not be possible to gauge it on the lattice,\footnote{~Our definition of~$\sf D$ involves a translation. There are some recent definitions of $e \leftrightarrow m$ duality symmetries of $H_{\rm TC}$ that involve finite-depth quantum circuits, but these symmetries are not on-siteable~\cite{Tu:2025bqf,Shirley:2025yji}. There are other lattice models in the toric code phase where $\mathsf{D}$ is on-site~\cite{Heinrich:2016wld}.} but we may certainly entertain doing so in the continuum $\bZ_2$ TQFT.

The result of this gauging~\cite{Teo:2015xla,Barkeshli:2014cna} is a non-Abelian TQFT we write as $Z({\rm Ising}) = {\rm Ising} \times \bar{\rm Ising}$, where ${\rm Ising}$ is the chiral TQFT with a fermion $\psi$ and a non-Abelian flux $\sigma$ of topological spin $e^{i\pi/8}$ (see e.g.~\cite{Kitaev:2005hzj} and references therein), and $\bar{\rm Ising}$ is its reflected anti-chiral partner. (We give a derivation of this fact in section \ref{secetainvariants}.) The anyons of $Z({\rm Ising})$ are ordered pairs. This theory has two bosons, $(\psi,\bar \psi)$ and $(\sigma,\bar \sigma)$. We can identify $(\psi,\bar \psi)$ with the~$\Z_2$ one-form symmetry that arises by gauging $\mathsf{D}$, since it is the only Abelian boson. Gauging this one-form symmetry returns us to the original~$\Z_2$ TQFT \cite{Burnell:2017otf}.

One can also show that $(\sigma,\bar \sigma)$ is a  condensable anyon, albeit a non-Abelian one, and if we algebraically condense it we obtain a trivial TQFT~\cite{Bais:2008ni}. Since this picture was the result of gauging $\mathsf{D}$, we recognize this trivial TQFT as the result of gauging $\mathsf{D}$ in a phase with two vacua, where~$\sf D$ is spontaneously broken. Thus the procedure of algebraically condensing the~$(\sigma,\bar \sigma)$ anyon relates the two phases of the~$\sf D$-gauged Fradkin-Shenker model adjacent to the multicritical point. In this sense the Fradkin-Shenker CFT realizes a dynamical version of non-Abelian anyon condensation (see footnote~\ref{fn:AC}). 

This observation becomes tantalizing when combined with the fact (also explained in section~\ref{secetainvariants}) that the Ising TQFT is the result of gauging fermion parity in a $p+ip$ superconductor~\cite{Kitaev:2005hzj,Bochniak:2021lzn}. The anyon $(\sigma,\bar\sigma)$ is thus a bound state of superconducting vortices in the $(p+ip) \times (p-ip)$ layers, each of which traps a Majorana zero mode. The Higgs-Yukawa-QED  (HYQED) Lagrangian~\eqref{intro:hyqedlag} is a natural QFT realization of these ingredients -- especially in its Higgs phase, which we analyze in sections~\ref{section:Z2Higgsphase} and~\ref{sec:SDTQFT}. 

One downside of HYQED from the point of view of the Fradkin-Shenker model is that it has to many global symmetries, in particular a continuous~$U(1)_e \times U(1)_m$ symmetry. We will now introduce a staggered generalization of the Fradkin-Shenker lattice model -- the SFS model -- with precisely such a global symmetry.

\section{The Staggered Fradkin-Shenker Lattice Model}\label{secU1FSmodel}

\subsection{Overview}

In this section we study a cousin of the Fradkin-Shenker model \eqref{eqnfsmodel} that has more symmetries, which we will refer to as the Staggered Fradkin-Shenker (SFS) model. In particular, it enjoys two $U(1)$ symmetries roughly corresponding to the conservation of $e$ and $m$ anyons (with a certain sublattice structure). The crystalline symmetries turn out to act non-trivially on the $U(1)$ charges, allowing us to identify some hidden symmetries and anomalies.

The staggered Fradkin-Shenker (SFS) Hamiltonian takes the form
\begin{equation}\label{eqnU1FSmodel}
H_{\rm SFS}(h_e,h_m) = H_{\rm TC} - h_e \sum_\ell \tilde Z_\ell - h_m \sum_\ell \tilde X_\ell \,,
\end{equation}
where $H_{\rm TC}$ is the toric code Hamiltonian \eqref{eqntoriccodeham}, $h_e,h_m \geq 0$, $\ell$ are the links of a square lattice, and we define
\begin{equation}
\tilde Z_\ell =  \frac{Z_\ell}{2}(1 + A_v A_{v'})~, \qquad
\tilde X_\ell =  \frac{X_\ell}{2}(1 + B_p B_{p'})~,
\end{equation}
where $v,v'$ are the vertices at the ends of $\ell$, and $p,p'$ are the plaquettes on either side of $\ell$. Its $U(1)_{e,m}$ symmetry charges are
\[\mathsf{Q}_e = \frac14 \sum_v (-1)^v A_v \,, \\
\mathsf{Q}_m = \frac14 \sum_p (-1)^p B_p \,,\]
where $A_v$ and $B_p$ are the toric code star and plaquette terms \eqref{eqnstarterm} and \eqref{eqnplaqterm}, whereas $(-1)^v$ and $(-1)^p$ denote $\pm 1$ depending on whether $v$ and $p$ belong to an even or odd sublattice, respectively, that we define in \eqref{eqnsublatticestructure} below. The normalization is the smallest such that a global~$2\pi$ rotation satisfies $\exp(2\pi i \mathsf{Q}_{e,m}) = 1$. In particular, local operators have integer charges under these generators. Meanwhile, a single $e$ (or $m$) anyon has $U(1)_e$ (or $U(1)_m$) charge $\pm 1/2$ (relative to the toric code ground state), depending on which sublattice it occupies.

\begin{table}
\begin{center}
\begin{tabular}{c c c c }
Lattice Symmetries & Description & $\mathsf{Q}_e$ & $\mathsf{Q}_m$ \\ 
\hline
$\mathsf{S}_{\hat x}, \mathsf{S}_{\hat y}$ & $\hat x$, $\hat y$ translation & $-$ & $-$ \\  
$\mathsf{R}_v^{\pi/2}$ & vertex-centered $\pi/2$ rotation & $+$ & $-$ \\
$\mathsf{R}_p^{\pi/2}$ & plaquette-centered $\pi/2$ rotation & $-$ & $+$ \\
$\mathsf{R}_l^{\pi}$ & link-centered $\pi$ rotation & $-$ & $-$ \\
$\mathsf{M}_v^H,\mathsf{M}_v^V$ & vertex-centered horizontal or vertical reflection & $+$ & $-$ \\
$\mathsf{M}_p^H,\mathsf{M}_p^V$ & plaquette-centered horizontal or vertical reflection & $-$ & $+$ \\
$\mathsf{T}$ & time reversal (complex conjugation) & $+$ & $+$ \\
$\mathsf{D}=\mathsf{S}_{\frac{1}{2}(\hat x+ \hat y)}\prod_\ell H_\ell$ & duality (Hadamard with $\frac{1}{2}(\hat x+ \hat y)$ translation) & $\mathsf{Q}_m$ & $\mathsf{Q}_e$
\end{tabular}
\end{center}
\caption{A summary of the symmetry generators of the Staggered Fradkin-Shenker model, described in section \ref{subseccryssymmetriesfsmodel}. On the right we indicate how these symmetries act on the $U(1)$ charges $\mathsf{Q}_e$ and $\mathsf{Q}_m$, with $+$ indicating that the charge is invariant, $-$ indicating that it is negated, and~$\mathsf{D}$ exchanging the two charges.}
\label{table:latticesymmetries}
\end{table}

For generic $h_e$ and $h_m$,  the SFS Hamiltonian~$H_{\rm SFS}(h_e,h_m)$ enjoys the~$U(1)_e \times U(1)_m$ symmetry, as well as all crystalline symmetries of the square lattice. Because of the sublattice structure in the definitions of $\mathsf{Q}_e$ and $\mathsf{Q}_m$, the crystalline symmetries have a non-trivial action on the charges, summarized in table \ref{table:latticesymmetries}. The theory also enjoys a time-reversal symmetry~$\sf T$, given by complex conjugation in the $Z$ basis. We may express the symmetry group as
\[\mathsf{G_{lat}} = (U(1)_e \times U(1)_m) \rtimes (\textbf{p4m} \times \bZ_2^\mathsf{T})\,,\]
where $\textbf{p4m}=\mathbb{Z}^2 \rtimes D_8$ is the symmetry of the square lattice ($\bZ^2$ are translations and $D_8$ is the dihedral group of degree 4 and order 8).

Along the line $h_e = h_m$ there is a further $\mathbb{Z}_2$ duality symmetry $\mathsf{D}$ acting as the $e \leftrightarrow m$ swap in the toric code, and generating the crystalline symmetries of a ``mini'' square lattice with basis $\frac{1}{2}(\hat x \pm \hat y)$, where $\hat x$ and $\hat y$ are the basis of the toric code square lattice. In particular,
\[\mathsf{D} H_{\rm SFS}(h_e,h_m) \mathsf{D}^{-1} = H_{\rm SFS}(h_m,h_e)~.\]
We may express this group as
\[\label{eqnlatticeselfdualsymmetry}\mathsf{G_{sd-lat}} = (U(1)_e \times U(1)_m) \rtimes (\textbf{p4mD} \times \bZ_2^\mathsf{T})\,,\]
where $\textbf{p4mD}$ is the symmetry of the mini square lattice, which is a $\bZ_2$ extension of $\textbf{p4m}$, i.e.~$\textbf{p4mD}/\textbf{p4m} = \bZ_2^{\mathsf D}$.

The fact that $e$ and $m$ anyons carry fractional charges under $U(1)_e$ and $U(1)_m$, respectively, implies that these symmetries have a fractional mutual Hall response corresponding to a mutual theta term with angle $\theta_{em} = \pi$ \cite{Barkeshli:2014cna}, corresponding to the 4d SPT action
\[\pi \int_{\cM_4} \frac{dA_e}{2\pi} \wedge \frac{dA_m}{2\pi}~.\] 
Any symmetry that pins $\theta_{em} = \pi$ thus has a mixed anomaly with $U(1)_e \times U(1)_m$.  
This includes time reversal $\mathsf{T}$ or the $\pi/2$ rotations $\mathsf{R}^{\pi/2}_{v,p}$. Thus there is no trivial phase in the phase diagram of $H_{\rm SFS}(h_e,h_m)$.

\begin{figure}[t]
    \centering
    \begin{tikzpicture}[x=4cm,y=4cm] 
  \draw[gray!60, line width=0.3pt] (0,0) rectangle (1.35,1.35);

  \node[below=8pt] at (0.64,0.02) {$h_m$};
  \node at (-0.15,0.64) {$h_e$};
  \draw[->] (0.49,-0.05) -- (0.79,-0.05);
  \draw[->] (-0.05,0.49) -- (-0.05,0.79);
  

  \node at (0.23,1.2) {$U(1)_e$};
  \node at (0.23,1.1) {SSB};

  \node at (1.12,0.28) {$U(1)_m$};
  \node at (1.12,0.18) {SSB};

  \node at (0.23,0.28) {Toric};
  \node at (0.23,0.20) {code};

    \node at (0.6,0.6) [rotate=45] {\large \textbf{???}};

\draw[<->] (-0.25,0.05) -- (0.05,-0.25);

\node at (-0.15,-0.15) {$\mathsf{D}$};

    \draw[line width=0.8pt,blue]
    (0.0,0.64) -- (0.2,0.64);
    \node at (0.2,0.72) {\textcolor{blue}{$O(2)^*_e$}};
    \draw[line width=0.8pt,blue]
    (0.64,0) -- (0.64,0.2);
    \node at (0.82,0.1) {\textcolor{blue}{$O(2)^*_m$}};

\end{tikzpicture}
    \caption{Phase diagram of the $U(1)_e \times U(1)_m$-symmetric, staggered Fradkin-Shenker model~\eqref{eqnU1FSmodel}, as a function of $h_e,h_m\geq 0$. We have analytic access only to the axes of this phase diagram, where it can be mapped to a $\bZ_2$-gauged spin-$1/2$ XY model. This model exhibits a second-order transition in the $O(2)^*$ universality class from the toric code phase to a $U(1)_e$ SSB or $U(1)_m$ SSB phase at some $h_e^*$ or $h_m^*$, respectively (solid blue lines meeting the axes). In the central region, there might be a multicritical point described by the HYQED. This would suggest that the central region where the three phases meet looks like the phase diagram in figure~\ref{figU1phasediagram} (rotated by 45 degrees). Unlike the Fradkin-Shenker model \eqref{eqnfsmodel}, we do not have analytic access to the large field limit. Between the $U(1)_e$ and $U(1)_m$ SSB phases there may be a direct transition, a coexistence phase where both symmetries are broken, or even something totally different. Note however that this phase diagram is constrained by a mixed anomaly involving $U(1)_e \times U(1)_m$ and time reversal; in particular there cannot be any trivial phase.}
    \label{figU1latticephasediagram}
\end{figure}
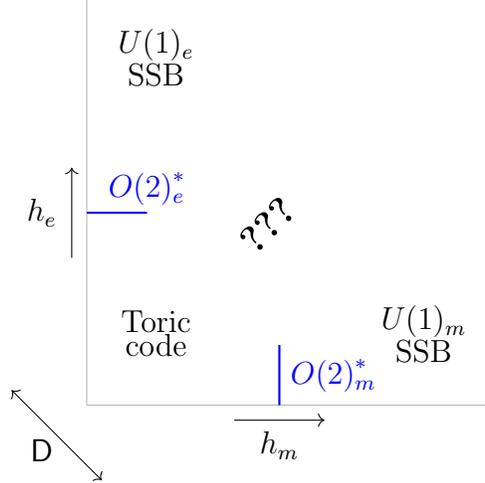

We will show that if $h_e = 0$ (resp.~$h_m=0$), $H_{\rm SFS}(h_e,h_m)$ reduces to a $\mathbb{Z}_2$-gauged XY model in a staggered chemical potential. This model has a toric code phase at small $h_m$ (resp.~$h_e$) and undergoes a continuous phase transition in the $\mathbb{Z}_2$-gauged $O(2)$ universality class, referred to as $O(2)^*$, to a phase where $U(1)_m$ (resp.~$U(1)_e$) is spontaneously broken \cite{Fisher:1989zza}. This is most clear in an equivalent presentation of the model, established in section~\ref{subsecgaugelatticemodel}, in a gauged Hilbert space with vertex, link, and plaquette degrees of freedom. This leads to the phase diagram in figure \ref{figU1latticephasediagram}.

Finally, we can relate the $U(1)_e \times U(1)_m$-symmetric SFS model to the original Fradkin-Shenker model via
\[\label{eqndeformingtheU1modeltotheoriginalmodel}H_{\rm FS}(h_e,h_m) = H_{\rm TC} - h_e \sum_\ell Z_\ell - h_m \sum_\ell X_\ell =H_{\rm SFS}(h_e,h_m) - h_e \sum_\ell \tilde Z_\ell^- - h_m \sum_\ell \tilde X_\ell^-~,\]
where
\[\tilde Z_\ell^- = \frac{Z_\ell}{2} (1-A_v A_{v'})~, \qquad
\tilde X_\ell^- = \frac{X_\ell}{2} (1-B_p B_{p'})~.\]
The operators $\t Z_\ell^-$ and $\t X_\ell^-$ have charge 1 under $Q_e$ and $Q_m$, respectively. They are also both time-reversal invariant. The uniform perturbations,
\[- h_e \sum_\ell \tilde Z_\ell^-  \,, \qquad -h_m\sum_\ell \tilde X_\ell^- \,,\]
preserve all crystalline symmetries of the original square lattice and are exchanged by duality~$\mathsf{D}$.

Before analyzing the staggered Fradkin-Shenker model in more detail, we note that similar Euclidean lattice models -- describing loop gases of conserved $e$ and $m$ worldlines with mutual $-1$ braiding statistics -- were studied in \cite{MotrunichPaper}. For both a softcore and a hardcore model, their numerical simulations suggest a phase diagram whose edges are similar to~\ref{figU1latticephasediagram}, but where the $O(2)^*_{e,m}$ lines turn into first-order lines away from the self-dual line, at a~duality-related pair of tricritical points. The first-order lines then merge with a self-dual first-order line at a three-fold-degenerate triple point (i.e.~not a multicritical point). The authors used a non-local formulation of the braiding statistics to produce a sign-problem-free model, so it is not clear how it is related to our Hamiltonian lattice model, which appears to have a sign problem.\footnote{~A Hamiltonian is sign-problem-free if in some basis all its off-diagonal terms are negative. At least in the $Z$ basis this is not true of our Hamiltonian.} It may be that the model of \cite{MotrunichPaper} did not include enough self-repulsion between particles of the same species to ensure that the~$O(2)_{e,m}^*$ transitions remain continuous. 

\subsection{$U(1)$ Symmetries of the Toric Code and Fractionalization}

Recall the toric code Hamiltonian \eqref{eqntoriccodeham},
\[H_{\rm TC} = - \sum_v A_v - \sum_p B_p \,.\]
Since all $A_v$ and $B_p$ are commuting, we can define two not-on-site $U(1)$ symmetries of $H_{\rm TC}$ as follows,
\[U(1)_e: \exp\left( \frac{i\theta}{4} \sum_v (-1)^v A_v\right) = \exp\left(i \theta \mathsf{Q}_e\right) \,, \\
U(1)_m: \exp\left( \frac{i\theta}{4} \sum_p (-1)^p B_p \right) = \exp\left(i \theta \mathsf{Q}_m\right) \,.\]
In this equation, the square lattice and its dual are bipartite, and we divide the vertices and plaquettes into even and odd, defining the parity $(-1)^v$ and $(-1)^p$ which are $\pm 1$ depending on whether $v$ or $p$ is part of the even or odd sublattice. This sublattice structure turns out to ensure that the phase diagram of~$H_{\rm SFS}$ has transitions described by $O(2)^*$ CFTs (see figure~\ref{figU1latticephasediagram} and section~\ref{o2starlat} below).

Thus, an $e$ anyon on the even sublattice has $U(1)_e$ charge $1/2$ and an $e$ anyon on the odd sublattice has $U(1)_e$ charge $-1/2$, and likewise for $m$ anyons. Note that this half-charge is acceptable since local operators only create $e$ or $m$ anyons in pairs, and thus have integer charges. The fractional charges of the anyons is an instance of symmetry fractionalization~\cite{Barkeshli:2014cna}.\footnote{~Note that this half charge is well-defined modulo~$\Z$ for all states in the same superselection sector as the elementary anyons above, since all such states differ by integer-charged local operators.}

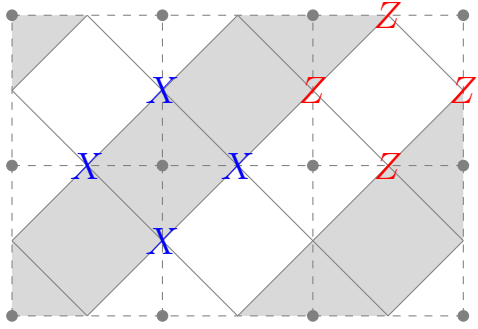
\begin{figure}[t]
    \centering
 \begin{tikzpicture}
    \color{gray}
    \def\length{4}
    \fill[gray!30] (-\length/2,-\length/4) -- (-\length/2,-\length/2) -- (-\length/4,-\length/2)  -- (3*\length/4,\length/2) -- (\length/4,\length/2) -- cycle;
    \fill[gray!30] (\length/4,-\length/2) -- (3*\length/4,-\length/2)  -- (\length,-\length/4) -- (\length,\length/4) -- cycle;
    \fill[gray!30] (-\length/2,\length/2) -- (-\length/4,\length/2)  -- (-\length/2,\length/4) -- cycle;

    \draw[-] (-\length/2, \length/4) -- (\length/4, -\length/2);
    \draw[-] (-\length/2, -\length/4) -- (-\length/4, -\length/2);
    \draw[-] (-\length/4, \length/2) -- (3*\length/4, -\length/2);
    \draw[-] (\length/4, \length/2) -- (\length, -\length/4);
    \draw[-] (3*\length/4, \length/2) -- (\length, \length/4);

    \draw[-] (-\length/2, \length/4) -- (-\length/4, \length/2);
    \draw[-] (-\length/2, -\length/4) -- (\length/4, \length/2);
    \draw[-] (-\length/4, -\length/2) -- (3*\length/4, \length/2);
    \draw[-] (\length/4, -\length/2) -- (\length, \length/4);
    \draw[-] (3*\length/4, -\length/2) -- (\length, -\length/4);
    
    \draw[dashed] (-\length/2, 0) -- (\length, 0);
    \draw[dashed] (-\length/2, -\length/2) -- (\length, -\length/2);
    \draw[dashed] (-\length/2, \length/2) -- (\length, \length/2);
    
    \draw[dashed] (0, -\length/2) -- (0, \length/2);
    \draw[dashed] (-\length/2, -\length/2) -- (-\length/2, \length/2);
    \draw[dashed] (\length/2, -\length/2) -- (\length/2, \length/2);
    \draw[dashed] (\length, -\length/2) -- (\length, \length/2);
    
    \filldraw (0, 0) circle (2pt);
    \filldraw (\length/2, 0) circle (2pt);
    \filldraw (-\length/2, 0) circle (2pt);
    \filldraw (0, \length/2) circle (2pt);
    \filldraw (0, -\length/2) circle (2pt);
    \filldraw (\length/2, \length/2) circle (2pt);
    \filldraw (-\length/2, \length/2) circle (2pt);
    \filldraw (-\length/2, -\length/2) circle (2pt);
    \filldraw (\length/2, -\length/2) circle (2pt);
    \filldraw (\length, -\length/2) circle (2pt);
    \filldraw (\length, 0) circle (2pt);
    \filldraw (\length, \length/2) circle (2pt);
    \color{blue}
    \node[] at (-\length/4, 0) {\large $X$}; 
    \node[] at (\length/4, 0) {\large $X$};  
    \node[] at (0, \length/4) {\large $X$};  
    \node[] at (0, -\length/4) {\large $X$}; 
    \color{red}

    \node[] at (\length, \length/4) {\large $Z$}; 
    \node[] at (3*\length/4, 0) {\large $Z$};  
    \node[] at (3*\length/4, \length/2) {\large $Z$};  
    \node[] at (\length/2, \length/4) {\large $Z$}; 
    \color{black}
    
\end{tikzpicture}
    \caption{The sublattice structure we use to define $(-1)^v$ and $(-1)^p$ in \eqref{eqnsublatticestructure}. We have drawn both the original square lattice of the toric code with lattice basis $\hat x$ and $\hat y$ (sites are the gray circles with dashed horizontal and vertical edges), and the mini square lattice with lattice basis $\frac12 (\hat x \pm \hat y)$ (sites at the midpoints of the dashed edges with solid gray edges connecting them). Plaquettes of the mini square lattice correspond to either vertices or plaquettes of the original square lattice of the toric code, and we have shaded them according to whether they belong to the even or odd sublattice. Along the self-dual line $h_e= h_m$, the model $H_{\rm SFS}(h_e,h_m)$ has the full crystalline symmetry of the mini sublattice.}
    \label{fig:sublatticestructure}
\end{figure}

There are four possible choices of the sublattice structure, which differ by crystalline symmetries of $H_{\rm SFS}(h_e,h_m)$ and are therefore physically equivalent. We show one such choice in figure \ref{fig:sublatticestructure}. It corresponds to 
\[\label{eqnsublatticestructure}(-1)^{v(x,y)} = (-1)^{x+y} \,, \qquad
(-1)^{p(x+1/2,y+1/2)} = (-1)^{x+y}\,,\]
where we labeled vertices $v$ by integer pairs $(x,y)$ and plaquettes $p$ by half-integer pairs $(x+1/2,y+1/2)$.

\subsection{The Staggered Fradkin-Shenker Model}

In the previous section, we identified two $U(1)$ symmetries of the toric code, with generators $\mathsf{Q}_e$ and $\mathsf{Q}_m$. The Fradkin-Shenker model,
\[H_{\rm FS}(h_e,h_m) = H_{\rm TC} - h_e\sum_\ell Z_\ell - h_m \sum_\ell X_\ell \,,\]
breaks these symmetries for all $h_e,h_m \neq 0$. In this section we define a related model which instead has these symmetries and shares some qualitative properties with the original Fradkin-Shenker model. We refer to it as the $U(1)_e \times U(1)_m$-symmetric, or staggered Fradkin-Shenker~(SFS) model.

Consider the operator $Z_\ell$. Acting on a toric code ground state, $Z_\ell$ produces two $e$ excitations, one at each end of the link $\ell$. Since these vertices always occur on distinct sublattices, these $e$'s have opposite $\mathsf{Q}_e$ charge. However, if there is already a single $e$ at one end, $Z_\ell$ will move the $e$ to the opposite sublattice, changing the $\mathsf{Q}_e$ charge by $\pm 1$. Therefore, we can define a $\mathsf{Q}_e$-preserving operator
\[\tilde Z_\ell = \frac{Z_\ell}{2}(1 + A_v A_{v'})\,,
\label{tildeZdefinition}\]
where $v,v'$ are the endpoints of $\ell$, so that the second factor projects onto states with either no $e$ excitations or two $e$ excitations at the endpoints of $\ell$. This operator is also automatically~$\mathsf{Q}_m$-preserving. We likewise define the $\mathsf{Q}_e$- and $\mathsf{Q}_m$-preserving operator
\[\tilde X_\ell = \frac{X_\ell}{2}(1 + B_p B_{p'})\,,
\label{tildeXdefinition}\]
where $p,p'$ are the plaquettes to either side of the link $\ell$.

We take the SFS Hamiltonian to be
\[\label{eqnU1FSmodel2}H_{\rm SFS}(h_e,h_m) = H_{\rm TC} - h_e \sum_\ell \tilde Z_\ell - h_m \sum_\ell \tilde X_\ell~.\]
The SFS model enjoys both $U(1)_{e,m}$ symmetries. It is a close cousin of the original Fradkin-Shenker model, which can be obtained by adding a deformation that explicitly breaks the $U(1)_e \times U(1)_m$ symmetry of the SFS model,
\begin{equation}
     H_{\rm SFS}(h_e,h_m) = H_{\rm FS}(h_e,h_m) + h_e \sum_\ell  \frac{Z_\ell}{2}(1 - A_v A_{v'}) + h_m \sum_\ell \frac{X_\ell}{2}(1 - B_p B_{p'}) \,.   
\end{equation}
For later convenience we define
\[\tilde X_\ell^- = \frac{X_\ell}{2}(1 - B_p B_{p'})~, \qquad
\tilde Z_\ell^- = \frac{Z_\ell}{2}(1 - A_v A_{v'})~,\]
so that the original and staggered Fradkin-Shenker models are simply related via
\[\label{eqndeformationtoFSmodel}H_{\rm FS}(h_e,h_m) = H_{\rm SFS}(h_e,h_m) - h_e \sum_\ell \tilde Z_\ell^- - h_m \sum_\ell \tilde X_\ell^- \,.\]

\subsection{Time Reversal and Mixed $U(1)_e \times U(1)_m$ Anomaly}\label{sec:latanomaly}

In addition to the $U(1)_{e,m}$ symmetries above, $H_{\rm SFS}(h_e,h_m)$ has an anti-unitary time-reversal symmetry $\mathsf{T}$ given by complex conjugation in the $Z$ basis, which acts on the Pauli operators as
\[\mathsf{T} = \begin{cases}X_\ell \mapsto X_\ell \\
Y_\ell \mapsto - Y_\ell \\
Z_\ell \mapsto Z_\ell
\end{cases}\]
This is clearly a symmetry of the model $H_{\rm SFS}(h_e,h_m)$, since all terms are made from~$X$ and~$Z$. For the same reason, both $U(1)$ charges defined above are also~$\mathsf{T}$-even.

Thanks to symmetry fractionalization, i.e.~the fact that $e$ anyons have $U(1)_e$ charge $1/2$, while~$m$ anyons have~$U(1)_m$ charge $1/2$, there is a mixed 't Hooft aanomaly with time reversal~\cite{Etingof:2009yvg,Barkeshli:2014cna,Chen:2016fxq}, which can be expressed as a (3+1)-dimensional theta term,
\[\omega_{\rm anomaly} = \pi \frac{dA_e}{2\pi} \wedge \frac{dA_m}{2\pi}~,\label{anomalystaggeredFS}\]
whose coefficient $\theta=\pi$ is pinned by time-reversal symmetry. For this quantization, it is important that both charges are preserved by $\mathsf{T}$, so that $\mathsf{T}$ (being anti-unitary) acts on the background gauge fields as $A_e \mapsto -A_e$ and $A_m \mapsto -A_m$. Likewise, any other symmetry that pins~$\theta=\pi$, such as several of the crystalline symmetries we consider below, has a mixed anomaly with~$U(1)_e \times U(1)_m$.

We can see this anomaly directly on the lattice by adapting the method of Else and Nayak \cite{Else:2014vma}. Let us truncate our $U(1)_e$ symmetry to act only in a region $R$,
\[U(1)_e^R: \exp\left(\frac{i\theta}{4} \sum_{v \in R} (-1)^v A_v\right)\,.\label{eq:truncR}\]
Although the original symmetry was correctly normalized when acting on the whole lattice, this operator is not $2\pi$ periodic when $R$ has boundary $\partial R$. Substituting~$\theta = 2\pi$ into~\eqref{eq:truncR} gives the operator
\[\prod_{v \in R} A_v = \prod_{\ell \in \partial R} X_\ell\,,\]
which transports an $m$ anyon around the boundary of $R$. We can think of this operator applied along the length of a cylinder as threading a $2\pi$ flux through the cylinder. Since the~$m$ anyon carries charge~$\pm 1/2$ under $U(1)_m$, this flux threading corresponds to a fractional Hall conductivity of $1/2$ (mod 1) in natural units, and indicates the presence of the $\theta=\pi$ term~\eqref{anomalystaggeredFS}.

As a consequence of this anomaly, the ground state of $H_{\rm SFS}(h_e,h_m)$ cannot be trivially gapped for any value of $h_e,h_m$. 

\subsection{Crystalline Symmetries}\label{subseccryssymmetriesfsmodel}

The SFS model $H_{\rm SFS}(h_e,h_m)$ enjoys all of the symmetries of the square lattice, generated by $\mathbb{Z}^2$ translations, vertex- and plaquette-centered $\pi/2$ rotations, edge-centered $\pi$-rotations, and reflections along the coordinate axes, all of these acting by simple permutations of the spins in the $Z$ basis. This wallpaper group is known as $\textbf{p4m}$ and is a symmorphic group of the form
\[\textbf{p4m} = \mathbb{Z}^2 \rtimes D_8 \,,\]
where $D_8$ is the dihedral group of order 8, which we can think of as the point group stabilizing any chosen vertex; it acts on the $\mathbb{Z}^2$ group of translations in the obvious way. Because of the sublattice structure shown in figure \ref{fig:sublatticestructure}, crystalline symmetries act non-trivially on the $U(1)$ charges, as summarized in table \ref{table:latticesymmetries}.

There is also a hidden crystalline symmetry $\mathsf{D}$. It is given by
\[\mathsf{D} = \mathsf{S}_{\frac12(\hat x+\hat y)}\prod_\ell H_\ell \,,\]
where $H_\ell$ is the Hadamard operator, which in the $Z$ basis is
\[H_\ell =  \frac{1}{\sqrt{2}}\begin{pmatrix}
    1 & 1 \\ 1 & -1
\end{pmatrix},\]
and $\mathsf{S}_{\frac12(\hat x+\hat y)}$ is a translation by $\frac12 (\hat x + \hat y)$.
Using~$H_\ell^2 = 1$, $H_\ell X_\ell H_\ell = Z_\ell$, and $H_\ell Z_\ell H_\ell = X_\ell$, we find the action of $\mathsf{D}$,
\[\mathsf{D} = \begin{cases}
    A_{v} \mapsto B_{v+(1/2,1/2)} \\
    B_{p} \mapsto A_{p+(1/2,1/2)}
\end{cases} \,,\]
where $v + (1/2,1/2)$ is the plaquette to the immediate northeast of $v$ and $p+(1/2,1/2)$ is the vertex to the immediate northeast of $p$. This implies that $\mathsf{D}$ exchanges $e$ and $m$ excitations of the toric code, so that we will refer to it as duality. With our chosen sublattice structure for the charges $\mathsf{Q}_e$ and $\mathsf{Q}_m$, it exchanges them without any signs,
\[\mathsf{D}:\mathsf{Q}_e \leftrightarrow \mathsf{Q}_m \,.\]
Likewise, we have
\[
\mathsf{D} = \begin{cases}
    X_\ell \mapsto Z_{\ell+(1/2,1/2)} \\
    Z_\ell \mapsto X_{\ell+(1/2,1/2)}
\end{cases} \,,\]
so that
\[\mathsf{D} H_{\rm SFS}(h_e,h_m) \mathsf{D}^{-1} = H_{\rm SFS}(h_m,h_e) \,.\]
In particular~$\mathsf{D}$ is a symmetry along the self-dual line $h_e = h_m$.

The duality symmetry is a crystalline symmetry that satisfies
\begin{equation}
   \label{dsqtrans}
\mathsf{D}^2 = \mathsf{S}_{\hat x+\hat y}~. 
\end{equation}
With it, the symmetry group \textbf{p4m} is extended to the symmetries of the mini square lattice with basis $\frac12(\hat x \pm \hat y)$, shown in figure \ref{fig:sublatticestructure}. Note that the original degrees of freedom live on the vertices of the mini square lattice. The wallpaper group of the mini square lattice is isomorphic to \textbf{p4m}, but it is a strictly larger group that includes the duality symmetry~$\mathsf{D}$, so we refer to it as \textbf{p4mD}. It resides in the non-split short exact sequence
\[1\to\textbf{p4m} \to \textbf{p4mD} \to \mathbb{Z}_2^\mathsf{D}\to 1~,\]
where the non-trivial element of the quotient is the equivalence class of all dualizing transformations mapping $H_{\rm SFS}(h_e,h_m) \to H_{\rm SFS}(h_m,h_e)$. The action of $\textbf{p4mD}$ is summarized in table~\ref{table:latticesymmetries}.

\subsection{Gauged Presentation}\label{subsecgaugelatticemodel}

Although the SFS model is defined in a tensor product Hilbert space, there is an illuminating way to embed it into a Hilbert space with gauge constraints. (See~\cite{Tupitsyn:2008ah} for a related discussion in the context of the original Fradkin-Shenker model.) To do this, we introduce spin-$\frac12$ degrees of freedom on vertices and plaquettes, with Pauli operators $X_v, Z_v$, $X_p, Z_p$ etc. We then impose the Gauss law constraints
\[\label{eqngaugeconstraint}Z_v =  (-1)^v A_v \,, \\
Z_p =  (-1)^p B_p \,.\]
These new degrees of freedom are completely redundant, in the following sense: there is a locality-preserving unitary map from this Hilbert space to the one where the constraints are simply $Z_v = 1$ and $Z_p = 1$, which is equivalent to our original Hilbert space. Therefore, we will be able to obtain a completely equivalent description of the model \eqref{eqnU1FSmodel2} in the gauged Hilbert space, i.e.~we are not gauging any symmetry; instead we are simply introducing some redundant degrees of freedom that will turn out to simplify the model.

The locality-preserving unitary map $E$ is a finite depth circuit,
\[ E = \prod_v E_v \prod_p E_p \,, \]
of the following commuting unitary operators, acting at each vertex $v$ and plaquette $p$,
\[\label{eqndisentangler}E_v = \frac12(1+X_v) + \frac12 (1-X_v) (-1)^v A_v \,, \\
E_p = \frac12(1+X_p) + \frac12 (1-X_p) (-1)^p B_p \,. \]
These satisfy $E_v Z_v E_v^\dagger = Z_v (-1)^v A_v$ and $E_p Z_p E_p^\dagger = Z_p (-1)^p B_p$. Thus, $E$ maps the constraints~$Z_v (-1)^v A_v = 1$ and $Z_p (-1)^p B_p = 1$ to $Z_v = 1$ and $Z_p=1$, respectively, which is what we wanted.

The signs in the Gauss laws \eqref{eqngaugeconstraint} are chosen so that we can express the $U(1)_e \times U(1)_m$ charges as the on-site terms
\[Q_e = \frac{1}{4}\sum_v Z_v \,, \\
Q_m = \frac{1}{4}\sum_p  Z_p \,.\]
Meanwhile, the toric code Hamiltonian can be thought of as a staggered chemical potential,
\[\label{eqnTCstaggeredpotential}H_{\rm TC} = -\sum_v (-1)^v Z_v - \sum_p (-1)^p Z_p \,.\]
Notice that the terms $Z_\ell$ and $X_\ell$ are not gauge-invariant, but applying the $E$ map we obtain
\[E Z_\ell E^\dagger = X_v Z_\ell X_{v'} \,, \qquad
E X_\ell E^\dagger = X_p X_\ell X_{p'} \,,\]
where $v,v'$ are the vertices of $\ell$, and $p,p'$ are the plaquettes bordering $\ell$.
This results in (see \eqref{tildeZdefinition} and \eqref{tildeXdefinition})
\[E \tilde Z_\ell E^\dagger = \frac{1}{2} Z_\ell(X_v X_{v'} + Y_v Y_{v'}) =  S^+_v Z_\ell S^-_{v'} + \text{h.c.} \,, \\
E \tilde X_\ell E^\dagger = \frac{1}{2} Z_\ell(X_p X_{p'} + Y_p Y_{p'}) =  S^+_p X_\ell S^-_{p'} + \text{h.c.}\,,\]
where we have defined the spin raising/lowering operators
\[S^\pm = \frac12(X\pm iY)~. \]

Thus, we can interpret the staggered Fradkin-Shenker model as a combination of two XY models in a staggered chemical potential. Equivalently, it is a combination of two hard-core boson models (namely, models where each site can be occupied by at most one boson), coupled to a $\bZ_2$ gauge field (the link degrees of freedom), such that one boson behaves like a charge and the other like a flux, and their hopping operators across a link anti-commute,
\[( S^+_v Z_\ell S^-_{v'} + \text{h.c.})( S^+_p X_\ell S^-_{p'} +  \text{h.c.}) = -( S^+_p X_\ell S^-_{p'} +  \text{h.c.})(S^+_v Z_\ell S^-_{v'} +  \text{h.c.})~.\]
In summary, the SFS model is equivalent to following Hamiltonian in the gauged presentation,
\begin{equation}
\begin{split}
\label{eqngaugeversion2}H_{\rm SFS, \, gauged}(h_e,h_m) = &-\sum_v  (-1)^v Z_v - \sum_p (-1)^p Z_p \\
&- h_e \sum_\ell \left( S^+_v Z_\ell S^-_{v'} +  \text{h.c.} \right) - h_m \sum_\ell \left( S^+_p X_\ell S^-_{p'} +  \text{h.c.} \right) \,.    
\end{split}
\end{equation}
The ordinary Fradkin-Shenker model can be obtained from this by adding the following unit-charge operators,
\[\label{eqndeftermsgauged}-h_e \sum_\ell \tilde Z_\ell^- =- h_e \sum_\ell \left( \frac{1}{2} S^+_v Z_\ell S^+_{v'} +  \text{h.c.} \right) \,, \\
-h_m \sum_\ell \tilde X_\ell^- =- h_m \sum_\ell \left( \frac{1}{2} S^+_p X_\ell S^+_{p'} +  \text{h.c.} \right) \,. \]
Note that in this presentation, crystalline symmetries that negate $(-1)^v$ and/or $(-1)^p$ must be combined with $\prod_v X_v$ and or $\prod_p X_p$ to preserve the Gauss laws.

\subsection{$O(2)^*$ Lines}\label{o2starlat}

The gauged presentation \eqref{eqngaugeversion2} of the staggered Fradkin-Shenker model makes it easier to analyze the axes of the phase diagram, where either $h_e = 0$ or $h_m = 0$. We will analyze $h_m = 0$, since the other axis $h_e = 0$ is related by the action of duality $\mathsf{D}$.

If~$h_m = 0$, then~$Z_p = (-1)^p B_p$ commutes with the Hamiltonian for each $p$, so that the model splits into sectors, each given by the XY model in a particular flux background, i.e.~there is a~$\pi$-flux wherever~$B_p = -1$. Activating any non-trivial such flux $B_p = -1$ increases the ground state energy; this essentially follows from the diamagnetic inequality of Simon~\cite{simon1976universal}, as we show in appendix~\ref{appdiamagneticinequality}. Moreover, the toric code terms/staggered chemical potential \eqref{eqnTCstaggeredpotential} give an extra energy cost $\Delta E = 2$ to any plaquette with $B_p = -1$, so for any states below an energy gap $\Delta E=2$, we have
\[Z_p = (-1)^p \qquad \text{when }\ h_m = 0~, \quad  \Delta E < 2~.\]
Inserting this into~\eqref{eqngaugeversion2}, we find the following low-energy Hamiltonian,
\[H_{\rm SFS, \, gauge}(h_e,0) = -\sum_v (-1)^v Z_v - h_e \sum_\ell \left( S^+_v Z_\ell S^-_{v'} + \text{h.c.} \right) \,, \]
with the constraint $B_p=1$ for all $p$.
We recognize this model as the spin-1/2 XY model in a staggered chemical potential, coupled to a flat $\bZ_2$ gauge field.

The spin-1/2 XY model with a staggered chemical potential is known to have a disordered phase at small $h_e$ and an ordered phase at large $h_e$ \cite{Aizenman_2004}. A direct transition between these phases has been observed numerically, and it is believed to be a continuous phase transition in the $O(2)$ Wilson-Fisher universality class \cite{Fisher:1989zza,hen2010phase}. What we have above is a $\bZ_2$ gauged version of this, where the disordered phase at small $h_e$ is the topologically ordered phase of the toric code, and the ordered phase at large $h_e$ is a phase where the $\bZ_2$ gauge field is Higgsed by the proliferation of bosons on the vertices, which we have identified with $e$ anyons. In this Higgs phase, the symmetry $Q_e$ is also spontaneously broken, giving rise to a gapless Nambu-Goldstone mode. We will refer to this $\bZ_2$-gauged $O(2)$ model as the $O(2)^*$ model.

\section{Continuum QED$_3$ with~$N_f=2$ Flavors}
\label{section:qed3review}

\subsection{Conventions}\label{sec:conventions}
We adopt Lorentzian signature with metric $\eta_{\mu\nu}=(-++)$ and Levi-Civita symbol $\varepsilon^{012}=+1$. The Dirac gamma matrices satisfy
\begin{equation}
    \lbrace \gamma^\mu, \gamma^\nu \rbrace = 2 \eta^{\mu\nu}~, \qquad \gamma^0 (\gamma^\mu)^\dagger \gamma^0 = \gamma^\mu~.
\end{equation}
We choose a Majorana basis in which they are all real,
\begin{equation}\label{majoranabasis}
\gamma^\mu = (i\sigma_y,\sigma_z,-\sigma_x) \,.
\end{equation}
The resulting representation~$D(\Lambda)$ of the Lorentz group is a 2-component Dirac column spinor~$\psi$, which acts from the left as~$\psi \to D(\Lambda)\psi$. We generally take~$\psi$ to be complex, and we typically suppress its spinor indices. When we choose to spell out the spinor indices explicitly, we use a 2+1 dimensional version of the two-component spinor formalism common in supersymmetric theories (see e.g.~\cite{Dumitrescu:2011iu}). The components of~$\psi$ are then denoted by
\begin{equation}
    (\psi)_\alpha = \psi_\alpha~, \qquad \alpha = 1,2~,
\end{equation}
and Lorentz transformations~${D(\Lambda)_\alpha}^\beta$ act from the left as real~$SL(2, \R)$ matrices, i.e.~the spinor indices~$\alpha = 1,2$ are fundamental~$SL(2, \R)$ indices.\footnote{~Thus the gamma matrices~\eqref{majoranabasis} have index structure~${(\gamma^\mu)_\alpha}^\beta$.} In quantum field theory, each component~$\psi_\alpha$ is an operator, whose Hermitian conjugate we denote by\footnote{~Note that we take Hermitian conjugation~$*$ to be order-reversing on quantum operators, e.g.~$(\psi_1 \chi_2)^* = \chi_2^* \psi_1^*$, but it does not act on the spinor indices. For this reason we do not use the more common notation~$\dagger$ to denote the Hermitian conjugate. In supersymmetric theories the Hermitian adjoint is often denoted by a bar, but we reserve this notation for the Dirac bar defined in~\eqref{Diracbar}.}
\begin{equation}
    (\psi_\alpha)^* = \psi^*_\alpha~.
\end{equation}
We are therefore free to impose an~$SL(2, \R)$-invariant Majorana condition~$\psi_\alpha = \psi_\alpha^*$, though we will almost exclusively use complex Dirac spinors throughout. Finally we note that a pair of up-down spinor indices can be contracted using the~$SL(2, \R)$ invariant~$\delta_\alpha^\beta$, while spinor indices can be raised and lowered from the left using the real and~$SL(2, \R)$-invariant Levi-Civita symbols,
\begin{equation}\label{raiselower}
    \psi^\alpha = \ep^{\alpha\beta} \psi_\beta = (i\sigma_y)^{\alpha\beta} \psi_\beta \qquad \psi_\alpha = \ep_{\alpha\beta} \psi^\beta = - (i\sigma_y)_{\alpha\beta} \psi^\beta~.
\end{equation}

In the usual Dirac spinor formalism, the Dirac conjugate~$\b \psi$ of~$\psi$ (which transforms from the right as~$\b \psi \to \b \psi D(\Lambda)^{-1}$) is taken to be the following row spinor, 
\begin{equation}\label{Diracbar}
\bar\psi \equiv \psi^\dagger \gamma^0 = \psi^{*\,t}\gamma^0 \,.
\end{equation}
Here~$\dagger$ denotes Hermitian conjugation~$*$ composed with transposition~$t$ of the spinor indices. Comparing with~\eqref{raiselower}, we see that the Dirac bar naturally has raised~$SL(2, \R$) indices,
\begin{equation}
    \b \psi^\alpha = - \psi^{* \alpha}~.
\end{equation}

The formulas above suffice for writing standard Dirac kinetic and mass terms, 
\begin{equation}\label{freeDirac}
\mathscr{L}_\text{Dirac} = - i \b \psi 
\dslash \psi + i m_D \b \psi \psi = i \psi^{*\alpha}{(\gamma^\mu)_\alpha}^\beta  \partial_\mu \psi_\beta - i m_D \psi^{*\alpha} \psi_\alpha~, \qquad m_D \in \R~,
\end{equation}
which preserve the~$U(1)$ phase-rotation symmetry acting on~$\psi_\alpha$. Here we have used the standard Dirac slash notation~$\dslash \equiv \gamma^\mu \d_\mu$, which we will also apply to other Lorentz vectors. The Lagrangian~\eqref{freeDirac} is (up to a total derivative) Hermitian, thanks to the following conjugation identities for the Lorentz-scalar and -vector bilinears constructed using two complex Dirac fermions $\psi$ and $\chi$,\footnote{~Here we use the fact that the fermion fields anticommute, while~$*$ is order-reversing on fields.}
\begin{equation}
(\bar\chi \psi)^* = - \bar\psi \chi~, \qquad
(\bar\chi \gamma^\mu \psi)^* =  \bar\psi \gamma^\mu \chi~.
\end{equation}

We will also need fermion mass terms of Majorana (or pairing) type, which break the~$U(1)$ phase-rotation symmetry of~$\psi$. These are easily expressed in two-component notation,
\begin{equation}\label{majorMass}
    \Delta \mathscr{L}_\text{Majorana} = - i m_M \ep^{\alpha\beta} \psi_\alpha \psi_\beta - i m_M^* \ep^{\alpha\beta} \psi^{*}_\alpha \psi^{*}_\beta~, \qquad m_M \in \C~,
\end{equation}
where Lorentz-invariance and hermiticity are manifest. To express~\eqref{majorMass} in Dirac notation, we need to use the charge-conjugation matrix~$C$, defined by~$C(\gamma^\mu)^tC^{-1}=-\gamma^\mu$, which is always an invariant tensor. In our basis we are free to take
\begin{equation}\label{C_on_psi}
  C \equiv \gamma^0 = i \sigma_y~, \qquad  \psi^C \equiv C \bar\psi^t = \psi^* \,.
\end{equation}
Here~$\psi^C$ is the charge-conjugated Dirac spinor, which has the same Lorentz-transformation properties as~$\psi$. Comparing with~\eqref{raiselower}, we see that the charge-conjugation matrix~$C$ is nothing but an~$SL(2, \R)$-invariant Levi-Civita symbol, so that the Majorana mass terms~\eqref{majorMass} can be converted to Dirac spinor notation as follows, 
\begin{equation}\label{majorMassDS}
     \ep^{\alpha\beta} \psi_\alpha \psi_\beta = \psi^t \gamma^0 \psi = \b {\psi^C} \psi~, \qquad   \ep^{\alpha\beta} \psi^*_\alpha \psi^*_\beta = \psi^\dagger \gamma^0 \psi^* = \b \psi \psi^C~.
\end{equation}
To verify the hermiticity of~\eqref{majorMass} it is useful to consider the following symmetric and Lorentz-invariant bilinear of the Dirac spinors~$\chi$ and~$\psi$,
\begin{equation}\label{Cbilin}
    \b {\chi^C} \psi = \chi^t \gamma^0 \psi = \ep^{\alpha\beta} \chi_\alpha \psi_\beta= \b {\psi^C} \chi~,
\end{equation}
whose Hermitian conjugate is given by
\begin{equation}\label{CHC}
    ( \b {\chi^C} \psi )^* =  - \b \psi \gamma^0 \b \chi^t= -\ep^{\beta\alpha} \psi^*_\beta \chi^*_\alpha = - \b {\psi} \chi^C~.
\end{equation}

Let us now discuss our conventions for currents and background fields.\footnote{~These differ from the conventions in~\cite{Dumitrescu:2024jko, Dumitrescu:2025vfp} by some signs.} The Dirac Lagrangian~\eqref{freeDirac} has a~$U(1)_A$ symmetry, and we will take the associated current and charge to be
\begin{equation}\label{Diracjq}
j^\mu_A = \b \psi \gamma^\mu \psi~, \qquad q_A = \int d^2 x\, j^0_A = - \int d^2 x \, \psi^\dagger \psi~.   
\end{equation}
Using the canonical anticommutation relations, it follows that
\begin{equation}
    [q_A, \psi(x)] = \psi(x)~, \qquad [q_A, \b \psi(x)] = - \b \psi(x)~,
\end{equation}
so that~$\psi$ has charge~$q_A = +1$ and~$\b \psi$ has charge~$q_A= -1$. We minimally couple the spin$_c$ connection~$A_\mu$ that serves as the background field for this~$U(1)_A$ symmetry to the Dirac Lagrangian as follows,
\begin{equation}\label{covDA}
 {\mathscr L}_\text{Dirac}[A] =  {\mathscr L}_\text{Dirac}[0] - A_\mu j_A^\mu = - i \b \psi \Dslash_A \psi + i m_D \b \psi \psi~, \qquad D_{A\,\mu} = \d_\mu - i A_\mu~.
\end{equation}
In other words, $j^0_A$ couples to~$-A_0 = A^0$ in the Lorentzian Lagrangian, while~$D_A$ is the conventionally normalized covariant derivatives on fields of charge~$q_A = +1$. Below we will use~$D_{-A}$ to denote the action of fields on charge~$q_A = -1$ etc. A properly quantized Chern-Simons term,\footnote{~More precisely, we must distinguish between the familiar case where~$A$ is a standard~$U(1)$ connection, and the case of interest here where~$A$ is a spin$_c$ connection. In the former case the Chern-Simons term is well-defined for even~$k$, but requires a choice of spin structure when~$k$ is odd. By contrast, if $A$ is a spin$_c$ connection we can make sense of the Chern-Simons term for any~$k \in \Z$ without choosing a spin structure, as long as we accompany it by a suitable gravitational Chern-Simons term (see section~\ref{secetainvariants} below). }
\begin{equation}\label{CSA}
   S_\text{CS}[A] =  {k \over 4 \pi } \int A \wedge dA = {k \over 4 \pi} \int d^3 x \, \ep^{\mu\nu\rho} a_\mu \d_\nu a_\rho~, \qquad k \in \Z~,
\end{equation}
thus assigns~$U(1)_A$ charge~$q_A = -k$ to a~$2\pi$ flux for~$A$ in the~$xy$-plane, i.e.~$\int (dA)_{12}= 2\pi$. 

For future reference, we note that the quantum contribution to the total effective Chern-Simons level~$k_\text{eff}$ that arises from integrating out the massive Dirac fermion is given by\footnote{~\label{fn:CSMatrix}If~$\psi$ carries charges~$q_I$ under~$\prod_I U(1)_I$, then the (not necessarily properly quantized) Chern-Simons term that arises from integrating out~$\psi$ is given by
\begin{equation}\label{genkmat}
{\half \sign(m_D) \over 4 \pi} \int \left(\sum_I q_I A_I\right) \wedge d \left(\sum_J q_J A_J\right)~.
\end{equation}}
\begin{equation}\label{deltakcs}
    \delta k_\text{eff} = \half \sign(m_D)~.
\end{equation}
This formula, though useful as a bookkeeping device, is not always sufficient to correctly capture the effect of integrating out massive fermions. A more precise discussion using~$\eta$-invariants can be found in section~\ref{secetainvariants}.

\subsection{Spontaneous Symmetry Breaking in QED$_3$}
\label{section:qed3SSB}

Here we briefly review (following the much more detailed presentation in~\cite{Dumitrescu:2024jko}) some facts about QED$_3$ with~$N_f = 2$ flavors of two-component Dirac fermions~$\psi^i~(i = 1,2)$ of electric charge $q_a=1$. In the absence of fermion mass terms, the Lagrangian is given by
\begin{equation}
\mathscr{L}_{\rm QED} = - \frac{1}{4e^2} f^{\mu\nu}f_{\mu\nu} - i\b\psi_i \gamma^{\mu}(\partial_\mu-ia_\mu) \psi^i \,.   
\label{QEDLagrangian}
\end{equation}
Several comments are in order:
\begin{itemize}
    \item The dynamical~$U(1)$ gauge field~$a_\mu$ is actually a spin$_c$ connection, so that the theory is bosonic: it has no gauge-invariant local operators that are fermions, and thus it can be formulated on an arbitrary three-manifold $\mathcal{M}_3$ without choosing a spin structure. In three dimensions, a free photon can be dualized to a compact scalar -- the dual photon.

    \item The strong-coupling scale of the theory is set by the gauge coupling $e^2$, whose mass dimension is one. The theory is weakly coupled at high energies (UV) and strongly coupled at low energies (IR).

    \item The continuous, unitary zero-form global symmetry of the theory is given by
    \begin{equation}\label{u2symm}
        U(2)_{\rm QED} = {SU(2)_f \times U(1)_\CM \over \Z_2}~.
    \end{equation}
    Here~$SU(2)_f$ is the flavor symmetry, under which the fermions~$\psi^i$ transform as a doublet. By contrast, we refer to~$U(1)_\CM$ as the magnetic or monopole symmetry and denote the corresponding monopole charge by~$q_\CM \in \Z$. The~$\Z_2$ quotient in~\eqref{u2symm} expresses the fact that gauge-invariant local operators with magnetic charge $q_\CM$ transform in an~$SU(2)_f$ representation of spin $j_f \in \half \Z$ that satisfies~$q_\CM \equiv 2j_f \mod 2 \mathbb{Z}$.

    \item The theory also enjoys a unitary charge-conjugation symmetry $\mathcal{C}$ and an anti-unitary time-reversal symmetry $\mathcal{T}$,\footnote{~By the relativistic~$\CC\CR\CT$-theorem, there is also a spatial reflection symmetry~$\CR$, which is redundant with~$\CC$ and~$\CT$. The fact that the theory has time-reversal symmetry requires the total effective Chern-Simons level for the dynamical gauge field~$a_\mu$ to vanish, as is implicit in~\eqref{QEDLagrangian}, see for instance~\cite{Wang:2017txt,Cordova:2017kue, Dumitrescu:2024jko, Dumitrescu:2025vfp}.} whose detailed action on the fields we will spell out below. There is a mixed 't Hooft anomaly between the unitary~$U(2)_{\rm QED}$ symmetry in~\eqref{u2symm} and any orientation-reversing symmetry -- such as $\mathcal{T}$ or $\mathcal{CT}$ -- that pins the following four-dimensional~$\theta$-angle to~$\theta = \pi$,
    \begin{equation}\label{QEDanomaly}
        S_{\rm anomaly} = \pi \int_{\mathcal{M}_4} c_2(U(2)_\text{QED}) \,.
    \end{equation}
    Here~$c_2(U(2)_\text{QED})$ is the second Chern class of the $U(2)_{\rm QED}$ background gauge fields. The anomaly-inflow (or SPT) action~\eqref{QEDanomaly} resides in the four-dimensional bulk~$\CM_4$, while the QED$_3$ theory of interest resides on its boundary~$\partial\CM_4=\CM_3$.  

\end{itemize}
The strongly-coupled IR behavior of this theory has remained controversial and inconclusive over the past decades. While the 't Hooft anomaly~\eqref{QEDanomaly} implies that the IR cannot be gapped, some lattice studies have indicated spontaneous symmetry breaking~\cite{Hands:2002dv, Hands:2002qt, Hands:2004bh, Strouthos:2007stc} while others have pointed to apparently conformal behavior without symmetry breaking~\cite{Karthik:2015sgq, Karthik:2016ppr, Karthik:2019mrr}. Relatively recently, conformal bootstrap studies~\cite{Li:2021emd} have indicated that~QED$_3$ with~$N_f =2$ flavors cannot flow to a symmetry-preserving CFT in the IR.\footnote{~By contrast, QED$_3$ with~$N_f = 4$ flavors appears to flow to a CFT~\cite{Albayrak:2021xtd}, suggesting that all theories with even~$N_f \geq 4$ are in the conformal window.} 

An enticing feature of the (apparently counterfactual) symmetry-preserving CFT scenario is that massless $N_f=2$ QED$_3$ can be argued to possess a self-duality that suggests an enhancement of the~$U(2)_{\rm QED}$ global symmetry to $O(4)$ in the deep IR \cite{Xu:2015lxa,Karch:2016sxi,Hsin:2016blu}. We refer to the reflection element of this~$O(4)$ as mirror symmetry, because it exchanges the flavor and the monopole symmetries, so that the latter is enhanced from~$U(1)_\CM$ to~$SU(2)_\CM$, much like in the~$\CN=4$ supersymmetric QED theories studied in~\cite{Intriligator:1996ex}. In section~\ref{sec:hyqed} below, we construct and analyze a multicritical deformation of QED (termed Higgs-Yukawa-QED) with an additional tuning parameter and reduced~$U(1)_f \subset SU(2)_f$ flavor symmetry. We argue in section~\ref{sec:emergemirr} that this theory actually has emergent mirror symmetry exchanging~$U(1)_f \leftrightarrow U(1)_\CM$ in the deep IR. 

Returning to the fate of massless $N_f=2$ QED$_3$, the proposal put forward in \cite{Chester:2024waw,Dumitrescu:2024jko} (see~\cite{Seiberg:1996nz, Senthil:2005jk} for earlier work on closely related models) is that this theory (with Lagrangian~\eqref{QEDLagrangian}), spontaneously breaks its global symmetry as follows,
\begin{equation}\label{QEDsymbreaking}
    U(2)_{\rm QED} \rightarrow U(1)_{\rm unbroken}~.
\end{equation}
The primary order parameter is the minimal monopole operator~$\mathcal{M}^i$, which is a Lorentz scalar of magnetic charge~$q_\mathcal{M}=1$ that transforms as an~$SU(2)_f$ doublet~\cite{Borokhov:2002ib}, and whose vev
\begin{equation}\label{Mvev}
    \langle \CM^i\rangle \neq 0~,
\end{equation}
triggers~\eqref{QEDsymbreaking}. Certain notions of charge conjugation and time reversal are also unbroken. This proposal is consistent with the Vafa-Witten theorems~\cite{Vafa:1983tf,Vafa:1984xg}. 

The symmetry-breaking pattern~\eqref{QEDsymbreaking} leads to three Nambu-Goldstone bosons that parameterize a sigma model into a squashed three-sphere~$\widetilde{S}^3$ with $U(2)$ isometry; it is the~$U(2)$ orbit of the monopole vev~\eqref{Mvev}, whose~$U(2)$-invariant norm is
\begin{equation}\label{eq:mmnorm}
v^2\equiv \langle (\CM^i)^*\rangle \langle \CM^i\rangle > 0~.   
\end{equation}
The squashed~$\t S^3$ can be viewed as a circle bundle that is Hopf-fibered over a~$\mathbb{CP}^1$ base. This~$\C\P^1$ can be described by an~$SU(2)_f$ triplet vector~$n^{a}$ of unit norm (neutral under~$U(1)_\CM$) that is determined by the monopole doublets~$\CM^i$ describing the full~$\t S^3$ via the Hopf map, 
\begin{equation}
\label{QED_monopole_vev}
\mathcal{M}^* \sigma^a \mathcal{M} = v^2 n^a~, \qquad n^a n^a = 1~, \qquad a = 1,2,3~, 
\end{equation}
where $\sigma^a$ are the Pauli matrices. Several comments are in order:

\begin{itemize}
\item Given $n^a$, the Hopf map determines~$\CM^i$ uniquely up to a phase, parametrized by the dual photon that shifts under~$U(1)_\CM$, which describes the $S^1$ fiber over each point of the base.  The unbroken subgroup in \eqref{QEDsymbreaking} is also non-trivially fibered over the base. 

\item The real~$SU(2)_f$ triplet fermion bilinear~$i\bar\psi \sigma^a\psi$ also gets a vev, which is (anti-) aligned with the monopole vev via the Hopf map, so that at low energies this operator flows to\footnote{~Our conventions in this formula differ from those used in~\cite{Dumitrescu:2024jko} by a sign.}
\begin{equation}
\label{QED_RG_triplet}
\CO^a \equiv i\bar\psi \sigma^a\psi \quad \xrightarrow{\rm IR} \quad -Cn^a~, \quad C>0~. 
\end{equation}
Thus~$\CO^a$ serves as a kind of secondary order parameter, which is sensitive to the breaking of the~$SU(2)_f$ symmetry to its Cartan subgroup (associated with the~$\C\P^1$ base of the Hopf fibration), but does not see the~$U(1)_\CM$ symmetry (associated with the fibers). 

\item In the low-energy~$\t S^3$ sigma model, the 't Hooft anomaly~\eqref{QEDanomaly} is matched by a conventional sigma-model~$\theta$-term, associated with its winding number~$\pi_3(\widetilde{S}^3)=\mathbb{Z}$, which is pinned at its non-trivial~$\mathcal{T}$-invariant value~$\theta=\pi$.
    
\end{itemize}

Evidence for the symmetry-breaking proposal~\eqref{QEDsymbreaking} can be gathered by examining its consistency with the phase diagram of the theory under various deformations. Mass deformations, reviewed below, were considered in~\cite{Chester:2024waw,Dumitrescu:2024jko}. More recently~\cite{Dumitrescu:2025vfp} analyzed the effect of a uniform magnetic field~$B$ and deduced the same symmetry-breaking pattern (and a non-relativistic variant of the~$\t S^3$ sigma model) in the strong-field regime~$B \gg e^4$.

The massless QED Lagrangian~\eqref{QEDLagrangian} admits two kinds of fermion mass deformations,
\begin{equation}\label{qedfm}
\mathscr{L}_{\psi~\text{masses}} = m \CO + m_a \CO_a~, \qquad m, m_a \in \R~.
\end{equation}
Here the~$SU(2)_f$ singlet and triplet fermion bilinears (defined to be Hermitian) are given by 
\begin{equation}\label{OOdef}
\mathcal{O}=i\bar\psi_i\psi^i~, \qquad \mathcal{O}_a = i\bar\psi_i{(\sigma_a)^i}_j\psi^j~.
\end{equation}
The former explicitly breaks time reversal~$\CT$, while the latter preserves a suitably defined time-reversal symmetry (discussed in more detailed below), but breaks~$SU(2)_f$ to a Cartan subgroup~$U(1)_f$. 

We will mostly be interested in the triplet mass~$m_a \CO_a$.\footnote{~The effect of the~$\CT$-breaking singlet mass~$m$ has been discussed in~\cite{Chester:2024waw}, where it was argued to deform both the metric and the~$\theta$-angle of the~$\t S^3$ sigma model. These authors conjectured that both smoothly vanish at two points of enhanced~$O(4) \supset U(2)_\text{QED}$ and time-reversal symmetry that are reached at some critical values~$m \sim \pm e^2$, described by the conventional $O(4)$ Wilson-Fisher CFT.} As long as~$|m_a| \ll e^2$ is small, it follows from~\eqref{QED_RG_triplet} that this mass term leads to a potential~$V = C m_a n_a$ in the~${\mathbb C \mathbb P}^1$ base of the~$\t S^3$ sigma model, which pins~$n_a \sim -m_a$ and lifts the two Nambu-Goldstone bosons that constitute the~$\C\P^1$. This leaves only the Hopf fiber over that base point, parameterized by the dual photon associated with the monopole operator that is aligned with~$-m_a \sim n_a$ via the Hopf map~\eqref{QED_monopole_vev}. Two special cases that will be important below are
\begin{equation}\label{m3monovev}
\begin{split}    
    & m_3 > 0 : \qquad n_3 = -1~, \qquad \langle \CM^2\rangle \neq 0~,\\
    & m_3 < 0 : \qquad n_3 = +1~, \qquad \langle \CM^1\rangle \neq 0~.
    \end{split}
\end{equation}
It was conjectured in~\cite{Chester:2024waw, Dumitrescu:2024jko} that this Coulomb phase extends all the way to the weakly-coupled regime at large fermion masses~$|m_a| \gg e^2$, without intervening phase transitions, and we will assume this below.

In the sections below, we will outline a proposal for a continuum QFT that describes the multicritical point and phase diagram of the Fradkin-Shenker model. In some parameter regimes, this theory effectively reduces to QED with $N_f=2$ Dirac fermions -- deformed by four-fermion and monopole operators. The fact that QED with $N_f=2$ does indeed spontaneously break its global symmetries as in~\eqref{QEDsymbreaking} via monopole condensation, leading to an~$\widetilde{S}^3$ sigma-model at low energies, plays an important role in correctly reproducing the Fradkin-Shenker phase diagram.

\section{Higgs-Yukawa-QED in the Continuum}\label{sec:hyqed}

\subsection{Lagrangian of Higgs-Yukawa-QED}\label{sec:hyqedlag}

We now add to the massless~$N_f=2$ QED$_3$ Lagrangian~\eqref{QEDLagrangian} a Lorentz scalar Higgs field~$\phi$ of electric charge~$q_a = 2$, and we couple it to the fermions via a Yukawa interaction. The resulting theory, which we refer to as Higgs-Yukawa-QED (HYQED), is described by the following Lagrangian 
\begin{equation}\label{LagrangianMinkowski}
\begin{split}
\mathscr{L}_{\rm HYQED} = &- \frac{1}{4e^2} f^{\mu\nu}f_{\mu\nu} - i \bar\psi_i \Dslash_a \psi^i  - |D_{2a} \phi|^2 -  \lambda_4 |\phi|^4 +   \mathscr{L}_\text{Yukawa}~, \qquad \lambda_4 > 0~.
\end{split}
\end{equation}
For future use we will write the Yukawa couplings using explicit 2-component spinor indices,\footnote{~Starting with a general complex~$y$, we can always achieve~$y > 0$ by rotating the phase of~$\phi$.}
\begin{equation}\label{yukawa2}
    \mathscr{L}_\text{Yukawa} =  i y \left( \phi^* \ep^{\alpha\beta} \psi_\alpha^1 \psi_\beta^2 +  \phi \ep^{\alpha\beta} (\psi_\alpha^1)^* (\psi_\beta^2)^*\right)~, \qquad y > 0~,
\end{equation}
as well as in Dirac notation (see~\eqref{Cbilin} and~\eqref{CHC}),
\begin{equation}\label{yukawaD}
    \mathscr{L}_\text{Yukawa} = i y  \left( \phi^* (\psi^{1})^t \gamma^0 \psi^2 +  \phi \b \psi^1 \gamma^0 (\b \psi^{2})^t\right)~,\qquad y>0~.
\end{equation}
This theory was recently discussed in~\cite{Jian:2017chw, Boyack:2018urb, Dupuis:2021flq}. 
Let us make a few comments:
\begin{itemize}
    \item The Lagrangian now contains three dimensionful constants~$e^2, \lambda_4, y^2$, all of mass dimension one, and hence its dynamics depends on two dimensionless ratios. Below we will outline a particular scenario for the dynamics, and we will argue that it is plausible as long as~$e^2 \ll \lambda_4, y^2$ is sufficiently small.
    
    As we will review, this is analogous to the Type II regime of Abelian Higgs models in 2+1 dimensions with a unit-charge Higgs field (a closely related variant with a Higgs field of charge two arises upon omitting the fermions from~\eqref{LagrangianMinkowski}). This regime is characterized by~$\lambda_4 \gtrsim e^2$, and the model (tuned to its massless point) flows to an interacting CFT, which is mapped by particle-vortex duality~\cite{Peskin:1977kp, Dasgupta:1981zz} to the~$O(2)$ Wilson-Fisher theory. This duality features prominently in our discussion below. 
    
    \item The structure of the Yukawa interaction, and the quantum numbers of the Higgs field~$\phi$, are chosen to allow for a simple, weakly-coupled description of pairing in the~$\ep^{\alpha\beta}\psi_\alpha^1 \psi_\beta^2$ channel, via a standard Higgs vev for~$\phi$ that can be arranged by dialing the mass of~$\phi$ (see below).\footnote{~See for instance~\cite{Seiberg:2016rsg, Cordova:2018acb, Dumitrescu:2023hbe, Dumitrescu:2024jko}, where this idea was used to analyze a variety of gauge theory phases.} Note that the theory remains bosonic, because~$\phi$ has charge two and thus respects the spin-charge relation enforced by the spin$_c$ connection~$a_\mu$.

    \item The quantum numbers of the monopole operators are the same as in~$N_f=2$ QED$_3$, because only the Dirac fermions carry zero modes in the monopole background. The monopole operators with the lowest magnetic charges are thus a doublet $\mathcal{M}^{i=1,2}$ with magnetic charge~$q_\CM = 1$, and a triplet~$\tilde{\mathcal{M}}^{a=1,2,3}$ with magnetic charge~$q_\CM = 2$, both of which are Lorentz scalars~\cite{Borokhov:2002ib}. By quantizing the fermion zero modes in the~$q_\CM = 2$ sector and applying the Gauss law, one also find an~$SU(2)_f$-singlet, Lorentz vector operator~$j_\mu \sim \ep_{ij} \CM^i \d_\mu \CM^j$. Similarly one can think of~$\t \CM^a \sim \CM^i {(\sigma^a)_i}^j \CM^j$.
\end{itemize}

\subsection{Symmetries of Higgs-Yukawa-QED}\label{sec:hyqedsym}

The presence of the Yukawa couplings~\eqref{yukawa2} in the Lagrangian~\eqref{LagrangianMinkowski} breaks some of the global symmetries of QED$_3$. See~\cite{Wang:2017txt, Cordova:2017kue, Dumitrescu:2024jko} and section~\ref{section:qed3review} above for further details. In this section we will describe the remaining symmetries of HYQED in detail. Before we do so, we give a brief summary of the main results for future reference. 

\subsubsection{Summary}

\noindent The symmetries of the HYQED Lagrangian~\eqref{LagrangianMinkowski} are given by
\begin{equation}\label{fullGhypreview}
    G_\text{HYQED} =  {\left( \text{Pin}^-(2)_f  \rtimes \Z_4^\CT\right)\times \text{Pin}^-(2)_\CM \over \Z_2 \times \Z_2}~.
\end{equation}
The~$(U(1)_{f} \times U(1)_{\CM})/\Z_2$ charges~$q_{f, \CM} \in \Z$ are congruent~$q_f \equiv q_\CM \mod 2$, so that $(-1)^{q_f} = (-1)^{q_\CM}$. We will also use the basis~$U(1)_e \times U(1)_m$, whose charges~$q_e = (q_f + q_\CM)/2$ and $q_m = (q_\CM - q_f)/2$ are unconstrained integers, $q_{e, m} \in \Z$. There are three more independent discrete symmetry generators: unitary duality~$\CD$ and charge conjugation~$\CC$, as well as anti-unitary time reversal~$\CT$, which satisfy the following relations,
\begin{equation}\label{CDTrels}
    \CC^2 = \CD^2 = 1~, \quad \CT^2 = (-1)^{q_f} = (-1)^{q_\CM}~, \quad \CC \CD = \CD \CC~, \quad \CC \CT = \CT \CC~, \quad \CD \CT = (-1)^{q_\CM} \CT\CD~. 
\end{equation}
The action of these discrete symmetries on the~$U(1)$ symmetries are fully captured by their actions on the charges, summarized in table \ref{tab:hyGonQ}. In table~\ref{tab:hyGonOps}  we list the action of the symmetries on some important gauge-invariant operators.
\begin{table}[t]
    \centering
    \begin{tabular}{ c | c c c c}
        Symmetry & $q_f$ & $q_\CM$ & $q_e$ & $q_m$ \\ \noalign{\vskip 2pt} \hline  \noalign{\vskip 2pt}
         $ \CD $ & $-q_f$ & $q_\CM$ & $q_m$ & $q_e$ \\
         $\CC$ & $-q_f$ & $-q_\CM$ & $-q_e$ & $-q_m$ \\
         $\CT$ & $q_f$ & $-q_\CM$ & $-q_m$ & $-q_e$
    \end{tabular}
    \caption{~Action of unitary duality and charge-conjugation symmetries~$\CD$ and~$\CC$, as well as anti-unitary time-reversal symmetry~$\CT$, on the~$U(1)$ charges~$q_f, q_\CM$, as well as their linear combinations~$q_e = \half (q_\CM + q_f)$, $q_m = \half (q_\CM - q_f)$.}
    \label{tab:hyGonQ}
\end{table}
\begin{table}[h]
    \centering
    \begin{tabular}{ c | c c c c c c }
        Symmetry & $\CM^1$ & $\CM^2$ & $\CO$ & $\CO_1$ & $\CO_2$ & $\CO_3$ \\ \noalign{\vskip 2pt} \hline  \noalign{\vskip 2pt}
        $q_f$ & $1$ & $-1$ & $0$ & d & d & $0$ \\
        $q_\CM$ & $1$ & $1$ & $0$ & $0$ & $0$ & $0$ \\
        $q_e$ & $1$ & $0$ & $0$ & d & d & $0$ \\
        $q_m$ & $0$ & $1$  & $0$ & d & d & $0$\\
         $ \CD $ & $\CM^2$ & $\CM^1$  & $\CO$ & $\CO_1$ & $-\CO_2$ & $-\CO_3$ \\
         $\CC$ & $(\CM^1)^*$ & $(\CM^2)^*$  & $\CO$ & $\CO_1$ & $-\CO_2$ & $\CO_3$ \\
         $\CT$ & $(\CM^2)^*$ & $-(\CM^1)^*$  & $-\CO$ & $-\CO_1$ & $\CO_2$ & $-\CO_3$
    \end{tabular}
    \caption{~The~$U(1)$ charges, as well as the action of the discrete symmetries~$\CD$, $\CC$, and~$\CT$, on some important gauge-invariant operators: the minimal monopole operators~$\CM^{i = 1,2}$, and the fermion bilinears~$\CO = i \b \psi_i \psi^i, \CO_a = i \b \psi_i {(\sigma_a)^i}_j \psi^j (a = 1,2,3)$ in~\eqref{OOdef}. The entries marked ``d'' indicate that~$\CO_1$ and~$\CO_2$ form a doublet under the~$U(1)_f$ symmetry, i.e.~$\CO_1 \pm i \CO_2$ have~$q_f = \mp 2$. Analogous statements hold for the~$q_{e, m}$ charges of these operators.}
    \label{tab:hyGonOps}
\end{table}

\subsubsection{Continuous Symmetries} 

\begin{equation}\label{globalsymHY}
\frac{U(1)_f \times U(1)_\mathcal{M}}{\mathbb{Z}_2}  \subset U(2)_\text{QED} = {SU(2)_f \times U(1)_\CM \over \Z_2}~.
\end{equation}
\begin{itemize}
    \item[($\CM$)] The $U(1)_\mathcal{M}$ magnetic symmetry of QED$_3$ is also a symmetry of HYQED. We normalize its conserved current as
    \begin{equation}\label{magnetic_current}
    j_\mathcal{M}^\mu = \frac{1}{4\pi} \varepsilon^{\mu\nu\rho} f_{\nu\rho} \,,
    \end{equation}
    so that $q_\mathcal{M} \in \mathbb{Z}$. Operators with~$q_\CM \neq 0$ are called monopole operators. An example is the unit-charge monopole~$\CM^i$ with~$q_\CM = 1$. 
     
    \item[($f$)] The Yukawa couplings preserve the~$U(1)_f \subset SU(2)_f$ Cartan subgroup of QED$_3$. We normalize~$U(1)_f$ so that~$\psi^1$ and $\psi^2$ have charges $+1$ and $-1$, respectively,
    \begin{equation}\label{U(1)factiononfermions}
    U(1)_f :\psi^i \rightarrow {\exp\left( i \alpha\sigma_z\right)^i}_j \psi^j \,, \qquad \alpha \sim \alpha + 2\pi~,
    \end{equation}
    while the Higgs field~$\phi$ is~$U(1)_f$ neutral. The~$U(1)_f$ current, given by 
    \begin{equation}\label{flavor_current}
    j_f^\mu = \bar\psi_i\gamma^\mu{(\sigma_z)^i}_j\psi^j~,  
    \end{equation}
is normalized so that the associated charges~$q_f \in \mathbb{Z}$ are integers.

    As in QED, a~$U(1)_f$ rotation by~$\alpha = \pi$ on the fermions~\eqref{U(1)factiononfermions} can be compensated by a gauge transformation, but it acts as~$(-1)^{q_\CM}$ on gauge-invariant monopole operators. This is responsible for the~$\Z_2$ quotient in~\eqref{globalsymHY}, which enforces the condition~$q_f \equiv q_\CM \mod 2$.\footnote{~The corresponding condition in QED was discussed below~\eqref{u2symm}.} For instance, the components of the monopole doublet~$\CM^i$ have charges
    \begin{equation}
        q_f(\CM^1) = 1~, \qquad q_f(\CM^2) = -1~, \qquad  q_\CM(\CM^{i}) =1~,
    \end{equation}
    and the components of the monopole triplet~$\tilde{\CM}^a$ have charges 
    \begin{equation}\label{charge2monopoles}
        q_f(\tilde{\CM}^1 \pm i\tilde{\CM}^2 ) = \mp 2~, \qquad q_f(\tilde{\CM}^3) = 0~, \qquad  q_\CM(\tilde{\CM}^a) =2~,
    \end{equation}
    while decomposing the~$SU(2)_f$ triplet fermion bilinear~$\CO_a = i \b \psi_i {(\sigma_a)^i}_j \psi^j$ leads to
    \begin{equation}\label{chargesofbilinears}
        q_f(\CO_1 \pm i \CO_2) = \mp 2~, \qquad q_f(\CO_3) = q_\CM(\CO_a) = 0~.
    \end{equation}

\end{itemize}

\subsubsection{Unitary Discrete Symmetries} 
\begin{itemize}
\item  A unitary $\mathbb{Z}_2^\mathcal{D}$ symmetry, generated by~$\CD$. Even though~$\Z_2^\CD$ is a subgroup of the~$U(2)_\text{QED}$ symmetry in~\eqref{globalsymHY}, we refer to its generator~$\CD$ as the duality symmetry of HYQED, because it will turn out to exchange mutually non-local electric and magnetic excitations in the topological~$\Z_2$ gauge theory phase of the model (see section~\ref{section:Z2Higgsphase}). This should not be confused with the genuine self-duality symmetry that emerges in the IR of HYQED, as discussed in section~\ref{sec:emergemirr}. In order to avoid confusion with~$\Z_2^{\CD}$, we refer to this emergent self-duality of HYQED as mirror symmetry. 

We define~$\CD$ to exchange the two fermion flavors~$\psi^1$ and $\psi^2$, 
\begin{equation}
\mathcal{D}: \psi^i\rightarrow {(\sigma_x)^i}_j\psi^j \,,    
\end{equation}
while leaving~$\phi$ and~$a_\mu$ invariant. It thus acts on the~$U(1)_{f}$ and~$U(1)_\CM$ charges via
\begin{equation}
    \CD : q_f \to - q_f~, \qquad q_\CM \to q_\CM~.
\end{equation}
Consistent with this, its action on the minimal monopole operator~$\CM^i$ and the fermion bilinears in~\eqref{OOdef} is given by 
\begin{equation}\label{action_of_D}
\mathcal{D} : \mathcal{M}^i \rightarrow {(\sigma_x)^i}_j\mathcal{M}^j \,, \quad 
\mathcal{O} \rightarrow \mathcal{O} \,, \quad \mathcal{O}^1 \pm i \CO^2 \rightarrow \mathcal{O}^1 \mp i \CO^2~,  \quad \mathcal{O}^3 \rightarrow -\mathcal{O}^3 \,.
\end{equation}

To embed~$\CD$ into the global~$U(2)_\text{QED}$ symmetry in~\eqref{globalsymHY} that acts on gauge-invariant local operators, we factorize~$\sigma_x$ into the product of an~$SU(2)_f$ Weyl reflection~$\CR_f = i \sigma_x$ and a~$U(1)_\CM$ transformation~$(-i)^{q_\CM}$,\footnote{~This factorization is non-unique, since we can multiply both factors by~$-1$. When acting on the fundamental fields~$\psi^i, \phi$ in the Lagrangian, $\sigma_x$ and~$i \sigma_x$ differ by a constant gauge transformation.} so that
\begin{equation}\label{Rfdef}
    \CD = \CR_f (-i)^{q_\CM}~, \qquad \CR_f \in SU(2)_f~, \qquad \CR_f^2 = (-1)^{q_f}~, \qquad \CD^2 = 1~.
\end{equation}
Thus, adjoining the generator~$\CD$ to the~$(U(1)_f \times U(1)_\CM)/\Z_2$ symmetry in~\eqref{globalsymHY} is equivalent to adjoining~$\CR_f$, which extends~$U(1)_f$ to the group~$\text{Pin}^-(2)_f$.

\item A unitary~$\Z_2^\CC$ charge-conjugation symmetry with generator~$\mathcal{C}$, which acts on the fields in the Lagrangian as follows,\footnote{~Recalling~\eqref{C_on_psi}, we can write~$\CC : \psi \to  \psi^C$.} 
\begin{equation}
    \mathcal{C} : \psi^i \rightarrow (\psi^i)^*~, \qquad a_\mu \rightarrow -a_\mu~, \qquad \phi \rightarrow \phi^*~,
\end{equation}
and thus it negates both global~$U(1)$ charges, 
\begin{equation}
    \CC : q_f \to -q_f~, \qquad q_\CM \to -q_\CM~.
\end{equation}
Its action on some gauge-invariant local operators of interest is
\begin{equation}
    \CC : \mathcal{M}^i \rightarrow (\mathcal{M}^i)^*~, \quad \mathcal{O} \rightarrow \mathcal{O} \,, \quad \mathcal{O}^1 \rightarrow \mathcal{O}^1 \,, \quad \mathcal{O}^2 \rightarrow -\mathcal{O}^2 \,, \quad \mathcal{O}^3 \rightarrow \mathcal{O}^3~. 
\end{equation}
Note that~$\CC$ commutes with~$\CD$, i.e.~they generate a~$\Z_2^\CC \times \Z_2^\CD$ product group,
\begin{equation}
    \CC \CD = \CD \CC~, \qquad \CD^2 = \CC^2 = 1~.
\end{equation}

In analogy with~\eqref{Rfdef}, we can define a unitary operator
\begin{equation}\label{RMdef}
    \CR_\CM \equiv \CC \CD i^{q_f} = \CC \CR_f i^{q_f - q_\CM}~, \qquad \CR_\CM^2 =  (-1)^{q_\CM}~,  
\end{equation}
which acts on the~$U(1)$ charges via
\begin{equation}
    \CR_\CM  : q_f \to q_f~, \qquad q_\CM \to - q_\CM~,
\end{equation}
so that~$\CR_\CM $ commutes with all of~$U(1)_f$, and in fact with all of~$\text{Pin}^-(2)_f$, thanks to
\begin{equation}\label{rfrmisc}
    \CR_f \CR_\CM  = \CR_\CM  \CR_f = \CC i^{q_f - q_\CM}~.
\end{equation}
It follows that~$\CR_\CM$ extends~$U(1)_\CM$ to~$\text{Pin}^-(2)_\CM$, so that the full unitary global symmetry of the theory is
\begin{equation}\label{globalsymHY2}
 G_\text{HYQED}^\text{unit.} =  {\text{Pin}^-(2)_f \times \text{Pin}^-(2)_\CM \over \Z_2}~.
\end{equation}
Thus the flavor and the monopole symmetries are exactly on the same footing. It follows from~\eqref{rfrmisc} that (up to~$U(1)$ rotations)~$\CC \sim \CR_f \CR_\CM$ is a simultaneous reflection in both factors.  

\item In section~\ref{sec:emergemirr} we will argue for the IR emergence of an additional unitary~$\Z_2^{{\sf C}_{m}}$ mirror symmetry -- not manifest at the level of the UV Lagrangian --  that exchanges~$\text{Pin}^-(2)_f$ and~$\text{Pin}^-(2)_\CM$. Since it exchanges monopole and non-monopole operators, it is an emergent self-duality symmetry of HYQED.  
\end{itemize}

\subsubsection{The~$U(1)_e \times U(1)_m$ Basis}

\noindent It is very convenient to define the diagonal and anti-diagonal $U(1)_e$ and~$U(1)_m$ symmetries, whose charges~$q_{e, m}$ are related to~$q_{f, \CM}$ via 
\begin{equation}\label{U(1)em_definition}
  q_e \equiv \half \left(q_\CM + q_f\right)~, \qquad q_m \equiv \half \left(q_\CM - q_f\right)~. 
\end{equation}
Note that~$q_e,q_m \in \mathbb{Z}$ are unconstrained, so that
\begin{equation}
    U(1)_e \times U(1)_m
\end{equation}
is a genuine product group. For instance, the gauge-invariant monopole operators with unit magnetic charge have
\begin{equation}
    (q_e, q_m)(\CM^1) = (1,0)~, \qquad (q_e, q_m)(\CM^2) = (0,1)~.
\end{equation}
By contrast, the fundamental fermions~$\psi^i$ are not gauge-invariant and carry fractional charges,
\begin{equation}
    (q_e, q_m)(\psi^1) = \left(\half,- \half\right)~, \qquad (q_e, q_m)(\psi^2)= \left(-\half, \half\right)~.
\end{equation}

Notice that charge conjugation $\mathcal{C}$ negates both $(q_e,q_m)$ charges, so it acts as a simultaneous reflection on both $U(1)_{e,m}$ factors, while duality $\mathcal{D}$ exchanges these charges, so it acts by swapping the two factors. Since $\mathcal{C}^2=\mathcal{D}^2=1$ and $\mathcal{CD}=\mathcal{DC}$, the unitary global symmetry \eqref{globalsymHY} of the theory in this basis can be written as
\begin{equation}
    G^{\rm unit.}_{\rm HYQED} = S(O(2)_e \times O(2)_m) \rtimes \mathbb{Z}_2^\mathcal{D} \,,
\end{equation}
where $S(O(2)_e \times O(2)_m)=\{(g_e,g_m)\in O(2)_e \times O(2)_m : (\det g_e )(\det g_m)=1\}$ is the special subgroup that includes rotations in either $O(2)_{e,m}$ factor and simultaneous reflections.

\subsubsection{Time-Reversal Symmetry}

\noindent There is an anti-unitary time-reversal symmetry~$\CT$, whose action on the fields takes the following form,\footnote{~Here~$\CT$ maps the spacetime point~$(t, \vec x)$ of every field to~$(-t, \vec x)$. Note that on fundamental fields $\mathcal{T}^2$ acts as fermion parity $(-1)^{F}$, which is a gauge transformation.}
\begin{equation}\label{Tdef}
\begin{split}
    \mathcal{T} : ~ &\psi^i \rightarrow \gamma^0 \psi^i \,, \qquad a_0 \rightarrow a_0 \,,  \qquad a_{1,2} \rightarrow -a_{1,2} \,, \qquad \phi \rightarrow -\phi~,
\end{split}
\end{equation}
so that
\begin{equation}\label{Tonqs}
    \CT : q_f \to q_f~, \qquad q_\CM \to -q_\CM~, \qquad q_e \to - q_m~, \qquad q_m \to - q_e~.
\end{equation}
Its action on some interesting gauge-invariant operators is\footnote{~The action of time reversal on monopole operators with unit magnetic charge can be read off from equation~(2.7) of~\cite{Cordova:2017kue}. From equations~(1.1) and~(1.3) there we see that our $\mathcal{T}$ is their $\mathcal{CT}$ and vice versa. We are free to choose the overall sign by composing with $(-1)^{q_\mathcal{M}}$.} 
\begin{equation}
\begin{split}
    \mathcal{T} &: \mathcal{M}^i \rightarrow {(i\sigma_y)^i}_j(\mathcal{M}^j)^*~, \quad \mathcal{O} \rightarrow -\mathcal{O}~, \quad \mathcal{O}^1 \rightarrow -\mathcal{O}^1 \,, \quad \mathcal{O}^2 \rightarrow \mathcal{O}^2 \,, \quad \mathcal{O}^3 \rightarrow -\mathcal{O}^3~.
\end{split}
\end{equation}
Note that~$\CT$ commutes with~$\CC$, and that it generates a~$\Z_4^\CT$ symmetry involving the central~$\Z_2$ subgroup generated by~$(-1)^{q_\CM} = (-1)^{q_f}$, 
\begin{equation}
    \mathcal{C}^2=1~,\qquad \CC \CT = \CT \CC~,\qquad \mathcal{T}^2=(\mathcal{CT})^2=(-1)^{q_\mathcal{M}}~.
\end{equation}
By contrast, $\CT$ does not commute with~$\CD$,
\begin{equation}
    \CD \CT = (-1)^{q_\CM} \CT \CD~, 
\end{equation}
but it does commute with~$\CR_f$ and~$\CR_\CM$,
\begin{equation}
    \CT \CR_{f, \CM} = \CR_{f, \CM} \CT~.
\end{equation}
Since~$\CT$ is anti-unitary, we know from~\eqref{Tonqs} that it complex conjugates~$U(1)_f$ rotations and leaves~$U(1)_\CM$ rotations invariant. Consequently, the addition of time-reversal symmetry~$\CT$ extends the unitary global symmetry~\eqref{globalsymHY2} of HYQED as follows,
\begin{equation}\label{fullGofHYQED}
    G_\text{HYQED} =  {\left( \text{Pin}^-(2)_f  \rtimes \Z_4^\CT\right)\times \text{Pin}^-(2)_\CM \over \Z_2 \times \Z_2}~.
\end{equation}
Here~$\Z_4^\CT$ is generated by~$\CT$, and the two~$\Z_2$ quotients enforce the relations
\begin{equation}
    \CT^2 = (-1)^{q_f} = (-1)^{q_\CM}~.
\end{equation}

\subsection{Background Fields and Anomalies}\label{sec:hyqed_anomalies}

We now couple the HYQED Lagrangian~\eqref{LagrangianMinkowski} to background gauge fields~$A_\CM$ and~$A_f$ for the $U(1)_\CM$ and $U(1)_f$ global symmetries, 
\begin{equation}\label{lagrangianHY_coupledtobackgrounds}
\begin{split}
    S_{\rm HYQED}[A_f,A_\CM] & =  \int d^3 x \, \left(\mathscr{L}_{\rm HYQED} - A_{f \mu} j^\mu_f - A_{\CM \mu} j_\CM^\mu\right) \\
    & =  \int d^3x \left( \mathscr{L}_{\rm HYQED} - \bar\psi_i \gamma^\mu {(\sigma_z)^i}_j\psi^j A_{f\,\mu}\right) - \frac{1}{2\pi} \int da \wedge A_\CM~, 
\end{split}
\end{equation}
Here we are using the conventions described below~\eqref{Diracjq}, together with the~$U(1)_\CM$ current~$j_\CM^\mu = {1 \over 4\pi} \ep^{\mu\nu\rho} f_{\nu\rho}$ in~\eqref{magnetic_current} and the~$U(1)_f$ current~$j_f^\mu = \b \psi_i \gamma^\mu {(\sigma_z)^i}_j \psi^i$ in~\eqref{flavor_current}. Note that~$\psi^1$ and~$\psi^2$ couple to~$a+A_f$ and~$a-A_f$ with unit charge,\footnote{~As we will review below, $a \pm A_f$ are both standard spin$_c$ connections.} so that the fermion kinetic terms in~\eqref{lagrangianHY_coupledtobackgrounds} can be written as
\begin{equation}\label{psi12charges}
    - i \b \psi_1 \Dslash_{a + A_f}\psi^1   - i \b \psi_2 \Dslash_{a - A_f}\psi^2~. 
\end{equation}

As reviewed in~\cite{Dumitrescu:2024jko}, both the dynamical gauge field~$a_\mu$ and the background fields~$A_{f, \CM}$ have modified flux quantization conditions on closed 2-cycles,\footnote{~Here~$w_2(T\mathcal{M}_3)$ is the second Stiefel-Whitney class of the tangent bundle of the spacetime manifold. All oriented three-manifolds are spin, so that~$\oint w_2(T\mathcal{M}_3)$ vanishes (mod 2). This is no longer the case when we extend~$a$ to four dimensions, e.g.~to analyze the anomalies of the theory, see~\cite{Dumitrescu:2024jko} and references therein for a detailed discussion.}
    \begin{equation}\label{modified_spinc_relation}
    \oint \frac{da}{2\pi} = \frac{1}{2} \oint w_2(T\mathcal{M}_3) + \oint \frac{dA_f}{2\pi} \text{ mod } \mathbb{Z} \,,    
    \end{equation}
and 
\begin{equation}\label{fluxquantizationAmAf}
\oint \frac{dA_\mathcal{M}}{2\pi} = \oint \frac{dA_f}{2\pi}  \text{ mod } \mathbb{Z}~.
\end{equation}
Let us make some comments:
\begin{itemize}
    \item In the absence of~$A_{f, \CM}$, the dynamical~$a$ is a standard spin$_c$ connection. 
    \item Since the global symmetry is~$(U(1)_f \times U(1)_\CM)/\Z_2$, the fluxes of~$A_f$ and~$A_{\CM}$ can be correlated half-integers. 
\item The appearance of~$A_f$ in the $a$-flux quantization rule captures the fact that a $\pi$-rotation in $U(1)_f$ acts on fundamental fermions as a gauge transformation.
\end{itemize}

The Higgs-Yukawa-QED theory inherits the mixed 't Hooft anomaly \eqref{QEDanomaly} of $N_f=2$ QED$_3$ as mixed anomaly between its $(U(1)_f \times U(1)_\mathcal{M})/\mathbb{Z}_2$ global symmetry and any other discrete symmetry -- for instance~$\CT$ or~$\CC\CT$ -- that pins the coefficient~$\theta = \pi$ in the following four-dimensional anomaly-inflow action,
\begin{equation}\label{mixedanomalyHYM}
S_{\text{anomaly}} = \pi \int_{\mathcal{M}_4} \left( \frac{dA_\mathcal{M}}{2\pi} \wedge \frac{dA_\mathcal{M}}{2\pi} - \frac{dA_f}{2\pi} \wedge \frac{dA_f}{2\pi}\right) = \pi \int_{\mathcal{M}_4} \frac{dA_e}{2\pi} \wedge \frac{dA_m}{2\pi}~. 
\end{equation}
Here~$A_e$ and $A_m$ are the $U(1)_e \times U(1)_m$ background gauge fields, whose relation with $A_\mathcal{M}$ and $A_f$ can be read off from \eqref{U(1)em_definition},
\begin{equation}\label{Aem_definition}
A_e = A_\mathcal{M} + A_f ~, \qquad A_m = A_\mathcal{M} - A_f~.
\end{equation}
It follows from~\eqref{fluxquantizationAmAf} that~$A_e$ and $A_m$ are standard~$U(1)$ connections, with independent, unconstrained fluxes, 
\begin{equation}
\oint \frac{dA_e}{2\pi} \in \Z~, \qquad \oint \frac{dA_m}{2\pi} \, \in \mathbb{Z}~.    
\end{equation}
Note that the~$\mathbb{Z}_2^\mathcal{D}$ duality symmetry acts on the background fields as follows,
\begin{equation}
\mathcal{D} : A_f \rightarrow -A_f~, \qquad A_\CM \rightarrow A_\CM~, \qquad A_e \leftrightarrow A_m~.
\end{equation}

\section{Phase Diagram of Higgs-Yukawa-QED}\label{subseccontinuumphasediagram}

\subsection{Choice of Tuning Parameters~$m_\phi^2$ and~$m_3$}\label{sec:tuningparams}

HYQED admits five gauge-invariant mass terms: the four fermion masses~\eqref{qedfm} already present in QED, and a mass for the Higgs field~$\phi$,
\begin{equation}\label{masslagrangian}
\mathscr{L}_{\text{masses}} = m \mathcal{O} + m_a \mathcal{O}^a - m_\phi^2|\phi|^2~, \qquad m, m_a, m_\phi^2 \in \R~,
\end{equation}
and it is interesting to consider the phase diagram of the theory as a function of these masses. 

We will focus on the two-dimensional parameter space that has a chance of matching the coupling constants~$h_e, h_m$ of the~$U(1)$-symmetric staggered Fradkin-Shenker model~\eqref{eqnU1FSmodel}. This model preserves all the symmetries in table~\ref{table:latticesymmetries}, except for the duality symmetry~$\sf D$, which exchanges~$h_e \leftrightarrow h_m$ and is therefore only a symmetry on the self-dual line~$h_e = h_m$; the orthogonal direction, with coupling~$\sim (h_m-h_e)$, is~$\sf D$-odd. See the phase diagram in figure~\ref{figU1latticephasediagram}.  

\begin{figure}[t]
    \centering
\begin{tikzpicture}[x=4cm,y=4cm] 
    \draw[gray!60, line width=0.3pt] (0,0) rectangle (1.35,1.35);
  \node at (0.675,-0.12) {$m_3$};
  \draw[->] (0.475,-0.05) -- (0.875,-0.05);
  
  \node at (-0.15,0.675) {$m^2_\phi$};
  \draw[->] (-0.05,0.475) -- (-0.05,0.875);

  \node at (1.1,0.80) {$U(1)_m$};
  \node at (1.1,0.68) {SSB};

  \node at (1.2,0.35) {\textcolor{blue}{$O(2)^*_m$}};

  \node at (0.3,0.80) {$U(1)_e$};
  \node at (0.3,0.68) {SSB};

  \node at (0.14,0.35) {\textcolor{blue}{$O(2)^*_e$}};

  \node at (0.675,0.25) {$\mathbb{Z}_2$ TQFT};

\node at (0.68,1.2) {First order};
\node at (0.687,1.0) {$S^1_e \sqcup S^1_m$};

\draw[<->] (0.55,1.5) -- (0.8,1.5);
\node at (0.675,1.6) {$\mathsf{D}$};

  \draw[line width=0.8pt,blue]
    (0.0,0.0) -- (0.675,0.675) -- (1.35,0);
  \draw[densely dotted, line width=0.8pt]
    (0.675,0.675) -- (0.675,1.35);
  \fill[color=purple] (0.675,0.675) circle (0.03);
\end{tikzpicture}
    \caption{Conjectured phase diagram of Higgs-Yukawa-QED as a function of the $\sf D$-even Higgs mass~$m_\phi^2$ and the $\sf D$-odd fermion mass~$m_3$ in~\eqref{intro:masses}. The multicritical point at the origin, described by the SFS CFT, is shown in red. The phase to its right ($m_3 > 0$) spontaneously breaks~$U(1)_m$, leading to an~$S^1_m$ circle of vacua. Its~$\sf D$-reflection at~$m_3 < 0$ spontaneously breaks~$U(1)_e$, with an~$S^1_e$ circle of vacua. These circles coexist on the~$m_\phi^2 > 0$ axis (indicated by the dotted black line), where the vacua are~$S_e^1 \sqcup S_m^1$ and the duality symmetry~${\sf D} : S_e^1 \leftrightarrow S_m^1$ is spontaneously broken. The phase at~$m_\phi^2 < 0$ is gapped, with a~$\Z_2$ TQFT in the IR, whose~$e$ and~$m$ anyons are exchanged by~$\sf D$. The transitions from the TQFT phase to the symmetry-breaking phases (indicated by blue lines) occur when either the~$e$ or the~$m$ anyon of the~$\Z_2$ TQFT becomes massless. The resulting second-order transitions are described by~$\Z_2$-gauged versions of the~$O(2)$ Wilson-Fisher CFT, which we denote by~$O(2)_{e,m}^*$, ending at the multicritical point.
    }
    \label{fig:HYQEDphasediagbis}
\end{figure}
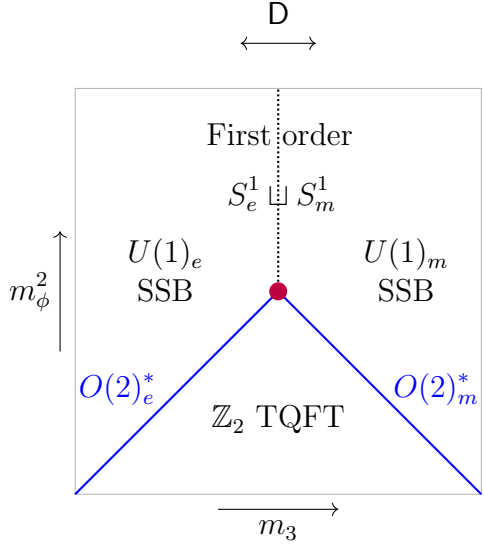

The mapping of the lattice symmetries to those of HYQED is explained in section~\ref{u1LatToCont}, and the results are summarized in table~\ref{tab:HYQEDlatticemapping}. Here we preview only the highlights that we will need in this section:
\begin{itemize}
    \item Both models have continuous~${(U(1)_f \times U(1)_\CM)} / \Z_2 = U(1)_e \times U(1)_m$ symmetries.  
    \item Lattice time-reversal symmetry~$\sf T$ acts on HYQED as 
    \begin{equation}\label{TisDCT}
        \sf T = \CD \CC \CT~.
    \end{equation}
    \item Full (i.e.~non-dualizing) lattice translations act on HYQED as 
    \begin{equation}\label{CisCm1}
        {\sf C} = \CC (-1)^{q_m}~.
    \end{equation}
    Given that~$U(1)_m$ is preserved, it follows that charge conjugation~$\CC$ is a symmetry. 
    \item Lattice duality~$\sf D$ acts on HYQED via
    \begin{equation}\label{DisDiqf}
        {\sf D} = \CD i^{q_f}~.
    \end{equation}
    Since~$U(1)_f$ is preserved everywhere, it follows that~$\CD$ is a symmetry if and only if~$\sf D$ is, i.e.~on the line~$h_e = h_m$. 
\end{itemize}
\noindent This leads to a natural identification of the lattice couplings~$h_e, h_m$ with HYQED mass terms:
\begin{itemize}
    \item The Higgs mass~$m_\phi^2 |\phi|^2$ preserves all global symmetries, and is thus identified with the duality-even direction, schematically
    \begin{equation}
        m_\phi^2 \sim h_e + h_m~. 
    \end{equation}
    \item The action of the global symmetries on the fermion mass operators~$\CO = i \b \psi \psi$ and $\CO_a = i \b \psi \sigma_a \psi$ is summarized in table~\ref{tab:hyGonOps}. We see that only~$\CO$ and~$\CO_3$ are~$U(1)_f$ singlets, and~$\CO$ is odd under~${\sf T} = \CD\CC\CT$ and hence ruled out. This leaves only~$\CO_3$, which preserves~$\CD\CC\CT$ and~$\CC$; it is also~$\CD$-odd so that we can schematically identify
    \begin{equation}
        m_3 \sim h_m - h_e~. 
    \end{equation}
\end{itemize}
In this section, we will therefore explore the phase diagram of HYQED~\eqref{LagrangianMinkowski} deformed by the following mass terms,\footnote{~The same phase diagram was previously considered in~\cite{Jian:2017chw}, under the assumption that QED$_3$ flows to an~$O(4)$-symmetric CFT. By contrast, we will assume the symmetry-breaking scenario for QED$_3$ reviewed in section~\ref{section:qed3review}. }
\begin{equation}\label{tuningparam}
    \mathscr{L}_{\rm masses} = m_3 \mathcal{O}^3- m_\phi^2|\phi|^2 = im_3\left(\bar\psi_1 \psi^1-\bar\psi_2\psi^2 \right)- m_\phi^2|\phi|^2~, \qquad m_3~, m_\phi^2 \in \R~.
\end{equation}
The salient features of this phase diagram are summarized in figure~\ref{fig:HYQEDphasediagbis}; they will established below.  Note that we are using a convenient regularization scheme in which the multicritical point occurs at~$m_\phi^2 = 0$, which is not pinned by any symmetries. By contrast, it necessarily lies on the~$m_3 = 0$ axis because it preserves the duality symmetry~$\sf D$. 

\subsection{Spontaneous~$U(1)_e$ and~$U(1)_m$ Breaking at Large~$|m_3|$}
\label{section: large m3}

In this section we will explain the~$U(1)_e$ and~$U(1)_m$ spontaneous symmetry breaking (SSB) phases in figure~\ref{fig:HYQEDphasediagbis}. As we will see, these are weakly coupled and can be analyzed reliably as long as the fermion mass~$|m_3 |$ is much larger than all other scales~$m^2_\phi,e^2,y^2,\lambda_4$ in the  problem. (Recall that we are assuming that $e^2 \ll y^2,\lambda_4$.) In our conjectured phase diagram these phases smoothly extend into the strong-coupling region at smaller~$|m_3|$, all the way to the~$O(2)^*$ lines when~$m_\phi^2 < 0$ (these are analyzed in section~\ref{section: particlevortex and O(2)*} below), to the positive~$m_\phi^2 > 0$ axis (discussed in section~\ref{firstorder} below), and to the multicritical point at~$m_\phi^2 = 0$.

\subsubsection{Effective Higgs Potential at Large $|m_3|$}

We claim that at sufficiently large fermion mass~$|m_3|$, the effective potential~$V_\text{eff}(\phi)$ for the Higgs field has a unique vacuum at~$\phi = 0$ in which the~$\phi$ particle is massive, i.e.~there is no Higgsing. This is clearly true when~$m_\phi^2 > 0$ is sufficiently large and positive. To see it more generally, we compute the leading one-loop correction to~$V_\text{eff}(\phi)$. 

Since we are assuming that~$y^2 \gg e^2$, this leading one-loop correction arises from integrating out the fermions in the presence of the Yukawa couplings~\eqref{yukawaD}, while ignoring the gauge interactions, i.e.~we start with the following Lagrangian for the fermions, 
\begin{equation}\label{lagrangian_fermions_intout}
\mathscr{L}_{\rm fermions} = -i \bar\psi_i \dslash \psi^i+im_3 \bar\psi_i {(\sigma_z)^i}_j \psi^j + iy\left( \phi^* ({\psi^1})^t \gamma^0 \psi^2 + \text{h.c.} \right)~. 
\end{equation}
In the absence of the gauge field, it is convenient to redefine~$\psi^2 \rightarrow (\psi^2)^{C}=(\psi^2)^*$, to obtain the following Lagrangian (see~\eqref{C_on_psi} and~\eqref{Cbilin}),
\begin{equation}
\mathscr{L}_{\rm fermions} = -i \bar\psi_i \dslash \psi^i+im_3 \bar\psi_i {(\sigma_z)^i}_j \psi^j + iy\left( \phi^* \bar\psi_2 \psi^1 + \phi \,\bar\psi_1 \psi^2 \right) \equiv \bar\psi_i {K^i}_j \psi^j  \,, 
\end{equation}
where the operator~${K^i}_j$ is defined as follows,
\begin{equation}\label{Kmatrix}
K=\begin{pmatrix}
        -i\dslash + i m_3 && iy\phi \\
         iy\phi^* && -i\dslash - i m_3
        \end{pmatrix} = V^\dagger
        \begin{pmatrix}
        -i\dslash + i \widetilde{m}(\phi) && 0 \\
         0 && -i\dslash - i \widetilde{m}(\phi)
        \end{pmatrix}
        V \equiv V^\dagger \widetilde{K} V~.
\end{equation}
Here~$V$ is an $SU(2)$ matrix that diagonalizes $K$, and $\widetilde{m}(\phi)$ is the field-dependent mass 
\begin{equation}\label{induced_fermion_mass}
    \t{m}(\phi) =\sqrt{m_3^2+y^2|\phi|^2} \,.
\end{equation}
Integrating out the fermions generates a one-loop contribution to the effective potential for $\phi$,
\begin{equation}
    V_{\rm 1-loop} = -i \Tr \log \widetilde{K} = -i \int \frac{d^3p}{(2\pi)^3} \, \tr \left( \log (-{p \!\!\!\hspace{1pt}\slash\,}+i\widetilde{m}) + \log (-{p \!\!\!\hspace{1pt}\slash\,}-i\widetilde{m}) \right)~,
\end{equation}
where~$\Tr (\cdots)$ is over spacetime, flavor, and spinor indices, whereas~$\tr(\cdots)$ is only over spinor indices. Performing the integral in dimensional regularization,\footnote{~This amounts to setting to zero the UV divergences~$\sim \Lambda_\text{UV}^3 + \Lambda_\text{UV}\t m^2$, the latter requiring an~$\CO(\Lambda_\text{UV} y^2)$ renormalization of~$m_\phi^2$, in line with the fact that we use a scheme in which the multicritical point is at~$m_\phi^2 = 0$.}  we find a positive-definite effective potential,\footnote{~An easy way to check this is to recall that in quantum mechanics, a single complex fermion of mass~$\mu$ has effective potential~$V_{\rm 1-loop} = - \half |\mu|$ (see e.g.~appendix B of~\cite{Dumitrescu:2025vfp}). It follows from~\eqref{Kmatrix} and~\eqref{induced_fermion_mass} that we have four quantum-mechanical complex fermion modes for every spatial momentum~$\vec p$, all of which have masses of absolute value~$\omega_{\vec p} = \sqrt{\t m^2 + |\vec p|^2}$, so that~$V_{\rm 1-loop} = -2 \int {d^2 \vec p \over (2 \pi)^2} \, \omega_{\vec p}$. This integral is divergent, but differentiating twice with respect to~$\t m^2$ leads to a finite result in exact agreement with~\eqref{v1loop},
\begin{equation}
    {\d^2 V_{\rm 1-loop} \over \d (\t m^2)^2} = {1 \over 4 \pi |\t m|}~.
\end{equation}
}
\begin{equation}\label{v1loop}
V_{\rm 1-loop} = \frac{1}{3\pi}|\widetilde{m}|^3 = \frac{1}{3\pi} \left(m_3^2+y^2|\phi|^2\right)^{3\over 2}~. 
\end{equation}
Expanding at large~$|m_3|$ and retaining only terms that depend on~$\phi$ we find 
\begin{equation}
V_{\rm 1-loop} \simeq \frac{y^2|m_3|}{2\pi} |\phi|^2 + \frac{y^4}{8\pi |m_3|} |\phi|^4~.   
\end{equation}
Adding this to the tree-level contribution, we obtain the following effective Lagrangian for the gauge and Higgs fields, 
\begin{equation}\label{Leff_scalar}
\mathscr{L}_{\rm eff} = - \frac{1}{4e^2} f^{\mu\nu}f_{\mu\nu}
- |D_{2a}\phi|^2 - V_\text{eff}(\phi)~, 
\end{equation}
with effective potential
\begin{equation}\label{veff}
    V_\text{eff}  = \left(m_\phi^2 + \frac{y^2|m_3|}{2\pi}\right)|\phi|^2 +  \left(\lambda_4 + \frac{y^4}{8\pi |m_3|}\right) |\phi|^4~.
\end{equation}
Thus a sufficiently large fermion mass~$|m_3|$ induces a positive effective Higgs mass-squared, regardless the sign of the tree-level~$m_\phi^2$, which in turn leads to a unique vacuum with~$\langle \phi\rangle = 0$ and a massive~$\phi$ field. Cleanly this is a Coulomb vacuum of the Abelian Higgs model~\eqref{Leff_scalar}, with a massless photon in the deep IR. We will now analyze it in more detail. 

\subsubsection{The Coulomb Phase at Large~$|m_3|$}

In a Coulomb phase, we expect the~$U(1)_\CM$ symmetry to be spontaneous broken by monopole operators, so that the (dual) photon plays the role of the gapless Nambu-Goldstone boson (NGB). In order to understand the quantum numbers of the symmetry-breaking monopole operators, we follow the discussion in~\cite{Dumitrescu:2024jko} and couple the theory to background gauge fields for the global symmetries. Integrating out massive fermions then leads to additional terms, which generally require a somewhat careful treatment that we postpone to section~\ref{secetainvariants}. Here we will focus on a particular mixed Chern-Simons involving the~$U(1)_f$ background gauge field~$A_f$ and the dynamical spin$_c$ connection~$a$ that can be understood by elementary means.  

Recalling the~$U(1)_a$ and~$U(1)_f$ charges of~$\psi^i$ in~\eqref{psi12charges}, as well as the form of the~$m_3$ fermion mass in~\eqref{tuningparam}, it follows from~\eqref{genkmat} that the induced Chern-Simons term that arises by integrating out the fermions at large $|m_3|$ takes the form
\begin{equation}
    S_\text{CS}^f = {\sign (m_3) \over 2\pi} \int A_f \wedge da~.
\end{equation}
Combining this with the bare Chern-Simons term~$-{1 \over 2\pi} \int A_\CM \wedge da$ in the Lagrangian~\eqref{lagrangianHY_coupledtobackgrounds}, and using the definitions of the~$U(1)_{e, m}$ background gauge fields~$A_{e, m}$ in~\eqref{Aem_definition}, we find
\begin{equation}\label{Mixed_CS_term}
S_{\rm CS}[A_\CM,A_f] = -\frac{1}{2\pi} \int \left(A_\CM -\text{sign}(m_3)A_f\right) \wedge da =
\begin{dcases}
    -\frac{1}{2\pi} \int A_m \wedge da~,  \quad &\text{if } m_3>0~,\\
    -\frac{1}{2\pi} \int A_e \wedge da~,  \quad &\text{if } m_3<0~.
\end{dcases}
\end{equation}
These terms have several important implications:
\begin{itemize}
    \item[(i)] They identify the spontaneously broken magnetic zero-form symmetry of the free Maxwell theory in the deep IR with $U(1)_m$ when $m_3 > 0$, and with $U(1)_e$ when $m_3 < 0$. Since the~$U(1)_e \times U(1)_m$ charges~$(q_e, q_m)$ of the HYQED monopole operators~$\CM^1$ and~$\CM^2$ are~$(1,0)$ and~$(0,1)$, respectively, we conclude that 
 \begin{equation}\label{largemassvacua}
    \begin{split}
    & m_3 > 0: \quad \braket{\cM^2} \neq 0~, \quad U(1)_m \text{ spontaneously broken}~, \\
    & m_3 < 0: \quad \braket{\cM^1} \neq 0~, \quad U(1)_e \text{ spontaneously broken}~,
    \end{split}
\end{equation}
in agreement with the QED result~\eqref{m3monovev}. Each of these phases describes a circle of vacua, referred to as $S^1_m$ and $S^1_e$, on which the respective spontaneously broken~$U(1)_m$ and~$U(1)_e$ symmetries act. 

The action of the discrete symmetries~$\CD, \CC, \CT$ (and hence the lattice symmetries~$\sf D, C, T$ summarized below~\eqref{masslagrangian} and in table~\ref{tab:HYQEDlatticemapping}) on~$S_{e,m}^1$ follow from their action on the monopole operators~$\CM^i$ in~table~\ref{tab:hyGonOps}. In particular, duality~$\CD$ exchanges~$m_3 \leftrightarrow -m_3$, but also $U(1)_e \leftrightarrow U(1)_m$ and hence~$\CM^1 \leftrightarrow \CM^2$, $S_e^1 \leftrightarrow S_m^1$, consistent with the above. 

\item[(ii)] They are responsible for 't Hooft anomaly matching. Even though~$A_{e,m}$ are standard~$U(1)$ connections, the dynamical gauge field is a generalized spin$_c$ connection, with modified flux quantization~\eqref{modified_spinc_relation} in the presence of background fields. As was shown in~\cite{Dumitrescu:2024jko}, extending the spacetime manifold~$\CM_3$ and the background fields to a bounding four-manifold~$\CM_4$ and using~\eqref{modified_spinc_relation}, \eqref{fluxquantizationAmAf}, \eqref{Aem_definition}  reproduces the~'t~Hooft anomaly in~\eqref{mixedanomalyHYM},
\begin{equation}\label{phases_anomaly_matching}
\frac{1}{2\pi} \int_{\mathcal{M}_4} da \wedge dA_{e,m} = \pi \int_{\mathcal{M}_4} \frac{dA_e}{2\pi} \wedge \frac{dA_m}{2\pi}~.    
\end{equation}

\item[(iii)] They imply at least one phase transition between the large~$m_3 > 0$ and large~$m_3 < 0$ Coulomb phases, roughly because it follows from~\eqref{Mixed_CS_term} that there is a jump in the Chern-Simons terms between the two regimes,
\begin{equation}
\Delta S_{\rm CS} = \frac{2}{2\pi} \int A_f \wedge da~.
\label{CSjump}
\end{equation}
Such a conclusion would be justified for Chern-Simons terms constructed only from background fields, but since~$a_\mu$ is dynamical we must be more careful. In the Coulomb phases above, $a_\mu$ describes a massless photon, and the fact that the symmetry-breaking patterns~\eqref{largemassvacua} at large $|m_3|$ depend on the sign of~$m_3$  shows that they are separated by at least one phase transition.\footnote{~In section~\ref{secetainvariants} we show that analogous reasoning fails when~$a_\mu$ is Higgsed by the charge-2 Higgs field~$\phi$.} In our proposed phase diagram (see figure~\ref{fig:HYQEDphasediagbis}), the intervening phase transition is first-order for~$m_\phi^2 > 0$ (see section~\ref{firstorder} below). By contrast, for~$m_\phi^2 < 0$ we argue for an entire intervening gapped phase, described by a~$\Z_2$ TQFT in the deep IR (see section~\ref{section:Z2Higgsphase} below), and separated from the symmetry-breaking phases by two lines of second-order transitions  (see section~\ref{section: particlevortex and O(2)*} below).
\end{itemize}

\subsubsection{Symmetry Fractionalization in the Coulomb Phase} \label{sec:SFcoul}

It is useful to interpret the way in which the 't Hooft anomalies are matched by the Chern-Simons terms~\eqref{phases_anomaly_matching} in terms of symmetry fractionalization (see for instance~\cite{Barkeshli:2014cna,Delmastro:2022pfo,Brennan:2022tyl, Brennan:2025acl,Seiberg:2025bqy} and references therein), i.e.~the phenomenon that line defects can transform projectively under the zero-form global symmetries, and that this can (and often must) lead to 't Hooft anomalies for those zero-form symmetries. 

To this end, consider bosonic Maxwell theory, formulated in terms of an ordinary~$U(1)$ connection~$b$, with fluxes in~$2 \pi \Z$. This theory has a magnetic~$U(1)_M^{(0)}$ zero-form symmetry and an electric~$U(1)_E^{(1)}$ one-form symmetry~\cite{Gaiotto:2014kfa} (as reviewed for instance in~\cite{McGreevy:2022oyu, Cordova:2022ruw}), with background fields~$A_M^{(1)}$ and~$B_E^{(2)}$, respectively, with action\footnote{~We could equivalently carry out the entire subsequent discussion in terms of the dual photon, i.e.~a compact boson with the same symmetries and anomalies.}
\begin{equation}\label{max}
    S_\text{Maxwell} =  -{1 \over 2e^2} \int \left|db - B_E^{(2)}\right|^2 - {1 \over 2\pi} \int A_M^{(1)} \wedge db~.
\end{equation}
Let us recall a few salient features of this theory:
\begin{itemize}
    \item The~$U(1)_E^{(1)}$ charged objects are Wilson lines of~charge~$n \in \Z$. In the presence of~$B_E^{(2)}$, a charge-$n$ Wilson line furnishes the boundary of a charge-$n$ Wilson surface~$\exp(i n \int B_E^{(2)})$.\footnote{~For our purposes it is sufficient to consider flat~$B_E^{(2)}$, so that this surface is topological, but this is not essential.}
    \item The~$U(1)_M^{(0)}$ charged objects are local monopole operators, already discussed at some length above. 
    \item There is a mixed 't Hooft anomaly characterized by the four-dimensional inflow action
    \begin{equation}\label{maxanom}
        S_\text{Maxwell anomaly} = {1 \over 2\pi} \int_{\CM_4} B_E^{(2)} \wedge dA_M^{(1)}~, \qquad \partial\mathcal{M}_4=\mathcal{M}_3 ~.
    \end{equation}
\end{itemize}

Let us now discuss how this theory and its anomaly arises in the Coulomb phase of HYQED at large~$m_3 > 0$.\footnote{~The large~$m_3 < 0$ regime is analogous, being related by the action of the duality symmetry~$\CD$.} As already discussed around~\eqref{Mixed_CS_term}, we should take~$A_M^{(1)} = A_m$, identifying the~$U(1)_M^{(0)}$ symmetry of Maxwell theory in the IR with the microscopic~$U(1)_m$ symmetry of HYQED. By contrast, HYQED has no exact one-form symmetries (see below). It does, however, have Wilson lines for the microscopic generalized spin$_c$ connection~$a$ that transform projectively under the microscopic zero-form symmetries of HYQED. Specifically, it follows from~\eqref{modified_spinc_relation} that the charge-1 Wilson line~$\exp(i \int a)$ is a fermion (i.e.~it transforms projectively under the bosonic~$SO(3)$ Lorentz group) of~$U(1)_f$ charge~$+1$ that is neutral under~$U(1)_\CM$.\footnote{~Note that this is a projective representation of~$(U(1)_f \times U(1)_\CM)/\Z_2$, whose faithful representations satisfy~$q_f = q_\CM \mod 2$.} This effect is captured in the IR Maxwell theory by setting\footnote{~A Wilson line defect~$\exp(i \int dt a_0(t))$ along the time direction is stabilized by the action of lattice time reversal~$\sf T = \CD \CC \CT$ and duality~$\CD$ (equivalently, $\CC\CT$), but not by charge conjugation~$\CC$ (or its lattice version~$\sf C$). It therefore makes sense to ask whether the line defect transforms projectively under~$\sf T$ and~$\sf D$ -- it does not, in accord with~\eqref{BEinflow}, which would otherwise contain additional inflow terms for those symmetries --  but it is not meaningful to ask this question for~$\CC$. }
\begin{equation}\label{BEinflow}
    B_E^{(2)} = \pi w_2(T\CM_3) + dA_f~.
\end{equation}
Since the 1d Wilson loop is the boundary of the 2d open surface~$\exp(i \int B_E^{(2)})$, the right-hand side of~\eqref{BEinflow} leads to anomaly inflow onto the loop and enforces its projective quantum numbers (see e.g.~\cite{Brennan:2022tyl} for a systematic exposition of this procedure). As already explained around~\eqref{phases_anomaly_matching}, substituting~\eqref{BEinflow} and~$A_M^{(1)} = A_m$ into~\eqref{maxanom} matches the 't Hooft anomaly~\eqref{phases_anomaly_matching} of HYQED. 

In fact, with this choice of background fields, the bosonic Maxwell theory~\eqref{max} only differs from the IR Coulomb phase of HYQED, which is formulated in terms of the generalized spin$_c$ connection~$a$, by a field redefinition:\footnote{~This holds up to local counterterms in the background fields.} let~$A_0$ by a fixed reference spin$_c$ structure, with~$\int dA_0 = \pi w_2(T\CM_3) \mod 2\pi$. Then~$B_E^{(2)} = d (A_0 + A_f)$, and we recognize~$a \equiv b - A_0 - A_f$ as our generalized spin$_c$ structure satisfying~\eqref{modified_spinc_relation}. Note that choosing a different reference spin$_c$ structure~$A_1$ leads to a shift by a~$U(1)$ connection~$A' = A_0 - A_1$ that can be absorbed by shifting~$b \to b + A'$.

Let us comment on the status of the~$U(1)_E^{(1)}$ form symmetry in the IR Maxwell theory. Most of it is an emergent symmetry that is violated by the dynamics of HYQED. However, its~$\Z_2^{(1)} \subset U(1)_E^{(1)}$ subgroup is special: according to~\eqref{BEinflow} its background fields can be accessed using only the exact zero-form symmetries of HYQED, and moreover this is crucial for matching the 't Hooft anomalies of these zero form symmetries~\cite{Delmastro:2022pfo,Brennan:2022tyl}. The~$\Z_2^{(1)}$ symmetry in the IR should thus be viewed as an exact consequence of the microscopic symmetries~\cite{Brennan:2025acl, Seiberg:2025bqy}. In the terminology of~\cite{Seiberg:2025bqy}, the Lorentz and~$U(1)_f$ symmetries are transmuted into the~$\Z_2^{(1)}$ one-form symmetry. We will see further examples of this phenomenon below. 

Symmetries and anomalies do not uniquely fix the symmetry fractionalization pattern. In fact, matching the Maxwell and HYQED anomalies~\eqref{maxanom} and~\eqref{phases_anomaly_matching} suggests another natural choice, 
\begin{equation}
    B_E^{(2)} = \half {dA_e}~, \qquad A_M = A_m~.
\end{equation}
The unit-charge Wilson line~$\exp(i \int b)$ of the IR Maxwell theory is a now boson (i.e.~it is not fractionalized under the Lorentz group) and has~$U(1)_e \times U(1)_m$ charges~$q_e = \half$, and~$q_m = 0$ (both mod~$\Z$). We claim that this is simply a different presentation of exactly the same theory,  which was formulated above using the generalized spin$_c$ connection~$a$. The two theories are dual, but there is a non-trivial map between their Wilson line defects. This self-duality will feature in sections~\ref{section: particlevortex and O(2)*} and~\ref{sec:easy} below. It has a direct analogue in~$\Z_2$ gauge theory, which is discussed in section~\ref{section:Z2Higgsphase}.

To understand why this is compatible with the distinct quantum numbers of the Wilson lines, observe that both theories have additional line defects: the topological symmetry defects of the~$U(1)_E^{(1)}$ one-form symmetry,
\begin{equation}\label{vortexop}
    \CV_\alpha = \exp\left({i \alpha \over e^2} \int_C * db\right)~, \qquad \alpha \sim \alpha + 2\pi~,
\end{equation}
where the angle~$\alpha$ specifies the~$U(1)_E^{(1)}$ transformation. Since the Wilson lines are charged under this symmetry, linking~$\CV_\alpha$ and~$\exp(i \int b)$  produces a phase~$e^{i\alpha}$. The operator~\eqref{vortexop} can thus be equivalently described as creating a fractional magnetic flux~$\alpha$ in the plane transverse to~$C$, i.e.~$C$ can be viewed as the worldline of a vortex with~$\alpha$-flux. (See\cite{Komargodski:2025jbu} for a recent detailed discussion of such vortex line defects.) Note that the vortex lines~\eqref{vortexop} do not braid with themselves, i.e.~they are uncharged under~$U(1)_E^{(1)}$, reflecting the absence of a  't Hooft self-anomaly for this symmetry. 

For our purposes, it will suffice to consider the vortex line~$\CV_{\pi}$ with~$\alpha = \pi$, whose braiding phase with~$\exp(i \int b)$ is~$-1$. Note that~$\CV_{\pi}$ is not fractionalized under the Lorentz and~$U(1)_e$ symmetries (in particular, it is a boson), but thanks to the second term in~\eqref{max} it carries~$U(1)_m$ charge~$q_m = \half \mod \Z$. We claim that the map of Wilson lines is given by
\begin{equation}\label{linesmatch}
    \exp\left(i \int a\right) \quad \longleftrightarrow \quad \CV_\pi \exp\left(i \int b\right)~,
\end{equation}
where the right-hand side denotes the fusion of~$\CV_\pi$ and the Wilson line of~$b$.\footnote{~Since~$\CV_\pi$ is topological in the IR Maxwell theory, this fusion does not lead to singularities.} Crucially, the quantum numbers of the lines identified in~\eqref{linesmatch} match: 
\begin{itemize}
\item Both sides have unit charge under~$U(1)_E^{(1)}$, since~$\CV_\pi$  is not charged under the one-form symmetry.  

    \item $\exp(i \int a)$ has~$q_f = 1$ and~$q_\CM = 0$, which amounts to~$q_e = q_m = \half \mod \Z$. This matches the right hand side, since~$\CV_\pi$ has~$(q_e, q_m) = (0, \half)$ and~$\exp(i\int b)$ has~$(q_e, q_m) = (\half, 0)$ (modulo~$\Z^2$). 
    \item Both~$\CV_\pi$ and~$\exp(i\int b)$ are bosons, but since they have braiding~$-1$ their fusion is a fermion, as is the case for~$\exp(i\int a)$.
\end{itemize}
When we Higgs the gauge group to~$\Z_2$ (as we do in section~\ref{section:Z2Higgsphase}), the discussion above reduces to the familiar statement that in topological~$\Z_2$ gauge theory the fermion line~$f$ can be viewed as the fusion of an electric charge~$e$ and a magnetic flux~$m$. 

The discussion above was phrased in terms of the symmetries of HYQED. More intrinsically, the self-duality~\eqref{linesmatch} of Maxwell theory~\eqref{max} can be effected by shifting
\begin{equation}\label{Bshiftmax}
    B_E^{(2)} \to B_E^{(2)} + \pi w_2(T\CM_3) + \half dA_M~.
\end{equation}
Since~$A_M^{(1)}$ is a standard~$U(1)$ connection, it follows that this shift preserves the 't Hooft anomaly~\eqref{maxanom}.\footnote{~Here we use the fact that~$w_2(T\CM_4) \cup x = x \cup x$ for any~$\Z_2$ class~$x$ on~$\CM_4$.}

\subsection{Spontaneous Duality Breaking at~$m_3 = 0$,~$m^2_\phi>0$}\label{firstorder}

In this section we will explain the region around the~$m_3 = 0$, $m_\phi^2 > 0$ axis of the phase diagram, indicated by a green double-line in figure~\ref{fig:HYQEDphasediagbis}. When~$m_\phi^2 > 0$ is sufficiently large, we can reliably integrate out the Higgs field, so that the low-energy dynamics is governed by the symmetry-breaking dynamics of~QED$_3$ (reviewed in section~\ref{section:qed3review}), deformed by irrelevant four-fermion operators that explicitly break the symmetries of QED$_3$ to those of HYQED. We argue that the~$U(1)_e$ and~$U(1)_m$ symmetry-breaking phases extend all the way to~$m_3 = 0$, which constitutes a first-order coexistence line along which the vacuum manifold is the disjoint union $S_e^1 \sqcup S_m^1$. The lattice duality symmetry~$\sf D$ in~\eqref{DisDiqf} (and similarly $\CD$) that is present at~$m_3 = 0$ exchanges~$S_e^1 \leftrightarrow S_m^1$, and is therefore spontaneously broken. We conjecture that this first-order line extends all the way to the multicritical point at the origin of figure~\ref{fig:HYQEDphasediagbis}. 

For sufficiently large~$m_\phi^2 > 0$, we can integrate out the Higgs field at tree level using its equations of motion,
\begin{equation}
\phi = \frac{iy}{m^2_\phi} (\psi^{1})^t \gamma^0 \psi^2 = {i y \over m^2_\phi} \ep^{\alpha\beta} \psi^1_\alpha \psi^2_\beta~.    
\end{equation}
Substituting back into the Lagrangian~\eqref{LagrangianMinkowski} and applying Fierz identities,\footnote{~These follow immediately from~$\ep^{\alpha[\beta} \ep^{\gamma\delta]} = 0$.} we obtain the following effective potential for the fermions,
\begin{equation}
\begin{split}\label{VeffHYM}
V_{\rm eff} &= -m_3 \mathcal{O}_3 - \frac{y^2}{m^2_\phi} \left( (\bar\psi_1 \psi^1)(\bar\psi_2\psi^2) + (\bar\psi_2\psi^1)(\bar\psi_1\psi^2) \right) \\
&= -m_3 \mathcal{O}_3 + \frac{y^2}{4m^2_\phi} \left( (\mathcal{O})^2-(\mathcal{O}_3)^2+(\mathcal{O}_1)^2+(\mathcal{O}_2)^2\right) \,.
\end{split}
\end{equation}
For future use in section~\ref{sec:emergemirr}, we point out that this effective potential can be expressed entirely in terms of~$\CO_3, \CO$, and the square of the~$U(1)_f$ flavor current~\eqref{flavor_current},\footnote{~Here we use the fact that
\begin{equation}
    {(\gamma^\mu)_\alpha}^\beta {(\gamma_\mu)_\gamma}^\delta = 2 \delta_\alpha^\delta \delta_\gamma^\beta - \delta_\alpha^\beta \delta_\gamma^\delta~.
\end{equation}}
\begin{equation}\label{jfsq}
    j_\mu^f = \b \psi_i \gamma_\mu {(\sigma_z)^i}_j \psi^j~, \qquad j_\mu^f j^{\mu f} = \CO^2 + 2 \CO_3^2 - \CO_1^2 - \CO_2^2~. 
\end{equation}

Recall from section~\ref{section:qed3review} that massless QED$_3$ spontaneously breaks its global symmetries, leading to an~${\t S}^3$ sigma model for the monopole vevs that are fibered over its~$\C\P^1$ base via~\eqref{QED_monopole_vev}.  In this sigma model, the singlet $\mathcal{O}$ flows to an irrelevant operator without a vev, whereas the triplet operator flows to~$\mathcal{O}^a \to - C n^a$ (with~$C > 0$), as explained around~\eqref{QED_RG_triplet}.  Up to a constant shift of the vacuum energy we thus find that~\eqref{VeffHYM} induces the following effective potential in the IR sigma model,
\begin{equation}\label{VeffHYM_2}
V_{\rm eff} = Cm_3 n^3 - \frac{C^2y^2}{2m^2_\phi} (n^3)^2~, \qquad C>0~.
\end{equation}
As long as this potential is sufficiently small, it can be reliably analyzed as a perturbation of the~${\t S}^3$ symmetry-breaking vacua of massless~$N_f = 2$ QED$_3$. Let us consider the vacua as a function of~$m_3$: 

\begin{itemize}
    \item[$m_3 > 0$:] In this case there is a unique circle~$S_m^1$ of vacua that matches the large $m_3> 0$ Coulomb phase in~\eqref{largemassvacua}, with spontaneously-broken~$U(1)_m$ symmetry,
    \begin{equation}\label{circlem}
        S^1_m : \qquad n^a=(0,0,-1)~, \qquad  \mathcal{M}^1 = 0~, \qquad \mathcal{M}^2=ve^{i\sigma_m}~.
    \end{equation}
    Here~$\sigma_m \sim \sigma_m + 2\pi$ is the dual photon, which parametrizes~$S_m^1$ and serves as the~$U(1)_m$ Nambu-Goldstone boson.
 \item[$m_3 < 0$:] As required by duality symmetry~$\CD$, which negates~$m_3$ and exchanges~$U(1)_e \leftrightarrow U(1)_m$, there is now an~$S_e^1$ circle of vacua matching the large $m_3 <  0$ Coulomb phase in~\eqref{largemassvacua}, with spontaneously-broken~$U(1)_e$ symmetry,
    \begin{equation}\label{circlee}
        S^1_e : \qquad n^a=(0,0,1)~, \qquad  \mathcal{M}^1 = ve^{i\sigma_e}~, \qquad \mathcal{M}^2=0~,
    \end{equation}
    where~$\sigma_e \sim \sigma_e + 2\pi$ is the dual photon NGB of the broken~$U(1)_e$ symmetry. 

\item[$m_3 = 0$:] The~$S_e^1$ and~$S_m^1$ vacua found above for~$m_3 \neq 0$ simply coexist, leading to a first-order phase transition as we dial through~$m_3 = 0$. The space of vacua there is the disjoint union~$S_e^1 \sqcup S_m^1$ of the two circles, and since the duality symmetry~$\CD$ in~\eqref{action_of_D} exchanges them, it is spontaneously broken. The same is true of any symmetry that swaps the two circles, including lattice~$\sf D$ in~\eqref{DisDiqf} and time reversal~$\CT$ (see table~\ref{tab:hyGonOps}). By contrast, any non-swapping symmetry, e.g.~charge conjugation~$\CC$ (see table~\ref{tab:hyGonOps}) and its lattice version~$\sf C$ in~\eqref{CisCm1}, as well as lattice~$\sf T = \CD \CC \CT$ in~\eqref{TisDCT} can be combined with the broken~$U(1)_{e,m}$ rotations on each circle to obtain unbroken symmetries. 

\end{itemize}

\noindent As was already discussed around~\eqref{m3monovev} in the context of QED$_3$, we will assume that the are no phase transitions as we increase~$|m_3|$ in the discussion above, as long as~$m_\phi^2 > 0$ is sufficiently large to ensure that we are in the QED$_3$ regime. 

What happens when we decrease~$m_\phi^2$, so that we exit the controlled QED$_3$ regime above? We know from the discussion around~\eqref{CSjump} that there must always be at least one phase transition between the large~$m_3 > 0$ and the large~$m_3 < 0$ Coulomb phases. The most economical conjecture is that the duality-breaking first-order line described above continues all the way to the multicritical point at~$m_\phi^2 = 0$. We will present evidence for this in section~\ref{sec:latcontmatch}, by relating HYQED deformed by monopole operators to the Fradkin-Shenker lattice model, which also has a~$\sf D$-breaking first-order line emanating from its multicritical point (see figure~\ref{figfsphasediagram}).

\subsection{The~$\Z_2$ Higgs Phase at~$m_\phi^2 < 0$}
\label{section:Z2Higgsphase}

When the Higgs mass~$m_\phi^2 < 0$ is negative and sufficiently large compared to the other scales~$|m_3|, e^2, y^2, \lambda_4$ in the problem, the theory is in a weakly-coupled Higgs phase, where $\langle \phi\rangle \neq 0$. Since~$\phi$ has charge two, it Higgses~$U(1)_a \to \Z_2$, while~$\phi$ and~$a_\mu$ become massive. Meanwhile, the fermions all acquire masses via the Yukawa couplings~\eqref{yukawa2}. A natural guess for the low-energy TQFT is therefore simply~$\Z_2$ gauge theory. Below we will establish this in detail and show how it is compatible with the microscopic symmetries and anomalies. This establishes the purple region in the phase diagram of figure~\ref{fig:HYQEDphasediagbis}, which we conjecture to extend all the way to the strong-coupling region and the multicritical point. We will present some evidence for this in section~\ref{sec:latcontmatch} by relating HYQED to the Fradkin-Shenker lattice model.

\subsubsection{Review of~$\Z_2$ Topological Gauge Theory}
\label{sec:z2gaugereview}

Let us briefly recall some aspects of pure~$\Z_2$ topological gauge theory, without a Dijkgraaf-Witten twist. This theory has many different, dual Lagrangian presentations. Here we begin with a description where the dynamical fields are~$\Z_2$ one-cochains~$e_1, m_1 \in C^1(\Z_2)$,\footnote{~It is often convenient to lift~$e_1, m_1$ to~$U(1)$ connections~$b^{(1)}$, $c^{(1)}$ and present the~$\Z_2$ TQFT as a BF/Chern-Simons theory with action (see e.g.~\cite{Kapustin:2014gua} and references therein),
\begin{equation}
    S_{\Z_2 \text{ TQFT}} = {1 \over 2\pi} \int b^{(1)} \wedge dc^{(1)}~,
\end{equation}
though we will not explicitly use this presentation here.
}
\begin{equation}\label{z2tqft}
    S_{\Z_2 \text{ TQFT}} =  \pi \int_{\CM_3} e_1 \cup \delta m_1~,
\end{equation}
which is invariant under gauge transformations that shift~$e_1, m_1$ by exact cochains. Moreover~$e_1, m_1$ are closed on shell and thus describe cohomology classes in~$H^1(\CM_3, \Z_2)$, i.e.~flat $\Z_2$ connections. Integrating out one of them, for instance~$m_1$, leads to a theory of a dynamical~$\Z_2$ gauge field~$e_1$ with vanishing action, and vice versa. The two pictures are related by a unitary duality symmetry~$\CD_\text{TQFT}$ (manifest in~\eqref{z2tqft}) that exchanges
\begin{equation}\label{Dtqft}
    \CD_\text{TQFT} : e_1 \leftrightarrow m_1~.
\end{equation}

All anyons in this theory are Abelian, and their worldlines~$C$ are described by Wilson lines of~$e_1$ and~$m_1$ along~$C$,
\begin{equation}
    e = \exp\left(i \pi \int_C e_1\right)~, \qquad m = \exp\left(i \pi \int_C m_1\right)~, \qquad f \equiv em = \exp\left(i \pi \int_C(e_1 + m_1)\right)~.
\end{equation}
Here the electric and magnetic anyons~$e$ and~$m$ are bosons of topological spin~$0$,\footnote{~In a bosonic TQFT such as ours the topological spin is meaningful mod~$1$.} while~$f$ is a fermion of topological spin~$\half$. The fusions rules are~$\Z_2^e \times \Z_2^m$, generated by~$e$ and~$m$, i.e.~$e^2 = m^2 = 1$. Since the TQFT is Abelian, it is completely specified by this data (see e.g.~\cite{Lee:2018eqa, Delmastro:2019vnj, Geiko:2022qjy} for a modern presentation with references).

Let us summarize the intrinsic global symmetries of the TQFT:
\begin{itemize}
    \item The theory has a~$(\Z_2^{(1)})_e \times (\Z_2^{(1)})_m$ one-form symmetry~\cite{Gaiotto:2014kfa}, generated by the~$m$ and~$e$ anyons. The charged lines are then the~$e$ and~$m$ anyons, respectively, which link non-trivially with the respective symmetry generating lines. The corresponding background gauge fields~$B^e_2, B_2^m \in H^2(\CM_3, \Z_2)$ couple to~\eqref{z2tqft} via
    \begin{equation}\label{Bcouplings}
        \Delta S[B^e, B^m]  = \pi \int_{\CM_3} \left( B_2^e \cup m_1 + B_2^m \cup e_1 \right) ~.
    \end{equation}
With this definition, the~$e$ anyon bounds a~$B_2^e$ surface, and the~$m$ anyon bounds a~$B_2^m$ surface. 

 \item   In the presence of~\eqref{Bcouplings}, the theory is invariant under dynamical, but not under background gauge transformations. This constitutes a  mixed 't Hooft anomaly between the electric and magnetic one-form symmetries, which is characterized by the following four-dimensional inflow action (or SPT) for the background fields,
    \begin{equation}\label{BBanom}
        S_\text{inflow} = \pi \int_{\CM_4} B_2^e \cup B_2^m~, \qquad \partial  \CM_4 = \CM_3~.
    \end{equation}

    \item The theory enjoys the unitary~$\Z_2$ duality symmetry~$\CD_\text{TQFT}$ in~\eqref{Dtqft}, which swaps $e \leftrightarrow m$ and $B_2^e \leftrightarrow B_2^m$, while fixing~$f$. 
    \item There is an anti-unitary~$\Z_2$ time-reversal symmetry~$\CT_\text{TQFT}$ that fixes all anyons~\cite{Lee:2018eqa, Delmastro:2019vnj, Geiko:2022qjy}.
\end{itemize}
Below we will identify~$\CD_\text{TQFT}$ and~$\CT_\text{TQFT}$  with specific UV symmetries of HYQED in the Higgs phase. By contrast, HYQED does not possess any microscopic one-form symmetries, but it has exact zero-form symmetries that fractionalize on the anyons of the~$\Z_2$ TQFT and are thus transmuted to its~$(\Z_2^{(1)})_e \times (\Z_2^{(1)})_m$ one-form symmetries.

\subsubsection{Symmetry Fractionalization in~$\Z_2$ Gauge Theory} \label{sec:SFtqft}

Whenever there are zero-form symmetries, their background fields can be substituted into the one-form symmetry background fields~$B_2^{e,m}$. This has the effect of making the anyons transform projectively under the zero-form symmetries -- a much-studied phenomenon broadly referred to as symmetry fractionalization that we already encountered in the Coulomb phases analyzed in section~\ref{section: large m3} above. For instance, substituting~$B_2^e = w_1^2(T \CM_3)$ (with $w_1(T\CM_3)$ the first Stiefel-Whitney class of the spacetime manifold~$\CM_3$) turns the $e$ anyon into a Kramers doublet under time reversal~$\CT_\text{TQFT}$. 

Similarly, setting~$B_2^e = w_2(T\CM_3)$  fractionalizes the Lorentz symmetry and turns the~$e$ anyon into a fermion (see for instance~\cite{Hsin:2019gvb} for a detailed discussion). As long as~$B_2^m = 0$ the~$m$ anyon, and hence the~$f = em$ anyon, are bosons. As explained in~\cite{Hsin:2019gvb}, the resulting theory is a dual description of the same untwisted~$\Z_2$ gauge theory~\eqref{z2tqft}, but the roles of the~$e$ and the~$f$ anyons are exchanged. Integrating out~$m_1$ in this description leads to a theory of the cochain~$e_1 \in C^1(\Z_2)$ subject to the constraint
    \begin{equation}
        \delta e_1 = w_2~,
    \end{equation}
    so that~$e_1$ is a dynamical spin structure. The TQFT of a dynamical spin structure (with vanishing action) is thus dual to the bosonic presentation of~$\Z_2$ gauge theory in~\eqref{z2tqft}. Note that the duality symmetry~$\CD_\text{TQFT}$ in~\eqref{Dtqft}, though of course present, is not manifest in the spin-structure presentation. In section~\ref{secetainvariants} we will write down an action for the~$\Z_2$ TQFT that depends on a dynamical spin structure and hast manifest duality symmetry~$\CD_\text{TQFT}$.

As we will show explicitly below, the~$\Z_2$ Higgs phase of our Higgs-Yukawa-QED theory is described by the~$\Z_2$ gauge theory Lagrangian~\eqref{z2tqft} without any symmetry fractionalization for the time-reversal or Lorentz symmetries, i.e~the anyons transform non-projectively under these symmetries. Indeed substituting such fractionalization patterns into~\eqref{BBanom} would generically lead to 't Hooft anomalies involving only~$w_1(T\CM_3)$ and~$w_2(T\CM_3)$ in the four-dimensional anomaly inflow SPT, which would be inconsistent with the 't Hooft anomaly~\eqref{mixedanomalyHYM} of the UV HYQED theory. 

However, we will also show that the anyons do transform projectively under the~$U(1)_{e, m}$ symmetries of HYQED:

\begin{itemize}
\item The~$e$ anyon has half-integer~$U(1)_e$ charge~$q_e = \half \mod \Z$ and integer~$U(1)_m$ charge $q_m = 0 \mod \Z$.\footnote{~Only the fractional part of these charges is meaningful, since the integer part can be shifted by well-defined background Wilson line counterterms the global~$U(1)_{e,m}$ symmetries.} We must therefore set
\begin{equation}\label{bmfrac}
    B_2^e = \left[{d A_e \over 2\pi}\right]_2 \in H^2(\CM_3, \Z_2)~.
\end{equation}
Here the subscript indicates that we should reduce the integer class represented by~$dA_e \over 2\pi$ to a~$\Z_2$ class.
\item Analogously, the~$m$ anyon has vanishing~$U(1)_e$ charge~$U(1)_m$ charge~$\half$ (modulo integer-charge counterterms), so we that we must set
\begin{equation}\label{befrac}
    B_2^m = \left[{d A_m \over 2\pi}\right]_2 \in H^2(\CM_3, \Z_2)~.
\end{equation}
\end{itemize}
Substituting~\eqref{bmfrac} and~\eqref{befrac} into~\eqref{BBanom}, we precisely match the 't Hooft anomaly~\eqref{mixedanomalyHYM} for the zero-form symmetries of HYQED.

\subsubsection{Symmetries and Particles in the~$\Z_2$ Higgs Phase of HYQED}

In the weakly-coupled Higgs regime, where~$m^2_\phi \ll 0$, the Higgs field~$\phi$ acquires a large vev that we take to be real and positive without loss of generality, $\langle \phi \rangle > 0$. This has several consequences:
\begin{itemize}
\item The physical Higgs particle, described by the radial mode of~$\phi$, is massive. Moreover, the fermions are also massive due to the Yukawa couplings~\eqref{yukawa2}, even when $m_3=0$.  

\item The~$U(1)_{e, m}$ symmetries under which~$\phi$ is neutral are unbroken. 

\item Since~$\phi$ has gauge charge~$q_a = 2$, the~$U(1)_a$ gauge group is Higgsed to its~$\Z_2^a$ subgroup, 
    \begin{equation}\label{Higgsing}
        \langle \phi \rangle > 0 \quad : \quad U(1)_a \to \Z^a_2~.
    \end{equation}
    Note that~$q_a(\psi^i) = 1$ and~$\Z_2^a$ is generated by~$(-1)^{q_a}$, which acts as fermion parity~$(-1)^F$ (i.e.~a $2\pi$ spatial rotation) on the fundamental fields in the Lagrangian. 

    As a consequence of Higgsing~\eqref{Higgsing}, the dynamical spin$_c$ connection~$a$ eats the angular mode of~$\phi$ to acquire a mass, leaving on only a topological~$\Z_2$ gauge theory (reviewed above) in the IR. Moreover, we naturally land on the description of that theory in terms of a dynamical spin structure: the ordinary~$U(1)$ connection~$2a$ is gauge-equivalent to zero in the Higgs phase. Up to gauge transformations, $a$ can thus be represented by a~$\Z_2$-valued cochain,\footnote{~This constitutes a partial gauge fixing; $\Z_2^a$ gauge transformations that shift~$[a]_2$ by an exact~$\Z_2$ cochain remain unfixed.}
    \begin{equation}\label{amod2}
        a \to \pi [a]_2~, \qquad [a]_2 \in C^1(\Z_2)~,
    \end{equation}
    whose torsion flux satisfies the mod-2 reduction of~\eqref{modified_spinc_relation},
    \begin{equation}\label{da2cond}
        \delta[a]_2 = \left[{da \over \pi}\right]_2 = w_2(T\CM_3) + \left[{d A_f \over \pi}\right]_2~.
    \end{equation}
    Thus~$[a]_2$ is a slight generalization of a spin structure (just as~$a$ is a slight generalization of a spin$_c$ structure), and it reduces to a spin structure as long as~$A_f$ only has integer fluxes~$\oint dA_f / 2 \pi \in \Z$. Recall from~\eqref{fluxquantizationAmAf} that its fluxes are in general half-integers.

While~$\phi$ is neutral under the unbroken~$\Z_2^a$ gauge symmetry and flows to the trivial anyon of the~$\Z_2$ TQFT at low energies, the massive fermions~$\psi^i$ are attached to~$\Z_2^a$ Wilson lines. They thus represent the fermionic quasiparticle~$f$ in the~$\Z_2$ TQFT.\footnote{~Note that~$\psi^1$ and~$\psi^2$ represent the same anyon superselection sector~$f$ in the~$\Z_2$ TQFT, since they differ by a gauge-invariant local operator~$\b \psi_1 \psi^2$. This is consistent with the fact that duality symmetry swaps~$\psi^1 \leftrightarrow \psi^2$, but fixes~$f$ , as discussed around~\eqref{Dhiggs} below.}
This is consistent with~\eqref{bmfrac}~and~\eqref{befrac}, which imply that the worldline of the fermion~$f = em$ resides on the boundary of an open Wilson surface for
\begin{equation}
    B_2^e + B_2^m = \left[{d (A_e + A_m)\over 2\pi}\right]_2 = \left[{d A_\CM \over \pi}\right]_2 = \left[{d A_f \over \pi}\right]_2~,
\end{equation}
where we have used~\eqref{fluxquantizationAmAf} and~\eqref{Aem_definition}. This agrees with the second term on the right-hand side of~\eqref{da2cond}, while the first term captures the fact that Wilson lines of~$[a]_2$ are fermions of topological spin~$\half \mod 1$. 

\item The~$\Z_2^{\sf D}$ lattice duality symmetry generated by~${\sf D} = \CD i^{q_f}$ in~\eqref{DisDiqf} leaves~$\phi, a_\mu$ invariant and is therefore unbroken. It acts non-projectively on the fermions,
\begin{equation}\label{Dhiggs}
    {\sf D} : \psi^i \to -{(\sigma_y)^i}_j \psi^j~, \qquad {\sf D}^2 = 1~.
\end{equation}
Thus~$\sf D$ leaves the~$f$-anyon of the~$\Z_2$ TQFT invariant, while swapping~$U(1)_e \leftrightarrow U(1)_m$. Comparing with~\eqref{bmfrac} and~\eqref{befrac}, we are led to identify~$\sf D$ with the duality symmetry~${\CD}_\text{TQFT}$ of the~$\Z_2$ gauge theory in~\eqref{Dtqft},
\begin{equation}
    {\sf D}= \CD i^{q_f} = \CD_\text{TQFT}~.
\end{equation}
We will give further arguments for this identification below when we discuss vortices.

\item The~$\Z_2^{\sf C}$ lattice charge-conjugation symmetry, given by ~${\sf C} = \CC (-1)^{q_m}$ in~\eqref{CisCm1}, sends
\begin{equation}
    {\sf C} : \phi \to \phi^*~, \qquad a_\mu \to -a_\mu~,
\end{equation}
and is thus unbroken by~$\langle \phi \rangle  > 0$. Note that the fermions transform non-projectively under~$\CC : \psi^i \to (\psi^i)^*$. Moreover, $\CC$ does not act on the IR~$\Z_2$ TQFT thanks to~\eqref{amod2}: the gauge field~$a$ is reduced to~$[a]_2 = -[a]_2 \mod 2$, so that~$[a]_2$ is~$\CC$-invariant.

However, the fermions and the lines of the TQFT do transform projectively under $(-1)^{q_m}$, which we will interpret as a~$\alpha = \pi$ rotation in~$U(1)_m$,\footnote{~If we chose~$\alpha = -\pi$ we will get a definition of~$(-1)^{q_m}$ on the fermions that differs by an unbroken~$\Z_2^a$ gauge transformation.}
\begin{equation}
    e^{i \pi q_m} : \psi^i \to e^{- i \pi q_f/2} \psi^i = -i {(\sigma_z)^i}_j \psi^j~.
\end{equation}
This notwithstanding, lattice~$\sf C$ also acts non-projectively on the fermions,
\begin{equation}
    {\sf C} : \psi^i \to -i {(\sigma_z)^i}_j (\psi^j)^*~, \qquad {\sf C}^2 \psi^i =  \psi^i~.
\end{equation}

    \item Since the~$\Z_2^{\sf T}$ lattice time-reversal symmetry generated by~${\sf T} = \CD \CC \CT$ in~\eqref{TisDCT} acts a~${\sf T} : \phi \to - \phi^*$, the unbroken time-reversal symmetry~${\sf T}'$ in the Higgs phase is obtained by combining~${\sf T}$ with a suitable~$U(1)_a$ gauge transformation,
    \begin{equation}
        {\sf T'} = {\sf T} e^{i \pi q_a/2}~. 
    \end{equation}
    The unbroken ${\sf T'}$ symmetry acts as follows on fields,
    \begin{equation}\label{ubtprime}
     {\sf T}': \phi \to \phi^*~, \qquad \psi^i \to -i (\sigma_x)_{ij} \gamma^0 (\psi^j)^*~, \qquad a_0 \to -a_0~, \qquad a_{1,2} \to a_{1,2}~.
    \end{equation}
    Note that~${\sf T'}^2 = 1$ on on all fields, so that the symmetry acts non-projectively. Below we will identify
    \begin{equation}
        {\sf T'} = \CD \CC \CT e^{i \pi q_a/2} \to  \CT_\text{TQFT}~. 
    \end{equation}
    Here~$\CT_\text{TQFT}$ is the time-reversal symmetry of the~$\Z_2$ gauge theory discussed below~\eqref{BBanom}, which fixes all the anyons and acts on them non-projectively. 
\end{itemize}

\subsubsection{Vortices in the~$\Z_2$ Higgs Phase of HYQED}\label{sec:vortices}

We will now deduce the fractionalized quantum numbers of the~$e$ and~$m$ anyons of the~$\Z_2$ TQFT, verifying~\eqref{bmfrac} and~\eqref{befrac}. In particular, we will find that the $e$ anyon carries $(q_e,q_m)$ charge $(1/2,0)$ and the $m$ anyon carries charge $(0,1/2)$ (all anyon charges are defined modulo the charge lattice $\bZ \times \bZ$ of local operators).

As we will see, these anyons describe the long-distance limit of vortex excitations, which are charged under the~$U(1)_\CM$ symmetry. In general, different vortices can represent the same anyon superselection sectors in the IR. For our purposes, it will suffice to consider a single vortex, at rest at the origin, and of minimal vorticity. 

As is well known (see e.g.~\cite{Dumitrescu:2025fme} for a recent review with references), the minimum energy vortex solution takes the following form, 
\begin{equation}\label{vortexbackground}
\phi(r,\theta) = \rho(r) e^{i\theta}~, \qquad \lim_{r\rightarrow\infty}\rho(r)=\langle \phi\rangle~.
\end{equation}
Here~$r, \theta$ are polar coordinates in the spatial~$xy$-plane. Note that the Higgs field winds precisely once around the circle at spatial infinity, as it approaches the Higgs vacuum. The energy is a sum of non-negative terms, including 
\begin{equation}
\int d^2x \, \left|(\partial_\mu-2ia_\mu)\phi\right|^2 \simeq 
2\pi \langle \phi\rangle^2 \int^\infty \frac{dr}{r}  \, \left(1-2a_\theta\right)^2 + (\text{finite})~.
\end{equation}
Since we would like to discuss finite-energy vortex excitations, the gauge field must be chosen to cancel this divergence, 
\begin{equation}
\lim_{r\rightarrow\infty}a_\theta=\frac{1}{2}~, 
\end{equation}
which in turn implies that the vortex is a~$\pi$-flux, i.e.~it carries~$U(1)_\CM$ charge~$q_\CM=\half$,
\begin{equation}
q_\CM=\frac{1}{2\pi}\int_{\mathbb{R}^2}f = \frac{1}{2\pi} \lim_{r\rightarrow\infty} \int_0^{2\pi} a_\theta \,d\theta = \frac{1}{2}~.
\end{equation}
We will not need the detailed form of the functions~$\rho(r), a_\theta(r)$ below (see~\cite{Dumitrescu:2025fme} and references therein for a detailed discussion of how to determine these functions). 

Let us summarize the symmetries that are unbroken in the presence of the vortex solution~$\phi(r, \theta)$ in~\eqref{vortexbackground}, together with~$a_\theta(r)$:
\begin{itemize}
    \item Rotations are only preserved by combining them with a constant~$U(1)_a$ gauge transformation. Since~$\phi$ has charge~$2$, the conserved angular momentum~$j'$ is therefore given in terms of the ordinary rotation generator~$j$ via
    \begin{equation}\label{jprime}
        j'= j - {q_a \over 2}~.
    \end{equation}
    Since the fermions~$\psi^i$ have~$q_a = 1$, they have integer spin in the vortex field (see below).  
    
    \item The~$(U(1)_\CM \times U(1)_f)/\Z_2=U(1)_e \times U(1)_m$ symmetry is preserved. 

\item The solution preserves the~${\sf T}'$ symmetry in~\eqref{ubtprime}.\footnote{~The fact that~${\sf T}'$ maps the operator~$\phi$ to its Hermitian conjugate implies that it preserves the~$c$-number vev~$\langle \phi \rangle$, as can be seen by writing~$\phi = \langle \phi\rangle {\mathds 1}$, with~${\mathds 1}$ the unit operator.}   Note that~${\sf T}'$ negates the conserved angular momentum~$j'$ in~\eqref{jprime}.

\item If~$m_3 = 0$, so that~$\sf D$ in~\eqref{Dhiggs} is a symmetry, than it is preserved by the vortex (and the same is true for~$\CD$). 
\end{itemize}
By contrast~${\sf C} : q_\CM \to -q_\CM$ negates the vorticity and is thus not preserved. 

We will now analyze the fermions in the vortex background. As is often the case, it suffices to consider the fermion zero modes to determine the quantum numbers of the vortex. See~\cite{Seiberg:2016rsg} for a detailed exposition with references, and~\cite{Jian:2017chw} for a discussion of the model studied here.

We begin our discussion on the self-dual line, where~$m_3 = 0$ and~$\sf D$ is preserved. (Below we will comment on the effect of turning on~$m_3 \neq 0$.) There the Dirac equation for~$\psi^i$ in the vortex background can be decoupled by using a basis that diagonalizes the Yukawa interaction~$\psi^1 \psi^2 \sim \psi_+^2 + \psi_-^2$,
\begin{equation}
\psi_+ \equiv \frac{1}{\sqrt{2}}\left(\psi^1+\psi^2\right)~, \qquad
\psi_- \equiv \frac{i}{\sqrt{2}}\left(\psi^1-\psi^2\right)~.   
\end{equation}
Let~$u(\vec x)$ be a classical (i.e.~$c$-number), static (i.e.~zero-energy) solution of this equation, 
\begin{equation}\label{ig0ofvort}
    \vec \gamma \cdot \left(\vec \partial - i \vec a(\vec x)\right) u(\vec x) = y \phi(\vec x) u^*(\vec x)~. 
\end{equation}
Here~$\phi(\vec x), \vec a(\vec x)$ denote the classical vortex solution discussed above. Since~$i\gamma^0$ is purely imaginary and anti-commutes with the Dirac operator on the left-hand side, it follows that~$i \gamma^0 u(\vec x)$ is another solution of the zero-mode equation.\footnote{~Note that~$i\gamma^0$ is essentially the action of the unbroken~$(\CC \CT)'$ symmetry (which differs from~$\sf T'$ by a duality swap~$\CD$) in the vortex background on the c-number zero modes.} We can thus choose a basis for the space of zero modes on which~$i\gamma^0$ (which is Hermitian and squares to unity) is diagonal, with eigenvalues~$\pm 1$. Let us denote the number of normalizable zero-mode solutions of each kind by~$N_\pm$. It was shown explicitly in~\cite{Jackiw:1981ee} that the minimal vortex of unit vorticity discussed above has a single zero mode solution satisfying
\begin{equation}
   q_\CM = \half \; \text{ vortex}: \quad N_+ = 1~, \qquad N_- = 0~, \qquad  i \gamma^0 u(\vec x) = u(\vec x)~.
\end{equation}
By contrast, the~$\sf C$-conjugate vortex with~$q_\CM = -1/2$ has~$N_+ = 0$ and~$N_- = 1$. More generally, the Callias index theorem~\cite{Callias:1977kg} (see appendix B of~\cite{Seiberg:2016rsg} for a modern exposition with references) states that the difference\footnote{~The sign in this equation can be fixed by considering a vortex of large, positive vorticity, whose interior harbors a nearly uniform, positive magnetic field~$B > 0$ (see e.g.~\cite{Dumitrescu:2025fme}) in which the zero-mode equation is straightforwardly solved (see e.g.~section 2.3 of~\cite{Dumitrescu:2025vfp} and references therein).}
\begin{equation}
    N_+ - N_- = \text{vorticity} = 2 q_\CM~,
\end{equation}
is a topological invariant that does not depend on the details of the vortex solution. A more refined version of the index theorem shows that the~$2 q_\CM$ zero modes all have distinct spins~$j' = q_\CM - 1/2, q_\CM - {3/2}, \ldots, -(q_\CM - 1/2)$. This is consistent with the statement above that the minimal vortices of vorticity~$\pm 1$ and~$q_\CM = \pm 1/2$ have spinless zero modes with~$j' = 0$.

Returning to the~$q_\CM = 1/2$ vortex, each fermion~$\psi_\pm$ has a single real zero mode and can therefore be expanded as follows, 
\begin{equation}
    \psi_\pm = \half u(\vec x)\gamma_\pm + (\text{non-zero modes})~, \qquad \int d^2 x \, u^\dagger(\vec x) u(\vec x) = 1~,
\end{equation}
where the operators~$\gamma_\pm$ are Majorana fermions, 
\begin{equation}
\gamma_\pm^\dagger = \gamma_\pm~,\qquad \gamma_\pm^2 = 1~, \qquad  \lbrace \gamma_+, \gamma_- \rbrace = 0~.
\end{equation}
In the more familiar complex basis,
\begin{equation}\label{ccomplex}
    c = \half \left(\gamma_+ + i \gamma_-\right)~, \qquad c^\dagger = \half \left(\gamma_+ - i \gamma_-\right)~,
\end{equation}
with anticommutation relations  
\begin{equation}\label{cacomm}
    \{c, c^\dagger\} = 1~, \qquad \{c,c\} = \{c^\dagger, c^\dagger\} = 0~.
\end{equation}
Note that
\begin{equation}\label{psitoc}
    \psi^1  = {1 \over \sqrt{2}} u(\vec x) c^\dagger + (\text{non-zero modes})~, \qquad \psi^2 =  {1 \over \sqrt{2}} u(\vec x) c + (\text{non-zero modes})~,
\end{equation}
and that substituting this expansion into the free Dirac Lagrangian for~$\psi^i$ leads to the following Lagrangian for the zero modes,
\begin{equation}
    L_0 = i c^\dagger \d_t c~,
\end{equation}
consistent with~\eqref{cacomm}.

As usual, this gives rise a two-dimensional Fock space. We will use the notation~$|m\rangle$ to denote the empty Fock vacuum, because this state will turn out to represent the~$m$ anyon of~$\Z_2$ TQFT at long distances (see below). By contrast, the occupied state will be denoted by~$\ket{e}$, because it resides in the~$e$ superselection sector of the TQFT. In formulas,
\begin{equation}
c\ket{m}=0~, \quad c^\dagger\ket{m}=\ket{e}~, \quad c\ket{e}=\ket{m}~, \quad c^\dagger \ket{e}=0~.    
\end{equation} 

Let us now discuss the action of the unbroken symmetries discussed around~\eqref{jprime} on the zero-mode Hilbert space. Here it is useful to to take the point of view (elaborated in the more involved context of section~\ref{u1LatToCont}) that the symmetries of the full QFT map into the symmetries~$G_0$ of the fermion zero modes via a group homomorphism (which in this case will be neither injective nor surjective) that is only defined up to conjugacy. Briefly,\footnote{~See appendix B of~\cite{Dumitrescu:2025vfp} for a recent detailed discussion.} the symmetries of the massless zero-mode quantum mechanics are
\begin{equation}
    G_0 = SO(3) \rtimes \Z_2^{\sf T_0}~.
\end{equation}
Here the unitary~$SO(3)$ symmetry is generated by
\begin{equation}
    J_3 = c^\dagger c  - \half~, \qquad J_1 + i J_2  = c^\dagger~, \qquad J_1 - i J_2 = c~,
\end{equation}
while the antiunitary~$\Z_2^{\sf T_0}$ time-reversal symmetry is generated by~${\sf T}_0 : c(t) \to c(-t)$. Note that~$\Z_2^{\sf T_0}$ extends~$SO(3)$ to~$O(3)$. 

Since~$SO(3)$ has rank one and there are three preserved~$U(1)$ charges in the presnce of the vortex -- namely~$q_f$, $q_\CM$, and~$j'$ -- only one of them can act non-trivially on the fermions, while the other two are necessarily central:
\begin{itemize}
\item[($j'$)] The fermion zero modes have vanishing spin~$j'$, and since the classical vortex solution also has~$j' = 0$, it follows that both vortex states are spin-$0$ bosons,\footnote{~Since the unbroken~$\Z_2^a$ gauge symmetry in the Higgs phase is generated by fermion number~$(-1)^F$, we conclude that~$e$ and~$m$ have vanishing~$\Z_2^a$ charge, and the same is true of the fermion zero modes themselves.}
\begin{equation}
    j' |e \rangle = j' |m\rangle = 0~.
\end{equation}

\item[($q_\CM$)] The fermion zero modes have vanishing~$U(1)_\CM$ charge, so that~$q_\CM$ is central. However, the bosonic vortex background has~$q_\CM = \half$, so that
\begin{equation}
    q_\CM |m \rangle = \half |m\rangle~, \qquad  q_\CM |e \rangle = \half |e\rangle~.
\end{equation}
\item[($q_f$)] We see from~\eqref{psitoc} that the fermion zero modes~$c^\dagger \sim \psi^1$ and~$c \sim \psi^2$ have~$U(1)_f$ flavor charges~$q_f = 1$ and $q_f = -1$, respectively. Up to conjugacy, we must therefore identify\footnote{~This expression also follows from substituting~\eqref{psitoc} into~$q_f = \int d^2 x \, j^0_f$, where~$j^\mu_f = \b \psi_i \gamma^\mu {(\sigma_z)^i}_j \psi^j$ is the~$U(1)_f$ current in~\eqref{flavor_current}. }
\begin{equation}\label{qfisj3}
    q_f = J_3 = c^\dagger c - \half ~, 
\end{equation}
which acts projectively on the vortices,
\begin{equation}
    q_f |e\rangle = \half |e\rangle~, \qquad q_f |m\rangle = -\half |m\rangle~.
\end{equation}
This reflects the fact that the Hilbert space transforms projectively under~$SO(3)$, which in turn signifies an 't Hooft anomaly. As we review below, this anomaly persists even if we restrict the~$SO(3)$ symmetry to its~$O(2)$ subgroup, which is a symmetry of HYQED. 
\end{itemize}
\noindent It is now straightforward to show that the discrete~$\Z_2^{\sf D} \times \Z_2^{\sf T'}$ symmetries of the vortex are represented on the zero modes via the following operators (unique up to conjugacy),
\begin{equation}
    {\sf D} = c + c^\dagger~, \qquad {\sf T'} = {\sf T}_0~.
\end{equation}
Here~$\sf D$ and~$\sf T'$ are characterized by their action on~$q_f$, 
\begin{equation}
    {\sf D} : q_f \to -q_f~, \qquad {\sf T'} : q_f \to q_f~,
\end{equation}
and they act as follows on the states,
\begin{equation}
    {\sf D} : |e\rangle \leftrightarrow |m\rangle~, \qquad {\sf T'} : |e\rangle \to |e\rangle~, \quad |m \rangle \to |m\rangle~.
\end{equation}
Notice that both~$\sf D$ and~$\sf T'$ act non-projectively. An important point is that the $O(2)_f = U(1)_f \rtimes \Z_2^{\sf D}$ symmetry of the QFT, which does not enjoy the full~$SO(3)$ symmetry of the zero modes, is sufficient to fix the c-number constant~$-\half$ in~\eqref{qfisj3}, and hence the fractional~$U(1)_f$ quantum numbers of the vortex states. Thus the~$SO(3)$ anomaly remains non-trivial under pull-back to~$O(2)_f$. 

It is now clear that the spin-0 vortices~$|e\rangle$ and~$|m\rangle$ discussed above are represented in the deep IR by the bosonic~$e$ and~$m$ anyons of the low-energy~$\Z_2$ TQFT~\eqref{z2tqft}. Moreover, we identify~$\sf D$ with the~$e \leftrightarrow m$ swapping~$\CD_{\text{TQFT}}$ symmetry discussed below~\eqref{BBanom}, while~$\sf T'$ is identified with~$\CT_\text{TQFT}$, which stabilizes all anyons. Moreover, both symmetries are represented non-projectively. Only the~$U(1)_\CM$ and~$U(1)_f$ symmetries are realized projectively on the vortices. Switching to the~$U(1)_e \times U(1)_m$ basis with charges~$q_e= \half (q_\CM + q_f)$ and $q_m= \half (q_\CM - q_f)$, we find that
\begin{equation}
    (q_e, q_m) |e\rangle = \left(\half, 0\right) |e\rangle~, \qquad (q_e, q_m) |m\rangle = \left(0, \half\right) |m\rangle~.
\end{equation}
This justifies the mapping of the background fields in~\eqref{bmfrac} and~\eqref{befrac}.

Finally, let us add a small~$\sf D$-breaking mass~$m_3 \in \R$, as in~\eqref{tuningparam}, and examine its effect on the zero modes by substituting~\eqref{psitoc}, 
\begin{equation}
    \delta {\mathscr L} =  i m_3 (\b \psi_1 \psi_1 - \b \psi_2 \psi^2) = - m_3 u^\dagger (i\gamma^0) u \left(c^\dagger c - \half\right)~.
\end{equation}
Since~$i \gamma^0 u  = u$, as discussed around~\eqref{ig0ofvort}, we conclude that the Hamiltonian of the zero modes is
\begin{equation}
    H_0 = m_3 \left(c^\dagger c - \half\right)~. 
\end{equation}
As expected, the degeneracy between the~$|e\rangle$ and the~$|m\rangle$ vortices is split, and the lightest vortex is determined by the sign of~$m_3$, 
\begin{equation}\label{lightvortex}
    \text{lightest~$q_\CM = \half$ vortex:} \quad \begin{cases} |m\rangle \quad \text{for} \quad m_3 > 0 \\ |e\rangle \quad\hskip4.5pt \text{for} \quad m_3 < 0 \end{cases} \ .
\end{equation}
As we will discuss in section~\ref{section: particlevortex and O(2)*} below, the fact that the~$|m\rangle$ vortex becomes light for~$m_3 > 0$ is consistent with the large~$m_3 > 0$ Coulomb phase we found in~\eqref{largemassvacua}, where the~$U(1)_m$ symmetry that acts on the~$|m\rangle$ vortex is spontaneously broken by the monopole operator~$\CM^2$, while the~$|e\rangle$ vortex is heavy and the~$U(1)_e$ symmetry is unbroken. Duality~$\sf D$ leads to analogous~$e \leftrightarrow m$ swapped picture at negative~$m_3 < 0$.

\subsection{Particle-Vortex Duality and the $O(2)_{e,m}^*$ Second-Order Lines}
\label{section: particlevortex and O(2)*}

In this section we consider the transitions between the~$\Z_2$ TQFT phase and the~$U(1)_{e,m}$ symmetry-breaking phases in figure~\ref{fig:HYQEDphasediagbis}. We will use a variant of particle-vortex duality to argue that the transitions are second-order and in the~$\Z_2$-gauged~$O(2)$ universality class, termed~$O(2)^*$. We will be able to argue this convincingly far away from the multicritical point, where the fermions are always heavy and can be integrated out. This is the regime of the~$U(1)$-symmetric Fradkin-Shenker lattice model that is under good theoretical control and leads to~$O(2)^*$ transitions, as discussed in section~\ref{o2starlat}. We conjecture that  the~$O(2)^*$ lines in HYQED remain second-order and continue all the way to the multicritical point, and it is the reason we take the Higgs quartic coupling~$\lambda_4$ to be large (see below). As we discuss in~\ref{FSviaMonos}, these transitions turn into the~Ising$^*$ transitions of the Fradkin-Shenker lattice model, which do indeed extend to its multicritical point. 

Let us take~$m_3 > 0$ and focus on the~$O(2)_m^*$ transition in figure~\ref{fig:HYQEDphasediagbis}. (The corresponding~$O(2)_e^*$ line at~$m_3 < 0$ is related by the action of the duality symmetry.) Let us now assume that~$m_3$ is large and dial the Higgs mass-squared~$m_\phi^2$ from large positive to large negative values. Recall from~\eqref{induced_fermion_mass} that the fermion mass is always bounded from below by~$m_3$, i.e.~the fermions remain heavy throughout. We are thus justified in integrating them out to the obtain the effective Lagrangian~\eqref{Leff_scalar}, which we repeat here, 
\begin{equation}\label{Leff_scalar_2}
\mathscr{L}_{\rm eff} = - \frac{1}{4e^2} f^{\mu\nu}f_{\mu\nu}
- |D_{2a}\phi|^2 - m_{\phi, \text{eff}}^2 |\phi|^2 -  \lambda_{4, \text{eff}} |\phi|^4 -\frac{1}{2\pi}  da \wedge A_m~.
\end{equation}
Here the effective Higgs mass and quartic coupling are given by
\begin{equation}
    m_{\phi, \text{eff}}^2 = m_\phi^2 + \frac{y^2|m_3|}{2\pi}~, \qquad \lambda_{4, \text{eff}} = \lambda_4 + \frac{y^4}{8\pi |m_3|}~.
\end{equation}

Several comments are in order:
\begin{itemize}
    \item The transition occurs at~$m_{\phi, \text{eff}}^2 = 0$, where~$m_\phi^2 \sim - y^2 |m_3|$, which is indeed far from the multicritical point. 

    \item The fact that~$\lambda_{4,\text{eff}} > \lambda_4 \gg e^2$ ensures that this transition is second-order, i.e.~the Abelian Higgs model~\eqref{Leff_scalar_2} describes a type II superconductor, whose vortices have repulsive interactions (see section~\ref{sec:pvreview} below). 

    \item Since we integrated out the unit-charge fermions and the bosonic model~\eqref{Leff_scalar_2} only has a charge-2 Higgs field, it possesses a~$\Z_2^{(1)}$ electric one-form symmetry that is not present in HYQED. However, as already discussed in other phases above, this symmetry is essential for matching the 't Hooft anomalies of HYQED via fractionalization of its zero-from symmetries (i.e.~it is the result of symmetry transmutation). 
\end{itemize}
Before we learn to apply particle-vortex duality to~\eqref{Leff_scalar_2}, which is formulated in terms of the spin$_c$ connection~$a$ and a charge-2 Higgs field, we review the more conventional case.

\subsubsection{Review of Particle-Vortex Duality}\label{sec:pvreview}

Particle-vortex duality~\cite{Peskin:1977kp,Dasgupta:1981zz}  (see~\cite{Seiberg:2016gmd} for a more recent review) relates the Abelian Higgs model (AHM), formulated in terms of a~$U(1)$ connection~$b$, with standard~$2 \pi \Z$ fluxes, and a single Higgs field~$h$ of charge~$+1$, to the~$O(2)$-symmetric theory of a single complex scalar field~$\Phi$ without gauge interactions, which we will refer to as the~$O(2)$ model.\footnote{~Whether or not this model flows to the~$O(2)$ Wilson-Fisher CFT in the IR depends on the choice of parameters, see below.} If we dial to the massless point, the duality takes the form 
\begin{equation}\label{fullPVlags}
\begin{split}
   \mathscr{L}_\text{AHM}& =   -{1 \over 2 e^2_b} |db|^2 -|D_b h|^2  - \lambda_{4, h}|h|^4 - \frac{1}{2\pi} A \wedge db~, \qquad \lambda_{4, h} > 0~, \\[5pt]
   & \text{Particle-Vortex Duality: } \bigg\updownarrow  \\[5pt]
   \mathscr{L}_{O(2)} & =  -|D_{A} \Phi |^2 - \lambda_{4, \Phi} |\Phi|^4 -  |\Phi|^6~, \qquad \lambda_{4, \Phi} \in \R~.
\end{split}
\end{equation}
Let us make several comments:
\begin{itemize}
    \item Both theories of have~$O(2) = U(1)_A\rtimes \Z_2^{\CC_A}$ global  symmetry, under which~$\Phi$ has charge~$+1$, while~$\CC_A : \Phi \to \Phi^*$. In the AHM,~$U(1)_A$ is the monopole symmetry, so that the charge~$+1$ monopole operator~$\CM_b$ creates a~$ 2\pi$ Dirac flux for~$b$ (and a corresponding vortex in the Higgs phase, see below).  Meanwhile, charge conjugation acts as~$\CC_A : b \to -b, h \to h^*$, and hence~$\CM_b \to \CM_b^*$. Thus the duality identifies
    \begin{equation}\label{Misphi}
        \CM_b \leftrightarrow \Phi~.
    \end{equation}
    This is the origin of the name particle-vortex duality. Note that the~$U(1)_A$ background gauge field~$A$ is a standard~$U(1)$ connection, with~$2 \pi \Z$ fluxes. 
  \item  The AHM has time-reversal symmetry,\footnote{~Since~$\CT_\text{AHM}$ preserves the gauge charge, we can absorb any phase in the~$h$-transformation by conjugating with the gauge symmetry.}
  \begin{equation}\label{T_AHM}
      \CT_\text{AHM} : \quad b_0 \to b_0~, \qquad b_{1,2} \to -b_{1,2}~, \qquad h \to h~, \qquad \CM_b \to \CM_b^*~,
  \end{equation} 
  whose action on the dual~$O(2)$ theory can be inferred from~\eqref{Misphi}. 
  \item The mapping of symmetry-preserving mass terms is
  \begin{equation}\label{massduality_PV}
    m^2_h |h|^2  \leftrightarrow -m^2_\Phi |\Phi|^2~.
\end{equation}
This is because the~$m_h^2 > 0$ phase of the AHM is a Coulomb phase, where~$\CM_b$ acquires a~$U(1)_A$ symmetry-breaking vev, corresponding to the~$m_\Phi^2 < 0$ phase in the~$O(2)$ model. By contrast, the AHM Higgs phase~$m_h^2 < 0$ is gapped and trivial in the IR, just as the~$m_\Phi^2 > 0$ phase in the~$O(2)$ model. In this phase the~$U(1)_A$ symmetry is unbroken, and~$\CM_b$ creates massive vortex excitations, which are described by~$\Phi$ in the~$O(2)$ dual.  

\item The~$O(2)$ model also has~$U(1)_A$-breaking deformations~$\Phi^n$. Thanks to~\eqref{Misphi} they correspond to deformations of the AHM Lagrangian by monopole operators~$\CM_b^n$, which will be important in section~\ref{FSviaMonos}. 

  \item The AHM depends on a dimensionless parameter~$\lambda_{4, h} / e^2_b$. Among other things, this parameter controls whether vortices in the Higgs phase attract (corresponding to a type I superconductor) or repel (corresponding to type II).\footnote{~See~\cite{Dumitrescu:2025fme} for a recent discussion with references.} Since the dual scalar~$\Phi$ is the vortex field, the sign of its quartic coupling~$\lambda_{4, \Phi}$ captures the same effect:
  
  In the type II regime, where~$\lambda_{4, h} \gtrsim e^2_b$ and the vortices repel, we have positive dual quartic coupling~$\lambda_{4, \Phi} > 0$, which leads to the interacting~$O(2)$ Wilson-Fisher CFT in the IR (upon fine-tuning the mass~\eqref{massduality_PV} to the critical point). In this regime the sextic coupling in the~$O(2)$ model can be safely omitted. 
  
  By contrast, the type~I regime has~$e_b^2 \gtrsim \lambda_{4, h}$ so that the vortices attract and~$\lambda_{4, \Phi} < 0$. This leads to a first-order phase transition as the mass~\eqref{massduality_PV} is dialed. In this case the sextic coupling~\eqref{fullPVlags} stabilizes the vacuum and cannot be neglected.  
\end{itemize}

\subsubsection{$\Z_2$ Gauging and the~$O(2)^*$ Theory}\label{sec:z2gaugeo2}

We define the~$O(2)^*$ theory to be the result of gauging a~$\Z_2 \subset U(1)_A$ symmetry in the~$O(2)$ model, which appears on one side of particle-vortex duality~\eqref{fullPVlags}. We thus decompose the background~$U(1)$ gauge field~$A$ into flat~$\Z_2$ connections $c_1 \in H^1(\cM_3,\Z_2)$, which are summed over, and a remaining rescaled~$U(1)$ connection~$\t A$ with standard $2\pi\mathbb{Z}$ fluxes,\footnote{~The constraint that~$c_1$ be a flat~$\Z_2$ chain can be enforced by a Lagrange multiplier, as in our discussion of~$\Z_2$ gauge theory (see section~\ref{sec:z2gaugereview}).}
\begin{equation}\label{AiscAt}
    A = \pi c_1 + \half \t A~.
\end{equation}
Substituting into the~$O(2)$ Lagrangian~\eqref{fullPVlags} leads to the Lagrangian of the~$O(2)^*$ model,\footnote{~Here we are writing the cup product for~$\Z_2$ classes with the understanding that it is integrated over all of spacetime~$\CM_3$.}
\begin{equation}\label{O(2)*Lagrangian}
\mathscr{L}_{O(2)^*}=-\left|D_{c_1 + \t A/2} \Phi\right|^2 - \lambda_{4, \Phi} |\Phi|^4 - |\Phi|^6 + \pi B_2^E \cup c_1~.
\end{equation}
Several comments are in order:
\begin{itemize}
    \item $\Phi$ is not gauge invariant and must be attached a Wilson line for the~$\Z_2$ gauge field~$c_1$.
    \item Relatedly, $\Phi$ carries charge~$\half$ under the conventionally normalized~$U(1)_{\t A}$ global symmetry, with background~$U(1)$ connection~$\t A$. All gauge-invariant operators, e.g.~$\Phi^2$, have integer~$U(1)_{\t A}$ charges.
    \item The flat~$\Z_2$ Wilson lines generate a~$\Z_2^{(1)}$ one-form symmetry, with background field $B_2^E \in H^2(\cM_3,\Z_2)$.
\end{itemize}

We can carry out the same gauging~\eqref{AiscAt} on the AHM side of particle-vortex duality~\eqref{fullPVlags}. Then integrating out~$c_1$ enforces the constraint that the fluxes of~$b$ are in~$4 \pi \Z$, rather than~$2 \pi \Z$. We thus introduce a conventionally normalized~$U(1)$ connection~$\t b = {b /2}$, in terms of which the~$\Z_2$ gauged AHM (or AHM$^*$) Lagrangian~\eqref{fullPVlags} becomes
    \begin{equation}\label{AHMstar}
        \mathscr{L}_{\text{AHM}^*} = -{2 \over e_b^2} \left|d \t b - \pi B_2^E \right|^2 - |D_{2 \t b} h|^2 - \lambda_{4, h} |h|^4 - {1 \over 2\pi} \t A \wedge d\t b~.
    \end{equation}
Note that the minimal local monopole operator~$\CM_{\t b}$ of the AHM$^*$ model has~$U(1)_{\t A}$ charge~$+1$, and it is dual to the gauge-invariant operator~$\Phi^2$ in the~$O(2)^*$ model,
\begin{equation}
    \CM_{\t b} \longleftrightarrow \Phi^2~.
\end{equation}
By contrast, the gauging procedure attaches a~$\Z_2$ Wilson line to the AHM monopole~$\CM_b$, which is now dual to the non-gauge-invariant~$O(2)^*$ field~$\Phi$. Finally, there is an electric~$\Z_2^{(1)}$ one-form symmetry, with background field~$B_2^E \in H^2(\cM_3,\Z_2)$, because~$h$ now has~$\t b$-charge~$2$. 

The local dynamics of the particle-vortex dual theories are not changed by the~$\Z_2$ gauging, but global features,  including symmetries and anomalies, are modified. In particular, both dual theories now have a mixed 't Hooft anomaly for the~$U(1)_{\t A}^{(0)}$ zero-form and~$\Z_2^{(1)}$ one-form symmetries, captured by 
\begin{equation}\label{o2*anomaly}
    S_\text{$O(2)^*$ Anomaly} = \pi \int_{\CM_4} B_2^E \cup \left[{d\t A \over 2\pi} \right]_2~, \qquad \partial\mathcal{M}_4=\mathcal{M}_3~, 
\end{equation}
so that no symmetry-preserving RG flow can lead to a trivially gapped phase.

As before, turning on~$m_h^2 = - m_\Phi^2 > 0$ leads to a gapless phase where~$U(1)_{\t A}$ is spontaneously broken. Using the AHM$^*$ description, this phase is Coulomb phase where the~$\Z_2^{(1)}$ one-form symmetry is enhanced to the electric~$U(1)_E^{(1)}$ one-form symmetry of Maxwell theory, and the anomaly~\eqref{o2*anomaly} is matched by the Maxwell anomaly~\eqref{maxanom}. 

By contrast, the~$m_h^2 = - m_\phi^2 < 0$ phase is gapped, but now harbors a~$\Z_2$ topological gauge theory (reviewed in section~\ref{sec:z2gaugereview}). Using the~$O(2)^*$ description, we drop the massive~$\Phi$ field in~\eqref{O(2)*Lagrangian} to find the~$\Z_2$ gauge field~$c_1$ coupled to the~$\Z_2^{(1)}$ background~$B_2^E$. However, the~$c_1$ Wilson line also carries fractional~$U(1)_{\t A}$ charge~$\half$, since it describes the massive~$\Phi$ particle. We thus identify the electric and magnetic one-form symmetry background gauge fields of~$\Z_2$ gauge theory in~\eqref{Bcouplings} as
\begin{equation}
    B_2^e = B_2^E~, \qquad B_2^m = \left[{d\t A \over 2\pi} \right]_2~.
\end{equation}
Substituting into the 't Hooft anomaly~\eqref{BBanom} of the~$\Z_2$ TQFT we match~\eqref{o2*anomaly}.

\subsubsection{The~$O(2)_{e,m}^*$ Transitions of HYQED}

Let us return to to the effective Lagrangian~\eqref{Leff_scalar_2}, 
\begin{equation}\label{Leff_scalar_3}
\mathscr{L}_{\rm eff} = - \frac{1}{2e^2} |da|^2 
- |D_{2a}\phi|^2 - m_{\phi, \text{eff}}^2 |\phi|^2 -  \lambda_{4, \text{eff}} |\phi|^4 -\frac{1}{2\pi}  da \wedge A_m~.
\end{equation}
We claim that it describes the~$O(2)_m^*$ transition at~$m_3 > 0$ in figure~\ref{fig:HYQEDphasediagbis}, being almost identical with the AHM$^*$ model in~\eqref{AHMstar}, with~$h = \phi$ and~$\t A = A_m$, except that the latter is formulated in terms of an ordinary~$U(1)$ connection~$\t b$, while~$a$ is a generalized spin$_c$ connection with fluxes satisfying~$\oint da = \pi \oint w_2(T\CM_3) + \oint dA_f \mod 2 \pi \Z$ (see~\eqref{modified_spinc_relation}). This can be captured by setting (see the related discussion around~\eqref{BEinflow})
\begin{equation}\label{BEfmla}
    B_2^E = w_2(T\CM_3) + {dA_f \over \pi}~, 
\end{equation}
in~\eqref{AHMstar}. Together with the $O(2)^*$ anomaly~\eqref{o2*anomaly} and~$\t A = A_m$, this matches the 't Hooft anomaly of HYQED.

In the particle-vortex dual~$O(2)^*$ description~\eqref{O(2)*Lagrangian} we find that the vortex field~$\Phi$, which is attached to a Wilson line of the~$\Z_2$ gauge field~$c_1$, carries~$U(1)_e \times U(1)_m$ charges~$(q_e, q_m) = (0, \half)$. It therefore precisely describes the~$|m\rangle$ vortex in the~$\Z_2$ Higgs phase of HYQED (see section~\ref{sec:vortices}), which becomes lighter as we approach the~$O(2)_m^*$ line at~$m_3 > 0$ in figure~\ref{fig:HYQEDphasediagbis} (see the discussion around~\eqref{lightvortex}). When this vortex condenses, i.e.~when~$\Phi$ acquires a vacuum expectation value~$\langle \Phi\rangle \neq 0$, then the~$\Z_2$ gauge field~$c_1$ is Higgsed and the~$U(1)_m$ symmetry is spontaneously broken; the gauge-invariant order parameter is~$\Phi^2 \sim \CM^{i = 2}$. This matches the~$U(1)_m$ spontaneous-symmetry-breaking Coulomb phase in figure~\ref{fig:HYQEDphasediagbis}. Since~$\lambda_{4, \text{eff}} \gg e^2$ in~\eqref{Leff_scalar_3}, i.e.~we are in the type II regime, the dual~$O(2)^*$ model~\eqref{O(2)*Lagrangian} has positive~$\lambda_{4, \Phi} >0$, so that the two phases are separated by a second-order transition described by the~$\Z_2$-gauged~$O(2)$ Wilson-Fisher CFT.  

Recall from section~\ref{sec:vortices} that the~$|m\rangle$ vortex described by~$\Phi$ is a boson, and corresponds to the~$m$ anyon of the low-energy~$\Z_2$ TQFT in the Higgs phase. Indeed the gauge field~$c_1$ to which it couples is an ordinary flat~$\Z_2$ connection and can be identified with the magnetic~$\Z_2$ gauge field~$m_1$ in the TQFT Lagrangian~\eqref{z2tqft}.

Let us now also discuss the fate of the~$e$ and~$f$ anyons in the TQFT phase, where the $\Z_2$ gauge field~$c_1$ is not Higgsed. These are described by~$\Z_2$ vortices of~$c_1$. It follows from the topological coupling in~\eqref{O(2)*Lagrangian} that such a vortex line is attached to a~$B_2^E$ surface, and thanks to the particular background~\eqref{BEfmla} this line is a fermion of~$U(1)_f$ charge~$+1$, i.e.~it describes the fermion~$f$. The~$e$ boson can then be viewed as an~$m f$ composite (see section~\ref{sec:z2gaugereview}). 

Recall from sections~\ref{sec:SFcoul} and~\ref{sec:SFtqft} that both the Maxwell and~$\Z_2$ gauge theories described there enjoy a self-duality under the relabeling~\eqref{linesmatch} of the line defects, which is captured by the shifting the background fields as in~\eqref{Bshiftmax} (with~$A_M = A_m)$,
\begin{equation}
    B_2^E \to B_2^E + w_2(T\CM_3) + {d A_m \over 2\pi}~.
\end{equation}
Adding this to~\eqref{BEfmla}, and using the fact that~$2 A_f = A_e - A_m$ (see~\eqref{Aem_definition}), we find that the~$O(2)_m^*$ transition is alternatively descried by an~AHM$^*$ model~\eqref{AHMstar} with
\begin{equation}\label{newB}
    B_2^E = {dA_f \over \pi} + {d A_m \over 2\pi} = {d A_e \over 2\pi} \mod 2~,
\end{equation}
i.e.~it is formulated in terms of a bosonic~$U(1)$ connection~$\t b$ that carries fractional~$U(1)_e$ charge~$\half$, which precisely describes the~$e$ anyon in the~$\Z_2$ TQFT phase. In the dual~$O(2)^*$ model~\eqref{O(2)*Lagrangian} the same anyon is now described by the vortex of the~$\Z_2$ gauge field~$c_1$, thanks to~\eqref{newB}. In this presentation the fermion must be viewed as the composite~$f = em$.

\medskip

\noindent Let us make some further comments:
\begin{itemize}
    \item So far we have described the~$O(2)_m^*$ transition at~$m_3 > 0$. It is related by duality~${\sf D}$ to the analogous~$O(2)_e^*$ transition at~$m_3 < 0$, where the~$|e\rangle$ vortex that is present in the~$\Z_2$ TQFT phase becomes massless. 
    \item Above we have used particle-vortex duality to describe the massless~$|m\rangle$ vortex on the~$O(2)_m^*$ line by the~$\Z_2$-gauged complex scalar field~$\Phi = \Phi_m$. Note that the~$|e\rangle$ vortex is heavy there. Similarly, applying particle-vortex duality to the~$O(2)_e^*$ transition (where~$|e\rangle$ is massless and~$|m\rangle$ is heavy) leads to a description in terms of a~$\Z_2$-gauged complex scalar~$\Phi_e$. These two description are mutually non-local, because they describe particles with~$-1$ mutual statistics in the TQFT phase. Equivalently, $\Phi_e$ couples to~$e_1$ and~$\Phi_m$ couples to~$m_1$ in the TQFT Lagrangian~\eqref{z2tqft}. Note that on the self-dual line~$m_3 = 0$ in the TQFT phase both~$\Phi_e$ and~$\Phi_m$ are massive, and they are exchanged by the duality symmetry~$\sf D$, i.e.~they are degenerate. 
    
    This picture has some similarities with~$SU(2)$ Seiberg-Witten theory~\cite{Seiberg:1994rs}, i.e.~$\CN=2$ supersymmetric~$SU(2)$ gauge theory in 3+1 dimensions. There is a continuous moduli space of vacua, all of which are described by~$U(1)$ gauge theory in the deep IR. There are two special monopole and dyon points, where one of these two mutually non-local particles becomes massless, while the other one remains heavy. The two points are exchanged by a global symmetry of the theory,\footnote{~It is a~$\Z_8$ zero-form $R$-symmetry, see~\cite{Seiberg:1994rs} and the more recent discussion in~\cite{Cordova:2018acb}.} which acts as electric-magnetic duality in the IR. At the self-dual point both particles are massive and degenerate.

    \item The discussion in this section is quite robust as long as we stay sufficiently far away from the multicritical point (see the discussion below~\eqref{Leff_scalar_2}). We conjecture that the second-order~$O(2)_{e,m}^*$ transition extend all the way to the multicritical point (see~\ref{FSviaMonos} for some evidence for this). At this point the mutually non-local~$\Phi_e$ and~$\Phi_m$ fields become massless simultaneously and the description in terms of~$O(2)_{e,m}^*$ degrees of freedom breaks down, so that we must use the full HYQED Lagrangian.
    
    Further extending the analogy to 3+1d~$\CN=2$ supersymmetric theories, the multicritical point is analogous to an Argyres-Douglas theory~\cite{Argyres:1995jj, Argyres:1995xn}, where the monopole and the dyon (and generally also other charged particles) become massless simultaneously. 
    
    \item Recall that we argued based on the jump~\eqref{CSjump} in the mixed Chern-Simons term for the flavor gauge field~$A_f$ and the dynamical gauge field~$a$ that there must be at least one phase transition separating the~$U(1)_e$ and~$U(1)_m$ phases at large $|m_3|$. At negative~$m_\phi^2 < 0$ there are the two~$O(2)_{e,m}^*$ transitions, with an intervening~$\Z_2$ TQFT phase. This picture is only consistent if the jump~\eqref{CSjump} trivializes in TQFT phase, where~$a \to  \pi[a]_2$ is reduced to a two-cochain, as discussed around~\eqref{amod2}. This leads to an apparent contradiction because~$A_f$ can have~$\pi \Z$ fluxes, whose resolution requires a more careful treatment of the fermion path integral -- and in particular its phase, which is related to Chern-Simons terms. In section~\ref{secetainvariants} we reformulate all Chern-Simons terms in terms of~$\eta$-invariants of suitable Dirac operators and show that the~$\Z_2$ Higgs phase is indeed smooth. 
\end{itemize}

\section{Fermion Path Integrals and~$\eta$-Invariants}\label{secetainvariants}

In this section, we explain how to carefully define the fermion path integral in $N_f = 2$ QED$_3$ and in HYQED using~$\eta$-invariants. (See for instance~\cite{Witten:2015aba, Seiberg:2016rsg,Seiberg:2016gmd} for related discussions.) As an application, we will write down a manifestly self-dual action for the $\bZ_2$ TQFT in the Higgs phase of HYQED. 

Recall that one Majorana fermion $\chi$ in 2+1 dimensions is a real, two-component spinor, which admits a single, real, Lorentz-invariant mass term $(im \chi^t \gamma^0 \chi+\text{c.c.})$. Its Euclidean partition function may be computed using a Pauli-Villars regulator field with a mass $M$. In this case, we find that the phase of the partition function is given by~\cite{Dai:1994kq,Witten:2015aba, Seiberg:2016rsg,Seiberg:2016gmd,Witten:2019bou}
\[Z_{\rm Majorana}(b) = |Z_{\rm Majorana}(b)|  \exp\left(i\frac{\pi}{4}(\sign(m) - \sign(M)) \eta(D_b)\right) \,,\]
where $D_b$ is the minimal (Euclidean, Hermitian) Dirac operator defined by the spin structure~$b$ that couples to $\chi$, which acts on two-component complex spinors.

Many familiar properties of induced Chern-Simons terms can be reformulated in terms of the $\eta$-invariant. For example, if we have a single Dirac fermion in 2+1 dimensions, now coupled to a spin$_c$ structure $b$, its partition function evaluates to
\[Z_{\rm Dirac}(b) = |Z_{\rm Dirac}(b)|
    \exp\left(i\frac{\pi}{2}(\sign(m) - \sign(M)) \eta(D_b)\right) \,, \]
where now $D_b$ is defined by the spin$_c$ structure $b$. In the non-trivial case, where~$m$ and~$M$ have opposite signs, we can express $e^{\pm i \pi \eta(\cD_b)}$ via the APS index theorem (see section~12.8.2 of~\cite{Nakahara:2003nw}),
\[\label{eqnAPSindex}\eta(D_b,\mathcal{M}_3) = \int_{\mathcal{M}_4} \left(\frac{db}{2\pi}\wedge\frac{db}{2\pi}-\frac{1}{8} \sigma\right) - 2{\rm Ind}(\tilde D_b) \,,\]
where the closed $3d$ spacetime~$\mathcal{M}_3$ is written as the boundary $\partial \mathcal{M}_4 = \mathcal{M}_3$ of a 4-manifold $\mathcal{M}_4$; the spin$_c$ structure $b$ is extended to $\mathcal{M}_4$, defining the Dirac operator $\tilde D_b$ and its index ${\rm Ind}(\tilde D_b) \in \bZ$; and $\sigma$ is a function of the metric curvature that integrates to the signature on a closed 4-manifold. The result is that we can express $e^{\pm i\pi \eta(D_b)}$ as a level $\pm 1$ Chern-Simons term for $b$ (plus a gravitational Chern-Simons term), which agrees with the familiar perturbative statement of the one-loop correction to the Chern-Simons level that we used in~\eqref{CSjump}. In particular, with our sign convention, for either sign of the regulator mass $M$, changing from~$m < 0$ to $m > 0$ increases the Chern-Simons level by $+1$.

\subsection{$p \pm i p$ Superconductors}

As a warm-up, let us consider the closely related example of a $p \pm ip$ superconductor, which we treat as a $\bZ_2$ Higgs phase of a dynamical spin$_c$ structure $b$.\footnote{~In many sources, the $p \pm ip$ superconductor is described as an invertible spin TQFT. Here we will obtain a non-invertible bosonic TQFT, which is related to the invertible theory by making the spin structure dynamical. This spin structure is the un-Higgsed part of the ``electromagnetic'' gauge field $b$.} We begin with a Dirac fermion $\psi$ carrying charge 1 under $b$. We also include a scalar $\phi$ carrying charge 2 under $b$. We will couple $\phi$ to $\psi$ while preserving gauge invariance. We can be concrete by writing $\psi$ in its real and imaginary parts,
\[\psi = \chi^+ + i \chi^- \,,\]
where $\chi^\pm$ are a pair of Majorana fermions, transforming in a real charge-1 $U(1)$ doublet.\footnote{~Besides the trivial representation, there is just one real 2d irrep of $SO(2)$ for each positive integer $n$, given by $e^{n \theta i \sigma_y}$. This representation is equivalent to $e^{-n \theta i \sigma_y}$ by conjugating by $\sigma_z$. When we complexify this representation, it splits as a sum of the complex 1d representations $e^{n i \theta}$ and $e^{-ni\theta}$ since $i\sigma_y$ has eigenvalues~$\pm i$.} We can write the following pair of Hermitian mass terms,
\[i(\bar \chi^+ \chi^+ - \bar \chi^- \chi^-) \,, \\
i(\bar \chi^+ \chi^- + \bar \chi^- \chi^+) \,,
\]
which transform as a real charge-2 doublet ($\bar \chi = \chi^t \gamma^0$). We couple these to the charge-2 complex scalar $\phi = \phi_1 + i \phi_2$ via
\[ i\phi_1 (\bar \chi^+ \chi^+ - \bar \chi^- \chi^-) + i\phi_2(\bar \chi^+ \chi^- + \bar \chi^- \chi^+) \,.\]

Now we drive the system into a regime where $\phi$ condenses, which leads to $b$ being Higgsed. This is different from the familiar Higgs mechanism, since $b$ is a spin$_c$ structure, and defines a principal bundle of spin$_c$ frames with structure group $({\rm Spin}(3) \times U(1)_b)/\bZ_2$, with the $\bZ_2$ quotient identifying the central $\bZ_2$ subgroup of each group. A non-zero section $\langle \phi \rangle$ of the complex line bundle with connection $2b$ allows us to reduce the structure group of the above frame bundle to $({\rm Spin}(3) \times \bZ_2)/\bZ_2 = {\rm Spin}(3)$. Thus, $b$ is Higgsed to a spin structure.

We may fix the gauge choosing $\phi_2 = 0, \phi_1>0$ and then integrate out the massive fermions. We need to use a $U(1)$-symmetric regulator to be compatible with gauge invariance, so we will include a Pauli-Villars regulator field for each $\chi^\pm$, and give all such regulators the same mass~$M$. With this choice, only the Majoranas above whose mass has the opposite sign of $M$ contribute to the universal part of the phase of the path integral, yielding the following effective path integral weight for the spin structure $b$,
\[Z(b) = \exp\left(-i \sign(M) \frac{\pi}{2} \eta(b)\right)|Z(b)|\,.\]
Only the phase is important for identifying which TQFT this describes. This phase is gauge invariant on a 3-manifold without boundary, but on a manifold with boundary it must be treated carefully. The Dai-Freed theorem \cite{Dai:1994kq,Witten:2019bou} explains how it becomes gauge invariant when combined with a certain Pfaffian, which is the partition function of a massless chiral Majorana fermion on the boundary, having odd charge under $b$ (see also section 3.6.4 of \cite{Seiberg:2016rsg}). From this observation, we can identify all the familiar features of this TQFT, which we refer to as the Ising TQFT for $M > 0$ and $\bar{\text{Ising}}$ for $M < 0$, which correspond to $\nu = 1$ and $-1$ respectively in Kitaev's 16-fold way \cite{Kitaev:2005hzj}. 

\subsection{QED$_3$ with $\eta$-Invariants}

Now we will explain how to give a definition of $N_f=2$ QED$_3$ in the UV using $\eta$-invariants. For each fermion, we introduce a corresponding Pauli-Villars regulator fermion transforming in the same way under the global symmetries, and we choose for each of these fermions the same regulator mass $M < 0$. This choice of regulator preserves all the global symmetries except time reversal. Indeed, when we apply a time-reversal transformation, the regulator mass $M$ flips, producing a ratio of determinants whose phase is $e^{-2\pi i \eta(a)}$. We can thus restore time reversal (in the absence of background fields) via the counterterm
\[\label{eqncounterterm}\exp\left(-S_{\rm counterterm}(a)\right) = \exp\left(-i\pi  \eta(a)\right) \,,\]
which is well-defined and local for the spin$_c$ structure $a$.

When we introduce background fields $\mathcal{A}_f=\mathcal{A}_f^a\,\sigma^a$ and $A_\cM$ for the $U(2) = (SU(2)_f \times U(1)_\cM)/\bZ_2$ global symmetry, the counterterm above is not invariant under gauge transformations because of the modified quantization rules
\begin{equation}
\begin{split}
\oint \frac{da}{2\pi} &= \frac12 \oint w_2(T\cM_3) + \oint \frac{dA_\cM}{2\pi} \mod \bZ \,, \\
\oint \frac{dA_\cM}{2\pi} &= \frac12 \oint w_2(SO(3)_f) \mod \bZ \,,
\end{split}
\end{equation}
which are the generalization of \eqref{modified_spinc_relation} (see~\cite{Dumitrescu:2024jko}). Instead we can use the counterterm
\[\label{eqncounterterm2}\exp\left(-S_{\rm counterterm}(a,A_\cM)\right) = \exp\left(-i\pi \eta(a+A_\cM)\right) \,, \]
which is gauge invariant since $a+A_\cM$ is a spin$_c$ structure by the above (recall that $2A_\mathcal{M}$ has integer fluxes). This choice is compatible with the full $U(2)$ symmetry. In general, we could also choose the combination $-i\pi \eta(a+kA_\cM)$ for any odd integer $k$. Different choices of $k$ amount to different definitions of the magnetic symmetry in this model, since by the APS index theorem $-i\pi \eta(a + k A_\cM)$ includes the crossterm $- \frac{ik}{2\pi} \int_{\cM_3} da \wedge A_\cM$. Thus we choose $k = 1$ to match \eqref{lagrangianHY_coupledtobackgrounds}.

We can compute the anomaly by applying a time-reversal transformation $\cC\cT$ (that commutes with $SU(2)_f$), which flips the sign of the regulator mass, as well as takes $A_f \mapsto A_f$, $A_\cM \mapsto -A_\cM$, and together with the counterterm produces a phase
\[\exp\left(-i \pi \eta(S_{\mathcal{A}_f+a}) + i\pi \eta(a+A_\cM) + i\pi \eta (a-A_\cM)\right) \,,\]
where $S_{\mathcal{A}_f,a}$ is the four-component complex spinor bundle associated with the Spin-$U(2)$ structure $\mathcal{A}_f + a$. Applying the APS index theorem, we can rewrite this as
\[\exp \Bigg( i\pi \int_{\cM_4} \left( -{\rm Tr} \Big( \frac{d\mathcal{A}_f}{2\pi} + \frac{da}{2\pi} \Big)^2 + \Big(\frac{da}{2\pi} + \frac{dA_\cM}{2\pi} \Big)^2 + \Big(\frac{da}{2\pi} - \frac{dA_\cM}{2\pi} \Big)^2 \right) \Bigg) \\
= \exp \left( i\pi\int_{\cM_4} \left( 2\Big(\frac{dA_\cM}{2\pi}\Big)^2 - {\rm Tr} \Big(\frac{d\mathcal{A}_f}{2\pi}\Big)^2\right)\right) = \exp\left( 2\pi i \int_{\mathcal{M}_4}c_2(U(2)) \right) \,, \]
where the trace is in the fundamental of $U(2)$. This is the Chern-Weil density corresponding to the second Chern class of $U(2)$. Thus we find the anomaly-inflow action of QED$_3$ to be $S=\pi \int c_2(U(2))$ as in \cite{Dumitrescu:2024jko}. When we reduce the symmetry to $U(1)_f \subset SU(2)_f$ by the Yukawa term, with background field $A_f=\mathcal{A}_f^{a=3}$, we get ${\rm Tr}(d\mathcal{A}_f/2\pi)^2 = 2 (dA_f/2\pi)^2$, reproducing the anomaly computed in \eqref{mixedanomalyHYM}.

\subsection{Self-Dual $\bZ_2$ TQFT Phase}\label{sec:SDTQFT}

Now we consider the $\bZ_2$ TQFT phase. As we will see, it looks like a combination of the $p+ip$ and $p-ip$ superconductor examples above, which was part of the motivation for introducing HYQED in section \ref{subsecsuperconductorintuition}.

It is convenient to express the $U(2)$ flavor symmetry of the two Dirac fermions (of which the central $U(1)$ is the gauge symmetry) as part of $SO(4)$ acting on four Majorana fermions $\lambda$. The Yukawa term breaks this symmetry down to $U(1) \times O(2)$, which we can express in a real basis of $\bR^4 = \bR^2 \otimes \bR^2$ as
\[U(1)_a: e^{i\theta \sigma_y} \otimes \mathbbm{1}_2 \,,\\
U(1)_f: \mathbbm{1}_2 \otimes e^{i\theta \sigma_y} \,, \\
\cD: \mathbbm{1}_2 \otimes \sigma_z \,. \]
In this basis, the Yukawa coupling is
\[i \phi_1 \bar \lambda (\sigma_z \otimes \mathbbm{1}_2) \lambda - i \phi_2 \bar \lambda (\sigma_x \otimes \mathbbm{1}_2) \lambda \,.\]
When we go to the $\bZ_2$ phase, where $\phi$ obtains a vev, we can apply the reasoning above, fixing the gauge to where $\phi$ is real and positive, which yields fermions with the mass term
\[\label{eqneffectivemasshiggs}i |\langle \phi\rangle| \bar \lambda (\sigma_z \otimes \mathbbm{1}_2) \lambda \,, \]
coupled to $a$, which is now a spin structure. In particular, we have two Majorana fermions of positive mass and two of negative mass. We will choose the same regulator mass $M <0$ for all four Pauli-Villars fields, which carry the same symmetry transformations as the corresponding components of $\lambda$. This regulator preserves all flavor symmetries but breaks time-reversal symmetry, yielding the effective action
\[Z_{\rm fermion}(a) =\exp\left(i \pi \eta(a)\right) \,.\]
The result when combined with the counterterm \eqref{eqncounterterm} above is a trivial effective action for the dynamical spin structure $a$,
\[Z(a) = Z_{\rm fermion}(a) \exp\left(-S_{\rm counterterm}\right)= 1~,\]
yielding an untwisted~$\Z_2$ TQFT equivalent to the toric code phase. This reflects the fact that our theory in the TQFT phase is equivalent to a stack $(p + ip) \times (p-ip)$, which gives a trivial superconductor.

Now let us consider turning on background fields. First we do so for the duality, introducing the $\bZ_2$ gauge field $A_\cD$. We see from the effective mass term \eqref{eqneffectivemasshiggs} that of the two positive mass Majorana fermions which contribute an $\eta$-invariant for regulator mass $M < 0$, one is $\cD$ even and the other is $\cD$ odd. With the counterterm, we thus obtain the effective path integral weight for $a$,
\[\label{eqnselfdualtqftaction}
Z(a,A_\cD)=\exp\left(i\frac{\pi}{2}(\eta(a+A_\cD) + \eta(a)) - i \pi \eta(a)\right) = \exp\left(i\frac{\pi}{2}(\eta(a+A_\cD) - \eta(a))\right)\,.\]
We can see from this expression that $\cD$ acts as the $em$ duality for the toric code phase by making $A_\CD$ dynamical. In that case, we have a path integral over two independent spin structures $a$ and $a + A_\cD$, and comparing with the $p \pm ip$ example above, we obtain the non-Abelian TQFT ${\rm Ising} \times \bar{\rm Ising}$, which is well-known to be the result of gauging the $em$ duality \cite{Burnell:2017otf,Barkeshli:2014cna}. Furthermore, we can only obtain a non-Abelian theory from an Abelian one by gauging a symmetry which permutes anyons. Note also that the expression above, once we sum over the spin structure $a$, has time-reversal symmetry when we combine it with the redefinition of the summed variable $a \mapsto a + A_\cD$, reflecting the fact that there is no $\cD$ anomaly with time reversal (cf.~\eqref{eqnparentanomaly}).

Now let us turn off the $A_\cD$ background and instead turn on the $U(1)_f$ and $U(1)_\cM$ background fields $A_f$, $A_\cM$, satisfying the quantization rule \eqref{modified_spinc_relation}, which makes $a \pm A_f$ a spin$_c$ structure. From the effective mass \eqref{eqneffectivemasshiggs}, we see that the model consists of a positive mass Dirac fermion charged under $a + A_f$ and a negative mass Dirac fermion with the same charge. With the regulator mass $M < 0$, integrating out this fermion contributes
\[Z_{\rm fermion}(a,A_f)=\exp\left(i \pi \eta(a+A_f)\right) \,,\]
which together with the counterterm \eqref{eqncounterterm2} gives
\[\label{eqntqftetaU1}Z(a,A_f,A_\cM)=\exp \left(i \pi \eta(a+A_f) -i \pi \eta(a+A_\cM)\right) \,.\]
This expression enjoys the global symmetries $\cD$ and $\cC$, since in 3d $\eta(b) = \eta(-b)$ for any spin$_c$ structure $b$,\footnote{~This follows because the spin-1/2 representation of ${\rm Spin}(3)$ is complex conjugate to itself. So if $D$ is the Dirac operator on $S \otimes L$ where $L$ is a complex line bundle and $S$ is the spinor bundle, then $D^*$ is the Dirac operator on $S \otimes L^*$, and has the same spectrum, and hence $\eta$-invariant.} and $a = -a$ in the Higgs phase.

Including both duality and $U(1)_f$, $U(1)_\cM$ backgrounds, we have a model consisting of a pair of Majorana $O(2)_{f,\cD,a}$ doublets, where the reflection element of $O(2)$ is the duality, $SO(2)$ rotations in $O(2)_{f,\cD,a}$ correspond to flavor rotations, and unfixed $\bZ_2$ gauge transformations act as a $\pi$ rotation. One of these doublets has positive mass and the other has negative mass. We thus obtain the following path integral weight for $a$,
\[\label{eqntqftO2}Z(a)=\exp \left(\frac{i\pi}{2} \eta(O(2)_{f,\cD,a}) -i \pi \eta(a+A_\cM)\right) \,.\]
Here $\eta(O(2)_{f,\cD,a})$ is the $\eta$-invariant of the 3d Euclidean Dirac operator with structure group $({\rm Spin}(3) \times O(2)_{f,\cD,a})/\mathbb{Z}_2$, with the quotient identifying the $\bZ_2$ centers of each group. Note these 3d fermions carry four complex components transforming in a complex tensor product of the 2d spin-1/2 representation of ${\rm Spin}(3)$ and the 2d charge-1 representation of $O(2)_{f,\cD,a}$.

We can check this reduces to \eqref{eqntqftetaU1} and \eqref{eqnselfdualtqftaction} when the appropriate backgrounds are turned off. First, turning off $A_\cD$, the four-component complex fermion can be split into two two-component complex fermions coupled to $a \pm A_f$. We thus obtain in this case $\frac{\pi}{2} \eta(O(2)) = \frac{\pi}{2}(\eta(a+A_f) + \eta(a-A_f))$. As noted above, $\eta(a-A_f) = \eta(a+ A_f)$ in the Higgs phase, so this agrees with \eqref{eqntqftetaU1}. Likewise, turning off the $U(1)_f$, $U(1)_\cM$ backgrounds, we can split the four-component complex fermion as one two-component complex fermion coupled to $a + A_\cD$ and another coupled to $a$, giving \eqref{eqnselfdualtqftaction}.

The anomaly we computed above implies the fractionalization pattern described in section \ref{section:Z2Higgsphase}. We can see it a bit more directly from \eqref{eqntqftetaU1} as follows. Consider gauging the $\bZ_2^b$ subgroup generated by $(-1)^f = (-1)^\cM$. We can express this model without background fields via a pair of spin structures $a_1 = a$, $a_2 = a+b$, with the weight
\[Z(a_1,a_2) = \exp\Big(i \pi \eta(a_1) - i\pi \eta(a_2) + \frac{i}{2\pi} \int B \cup (a_1 + a_2)\Big) \,,\]
where $B$ is a background two-form $\bZ_2$ gauge field which we will later gauge to give us the original theory again. The theory of $a_1$ and $a_2$ separately describe $\nu = \pm 2$ in Kitaev's 16-fold way \cite{Kitaev:2005hzj}, which are equivalent to $U(1)_4$ and $U(1)_{-4}$ respectively. Promoting $B$ to a dynamical field corresponds to algebraic condensation of the boson $(2,2)$, which reduces $U(1)_4 \times U(1)_{-4}$ to the toric code, in agreement with our analysis above. In particular we may identify the orbit $(2,0) \sim (0,2)$ with the fermion and $(1,1) \sim (3,3)$, $(1,3) \sim (3,1)$ with the two bosons. The problem is thus reduced to understanding the fractionalization in this theory.

Upon gauging this subgroup, the global symmetry becomes $U(1)_{f'} \times U(1)_{\cM'}$ with integer charges $q_f' = q_f/2$ and $q_\cM' = q_\cM/2$. These are extended by $\bZ_2^{a_1}$ and $\bZ_2^{a_2}$, respectively. This means that the unit Wilson lines for $a_1$ and $a_2$, which are the anyons $(2,0)$ and $(0,2)$, carry $(q_f',q_\cM')$ charge $(1/2,0)$ and $(0,1/2)$, respectively. It follows (up to relabeling) that $(1,0)$ carries charge $(1/4,0)$ and $(0,1)$ carries charge $(0,1/4)$. When we gauge the one-form symmetry $(2,2)$, and recover the charges $q_e = q_\cM' + q_f'$, $q_m = q_\cM' - q_f'$, we find the bososn $(1,1) \sim (3,3)$ has $(q_e,q_m)$ charge $(1/2,0)$ and $(1,3) \sim (3,1)$ has $(q_e,q_m) = (0,1/2)$. This is the claimed fractionalization pattern of section \ref{section:Z2Higgsphase}.


\section{Matching Lattice and Continuum Descriptions}\label{sec:latcontmatch}

\subsection{Mapping Lattice Symmetries to the Continuum}\label{u1LatToCont}

In this section, we establish the map from the symmetries of the staggered Fradkin-Shenker (SFS) model (see section \ref{secU1FSmodel}) into those of the Higgs-Yukawa-QED theory (HYQED, see section \ref{sec:hyqed}). We will find that -- with some input from matching the phase diagram in the symmetry-breaking phases, as well as the existence of a certain lattice operator giving a deformation to the Fradkin-Shenker model -- this mapping is unique.

Generally speaking, the lattice-continuum correspondence at the level of symmetries is defined by a conjugacy class of group homomorphisms,
\begin{equation}
   \rho:G_\text{UV} \to G_\text{IR}~, 
\end{equation}
where $G_\text{UV}$ is the symmetry of the lattice model, and $G_\text{IR}$ is the symmetry of the continuum QFT.
This homomorphism is only defined up to conjugacy classes in $G_\text{IR}$, because one can always redefine the QFT operators $\mathcal{O}$ by the action of a symmetry of the continuum theory, $\mathcal{O} \rightarrow g\,\mathcal{O}$, where $g\in G_\text{IR}$ is a fixed element. Thus, if a lattice symmetry element $h\in G_{\text{UV}}$ is represented in the continuum as $\rho(h)$, after the redefinition above it acts as $\rho'(h)=g \rho(h) g^{-1}$. We must then regard $\rho(h)$ to be equivalent to the conjugate homomorphism defined by
\[h \mapsto \rho'(h)=g \rho(h) g^{-1} \,.\]
This conjugation equivalence will play an important role below. For simplicity of exposition, will only consider $G_\text{IR}$ as the quotient of the full symmetry group by the connected component of the identity of its Lorentz/conformal symmetries and the canonical CRT symmetry (all of these symmetries form a normal subgroup of the full symmetry group).

In our case $G_\text{UV} = {\sf G}_{\sf sd-lat}$ is the symmetry of the SFS model along the self-dual line, and $G_\text{IR}$ is some group $G_\text{SFS CFT}$ containing $G_\text{HYQED}$ as a subgroup, as well as some symmetries that emerge at the multicritical point. We will argue below that a minimal such extension of $G_\text{HYQED}$ is
\[\label{eqnIRsymmetry}G_\text{SFS CFT} = (O(2)_e \times O(2)_m) \rtimes (\bZ_2^\mathsf{D} \times \bZ_2^\mathsf{T})~.\]
In particular, mirror elements such as $\mathsf{R}_v^{\pi/2}$ (see table \ref{table:latticesymmetries}), which negate precisely one sublattice charge, must show up in $G_\text{SFS CFT}$ because they act non-trivially on conserved currents, but on the other hand from the point of view of HYQED they exchange flavor and magnetic charges,\footnote{~We will call symmetries acting on the charges this way mirror symmetries, in analogy with similar symmetries that arise in supersymmetric dualities~\cite{Intriligator:1996ex}.} and thus cannot be manifest in HYQED. We must assume for consistency of our proposal that they emerge at the multicritical point. We will present some evidence for this in section~\ref{sec:emergemirr}. 

\begin{table}[t!]
    \centering
        \begin{tabular}{c | c | c c c c c}
        SFS Model  & HYQED  & $\CM^1$ & $\CM^2$ & $\CO$ & $\CO_1-i\CO_2$ & $\CO_3$ \\ \noalign{\vskip 2pt} \hline  \noalign{\vskip 2pt}
         $\mathsf{Q}_e$ & $q_e = \half (q_\CM + q_f)$ & $1$ & $0$ & 0 & 1 & 0 \\[2pt]
         $\mathsf{Q}_m$ & $q_m = \half (q_\CM - q_f)$ & $0$ & $1$  & 0 & $-1$ & 0 \\[2pt]
         $\mathsf{T}$ & $\CD\CC\CT$ & $\CM^1$ & $-\CM^2$ & $-\CO$ & $-\CO_1+i\CO_2$ & $\CO_3$ \\[2pt]
         $\mathsf{S}_{\hat{x}}, \mathsf{S}_{\hat{y}}$ & $\CC (-1)^{q_m}$ & $(\CM^1)^*$ & $-(\CM^2)^*$  & $\CO $ & $-\CO_1-i\CO_2$  & $\CO_3$ \\[2pt]
         $\mathsf{D}$ & $\CD i^{q_f}$ & $i \CM^2$ &$-i \CM^1$ & $\CO$ & $-\CO_1-i\CO_2$  & $-\CO_3$ \\[2pt]
        $\mathsf{C}_e$, $\mathsf{R}_{v}^{\pi/2}$ & \text{emergent} & $(\cM^1)^*$ & $\cM^2$ & $-\cO$ & $\tilde{\CM}^*$ & $\cO_3$ \\[2pt]
        $\mathsf{C}_m$, $\mathsf{R}_{p}^{\pi/2}$ & \text{emergent} & $\cM^1$ & $-(\cM^2)^*$ & $-\cO$ & $\tilde{\CM}$ & $\cO_3$
    \end{tabular}
    \caption{The mapping of symmetries from the Staggered Fradkin-Shenker (SFS) lattice model (with Hamiltonian~\eqref{eqnU1FSmodel} in section~\ref{secU1FSmodel}) to the Higgs-Yukawa-QED continuum field theory (with Lagrangian~\eqref{LagrangianMinkowski} in section~\ref{sec:hyqed}). This mapping is determined (up to conjugacy) by matching the phases where~$U(1)_{e, m}$ are spontaneously broken; and by demanding the existence of local Hamiltonian/Lagrangian deformations that have unit charges~$q_{e, m} = \pm 1$, but preserve all discrete lattice symmetries. We display the action of the symmetries on continuum monopole and fermion-mass operators (see table~\ref{tab:hyGonOps}). We denote the monopole operator with $q_\mathcal{M}=2$ and $q_f=0$ as $\tilde{\mathcal{M}}=\tilde{\mathcal{M}}^{a=3}$ (see \eqref{charge2monopoles}). Note that symmetries of the lattice model that do not act on the charges~${\sf Q}_{e, m}$ act trivially in continuum HYQED. By contrast, mirror symmetries that negate precisely one of the charges~(e.g.~$\mathsf{R}_v^{\pi/2}$ in table~\ref{table:latticesymmetries}), whose equivalence classes in $\mathsf{G_{sd-lat}/K}$ we express as $\mathsf{C}_e$ and $\mathsf{C}_m$, are not manifest in the HYQED Lagrangian in the UV and must emerge as accidental symmetries of the multicritical point in the IR (see section~\ref{sec:emergemirr}). The question marks in the last two rows are charge 2 monopole operators. The action of the mirror symmetries on $\cO$ and $\cO_3$ are determined by the duality in section \ref{sec:easy}, and the rest of the action by the mapping of lattice symmetries.}
    \label{tab:HYQEDlatticemapping}
\end{table}

In table \ref{tab:HYQEDlatticemapping}, we describe the images of a generating set of the subgroup $\mathsf{G}_\mathsf{sd-lat}' \subset \mathsf{G}_\mathsf{sd-lat}$ of non-mirror symmetries (see \eqref{eqnnonmirrorlatticesymm}) in $G_{\rm HYQED}$ under our chosen conjugacy class of $\rho$. In particular, the subgroup $\mathsf{K} \subset \mathsf{G}_\mathsf{sd-lat}$ of crystalline symmetries of the lattice model that preserve the $U(1)$ charges, equivalently those which preserve the sublattice structure of vertices and plaquettes shown in figure \ref{fig:sublatticestructure}, all act trivially in $G_{\rm HYQED}$. The full map $\rho:\mathsf{G}_\mathsf{sd-lat} \to G_\text{SFS CFT}$ extends the map above and gives an isomorphism
\[\mathsf{G}_\mathsf{sd-lat}/\mathsf{K} \cong G_\text{SFS CFT}\,.\]

Recall from the discussion around~\eqref{eqnlatticeselfdualsymmetry} that the SFS model has symmetry group along its self-dual line (i.e.~$h_e = h_m$~in~\eqref{eqnU1FSmodel}) given by
\begin{equation}
    {\sf G}_\text{\sf sd-lat} = (U(1)_e \times U(1)_m) \rtimes (\textbf{p4mD} \times \mathbb{Z}_2^\mathsf{T})~.
\end{equation}
By contrast, the manifest UV symmetry of the massless Higgs-Yukawa-QED (HYQED) Lagrangian~\eqref{LagrangianMinkowski} is given by~\eqref{fullGhypreview}, which we repeat here,
\begin{equation}\label{fullGhybis}
    G_\text{HYQED} =  {\left( \text{Pin}^-(2)_f  \rtimes \Z_4^\CT\right)\times \text{Pin}^-(2)_\CM \over \Z_2 \times \Z_2}~.
\end{equation}
Here~$\text{Pin}^-(2)_{f, \CM}$ is generated by~$U(1)_{f,\CM}$ (with charges~$q_{f, \CM} \in \Z$), together with reflection elements~$\CR_{f, \CM}$ satisfying~$\CR_{f, \CM}^2 = (-1)^{q_f, q_\CM}$. The two~$\Z_2$ quotients enforce the relations~$q_f \equiv q_\CM \mod 2$ and~$\CT^2 = (-1)^{q_f} = (-1)^{q_\CM}$. As explained around~\eqref{Rfdef} and~\eqref{RMdef}, the reflections~$\CR_{f, \CM}$ can be traded (up to~$U(1)$ rotations) for the unitary duality and charge-conjugation symmetries~$\CD$ and~$\CC$, which we will use below. Together with time reversal~$\CT$, they satisfy the relations~\eqref{CDTrels}, which we repeat here,
\begin{equation}\label{CDTrelsbis}
    \CC^2 = \CD^2 = 1~, \quad \CT^2 = (-1)^{q_f} = (-1)^{q_\CM}~, \quad \CC \CD = \CD \CC~, \quad \CC \CT = \CT \CC~, \quad \CD \CT = (-1)^{q_\CM} \CT\CD~. 
\end{equation}
We will also need the action of these symmetries on the~$U(1)$ charges in table~\ref{tab:hyGonQ}, 
\begin{equation}\label{CDTonqsbis}
    \CD~: (q_f, q_\CM) \to (-q_f, q_\CM)~, \quad \CC : (q_f, q_\CM) \to (-q_f, -q_\CM)~, \quad \CT : (q_f, q_\CM) \to (q_f, -q_\CM)~.
\end{equation}

\subsubsection{Mapping the~$U(1)_e \times U(1)_m$ Symmetries}

We want to identify the $U(1)_e$ and $U(1)_m$ SSB phases found in HYQED in section~\ref{subseccontinuumphasediagram} (see figure \ref{fig:HYQEDphasediagbis}) with the corresponding $U(1)_e$ and $U(1)_m$ SSB phases found on the lattice in section~\ref{o2starlat}. Note that these regions extend to the axes of the phase diagram, which are understood from both the lattice and the continuum point of view, where they correspond to the limit of large $h_{e,m}$ and large $|m_3|$, respectively. We thus choose to identify the $U(1)$ charges/generators as follows,
\[\label{Qqmap}
\mathsf{Q}_e \mapsto q_e = \half (q_\CM+q_f)~, \\
\mathsf{Q}_m \mapsto q_m = \half (q_\CM-q_f)~.\]
Recall that~$q_{e, m}$ are unconstrained integers, thanks to the relation~$q_f = q_\CM \mod 2$, reviewed below~\eqref{fullGhybis}. 

Different choices of images of the $U(1)$ generators $\mathsf{Q}_e$, $\mathsf{Q}_m$ sending $\pm 1$ charges on the lattice to $\pm 1$ charges in the continuum must be all physically equivalent, since they are conjugate by the UV symmetries $\mathsf{D}$, $\mathsf{T}$, $\mathsf{R}_{v,p}^{\pi/2}$ (see table \ref{table:latticesymmetries}). In terms of HYQED we can conjugate by $\CD$, $\CC$, and $\CT$ (but we are missing one conjugation by a ``mirror symmetry'', see below). Once we have made the choice above, we have fixed the freedom to conjugate by those symmetries that act on the charges. However, we are still free to conjugate by~$\mathcal{DCT}$, which is their unique combination that preserves both charges~$q_{e, m}$. We can further conjugate by the connected part of the symmetry group,
\begin{equation}
    {U(1)_f \times U(1)_\CM \over \Z_2}  = U(1)_e \times U(1)_m~,
\end{equation}
which also preserves the charges. Below we will proceed step-by-step to construct a representative of the conjugacy class of~$\rho$.

\subsubsection{Mirror Symmetries}\label{subsubsecmirrorsymmmatching1}

At this stage, we can already recognize an issue. On the lattice there are symmetries like $\mathsf{R}^{\pi/2}_v$ which negate only one of the charges, in this case $\mathsf{Q}_m$ (see table~\ref{table:latticesymmetries}). Such a symmetry in the continuum would necessarily act to exchange $q_f$ and $q_\CM$. But this symmetry would exchange monopoles of $(q_f,q_\CM)$ charge $(0,2)$ with fermion bilinears of charge $(2,0)$. Such a symmetry cannot be manifest in HYQED, i.e.~it must be an emergent self-duality of the theory at the multicritical point in the IR.  As already mentioned above, we will give some evidence for the emergence of mirror symmetry in sections~\ref{sec:emergemirr}, \ref{sec:easy}, and~\ref{sec:largeN} below. 

The subgroup of lattice symmetries ${\sf G}'_{\sf sd-lat} \subset {\sf G}_{\sf sd-lat}$ that maps into~$G_\text{HYQED}$ are those which are not mirror symmetries, which we will denote by
\[\label{eqnnonmirrorlatticesymm}{\sf G}'_{\sf sd-lat} = (U(1)_e \times U(1)_m) \rtimes (\sf{pmm} \times \mathbb{Z}_2^\mathsf{T})~.\]
Here $\sf{pmm}$ is the symmetry of the mini square lattice preserving the orientation of the stripe pattern in figure \ref{fig:sublatticestructure}. This group is generated by translations $\mathsf{S}_{\hat x}$ and $\mathsf{S}_{\hat y}$, $\mathsf{D}$, and the point group symmetries~$\mathsf{R}_v^\pi$, $\mathsf{R}_v^{\pi/2} \mathsf{M}_v^H$, and $\mathsf{R}_v^{\pi/2} \mathsf{M}_v^V$. In short, we will obtain a commutative diagram
\begin{equation}
\begin{tikzcd}
    {\sf G}_{\sf sd-lat} \arrow[r,"\rho"] & G_\text{SFS CFT} \\ {\sf G}_{\sf sd-lat}' \arrow[u] \arrow[r,"\rho'"] & G_\text{HYQED} \arrow[u]
\end{tikzcd}
\end{equation}
where the vertical maps are inclusions. The full symmetry~$G_\text{IR}$ of the multicritical CFT in the IR will  be a~$\bZ_2$ extension of $G_\text{HYQED}$ by an emergent mirror symmetry.

\subsubsection{The Kernel of the Symmetry Map}

Above we defined the subgroup ${\sf K} \subset {\sf G_{sd-lat}}$ of crystalline symmetries acting trivially on the $U(1)$ charges ${\sf Q}_{e, m}$. We consider also its subgroup ${\sf K}' \subset {\sf K}$ of \emph{orientation-preserving} symmetries. These are related by ${\sf K}' = {\sf K} \rtimes \bZ_2^{\mathsf{R}}$ where $\bZ_2^\mathsf{R}$, which is the equivalence class of $\mathsf{M}_v^H \mathsf{R}_v^{\pi/2}$, corresponding to reflection over the $\hat x  + \hat y$ axis.

We now argue that ${\sf K}'$ symmetries should act trivially in~$G_\text{SFS CFT}$, i.e.~they are in the kernel of $\rho$, assuming that there are no further emergent symmetries beyond mirror. Under this assumption, since they are not mirror symmetries, $\rho$ takes $\mathsf{K}$ into $G_\text{HYQED}$. On the other hand, in $G_\text{HYQED}$ the only spacetime-orientation-preserving symmetries acting trivially on the $U(1)$ charges are the $U(1)$ symmetries themselves. These symmetries are spontaneously broken in the $U(1)_e$ and $U(1)_m$ SSB phases, while the symmetries in $\mathsf{K}'$ are not. Thus, all symmetries in $\mathsf{K}'$ must map to trivial symmetries.

It is thus sufficient to consider conjugacy classes of homomorphisms 
\begin{equation}
    \begin{split}
        \bar \rho &: \bar {\sf G} = {\sf G}_{\sf sd-lat}/{\sf K}' \to G_\text{SFS CFT}~, \\
        \bar \rho' & : \bar {\sf G}' = {\sf G}_{\sf sd-lat}'/{\sf K}' \to G_\text{HYQED}~.
    \end{split}
\end{equation}
from the quotient groups,  
\begin{equation}
    \begin{split}
        & \bar {\sf G} = (O(2)_e \times O(2)_m) \rtimes ( \Z_2^{{\mathsf{D}}} \times \Z_2^{{\mathsf{T}}}) \times \bZ_2^{\mathsf{R}}~, \\
        & \bar {\sf G}' = (U(1)_e \times U(1)_m) \rtimes (\bZ_2^\mathsf{C}  \times \bZ_2^\mathsf{D} \times \bZ_2^\mathsf{T}) \times \bZ_2^{\mathsf{R}} \subset \b {\sf G} ~.
    \end{split}
\end{equation}
Here (by a slight abuse of notation) ${\mathsf{T}}$ and ${\mathsf{D}}$ denote the equivalence classes of $\mathsf{T}$ and $\mathsf{D}$ in the quotient groups. The groups~$O(2)_{e,m} = U(1)_{e,m} \rtimes \mathbb{Z}_2^{\mathsf{C}_{e,m}}$ include the mirror elements ${\mathsf{C}}_e$ and ${\mathsf{C}}_m$, which are the equivalence classes~of $\mathsf{R}_p^{\pi/2}$ and $\mathsf{R}_v^{\pi/2}$, respectively, in the quotient~$\b {\sf G}$. Their non-mirror product~$\mathsf{C} = \mathsf{C}_e \mathsf{C}_m$ is the equivalence class of the translations $\mathsf{S}_{\hat x}$, $\mathsf{S}_{\hat y}$ in both quotient groups. As above, $\mathsf{R}$ is the equivalence class of $\mathsf{M}_v^H \mathsf{R}_v^{\pi/2}$, corresponding to reflection over the $\hat x  + \hat y$ axis, which preserves both $\mathsf{Q}_e$ and $\mathsf{Q}_m$, and commutes with all the other symmetries in the quotient.

The quotient group~$\b {\sf G}$ is then characterized by the following relations, which follow from~table~\ref{table:latticesymmetries} upon trivializing those lattice symmetries in the kernel~$\sf K'$,\footnote{~For instance, it follows from~\eqref{dsqtrans} that
\begin{equation}
    {\sf D}^2 = {\sf S}_{\hat x + \hat y} \in {\sf K}'~,
\end{equation}
so that the image of~$\sf D$ in the quotient group~$\bar {\sf G}$ squares to the identity. 
} 
\begin{equation}\label{quotrels}
\begin{split}
    & {\mathsf{T}}^2 = {\mathsf{C}}_e^2 = {\mathsf{C}}_m^2 = {\mathsf{D}}^2 = \mathsf{R}^2 = 1~, \\
    & {\sf T  C}_{e, m} = {\sf C}_{e,m} {\sf T}~, \quad {\sf T  D} = {\sf D} {\sf T}~, \quad {\sf C}_e {\sf C}_m =  {\sf C}_m {\sf C}_e~, \quad {\sf D C}_e = {\sf C}_m {\sf D}~,\\
    & {\mathsf{T}} \mathsf{Q}_e {\mathsf{T}} = \mathsf{Q}_e~, \qquad\quad \,\; {\mathsf{T}} \mathsf{Q}_m {\mathsf{T}} = \mathsf{Q}_m~, \\
& {\mathsf{C}}_e \mathsf{Q}_e {\mathsf{C}}_e = -\mathsf{Q}_e~, \qquad {\mathsf{C}}_e \mathsf{Q}_m {\mathsf{C}}_e = \mathsf{Q}_m~, \\
& {\mathsf{C}}_m \mathsf{Q}_e {\mathsf{C}}_m = \mathsf{Q}_e~, \qquad \, {\mathsf{C}}_m \mathsf{Q}_m {\mathsf{C}}_m = -\mathsf{Q}_m~, \\
& {\mathsf{D}} \mathsf{Q}_e {\mathsf{D}} = \mathsf{Q}_m~, \qquad \;\;\;\;{\mathsf{D}} {\mathsf{C}}_e {\mathsf{D}} = {\mathsf{C}}_m~.
\end{split}
\end{equation}
It follows from these relations that the non-mirror charge-conjugation symmetry~$\sf C$ satisfies
\begin{equation}\label{cquotrel}
\begin{split}
    & {\sf C} = {\sf C}_e {\sf C}_m~, \quad {\sf C}^2 = 1~, \quad {\sf C}{\sf T} = {\sf T} {\sf C}~, \quad {\sf C}{\sf D} = {\sf T} {\sf D}~,\\
    & {\sf C} {\sf Q}_e{\sf C} = - {\sf Q}_e~, \qquad  {\sf C} {\sf Q}_m {\sf C} = - {\sf Q}_m~.
\end{split}
\end{equation}

\subsubsection{Mapping the Discrete Symmetries}

Now we will consider the mapping of the discrete symmetries. As explained below~\eqref{Qqmap}, after choosing the mapping of continuous symmetries, we are only free to conjugate by~$\CD \CC \CT$, as well as by~$U(1)_{e,m}$ rotations.

Let us now analyze the continuum mapping of the anti-unitary~$\mathsf{T}$ symmetry on the lattice. Its image must preserve both $U(1)$ charges, hence it must satisfy $\b \rho(\mathsf{T})=\CD \CC\CT e^{i\alpha q_e + i\beta q_m}$ for some $\alpha,\beta$. These are all conjugate by $U(1)$ rotations, since $e^{i\alpha q_e} \CD\CC\CT e^{-i\alpha q_e} = \CD\CC\CT e^{-2i\alpha q_e}$ (and similarly for~$U(1)_m$ rotations). Thus, we set
\[\b \rho(\mathsf{T}) = \CD\CC\CT~,\]
which fixes the ambiguity up to the $\bZ_{2, e} \times \bZ_{2, m}$ subgroups of~$U(1)_e \times U(1)_m$, as well as~$\mathcal{DCT}$. Note that $(\mathcal{DCT})^2=1$ as required to match $\mathsf{T}^2=1$.

Next we consider the unitary~$\mathsf{D}$ symmetry on the lattice. We need the image of $\mathsf{D}$ to exchange the $U(1)$ charges~${\sf Q}_e \leftrightarrow {\sf Q}_m$. This means $\b \rho(\mathsf{D}) = \CD e^{i\alpha q_e + i\beta q_m}$ for some $\alpha$, $\beta$. Using that $\CD e^{i\alpha q_e + i\beta q_m} \CD e^{i\alpha q_e + i\beta q_m} = e^{i(\alpha+\beta)(q_e + q_m)}$ and imposing $\mathsf{D}^2=1$, we find $\beta = -\alpha \mod 2 \pi$. We also need it to commute with the image of $\mathsf{T}$, namely $\mathcal{DCT}$, which is equivalent to imposing $(\mathsf{DT})^2=1$. We then compute
\[\CD e^{i\alpha(q_e -q_m)} \mathcal{DCTD} e^{i\alpha(q_e-q_m)} \mathcal{DCT} =  e^{i\pi(q_e+q_m)-2i\alpha(q_e -q_m)}~,\]
where we used $\mathcal{DCTD} = \mathcal{CT}(-1)^{q_\CM}$ and~$(-1)^{q_\CM} = (-1)^{q_e + q_m}$. Setting this to unity, we find two possible solutions, $\alpha = \pm \pi/2$ mod $2\pi$. These two choices are related by conjugation in~$\Z_{2, e}$, since 
\[e^{i\pi q_e} \CD e^{i\pi q_e} = \CD e^{i\pi(q_e+q_m)}~,\]
and in fact the same is true of we use~$\Z_{2, m}$. 
We thus choose $\alpha=\pi/2$, so that
\[\b \rho(\mathsf{D}) = \CD e^{i(\pi/2) (q_e-q_m)} = \CD i^{q_f}~,\]
which fixes this ambiguity. The only remaining ambiguity is conjugation by $\CD\CC\CT$ and by the diagonal~$\Z_2 \subset \Z_{2, e} \times \Z_{2, m} $, which is generated by~$e^{i\pi (q_e + q_m)}=(-1)^{q_\mathcal{M}}=(-1)^{q_f}$.

The equivalence class of unit translations $\mathsf{S}_{\hat x}, \mathsf{S}_{\hat y}$ is $\mathsf{C}$. This must be sent to a unitary symmetry negating both charges, so $\b \rho(\mathsf{C}) = \CC e^{i\alpha q_e + i\beta q_m}$. Note that all of these square to the identity. We must now enforce that~$\mathsf{C}$ commutes with $\mathsf{T}$, so we compute $(\mathsf{CT})^2$ as
\[\CC e^{i\alpha q_e + i\beta q_m} \mathcal{DCT} \CC e^{i\alpha q_e + i\beta q_m} \mathcal{DCT}
= e^{-2i(\alpha q_e + \beta q_m)}~.\]
Setting this to the identity, we must have $\alpha = 0,\pi$ and $\beta = 0,\pi$, mod $2\pi$. $\mathsf{C}$ also commutes with $\mathsf{D}$, so we do a similar computation of $(\mathsf{DC})^2$ as
\[\CD e^{i(\pi/2) (q_e-q_m)} \CC e^{i\alpha q_e + i\beta q_m}\CD e^{i(\pi/2) (q_e-q_m)} \CC e^{i\alpha q_e + i\beta q_m}
= e^{i\pi (q_e-q_m)} e^{i(\alpha-\beta)(q_e-q_m)}~,\]
so $\alpha-\beta = \pi$ mod $2\pi$. Thus we find two solutions $(\alpha,\beta) = (0,\pi)$ and $(\pi,0)$.

These are in fact two physically distinct solutions, since they are not related by conjugation by the remaining elements $\CD\CC\CT$ and $e^{i\pi (q_e + q_m)}$. To distinguish them we will need some extra physical input from the lattice. We will use the existence of the lattice operators $\tilde Z_\ell^-$ and $\tilde X_\ell^-$ defined in \eqref{eqndeformingtheU1modeltotheoriginalmodel}, which can be combined to a deformation
\[-\sum_\ell \left(\tilde Z_\ell^- + \tilde X_\ell^- \right) \,,\]
which is a combination of charge~$\pm1$ operators under $U(1)_e$ and $U(1)_m$ that preserves $\mathsf{T}$ and all crystalline symmetries \textbf{p4mD}, including $\mathsf{D}$ and $\mathsf{C}$ (translations). Note that this is precisely the deformation that connects the staggered Fradkin-Shenker model to the original Fradkin-Shenker model along the self-dual line~$h_e = h_m$, as explained around \eqref{eqndeformingtheU1modeltotheoriginalmodel}.

This deformation in the field theory must correspond to a Hermitian combination of $q_\CM=\pm1$ monopole operators that is invariant under the images of $\mathsf{T}$, $\mathsf{D}$, and $\mathsf{C}$. A general Hermitian combination of $q_\CM=\pm1$ monopole operators takes the form
\[\CM_{\rm sym}=\alpha \CM^1 + \beta \CM^2 + \alpha^* (\CM^1)^* + \beta^* (\CM^2)^*~,\]
where $\alpha,\beta \in \bC$. $\b \rho(\mathsf{T}) = \mathcal{DCT}$ acts as
\[\b\rho(\mathsf{T})=\mathcal{DCT} = \begin{cases} \CM^1 \mapsto \CM^1 \\ \CM^2 \mapsto - \CM^2 \end{cases}\]
The operator $\CM_{\rm sym}$ must therefore have $\alpha$ real and $\beta$ imaginary. $\b \rho(\mathsf{D}) = \mathcal{D} i^{q_f}$ acts as
\[\b \rho(\mathsf{D})=\mathcal{D} i^{q_f} = \begin{cases} \CM^1 \mapsto i\CM^2 \\ \CM^2 \mapsto - i\CM^1 \end{cases}\]
This implies $\beta = i\alpha$. Therefore our operator takes the form
\[\label{eqnsymmetricmonopole}\CM_{\rm sym}=\alpha \left( \CM^1 + i \CM^2 + (\CM^1)^* - i (\CM^2)^* \right) \,,\]
with $\alpha \in \bR$, meaning there is a unique such operator up to a constant. We have
\[\mathcal{C} = \begin{cases} \CM^1 \mapsto (\CM^1)^* \\ \CM^2 \mapsto (\CM^2)^* \end{cases}\]
Thus there is only one choice for $\mathsf{C}$ that works, namely
\[\b \rho(\mathsf{C}) = \CC (-1)^{q_m}\,,\]
preserving $\CM_{\rm sym}$.

Although we cannot identify the mirror symmetries acting in $G_\text{HYQED}$, we can in fact identify their action on the monopole operators. This is because the only charge $\pm 1$ operators for $U(1)_e$ and $U(1)_m$ are unit monopole operators. Let $\mathsf{C}_e$ be the equivalence class of $\mathsf{R}^{\pi/2}_p$ in the quotient group. The symmetry $\b \rho(\mathsf{C}_e) = \CC_e$ must negate $q_e$ and preserve $q_m$, commute with $\b \rho(\mathsf{C})$ and $\b \rho(\mathsf{T})$, and preserve the monopole operator above, since it preserves the corresponds lattice deformation $\sum_\ell (\tilde Z_\ell^- + \tilde X_\ell^-)$. It must therefore act as
\[\label{eqnCemirror}\b \rho(\mathsf{C}_e) = \CC_e = \begin{cases}
    \CM^1 \mapsto (\CM^1)^* \\
    \CM^2 \mapsto \CM^2
\end{cases}\]
Likewise, let $\mathsf{C}_m$ be the equivalence class of $\mathsf{R}^{\pi/2}_v$. It must preserve the monopole operator above, so we find
\[\label{eqnCmmirror}\b \rho(\mathsf{C}_m) = \CC_m = \begin{cases}
    \CM^1 \mapsto \CM^1 \\
    \CM^2 \mapsto -(\CM^2)^*
\end{cases}\]
Note that $\b \rho(\mathsf{C}_e) \b  \rho(\mathsf{C}_m) = \b  \rho(\mathsf{C})$, yielding an independent check of the consistency of this map.

Finally, let us consider the reflection symmetry $\mathsf{R}$, which is the equivalence class of the lattice symmetry which reflects over the axis $\hat x + \hat y$. This preserves both $U(1)$ charges and moreover is unbroken in the $U(1)$ SSB phases. Consider the canonical CRT transformation $\Theta_x$ of HYQED, acting on spacetime coordinates as $t \mapsto -t$, $x \mapsto -x$, $y \mapsto y$. This transformation commutes with all internal symmetries (see~e.g.~\cite{Hason:2020yqf} and references therein), and so on monopole operators acts as
\[\Theta_x: \begin{cases}
    \cM^1 \mapsto (\cM^1)^* \\
    \cM^2 \mapsto (\cM^2)^*
\end{cases}\]
We find that the unique reflection symmetry $\bar \rho(\mathsf{R})$ preserving the monopole operators in the continuum is (up to conjugation by Lorentz transformations)
\[\bar \rho(\mathsf{R}) = \bar\rho(\mathsf{C})\bar\rho(\mathsf{T}) \Theta_x \]
Note this implies the pleasant formula $\bar\rho(\mathsf{CRT})=\Theta_x$. As mentioned above we identify $\Theta_x = 1$ in $G_\text{SFS CFT}$,  so that~${\sf CRT}$ is also in the kernel of $\b \rho$.

It is easy to see that $\b \rho(\mathsf{D})$, $\b \rho(\mathsf{T})$, $\b \rho(\mathsf{C})$, and $U(1)_e$, $U(1)_m$ generate $G_\text{HYQED}$. In other words the restriction~$\bar \rho':\bar {\sf G}'/\bZ_2^{{\sf CRT}} \to G_\text{HYQED}$ of the map~$\b \rho$ we have constructed to~${\sf G}' \subset {\sf G}$ is an isomorphism. In turn, the minimal extension~$G_\text{IR} \supset G_\text{HYQED}$ that includes the emergent mirror symmetry and nothing else is thus $\bar {\sf G}$, so we may write
\[G_\text{SFS CFT} = (O(2)_e \times O(2)_m) \rtimes (\bZ_2^\mathsf{D} \times \bZ_2^\mathsf{T})~,\]
where $O(2)_{e,m}=U(1)_{e,m}\rtimes\mathbb{Z}_2^{\mathsf{C}_{e,m}}$ and $\mathsf{C}=\mathsf{C}_e\mathsf{C}_m$. From now on we identify $\mathsf{D}$ etc. with their images in~$G_\text{IR}$. We stress that although mirror symmetry can only emerge in HYQED near the multicritical point, it is an exact symmetry of the SFS and ordinary Fradkin-Shenker lattice models.

\subsection{Emergence of Mirror Symmetry in Higgs-Yukawa-QED}\label{sec:emergemirr}

In this section, we will propose an intuitive physical picture for the emergence of~$\mathbb{Z}_2$ mirror symmetry at the multicritical point of HYQED, by pointing out its relation to previously conjectured self-dualities of QED$_3$ (reviewed below). In later sections we will present two further, sharper pieces of evidence for the emergence of mirror symmetry:
\begin{itemize}
    \item[(i)] In section~\ref{sec:easy} we present a continuum QFT dual of HYQED, namely the (multicritical) easy-plane $\mathbb{CP}^1$ model, where mirror symmetry is manifest, while duality~$\mathsf{D}$ is emergent.
    \item[(ii)] In section~\ref{sec:largeN} we give some computational evidence for the emergence mirror symmetry at the multicritical point of HYQED, by estimating the scaling dimensions of the lightest charged operators that participate in its selection rules.
\end{itemize}

Let us begin by reviewing how mirror symmetry was conjectured to emerge in the (now unlikely) scenario that $N_f=2$ QED$_3$ flows to a symmetry-preserving CFT at low energies. In this case,  a fermionic variant of particle-vortex duality~\cite{Wang:2015qmt,Metlitski:2015eka,Seiberg:2016gmd} gives rise to a self-duality of QED$_3$ in the IR, under which flavor and monopole symmetries are exchanged~\cite{Xu:2015lxa,Karch:2016sxi, Hsin:2016blu, Wang:2017txt}. Schematically, and up to terms that only depend on background fields, the self-duality reads
\begin{equation}\label{eq:qed3SelfD}
\begin{split}
-i\bar\psi_1\slashed{D}_{a+A_f}\psi^1 - i \bar\psi_2&\slashed{D}_{a-A_f}\psi^2 - \frac{1}{2\pi}da\wedge A_\CM \\
\text{Self-Duality : } \ &\Big\updownarrow \\
-i\bar\chi_1\slashed{D}_{\tilde{a}+A_\CM}\chi^1 - i \bar\chi_2&\slashed{D}_{\tilde{a}-A_\CM}\chi^2 - \frac{1}{2\pi}d\tilde{a}\wedge A_f \,.
\end{split}
\end{equation}
In the first line, we have $N_f=2$ QED$_3$ coupled to background fields $A_\CM$ and $A_f$ for the $U(1)_\CM$ magnetic symmetry and for the $U(1)_f$ Cartan subgroup of the $SU(2)_f$ flavor symmetry, respectively. In the second line, we have the same theory, but with the roles of $U(1)_f$ and $U(1)_\CM$ reversed. In the first description, the manifest UV symmetry contains $(SU(2)_f \times U(1)_\cM)/\bZ_2$, whereas in the second description it contains $(SU(2)_\cM \times U(1)_f)/\bZ_2$. A putative IR CFT would therefore have the following emergent symmetry,
\begin{equation}
    O(4) = \frac{SU(2)_f \times SU(2)_\CM}{\mathbb{Z}_2} \rtimes \mathbb{Z}^{\rm mirror}_2 \,,
\end{equation}
where the $\mathbb{Z}_2^{\rm mirror}$ symmetry swapping the two $SU(2)$ factors arises from the self-duality in~\eqref{eq:qed3SelfD}. Mirror symmetry acts as follows on the Hermitian fermion bilinears~$\CO = i \b \psi \psi$ and~$\CO_3 = i \b \psi \sigma_z \psi$,
\begin{equation}\label{mirror_on_bilinears}
\mathbb{Z}^{\rm mirror}_2 : \quad \CO \leftrightarrow -\t\CO \,, \qquad {\CO}_3 \leftrightarrow \t\CO_3 \,,     
\end{equation}
where $\tilde{O}$ and $\tilde{O}_3$ are the corresponding bilinear operators built out of the dual fermions $\chi^{1,2}$. The bilinear~$\CO_1 - i \CO_2$, which carries~$U(1)_f$ charge~$q_f = 2$, is exchanged with a monopole operator of~$U(1)_\CM$ charge~$q_\CM = 2$.

It is not obvious how to extend the self-duality of QED$_3$ reviewed above to full HYQED \eqref{LagrangianMinkowski},  because the QED$_3$ operators that couple to the Higgs field~$\phi$ via the Yukawa couplings~\eqref{yukawa2} are not gauge invariant. 

We sidestep this issue by focusing on the first-order line~$m_3 = 0$, $m_\phi^2 > 0$ discussed in section~\ref{firstorder}. At least for sufficiently large~$m_\phi^2 > 0$, the effective theory is $N_f=2$ QED$_3$ (with a renormalized effective gauge coupling $e^2_{\rm eff}$) deformed by the gauge-invariant four-fermion effective potential~$V_\text{eff}$ in~\eqref{VeffHYM} that  explicitly breaks~$SU(2)_f\rightarrow U(1)_f$, obtained by integrating out the Higgs field. Using the fact that the~$U(1)_\CM$ current is~$j_\CM^\mu ={1 \over 4 \pi} \ep^{\mu\nu\rho} f_{\nu\rho}$ (see~\eqref{magnetic_current}) and the identity~\eqref{jfsq} for the~$U(1)_f$ current~$j^\mu_f$, we can express the effective Lagrangian for this deformed $N_f=2$ QED$_3$ theory as follows,
\begin{equation}
\begin{split}
\mathscr{L}_{\rm eff} &= -\frac{1}{4e^2_{\rm eff}} f_{\mu\nu}f^{\mu\nu} - i\bar\psi_i\slashed{D}_a\psi^i- V_{\rm eff} \\
&= - i\bar\psi_i\slashed{D}_a\psi^i +\frac{2\pi^2}{e^2_{\rm eff}} j_\mu^\CM j^{ \CM \mu} + \frac{y^2}{4m^2_\phi}j_\mu^f j^{f\mu} -
\frac{y^2}{4m^2_\phi} \left( 2(\mathcal{O})^2+(\mathcal{O}_3)^2\right) \,.
\end{split}
\end{equation}
Consider now the action of the $\mathbb{Z}_2^{\rm mirror}$. It follows from \eqref{mirror_on_bilinears} that $(\mathcal{O})^2$ and $(\mathcal{O}_3)^2$ are invariant, while the currents are swapped, $j_\mu^\CM \leftrightarrow j_\mu^f$. So mirror symmetry requires the equality of the effective gauge and four-fermion couplings, 
\begin{equation}\label{mirror_relation}
    \frac{1}{e^2_{\rm eff}} \overset{?}{=} \frac{y^2}{8\pi^2m^2_\phi} \,.
\end{equation}
Far away from the multicritical point, at sufficiently large~$m_\phi^2$, we have~$e_\text{eff}^2 \simeq e^2$, and this condition is clearly not satisfied.

Beyond this controlled regime, both effective couplings in~\eqref{mirror_relation} will be replaced by non-trivial (and incalculable) functions of~$m_\phi^2$ and the other dimensionful parameters of the theory, and two such functions might be equal at a specific value of the tuning parameter~$m_\phi^2$. We expect precisely this to happen at the HYQED multicritical point. Of course this is only a plausibility argument that a mirror symmetry that acts as in~\eqref{mirror_on_bilinears} might emerge there. As already mentioned above, we will give sharp, quantitative arguments for this in sections~\ref{sec:easy} and~\ref{sec:largeN} below.

\subsection{Recovering the Fradkin-Shenker Phase Diagram}\label{FSviaMonos}

\subsubsection{Monopole Deformations of HYQED}

\begin{figure}[t]
    \centering   
\begin{tikzpicture}[x=4cm,y=4cm] 
  \draw[gray!60, line width=0.3pt] (0,0) rectangle (1.35,1.35);

  \node[below=8pt] at (0.64,0.02) {$h_m$};
  \node at (-0.15,0.64) {$h_e$};
  \draw[->] (0.49,-0.05) -- (0.79,-0.05);
  \draw[->] (-0.05,0.49) -- (-0.05,0.79);

  \node at (0.23,1.06) {};
  \node at (0.3,0.96) {Higgsing};
  \node at (0.23,0.72) {\textcolor{blue}{Ising$^*_e$}};

  \node at (0.81,0.43) {\textcolor{blue}{Ising$^*_m$}};
  \node at (1.0,0.25) {Confinement};
  \node at (1.05,0.18) {};

  \node at (0.23,0.28) {Toric};
  \node at (0.23,0.18) {code};

\draw[<->] (-0.25,0.05) -- (0.05,-0.25);

\node at (-0.15,-0.15) {$\mathsf{D}$};

  \draw[line width=0.8pt,blue]
    (0.0,0.64) -- (0.64,0.64) -- (0.64,0.00);

  \draw[densely dotted, line width=0.8pt]
    (0.64,0.64) -- (1,1);
  \node at (0.91,0.85) {\hskip-12pt $\mathsf{D}$ SSB};  
  \fill[color=teal] (1,1) circle (0.03);
  \fill[color=purple] (0.640,0.640) circle (0.03);
  \node at (0.9,0.640) {\textcolor{purple}{FS CFT}};
  \node at (1.1,1.1) {\textcolor{teal}{Ising}};
\end{tikzpicture}
    \caption{Schematic phase diagram of the Fradkin-Shenker (FS) model~\eqref{eq:Hfsintro}, reproduced from \cite{Wu:2012cj}, as a function of~$h_e,h_m \geq 0$. The~$\Z_2^{\sf D}$ duality symmetry acts as~${\sf D}: h_e \leftrightarrow h_m$ and is preserved along the self-dual diagonal~$h_e = h_m$. In the lower left region there is the toric code phase, described at long distances by an emergent~$\Z_2$ gauge theory, which couples electrically and magnetically to dynamical, massive~$e$ and~$m$ anyons that are exchanged by~$\sf D$. These become massless on the second-order~$\Z_2$-gauged Ising lines (indicated in blue and labeled~Ising$^*_e$ and Ising$^*_m$, respectively), which describe electric and magnetic Higgs transitions into the trivial Higgs/confined phase. The Ising$^*_{e,m}$ lines meet at a multicritical point (indicated in red) on the self-dual diagonal, which is described by the FS CFT. Continuing further along this diagonal into the Higgs/confined phase, we find a first-order (dotted) line on which~$\sf D$ is spontaneously broken, and which ends at a conventional Ising critical point (indicated in green).}
    \label{figfsphasediagrambis}
\end{figure}
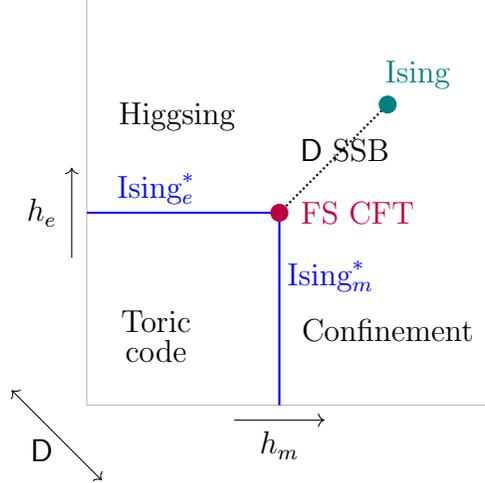

We have explored the phase diagram of HYQED in section \ref{subseccontinuumphasediagram} and in figure \ref{fig:HYQEDphasediagbis}. Under the assumption that it captures the full phase diagram of the SFS lattice model, we will now argue that it reproduces the phase diagram of the original Fradkin-Shenker model in figure~\ref{figfsphasediagrambis}, lending additional credence to our conjectured duality web. In important role will be played by the symmetry-breaking dynamics of QED$_3$, reviewed in section~\ref{section:qed3review}.

Recall from \eqref{eqndeformationtoFSmodel} that the SFS and the Fradkin-Shenker models are related as follows,
\[H_{\rm FS}(h_e,h_m) = H_{\rm SFS}(h_e,h_m) - h_e \sum_\ell \tilde Z_\ell^- - h_m \sum_\ell \tilde X_\ell^- \,. \]
To continuously interpolate between them, we introduce the four-parameter model
\[\tilde H_{\rm FS}(h_e,h_m,\epsilon_e,\epsilon_m) = H_{\rm SFS}(h_e,h_m) - \epsilon_e \sum_\ell \tilde Z^-_\ell - \epsilon_m \sum_\ell \tilde X^-_\ell \,, \]
where $\epsilon_e = \epsilon_m = 0$ for the SFS model and $\epsilon_e = h_e$, $\epsilon_m = h_m$ in the Fradkin-Shenker model. Duality $\mathsf{D}$ acts on the couplings as $h_e \leftrightarrow h_m$, $\epsilon_e \leftrightarrow \epsilon_m$.

Assuming that HYQED correctly describes the multicritical point of the SFS model, we can study perturbations with small $\epsilon_{e,m}$. In \eqref{eqnsymmetricmonopole} above, we identified a unique monopole operator $\CM_{\rm sym}$ with the right quantum numbers to match the lattice deformation to the Fradkin-Shenker model along the self-dual line. With the same argument we can identify
\begin{equation}\label{deformationmap}
    \begin{split}
    \tilde Z_\ell^- &\sim \left( \CM^1 + (\CM^1)^* \right) + \cdots \,, \\ 
    \tilde X_\ell^- &\sim \left( i\CM^2 - i (\CM^2)^* \right) + \cdots \,,
    \end{split}
\end{equation}
where the ellipses denote other operators with the same charges. We will assume that these are the leading contributions near the multicritical point, i.e.~that the other operators contributing to \eqref{deformationmap} have higher scaling dimension. As we will show in section \ref{subseclargeNextrap}, the perturbations~\eqref{deformationmap} are relevant, since~$\Delta[\CM^{1,2}] \simeq 0.63$ in the SFS CFT. Thus, turning on small $\epsilon_e$, $\epsilon_m$ amounts to adding the following relevant operator to HYQED,
\[\label{eqnmonopoleperturbation} \Delta {\mathscr L} = -\varepsilon_e(\CM^1 + (\CM^1)^*)-\varepsilon_m (i\CM^2 - i(\CM^2)^*) \,,\]
where $\varepsilon_e/\varepsilon_m=\epsilon_e/\epsilon_m$ and we take $\varepsilon_{e,m}>0$ without loss of generality. This deformation is invariant under $\mathsf{T}$ and $\mathsf{C}_{e,m}$, as well as $\mathsf{D}$ when $\varepsilon_e=\varepsilon_m$.

Strictly speaking, the description~\eqref{eqnmonopoleperturbation} is valid near the SFS multicritical point, and for small~$\ep_{e, m}$. However, we will allow ourselves to explore these deformations in the entire phase diagram of HYQED (see section \ref{subseccontinuumphasediagram} and figure \ref{fig:HYQEDphasediagbis}), i.e.~for all~$m_\phi^2$, $m_3$ and for sufficiently large~$\ep_{e,m}$ to reach the Fradkin-Shenker model, while continuing to treat~\eqref{eqnmonopoleperturbation} as a leading-order perturbation. This assumes the commutativity of the RG flows involved, i.e.~the absence of accidental phase transitions that we cannot control. With these assumptions we recover the phase diagram of the Fradkin-Shenker model (depicted in figure~\ref{figfsphasediagrambis}) from HYQED.

\subsubsection{Matching the Phases at $m_3\neq 0$}
Let us start by considering how the monopole deformation~\eqref{eqnmonopoleperturbation} affects the vacua of HYQED at $m_3\neq 0$, where $\mathsf{D}$ is explicitly broken.

For sufficiently negative~$m_\phi^2 < 0$, HYQED flows to the Higgs phase described by $\mathbb{Z}_2$ TQFT (see section \ref{section:Z2Higgsphase}). This gapped phase is stable to all sufficiently small local perturbations, so it is unchanged when adding the monopole operator \eqref{eqnmonopoleperturbation}. It thus gives rise to the toric code phase of the Fradkin-Shenker model.

Net, let us study the effect of the monopole perturbation~\eqref{eqnmonopoleperturbation} on the $U(1)_e$ and $U(1)_m$ symmetry-breaking Coulomb phases of HYQED (see section \ref{section: large m3}). The corresponding order parameters are the monopole operators $\CM^1 = v e^{i \sigma_e}$ and $\CM^2 = v e^{i \sigma_m}$, respectively. Here~$\sigma_{e,m} \sim \sigma_{e, m} + 2 \pi$ are the dual photons parametrizing the ground-state circles $S^1_{e,m}$ in~\eqref{circlem} and~\eqref{circlee}. The monopole deformation~\eqref{eqnmonopoleperturbation} gives rise to the following scalar potential,
\begin{equation}\label{eq:perturbgeneric}
2\varepsilon_e {\rm Re}(\CM^1) - 2\varepsilon_m {\rm Im}(\CM^2) =
\begin{dcases}
+2\varepsilon_e v \cos\sigma_e \quad &\text{on } S^1_e \quad (\text{for } m_3 < 0)~, \\
-2\varepsilon_m v\sin\sigma_m \quad &\text{on } S^1_m  \quad(\text{for } m_3 > 0)~.
\end{dcases}    
\end{equation}
This picks out a point on each circle, namely~$\sigma_e=\pi$ and $\sigma_m=\pi/2$, leading to a trivially gapped vacuum (preserving $\mathsf{T}$ and $\mathsf{C}_{e,m}$) for either sign of~$m_3$. Note that the unique vacuum on~$S^1_e$ describes the Higgs regime of the Fradkin-Shenker model, while the unique vacuum on~$S^1_m$ describes the confining regime (see figure~\ref{figfsphasediagrambis}).

Let us now describe the effects of the monopole deformation~\eqref{eqnmonopoleperturbation} on the $O(2)^*_{e,m}$ models (with Lagrangian~\eqref{O(2)*Lagrangian}) that describe the transition between the toric code and~$U(1)_{e,m}$ symmetry-breaking phases of HYQED. It suffices to to consider the case~$m_3<0$, i.e.~the~$O(2)_e^*$ transition, since the~$O(2)_m^*$ transition at $m_3>0$ is related to it by the action of duality $\mathsf{D}$. 

In the~$O(2)_e^*$ model, the UV operator~$\mathcal{M}^1$ is represented by~$\Phi^2$, while~$\CM^2$ is trivial in the IR. We thus find that the monopole deformation~\eqref{eqnmonopoleperturbation} amounts to adding the gauge-invariant, relevant operator~$\text{Re}(\Phi^2)$ to the potential of the $O(2)^*_e$ model, leading to
\begin{equation}
V_{O(2)^*_e + \text{monopoles}} = \lambda_{4,\Phi}|\Phi|^4 + m^2_\Phi |\Phi|^2  + \frac{m^2_\varepsilon}{2} \left( \Phi^2+(\Phi^*)^2\right) \,, \qquad m^2_\varepsilon >0~,
\end{equation}
where $\lambda_{4,\Phi}>0$ and $m^2_\Phi$ are the effective couplings of the $O(2)^*_e$ model, which depend on $m_3$, and $m^2_\varepsilon \sim \varepsilon_e$.
In terms of the real and imaginary parts of $\Phi \equiv \frac{1}{\sqrt{2}}(\varphi_1+i\varphi_2)$,
\begin{equation}
V_{O(2)^*_e + \text{monopoles}} = \frac{\lambda_{4,\Phi}}{4} (\varphi_1^2+\varphi_2^2)^2 + \frac{1}{2}(m^2_\Phi+m^2_\varepsilon)\varphi_1^2+\frac{1}{2}(m^2_\Phi-m^2_\varepsilon)\varphi_2^2~, \qquad m^2_\varepsilon >0~.    
\end{equation}
This potential explicitly breaks the $U(1)_e$ symmetry, while preserving~$\mathsf{T}$, $\mathsf{C}_{e,m}$. Let us study its semiclassical phases.
\begin{itemize}
    \item When $m^2_\Phi>m^2_\varepsilon$, we have $\braket{\varphi_1}=\braket{\varphi_2}=0$. Both $\varphi_1$ and $\varphi_2$ have a positive mass and can be integrated out.  We are thus left with a~$\Z_2$ TQFT at low energies. 
    
    \item When $m^2_\Phi<m^2_\varepsilon$, we have that $\braket{\varphi_1}=0$ and $\braket{\varphi_2}= ((m^2_\varepsilon-m^2_\Phi)/\lambda_{4,\Phi})^{1/2}$. Thus, the $\mathbb{Z}_2$ gauge field is Higgsed, and we are left with a unique, trivially gapped vacuum. This continues to holds even if we further lower the value of $m^2_\Phi$; in particular there is no further transition at $m^2_\Phi=-m^2_\varepsilon$. Indeed, once $\varphi_2$ has condensed, both $\widehat\varphi_2 \equiv \varphi_2-\braket{\varphi_2}$ and $\varphi_1$ have positive masses-squared, given by $3\lambda_{4_\Phi}\braket{\varphi_2}^2+m_\Phi^2-m_\varepsilon^2=2(m^2_\varepsilon-m^2_\Phi)$ and $\lambda_{4_\Phi}\braket{\varphi_2}^2+m_\Phi^2+m_\varepsilon^2 =2m^2_\varepsilon$, respectively.

    \item When $m^2_\Phi=m^2_\varepsilon$, we have a transition between the phases described above, driven by the condensation of $\varphi_2$, which is massless at the transition and subject to the gauge identification $\varphi_2 \sim -\varphi_2$. This implies that the $O(2)^*_e$ phase transition of HYQED is reduced to a continuous~$\mathbb{Z}_2$-gauged Ising transition, which we denote by Ising$^*_e$, in agreement with figure~\ref{figfsphasediagrambis}. 
    
\end{itemize}

\noindent Note that~$\mathsf{T}$ and $\mathsf{C}_{e,m}$ are not spontaneously broken in any of these phases.

\subsubsection{Matching the Phases at $m_3= 0$}

Let us now consider the self-dual line~$m_3=0$, where $\mathsf{D}$ is a symmetry of the theory. For $m^2_\phi<0$, HYQED flows to the $\mathbb{Z}_2$ gauge theory phase, which is stable to small local perturbations and thus is unchanged, as already explained above.

For $m^2_\phi>0$ we have the first-order line where the two circles of vacua $S^1_e$ and $S^1_m$ are degenerate (see section~\ref{firstorder}). We now show that for small enough $m^2_\phi$, the monopole perturbation~\eqref{eqnmonopoleperturbation} lifts these circles and leaves two gapped, degenerate vacua (one on each circle) that preserve $\mathsf{T}$ and $\mathsf{C}_{e,m}$, but are exchanged by the duality symmetry $\mathsf{D}$, which is therefore spontaneously broken. At larger values of $m^2_\phi$, it will instead give rise to a unique gapped vacuum that also preserves~$\mathsf{D}$. The two are separated by a conventional second-order Ising transition. 

To show this explicitly, let us recall from section~\ref{section:qed3review} that~$N_f = 2$ QED$_3$ (without a Higgs field~$\phi$) spontaneously breaks its~$U(2)$ global symmetry via the monopole vev~$\langle \CM^i\rangle \neq 0$ in~\eqref{Mvev}. This leads to a squashed~${\t S}^3$ sigma model with~$U(2)$ symmetry for the NGBs, described by the monopole field~$\CM^i$ subject to the constraint~\eqref{eq:mmnorm}, which we repeat here,
\begin{equation}\label{eq:mmconsbis}
   {\t S}^3 : \qquad  (\CM^i)^* \CM^i = v^2~.
\end{equation}
If we now add back in the Higgs field~$\phi$ of HYQED and assume that its mass~$m_\phi^2 > 0$ is sufficiently large and positive, we obtain the effective potential~\eqref{VeffHYM_2} on~${\t S}^3$ (with~$m_3 = 0$), to which we must now also add the monopole perturbation~\eqref{eqnmonopoleperturbation} (with~$\varepsilon_e=\varepsilon_m\equiv \varepsilon > 0$),
\begin{equation}\label{effpotentialwithmonopoles}
    V_{{\t S}^3 \text{ eff}} = - \beta \left(|\CM^1|^2- |\CM^2|^2\right)^2 + 2\varepsilon \left( {\rm Re}(\CM^1) - {\rm Im}(\CM^2) \right)~, \qquad \beta = \frac{C^2y^2}{2m_\phi^2v^4}~.
\end{equation}

Let us first consider the case $\beta \gg \varepsilon$, where we can treat $\varepsilon$ as a small perturbation around the two circles of vacua $S^1_{e,m}$ in \eqref{largemassvacua}. The energy shifts due to the perturbation are
\begin{equation}\label{energyshiftsvacua}
\begin{split}
\Delta V^e_{\rm eff} &= + 2\varepsilon v \cos\sigma_e \,, \\
\Delta V^m_{\rm eff}  &= -2\varepsilon v \sin\sigma_m \,,
\end{split}
\end{equation}
which is the same as~\eqref{eq:perturbgeneric} with~$\ep_{e,m} = \ep$. These potentials have unique minima, at $\sigma_e=\pi$ and $\sigma
_m=\pi/2$, respectively, with~$\Delta V^e_{\rm eff} = \Delta V^m_{\rm eff}$. The perturbation picks a single vacuum on each circle, and these vacua are degenerate. They preserve $\mathsf{T}$ and $\mathsf{C}_{e,m}$, while they are mapped into each other by $\mathsf{D}$, which is thus spontaneously broken. Thus, for large values of~$\beta$, i.e.~when $m^2_\phi$ is small and we are close to the multicritical point, we still have a line of first-order phase transitions, with two degenerate, gapped vacua exchanged by the spontaneously broken~$\mathsf{D}$ symmetry. This precisely matches the dotted first-order line in the Fradkin-Shenker phase diagram (see figure~\ref{figfsphasediagrambis}). 

As we increase $m^2_\phi$, moving further away from the multicritical point, $\beta$ gets smaller, and the effect of the monopole perturbation becomes more pronounced. For $\beta \ll \varepsilon$ we only retain the second term in~\eqref{effpotentialwithmonopoles}, which is linear in the monopole fields and leads to a unique vacuum on~${\t S}^3$,
\begin{equation}\label{uniquevacuum}
\left( \mathcal{M}^1= -\frac{v}{\sqrt{2}}  \,,  \mathcal{M}^2= \frac{iv}{\sqrt{2}}\right) \,~
\end{equation}
which necessarily preserves~$\mathsf{D}$. For sufficiently large~$m^2_\phi>0$, we thus find a single, trivially gapped vacuum that preserves all global symmetries. This matches the trivial Higgsed/confined phase of the Fradkin-Shenker model in the top-right corner of the phase diagram in figure~\ref{figfsphasediagrambis}. As we will show below, this vacuum is continuously connected to the trivially gapped vacua described below~\eqref{eq:perturbgeneric}, which occur for sufficiently large~$m_3 \neq 0$, i.e.~sufficiently far away from the self-dual line. 

Let us explicitly analyze the transition between the~$\Z_2^{\sf D}$ breaking and preserving vacua at small and large~$m_\phi^2$ described above. The problem can be simplified by noting that all extrema of the effective potential \eqref{effpotentialwithmonopoles} with $\varepsilon\neq 0$ have
\begin{equation}
\mathrm{Im}(\mathcal{M}^1) = \mathrm{Re}(\mathcal{M}^2) = 0~, 
\end{equation}
which preserves $\mathsf{T}$, $\mathsf{C}_{e,m}$, and $\mathsf{D}$. 

This allows us to impose the $\widetilde{S}^3$ constraint~\eqref{eq:mmconsbis} as
\begin{equation}\label{eq:mviaalpha}
\mathrm{Re}(\mathcal{M}^1) = v \cos\alpha \,, \qquad \mathrm{Im}(\mathcal{M}^2) = v \sin\alpha \,,   
\end{equation}
which reduces the problem to minimizing the following effective potential in one variable,
\begin{equation}
V_{{\t S}^3 \text{ eff}}(\alpha) = - \beta v^4 \cos^2(2\alpha)+2\varepsilon v(\cos\alpha-\sin\alpha) \,.    
\end{equation}
Its critical points satisfy 
\begin{equation}
\sin(4\alpha) =  \frac{\varepsilon}{\beta v^3} (\sin\alpha+\cos\alpha) \,.   
\end{equation}
When $\varepsilon/\beta<4v^3/\sqrt{2}$ there are two isolated, degenerate minima at $\alpha_\pm=3\pi/4 \pm \hat\alpha$, where $\hat\alpha\in[0,\pi/4)$ is a monotonically decreasing function of $\varepsilon/\beta$. These minima preserve $\mathsf{T}$ and $\mathsf{C}_{e,m}$, but are exchanged by $\mathsf{D}$. In the limit $\varepsilon\rightarrow 0$, one gets $\hat{\alpha}\rightarrow \pi/4$, and the minima approach the ones discussed below \eqref{energyshiftsvacua}. At the critical value $\varepsilon/\beta=4v^3/\sqrt{2}$ one finds $\widehat\alpha=0$, so that the two minima merge continuously into the unique vacuum~\eqref{uniquevacuum}, described by the~$\mathsf{D}$-preserving minimum $\alpha=3\pi/4$. Thus the first-order lines ends in a second-order Ising transition, as in figure~\ref{figfsphasediagrambis}, which is described by the Ising order parameter~$\alpha- 3 \pi/4$.

\subsubsection{Interpolating From Higgsing to Confinement on ${\t S}^3$}

It is instructive to understand how the unique Higgs/confining vacuum~\eqref{uniquevacuum} along the self-dual~$m_3 = 0$ line evolves into the unique Higgs or confined vacua at sufficiently large~$m_3 < 0$ or~$m_3 >0$, respectively, that we found below~\eqref{eq:perturbgeneric}. This closely parallels our discussion of Higgs-confinement continuity in the Fradkin-Shenker model in section~\ref{sec:FSdetail}. 

To this end, consider the full effective potential on~${\t S}^3$, including both the effects of different $\varepsilon_{e,m}$ in the monopole perturbation~\eqref{eqnmonopoleperturbation} and the mass term $m_3 \neq 0$ in \eqref{VeffHYM_2},
\begin{equation}
    V_{{\t S}^3 \, {\rm eff}} = - \beta \left(|\CM^1|^2- |\CM^2|^2\right)^2 + 2\varepsilon_e {\rm Re}(\CM^1) - 2\varepsilon_m{\rm Im}(\CM^2) + \frac{Cm_3}{v^2} \left(|\CM^1|^2- |\CM^2|^2\right)~.
\end{equation}
As before, all extrema must satisfy $\mathrm{Im}(\mathcal{M}^1) = \mathrm{Re}(\mathcal{M}^2) = 0$, so we can use the parametrization~\eqref{eq:mviaalpha} to reduce the effective potential to
\begin{equation}
V_{{\t S}^3 \, {\rm eff}}(\alpha) = - \beta v^4 \cos^2(2\alpha)+2\varepsilon_e v \cos\alpha - 2\varepsilon_m v \sin\alpha + Cm_3 \cos(2\alpha) \,.    
\end{equation}
It is convenient to define parameters~$\nu$ and~$\theta$ via
\begin{equation}
2\varepsilon_e v \cos\alpha - 2\varepsilon_m v \sin\alpha = \nu \cos(\alpha+\theta) \,, \quad \nu = 2v\sqrt{\varepsilon_e^2+\varepsilon_m^2} > 0\,, \quad \theta=\arctan\left(\frac{\varepsilon_m}{\varepsilon_e}\right) \in (0,\pi/2) \,,
\end{equation}
and rewrite the effective potential as
\begin{equation}\label{eq:fullvs3eff}
V_{{\t S}^3 \, {\rm eff}}(\alpha) = - \beta v^4 \cos^2(2\alpha)+\nu \cos(\alpha+\theta) + Cm_3 \cos(2\alpha) \,.    
\end{equation}

For our purposes, and paralleling the discussion in section~\ref{sec:FSdetail}, it will suffice to stay far away from the multicritical point by taking~$m_\phi^2 > 0$ to be very large and~$\beta \simeq 0$ in~\eqref{eq:fullvs3eff}. It can be checked that the resulting effective potential has a unique, gapped global minimum as a function of~$\theta$ and~$m_3$ that continuously interpolates between the limits of interest:
\begin{itemize}
    \item On the self-dual line~$m_3 = 0$, $\theta = \pi/4$, there is a unique minimum at~$\alpha = \pi - \theta = 3\pi/4$, in agreement with~\eqref{uniquevacuum}. 
    \item As we go into the Higgsing region, $m_3 < 0$ is large and negative, while~$\theta \to 0$ and~$\nu > 0$ is large. In this case we find a unique minimum at~$\alpha = \pi$, which coincides with the Higgs vacuum found below~\eqref{eq:perturbgeneric}.

    \item In the confining region we have~$m_3 > 0$ and~$\theta \to \pi/2$ with large~$\nu > 0$, leading to a unique minimum at~$\alpha = \pi/2$. This precisely agrees with the unique confining vacuum found below~\eqref{eq:perturbgeneric}. 
\end{itemize}
Thus the unique gapped vacuum continuously rotates from~$\alpha = \pi$ to~$\alpha = \pi/2$ in the~$\text{Re}(\CM^1)$--$\text{Im}(\CM^2)$ plane on~${\t S}^3$, as we interpolate from Higgsing to confinement.

\section{Multicritical Duality with the Easy-Plane~$\C\P^1$ Model}\label{sec:easy}

In section \ref{u1LatToCont} we studied the lattice-continuum correspondence and found that the symmetry of the SFS multicritical point is strictly larger that of HYQED. In particular, the mirror symmetry $\mathsf{C}_e$ (see \eqref{eqnCemirror}) must emerge in the IR from the point of view of HYQED. 

In this section, we will argue that the easy-plane $\mathbb{CP}^1$ model (EP$\mathbb{CP}^1$), i.e.~scalar QED$_3$ with two charge-1 complex scalars and a suitable scalar potential (see below), provides another, dual continuum QFT description of the same physics. In this~EP$\mathbb{CP}^1$ dual, mirror symmetry~$\mathsf{C}_e$ is manifest, while~duality symmetry $\mathsf{D}$ is not and must now emerge in the deep IR. More precisely, we have the following diagram of inclusions (see \eqref{eqnIRsymmetry}),
\begin{equation}\label{eqnsymmetrydualitycomparison}
    \begin{tikzcd}[column sep=small, row sep=small]
    G_{\text{EP}\mathbb{CP}^1}= (O(2)_e \times O(2)_m) \rtimes \bZ_2^{\mathsf{T}} \arrow[r,hook] & G_{\text{SFS CFT}} =  (O(2)_e \times O(2)_m) \rtimes (\bZ_2^\mathsf{D} \times \bZ_2^\mathsf{T})  \\
     G_{\text{EP}\mathbb{CP}^1} \cap G_{\rm HYQED}=S(O(2)_e \times O(2)_m) \rtimes \bZ_2^{\mathsf{T}} \arrow[u,hook] \arrow[r,hook] & G_{\rm HYQED} = S(O(2)_e \times O(2)_m) \rtimes (\bZ_2^\mathsf{D} \times \bZ_2^\mathsf{T})  \arrow[u,hook]
    \end{tikzcd}
\end{equation}
Here we have expressed the group $G_{\text{EP}\mathbb{CP}^1}$ in terms of identifications with the symmetries in $G_{\text{SFS CFT}}$, whose action on the fields of the EP$\mathbb{CP}^1$ model are discussed below.
Notice that $G_{\text{SFS CFT}}$ is the minimal group that contains the union $G_{\text{EP}\mathbb{CP}^1} \cup G_{\rm HYQED}$. 

Numerical evidence for the emergence of mirror symmetry of HYQED is given in section~\ref{subseclargeNextrap}, and we regard this as evidence for this proposed duality, since mirror symmetry is manifest in the EP$\mathbb{CP}^1$ duality frame. Crucially, the proposed duality is a multicritical duality, i.e.~it holds in the vicinity of the multicritical point of each model, described by the SFS CFT, which is reached by tuning two relevant operators. 

This differs sharply from previous proposals that assumed the~EP$\mathbb{CP}^1$ model would flow to a CFT with the tuning of a single relevant operator. In this scenario, the theory would have been IR dual to ordinary QED$_3$ with $N_f =2$ fermions (without a Higgs field) \cite{Karch:2016sxi, Wang:2017txt, Benini:2017aed}. Taken together, this would have implied the enhancement of the unitary symmetry in the IR CFT to~$O(4)$, which is incompatible with conformal bootstrap results~\cite{Li:2018lyb,Li:2021emd}. 

On the other hand our proposal is also a natural extension of these ideas, e.g.~the matching of symmetries and anomalies (and in part also phases) follows almost immediately from considerations in the literature, and the mechanism by which the~$\Z_2$ mirror and duality symmetries emerge at the multicritical point described by the SFS CFT involves self-dualities of HYQED and the EP$\C\P^1$ model.

\subsection{Lagrangian, Symmetries, and Anomalies}
\label{sec:EPCP1_lag}

The easy-plane $\mathbb{CP}^1$ model is scalar QED$_3$ with two flavors of complex scalars $z_{i}$ ($i=1,2$) with electric charge $q_b=1$ under a $U(1)$ gauge field $b$ (with conventionally normalized fluxes in the absence of background fields, $\oint db \in 2\pi\bZ$), and a suitable scalar potential. Including a scalar mass term~$m_z^2 \in \mathbb{R}$, the Lagrangian is
\begin{equation}\label{eqnCP1lagrangian}
\mathscr{L}_{\text{EP}\mathbb{CP}^1} = -\frac{1}{4\tilde{e}^2} f^{\mu\nu} f_{\mu\nu} - |D_b z_i|^2 - m_z^2|z_i|^2  - V_{\rm EP} \,,
\end{equation}
where $\tilde{e}^2$ is the gauge coupling, $f=db$ is the field strength, and $V_{\rm EP}$ is the easy-plane scalar potential,
\begin{equation}
\label{EP_potential}
V_{\text{EP}} = \lambda_{SO(3)}\left(|z_1|^2+|z_2|^2\right)^2 + \lambda_{\rm EP} |z_1|^2 |z_2|^2  \,, \qquad \lambda_{SO(3)}>0~, \qquad -4 < \lambda_{\rm EP}/\lambda_{SO(3)} < 0~.
\end{equation}
The stability conditions $\lambda_{SO(3)}>0$ and $\lambda_{\rm EP}/\lambda_{SO(3)}>-4$ ensure that the classical potential is bounded from below. The further requirement $\lambda_{\rm EP}<0$ follows by matching the phases of this theory with those of HYQED, as we will show in section \ref{sec:EPCP1_phases}.

In the isotropic case where $\lambda_{\rm EP}=0$, the continuous part of the global symmetry is $SO(3)_e \times U(1)_m$. Here~$U(1)_m$ factor is the usual magnetic/monopole symmetry of Abelian gauge theories, while $SO(3)_e$ is the flavor symmetry under which gauge-invariant local operators transform faithfully; the gauge-charged scalars~$z_i$ transform projectively as a doublet.\footnote{~See e.g.~\cite{Komargodski:2017dmc, Wang:2017txt} and references therein for a discussion of the symmetries and anomalies at $\lambda_{\rm EP}=0$.} Note that in this theory the monopole operators are not dressed by the matter fields, and hence they do not carry flavor charge. 

In the presence of the easy-plane deformation, $\lambda_{\rm EP}\neq 0$, the $SO(3)_e$ flavor symmetry is explicitly broken to its $O(2)_e$ subgroup, so that the full UV global symmetry of the EP$\C\P^1$ model is given by
\begin{equation}
G_{\text{EP}\mathbb{CP}^1}= (O(2)_e \times O(2)_m) \rtimes \bZ_2^{\mathsf{T}}~.
\end{equation}
Below we will explicitly write~$O(2)_{e,m}=U(1)_{e,m} \rtimes \mathbb{Z}_2^{\mathsf{C}_{e,m}}$. In our proposed duality with HYQED, the symmetries in~$G_\text{EP$\C\P^1$} \cap G_\text{HYQED}$ (see~\eqref{eqnsymmetrydualitycomparison}) will be explicitly be identified below, while mirror symmetry (which can be viewed as either one of~${\sf C}_{e, m}$) is manifest in the EP$\C\P^1$ model, and duality~$\sf D$ is manifest in HYQED (see section~\ref{sec:EPO5} below for more detail). 

Let us spell out the manifest symmetries of the  EP$\C\P^1$ model:
\begin{itemize}
    \item The $U(1)_e \subset SO(3)$ symmetry is the flavor Cartan,
    \begin{equation}
    U(1)_e : z_i \rightarrow {\exp\left(i\alpha \frac{\sigma_z}{2}\right)_i}^{\, j} \, z_j~, \qquad \alpha \sim\alpha+2\pi \,.   
    \end{equation}
    Here~$U(1)_e$ is normalized so that a $2\pi$ rotation is a gauge transformation, i.e.~$z_{1,2}$ have charge~$q_e=\pm 1/2$, respectively. Gauge invariant operators have $q_e \in \mathbb{Z}$.

    \item The current of the magnetic~$U(1)_m$ symmetry is 
    \begin{equation}
    j_m^\mu = \frac{1}{4\pi} \varepsilon^{\mu\nu\rho} f_{\nu\rho}~.    
    \end{equation}
    We denote the unit-charge monopole operator (with~$q_m = 1$) by~$\mathcal{M}_b$, and all gauge invariant operators have $q_m \in \mathbb{Z}$. Clearly~$U(1)_e$ and $U(1)_m$ commute, so the continuous symmetry of the model is simply $U(1)_e \times U(1)_m$.

    \item The $\bZ_2^{\mathsf{C}_e} \subset SO(3)$ symmetry is a Weyl reflection that swaps the two scalars,
    \begin{equation}
    \mathsf{C}_e : z_1 \leftrightarrow z_2 \,, \quad b_\mu \rightarrow b_\mu \,, \quad \CM_b \rightarrow \cM_b \,.
    \end{equation}
    It acts on the $U(1)_{e,m}$ charges as follows,
    \begin{equation}
    \mathsf{C}_e : q_e \rightarrow -q_e \,, \quad q_m \rightarrow q_m~,
    \end{equation}
    which immediately identifies it as a mirror symmetry (see~e.g.~table~\ref{table:latticesymmetries}). Thus, it commutes with $U(1)_m$ and combines with $U(1)_e$ to form $O(2)_e=U(1)_e \rtimes \mathbb{Z}_2^{\mathsf{C}_e}$. 

    \item The $\bZ_2^{\mathsf{C}_m}$ symmetry is obtained by combining the $\mathbb{Z}_2^{\mathsf{C}_e}$ mirror symmetry above with a conventional~$\mathbb{Z}_2^{\tilde\cC}$ charge-conjugation symmetry,
    \begin{equation}
    \tilde{\mathcal{C}} : z_i \rightarrow z_i^* \,, \quad b_\mu \rightarrow -b_\mu \,, \quad \CM_b \rightarrow \cM_b^* \,, \quad q_e \rightarrow -q_e \,, \quad q_m \rightarrow -q_m \,.
    \end{equation}
    Notice that $\mathsf{C}_e$ and $\tilde{\mathcal{C}}$ commute. We define $\mathsf{C}_m=\tilde{\mathcal{C}}\mathsf{C}_e(-1)^{q_m}$, which acts as
    \begin{equation}
    \mathsf{C}_m : z_1 \leftrightarrow z_2^* \,, \quad b_\mu \rightarrow -b_\mu \,, \quad \CM_b \rightarrow -\cM_b^* \,.
    \end{equation}
    It acts on the $U(1)$ charges as
    \begin{equation}
    \mathsf{C}_m : q_e \rightarrow q_e \,, \quad q_m \rightarrow -q_m~,
    \end{equation}
    and thus it also constitutes a mirror symmetry, which differs from~$\sf C_e$ by a non-mirror symmetry. It follows that~$\sf C_e$ commutes with $U(1)_e$ and combines with $U(1)_m$ to form $O(2)_m=U(1)_m \rtimes \mathbb{Z}_2^{\mathsf{C}_m}$.

    \item The $\mathbb{Z}_2^\mathsf{T}$ anti-unitary time-reversal symmetry is the combination of the $\mathbb{Z}_2^{\mathsf{C}_m}$ symmetry defined above with the usual $\mathbb{Z}_2^{\tilde{\mathcal{T}}}$ anti-unitary time-reversal symmetry of the Abelian Higgs model (see e.g.~\eqref{T_AHM}),
    \begin{equation}
    \tilde{\mathcal{T}} : z_i \rightarrow z_i \,, \quad b_0 \rightarrow b_0 \,, \quad b_{1,2} \rightarrow - b_{1,2} \,, \quad \CM_b \rightarrow \cM_b^* \,, \quad q_e \rightarrow q_e \,, \quad q_m \rightarrow -q_m \,.
    \end{equation}
    Note that~$\tilde{\mathcal{T}}$ commutes with both $\mathcal{C}$ and $\mathsf{C}_e$, hence with $\mathsf{C}_m$. We define $\mathsf{T}=\tilde{\mathcal{T}}\mathsf{C}_m$, which acts as follows,
    \begin{equation}
    \mathsf{T} : z_1 \leftrightarrow z_2^* \,, \quad b_0 \rightarrow -b_0 \,, \quad b_{1,2} \rightarrow b_{1,2} \,, \quad \CM_b \rightarrow -\cM_b \,, \quad q_e \rightarrow q_e \,, \quad q_m \rightarrow q_m \,.
    \end{equation}
    Notice that $\mathsf{T}$ preserves both $U(1)_{e,m}$ charges and satisfies $\mathsf{T}^2=1$.
\end{itemize}

Let us discuss the 't Hooft anomalies of the global symmetries above, and how they match the anomaly of HYQED in \eqref{mixedanomalyHYM}. The theory at $\lambda_{\rm EP}=0$ has a mixed anomaly between the $SO(3)_e$ flavor symmetry and the $U(1)_m$ magnetic symmetry, with inflow action \cite{Benini:2017dus}
\begin{equation}
S^{\mathbb{CP}^1}_{\rm anomaly} = \pi \int_{\mathcal{M}_4} w_2(SO(3)_e) \cup \left[\frac{dA_m}{2\pi}\right]_2 \,, \qquad \partial\mathcal{M}_4=\mathcal{M}_3 \,.  
\end{equation}
where $w_2(SO(3)_e)$ is the second Stiefel-Whitney class of the $SO(3)_e$ bundle and $[-]_2$ indicates that this anomaly only depends on the integer class $dA_m/2\pi$ mod 2. This anomaly immediately follows from the 't Hooft anomaly of pure Maxwell theory (see e.g.~\eqref{maxanom}) by substituting~$B_E^{(2)} \to \pi w_2(SO(3)_e)$, which captures the projective~$SO(3)_e$ transformation of the unit-charge Wilson line and its~$z_i$ endpoints. This is a prototypical example of an~'t Hooft anomaly that arises from symmetry fractionalization (see e.g.~\cite{Barkeshli:2014cna,Delmastro:2022pfo,Brennan:2022tyl, Brennan:2025acl,Seiberg:2025bqy}). Upon explicit breaking~$SO(3)_e$ to its~$U(1)_e$ Cartan subgroup, the class~$w_2(SO(3)_e)$ reduces to the first Chern class of the $U(1)_e$ bundle, leading to
\begin{equation}
S^{\text{EP}\mathbb{CP}^1}_{\rm anomaly} = \pi \int_{\mathcal{M}_4} \frac{dA_e}{2\pi} \wedge \frac{dA_m}{2\pi} \,, \qquad \partial\mathcal{M}_4=\mathcal{M}_3~.
\end{equation}
This precisely matches the 't Hooft anomaly of HYQED in~\eqref{mixedanomalyHYM}.

\subsection{Operator Mapping and Phases}
\label{sec:EPCP1_phases}

\begin{table}[t]
    \centering
    \begin{tabular}{c | c}
    HYQED &  EP$\mathbb{CP}^1$ \\
    $|\phi|^2$ & $|z_1|^2 |z_2|^2$ \\
    $-i \bar \psi \sigma_z \psi$ & $|z_1|^2 + |z_2|^2$ \\
    $-i \bar \psi \psi$ & $|z_1|^2 - |z_2|^2$ \\
    $\cM^1$ & $z_1 z_2^*$ \\
    $\cM^2$ & $\cM_b$ \\
    $\mathcal{O}_1+i\mathcal{O}_2$ & $\cM_b z^*_1 z_2$
\end{tabular}
    \caption{Proposed operator mapping between Higgs-Yukawa-QED and the easy-plane $\mathbb{CP}^1$ model at the multicritical point.}
    \label{tabCP1dualityoperatormapping}
\end{table}

Given the identification of the symmetries of the EP$\mathbb{CP}^1$ model and HYQED reviewed above, we can also map some of the gauge-invariant operators (see table~\ref{tabCP1dualityoperatormapping}). In particular, we identify
\begin{equation}\label{EPCP1_map_def}
 |\phi|^2 \leftrightarrow |z_1|^2 |z_2|^2 \,, \qquad
-i \bar\psi\sigma_z\psi \leftrightarrow |z_1|^2 + |z_2|^2~.
\end{equation}
Note that the easy-plane anisotropy~$\lambda_\text{EP}$ is mapped to the Higgs mass~$m_\phi^2$ of HYQED, and both of them are invariant under all symmetries, whereas the easy-plane scalar mass~$m_z^2$ is mapped to the $\mathsf{D}$-odd fermion mass~$m_3$ of HYQED. Since the scalar mass~$m_z^2$ in the EP$\mathbb{CP}^1$ model is invariant under all of~$G_{\text{EP}\mathbb{CP}^1}$, it follows that the duality symmetry $\mathsf{D}$ must emerge at the multicritical point in order for the proposed duality to work. At the multicritical point, the self-duality $\mathsf{D}$ may be derived for the EP$\mathbb{CP}^1$ model by applying the usual particle-vortex duality to both scalars~\cite{Motrunich:2003fz}, which shows that~$m_z^2$ is indeed~$\sf D$-odd and $\lambda_{\rm EP}$ is~$\sf D$-even. 

Let us discuss the phase diagram of the EP$\mathbb{CP}^1$ model. We propose that the theory flows to a multicritical CFT at low energies (with $\lambda_{SO(3)}$ and $\tilde{e}^2$ flowing to their IR fixed point and thus corresponding to irrelevant operators of the IR CFT) by tuning two relevant parameters: the $\mathsf{D}$-odd mass~$m_z^2$ and the $\mathsf{D}$-even easy-plane anisotropy~$\lambda_{\rm EP}$. This has to be contrasted with previous proposals, where only the mass term was assumed to correspond to a relevant deformation of the IR CFT.\footnote{~In \cite{Wang:2017txt}, the authors mentioned that ordinary QED$_3$ with $N_f=2$ fermions (and no scalar fields) does not have an operator dual to the anisotropy $|z_1|^2|z_2|^2$, and so the critical duality with the EP$\mathbb{CP}^1$ model may not be correct. They also noted that a direct simulation of the easy-plane transition in magnetic systems generically finds a first-order transition. This is all consistent with our multicritical proposal.} 

The duality-odd direction is the only one that can be convincingly analyzed from the EP$\mathbb{CP}^1$ point of view:\footnote{~This is the original direction considered for the easy-plane N\'eel-VBS transition (see \cite{Wang:2017txt} and references therein).} 

\begin{itemize}

    \item When $m_z^2 > 0$ is large and positive, the scalars $z_{1,2}$ can be reliably integrated out, and we are left with a pure Maxwell phase at low energies. This preserves the $U(1)_e$ flavor symmetry, while spontaneously breaking the $U(1)_m$ magnetic symmetry via monopole condensation $\braket{\mathcal{M}_b} \neq 0$. This gives rise to an $S^1_m$ space of vacua, which matches the SSB phase of HYQED reached upon deforming with a large positive fermion mass $m_3 > 0$, hence the sign in the operator map \eqref{EPCP1_map_def}.
    
    \item When $m_z^2 < 0$ is large and negative, the gauge field $b$ is completely Higgsed and the photon gets a mass. The $U(1)_m$ magnetic symmetry is preserved, while the $U(1)_e$ flavor symmetry is spontaneously broken by the condensation of the gauge-invariant order parameter $\braket{z_1 z_2^*} \neq 0$. This gives rise to an $S^1_e$ space of vacua, which corresponds to the SSB phase of HYQED reached upon deforming with a large negative fermion mass $m_3 < 0$. 
    
    In order to have this pattern of spontaneous symmetry breaking with the classical potential $V_{\rm EP}$ in \eqref{EP_potential}, it is necessary to take~$\lambda_{\rm EP}<0$, together with the stability conditions $\lambda_{\rm SO(3)}>0$ and $\lambda_{\rm EP}/\lambda_{SO(3)}>-4$. This is because the classical vacua of $V_{\rm EP}$ in the stability region and for $m_z^2<0$, after quotienting by the $U(1)$ gauge redundancy, depend on the sign of $\lambda_{\rm EP}$ as follows (see e.g.~\cite{Benini:2017aed}),
    \begin{itemize}[leftmargin=5em,labelsep=0.5em]
        \item[$\lambda_{\rm EP}<0$] : $|z_1|^2=|z_2|^2 =  \dfrac{|m_z^2|}{4\lambda_{SO(3)}+\lambda_{\rm EP}}$ $\quad \Rightarrow \quad S^1$ vacua~. 
        \item[$\lambda_{\rm EP}=0$] : $|z_1|^2+|z_2|^2 = \dfrac{|m_z^2|}{2\lambda_{SO(3)}}$ $\quad \Rightarrow \quad S^2$ vacua~. 
        \item[$\lambda_{\rm EP}>0$] : $\left\{z_1=0,|z_2|^2 = \dfrac{|m_z^2|}{2\lambda_{SO(3)}} \right\} \sqcup \left\{|z_1|^2 = \dfrac{|m_z^2|}{2\lambda_{SO(3)}},z_2=0\right\}$ $\quad \Rightarrow \quad 2$ vacua~.
    \end{itemize}
    Since we must have an $S^1$ space of vacua to match the corresponding phase of HYQED, we must restrict  the quartic couplings as in~\eqref{EP_potential}.\footnote{~Notice that $\lambda_{\rm EP}=0$ is a point of enhanced $SO(3)$ flavor symmetry of \eqref{eqnCP1lagrangian}, so that generically we do not expect the $\lambda_{\rm EP}=0$ separatrix in parameter space to be crossed along the RG flow. The regime $\lambda_{\rm EP} > 0$ is sometimes referred to as the easy-axis regime.}
\end{itemize}

The duality-even direction cannot be directly analyzed by examining the classical potential~$V_{\rm EP}$. Let us offer some intuition about the possible quantum behavior of the model as follows: let us adopt a regularization scheme where the multicritical point of the EP$\mathbb{CP}^1$ model is reached when $m_z^2=0$ and $\lambda_{\rm EP}=-2\lambda_{SO(3)}$, and rewrite
\begin{equation}
V_{\rm EP} = \lambda_{SO(3)} \left( |z_1|^4 + |z_2|^4 \right) + \lambda_{SO(3)} \tilde\lambda_{\rm EP} |z_1|^2 |z_2|^2 \,, \qquad \tilde\lambda_{\rm EP} \equiv 2+\lambda_{\rm EP}/\lambda_{SO(3)} \,,  
\end{equation}
which we take in the range $-2<\tilde{\lambda}_{\rm EP}<2$, the critical point being at $\tilde{\lambda}_{\rm EP}=0$.

When $-2<\tilde{\lambda}_{\rm EP}<0$, the easy-plane potential is (relatively speaking) attractive between the two scalars $z_1$ and $z_2$. Eventually, this attraction may overcome their electric repulsion  (at least if~$\tilde{e}^2$ is not too large), allowing the condensation of a flavor-neutral charge-2 bound state $z_1 z_2$. The electrically-charged condensate induces the Higgsing of the $U(1)$ gauge group to $\mathbb{Z}_2$, leading to a $\mathbb{Z}_2$ TQFT at low energies where $U(1)_e$ and $U(1)_m$ are unbroken and all particles are massive. This is to be identified with the toric code phase of HYQED, which occurs when~$m_\phi^2<0$ (hence the sign in the operator map \eqref{EPCP1_map_def}).\footnote{~Similar proposals have been made to describe the formation of spin liquid states in lattice systems of competing order, such as \cite{Chatterjee:2017pqv}, although their Higgs condensate has different quantum numbers than ours.}

Instead, when $\tilde{\lambda}_{\rm EP} > 0$ we assume that the theory exhibits a first-order transition, consistent with numerical investigations of magnetic systems that generically find a first-order transition in the EP$\mathbb{CP}^1$ model by dialing $m_z^2$ \cite{Desai:2019tel}. Restricting $0<\tilde{\lambda}_{\rm EP}<2$, this is a transition between the $S^1_e$ and $S^1_m$ circles of vacua, where both coexist and the (emergent) duality symmetry exchanging them is spontaneously broken. This matches the first-order self-dual line of HYQED obtained when $m_\phi^2>0$.

Note that we can use the operator map in table~\ref{tabCP1dualityoperatormapping} to express the unit-charge monopole deformation~\eqref{eqnmonopoleperturbation} that interpolates between HYQED an the Fradkin-Shenker model in the EP$\C\P^1$ duality frame. Finally, we can add charge-four monopoles to discuss the N\'eel-VBS transition for square-lattice spin-1/2 anti-ferromagnets, as we did in section~\ref{intro:NVBS}.

\subsection{Embedding $G_{\rm SFS~CFT}$ in $O(5)$}\label{sec:EPO5}

A convenient way to describe the symmetry map and the anomaly matching between HYQED and the EP$\mathbb{CP}^1$ model is to embed their unitary symmetry, which includes emerging mirror and duality respectively, $G_\text{SFS CFT}^{\rm unit.}=(O(2)_e \times O(2)_m) \rtimes \mathbb{Z}_2^\mathsf{D}$ in $O(4)$ as follows (see for instance~\cite{Wang:2017txt} and references therein),
\[\begin{pmatrix}
    O(2)_e & 0 \\ 0 & O(2)_m
\end{pmatrix} \subset O(4)~,\]
with the $\mathbb{Z}_2^{\mathsf{D}}$ duality symmetry acting as
\[\mathsf{D}= \mathcal{D}e^{\frac{i\pi}{2}(q_e-q_m)}=\begin{pmatrix}
    0 & \mathbbm{1}_2 \\ \mathbbm{1}_2 & 0
\end{pmatrix}\begin{pmatrix}
    -i\sigma_y & 0 \\ 0 & i\sigma_y
\end{pmatrix} = \begin{pmatrix}
     0 & i\sigma_y \\ -i\sigma_y & 0
\end{pmatrix} \in O(4) \,.\]
For instance, charge-conjugation and mirror symmetries act as the $O(4)$ elements
\begin{equation}
\begin{split}
  \mathsf{C}=\mathcal{C}&e^{i\pi q_m} = \begin{pmatrix}
    \sigma_z & 0 \\ 0 & \sigma_z
\end{pmatrix}\begin{pmatrix}
    \mathbbm{1}_2 & 0 \\ 0 & -\mathbbm{1}_2
\end{pmatrix}=\begin{pmatrix}
    \sigma_z & 0 \\ 0 & -\sigma_z
\end{pmatrix} \,, \\
  &\mathsf{C}_e = \begin{pmatrix}
    \sigma_z & 0 \\ 0 & \mathbbm{1}_2
\end{pmatrix}\,, \qquad 
  \mathsf{C}_m = \begin{pmatrix}
    \mathbbm{1}_2 & 0 \\ 0 & -\sigma_z
\end{pmatrix} \,.
\end{split}
\end{equation}
This embedding was previously used to support the scenario of a continuous symmetry enhancement to $O(4)$; here it will be a mathematical convenience. We can further embed $O(4)$ in $SO(5)$ as
\begin{equation}
 \begin{pmatrix}
    O(4) & 0 \\ 0 & \pm 1
\end{pmatrix}  \subset SO(5) \,, 
\end{equation}
where the $\pm 1$ entry is fixed for each $O(4)$ transformation by its determinant. This allows us to also include anti-unitary symmetries as $O(5)$ transformations with determinant $-1$, e.g.~the $\mathbb{Z}_2^\mathsf{T}$ time-reversal symmetry is represented by
\[\mathsf{T}= \mathcal{DCT}=\begin{pmatrix}
    \sigma_z & 0 & 0 \\
    0 & -\sigma_z & 0 \\
    0 & 0 & -1
\end{pmatrix}\in O(5) \,.\]
In the basis of operators of HYQED, this embedding can be conveniently described by taking the fundamental $O(5)$ vector to be comprised of the following operators,
\[\left({\rm Re}(\CM^1),{\rm Im}(\CM^1),{\rm Re}(\cM^2),{\rm Im}(\cM^2),i \bar \psi \psi\right) \,.\]
For instance, it is obvious in this embedding that the only unitary symmetries under which the fermion bilinear $i\bar\psi\psi$ is odd are the ones that have negative $O(4)$ determinant, such as mirror symmetries $\mathsf{C}_{e,m}$.
Upon using the duality maps, this defines the embeddings into $O(5)$ of the global symmetry groups of all the models we are discussing, including $G_\text{SFS CFT}$. The anomaly for each of these groups is given by pulling back the Euler class
\[\label{eqnparentanomaly}e_5 \in H^5(BO(5),\bZ^{\rm det})~,\]
with coefficients twisted by the determinant line of $O(5)$. Note that this (twisted) integer class should be viewed as the gauge-invariant anomaly polynomial, which in turn defines a Chern-Simons-like anomaly-inflow action in 3+1d. It also describes the~$O(5)$ anomaly of the~$S^4$ sigma model with Wess-Zumino term considered in~\cite{Senthil:2005jk}.

For the unitary symmetries that form a subgroup of $SO(5)$, this amounts to the anomaly inflow action
\begin{equation}
    S_{\rm anomaly} = \pi \int_{\mathcal{M}_4} w_4(SO(5)) \,, \qquad \partial\mathcal{M}_4=\mathcal{M}_3 \,,
\end{equation}
where $w_4(SO(5))$ is the fourth Stiefel-Whitney class. This can be derived by noting that for~$SO(5)$, the Bockstein of $w_4$ is the Euler class. For the $G_\text{SFS CFT}=(O(2)_e \times O(2)_m) \rtimes \mathbb{Z}_2^\mathsf{D}$ subgroup, the anomaly can be written as
\[ S_{\rm anomaly} = \pi \int_{\cM_4} w_2(O(2)_e) \cup w_2(O(2)_m) \,, \qquad \partial\mathcal{M}_4=\mathcal{M}_3 \,, \]
by the Whitney sum formula \cite{milnor1974characteristic}.

Let us stress again thatHYQED and the EP$\mathbb{CP}^1$ model only show a discrete $\mathbb{Z}_2$ enhancement of their respective global symmetries to $G_\text{SFS CFT}$, without further continuous enhancement of the unitary symmetry to $O(4)$ or $SO(5)$. In fact, the large-$N$ estimates of anomalous dimensions we obtain in section~\ref{sec:largeN} show that the singlet fermion bilinear $i\bar \psi \psi$ is highly irrelevant, with~$\Delta\simeq 4.70$, while unit-charge monopole operators are highly relevant, with~$\Delta\simeq 0.63$, so the SFS CFT is very far from having an $SO(5)$ global symmetry.  Nevertheless, the embedding described above is a convenient way to encode the anomaly and to make all symmetries manifest. It is also consistent with the existence of an $SO(5)$ fixed point that flows to the SFS CFT upon adding relevant, $SO(5)$-breaking deformations (see~\cite{Wang:2017txt} for a related discussion).

\section{Large-$N$ Generalization of Higgs-Yukawa-QED}\label{sec:largeN}

In this section, we use the large-$N$ expansion to compute the scaling dimensions of the relevant operators $\bar\psi_i\psi^j$ and $\phi^*\phi$ of the $U(1)_f\times U(1)_\mathcal{M}$-preserving model at the tricritical point. To this end, we generalize our model \eqref{LagrangianMinkowski} to include $2N$ Dirac fermions $\psi_i$, where $i=1,\dots,2N$, with unit electric charge.
In Euclidean signature, the Lagrangian is\footnote{~We perform the Wick rotation $t=-i\tau \,, \partial_t=i\partial_\tau \,, a_t = i a_\tau \,, \gamma^\tau = i\gamma^t$. The Euclidean gamma matrices are then the Hermitian matrices $\gamma^\mu=\lbrace -\sigma_y,\sigma_z,-\sigma_x \rbrace$. Note that $(\gamma^\mu)^2=1$ for all $\mu$, and the Euclidean $\gamma^0$ is anti-symmetric and imaginary. We take $\bar\psi=\psi^\dagger \gamma^0$, and we choose normalization factors in the couplings for future convenience.
The $i,j$ indices are flavor labels and their up/down position does not matter.}
\begin{equation}\label{largeNLagrangian}
\begin{split}
\mathscr{L} &= \frac{1}{4e^2} f^{\mu\nu} f_{\mu\nu} + {\bar\psi}_i \gamma^\mu \left( \partial_\mu - i a_\mu\right) \psi_i+|(\partial_\mu-2ia_\mu)\phi|^2 \\
&+ \frac{\lambda_4}{N} (\phi^*\phi)^2 + \frac{i}{2\sqrt{N}}\left( \mathscr{Y}_{ij}\phi^*\psi_i^t\gamma^0\psi_j + \mathscr{Y}^*_{ij} \phi\, \bar\psi_i \gamma^0 \bar\psi^t_j \right) \,.
\end{split}
\end{equation}
In the Lagrangian above, $\mathscr{Y}_{ij}$ is the symmetric matrix of Yukawa couplings,\footnote{~The matrix $\mathscr{Y}_{ij}$ is symmetric because $\psi^t_i \gamma^0 \psi_j = (\psi^t_i \gamma^0 \psi_j)^t = - \psi^t_j (\gamma^0)^t \psi_i = \psi^t_j \gamma^0 \psi_i$. The complex conjugate term follows from the property $(\psi^t_i \gamma^0 \psi_j)^\dagger = (\psi^t_j \gamma^0 \psi_i)^\dagger =\psi_i^\dagger (\gamma^0)^\dagger \psi_j^*= -\bar\psi_i \gamma^0 \bar\psi^t_j$.} which as a spurion transforms as $\mathscr{Y} \rightarrow U^T \mathscr{Y} U$ under an $SU(2N)$ transformation $U$. For generic $N$, we take
\begin{equation}\label{YukawacouplingN}
\mathscr{Y} = y \, \sigma_x \otimes \mathbbm{1}_N = y \begin{pmatrix}
0 & \mathbbm{1}_N \\
\mathbbm{1}_N & 0
\end{pmatrix} \,,
\end{equation}
which satisfies $\mathscr{Y}^\dagger \mathscr{Y} = |y|^2 \mathbbm{1}_{2N}$. Thus, up to discrete quotients, the Lagrangian \eqref{largeNLagrangian} preserves the $U(1)_\mathcal{M}$ magnetic symmetry and a subgroup $G_f$ of the $SU(2N)$ flavor symmetry of ordinary QED with $2N$ Dirac fermions,\footnote{~The condition $U^T \mathscr{Y} U= \mathscr{Y}$ can be rewritten as $V^TV=\mathbbm{1}_{2N}$, where $V=P^{-1}UP$ and $P$ is the unitary matrix such that $P^T \mathscr{Y}P=\mathbbm{1}_{2N}$, namely
\begin{equation}
    P = \frac{1}{\sqrt{2}}\begin{pmatrix}
\mathbbm{1}_N & i\mathbbm{1}_N \\
\mathbbm{1}_N & -i\mathbbm{1}_N
\end{pmatrix} \,.
\end{equation}
Imposing that $U^\dagger U=\mathbbm{1}_{2N}$ and $\det U=1$ further constrains $V^\dagger V=\mathbbm{1}_{2N}$ (which implies $V=V^*$) and $\det V=1$. Thus $G_f \simeq \{V \in GL(2N,\mathbb{C}):V=V^*,V^TV=\mathbbm{1}_{2N},\det V=1\}=SO(2N,\mathbb{R})$.
\label{footnote:unbrokensubgroup}}
\begin{equation}
G_f = \{ U \in SU(2N) : U^T \mathscr{Y} U = \mathscr{Y}\} \simeq SU(2N) \cap O(2N,\mathbb{C}) \simeq SO(2N) \,.   
\end{equation}
The only gauge-invariant Yukawa coupling that preserves $G_f$ is then $\mathscr{Y}$ as in \eqref{YukawacouplingN}, up to rescaling.\footnote{~Let $\mathscr{Y}'$ be a symmetric matrix such that $U^T \mathscr{Y}'U=\mathscr{Y}'$ for all $U \in G_f$. Writing $U=PVP^{-1}$ as in footnote \ref{footnote:unbrokensubgroup}, this condition is equivalent to $[V,P^T\mathscr{Y}'P]=0$ for all $V\in SO(2N)$. If $N>1$, this implies that $P^T \mathscr{Y}'P$ is a multiple of the identity, by Schur's lemma. If $N=1$, instead, it implies that $P^T\mathscr{Y}'P$ is a linear combination of the identity and the Levi-Civita symbol, but the latter contribution must vanish because $\mathscr{Y}'$ is symmetric. In both cases we get $P^T \mathscr{Y}'P = c\mathbbm{1}_{2N}$, namely $\mathscr{Y}'=c\,\mathscr{Y}$.
\label{footnote:preservingY}}
The Higgs-Yukawa-QED model is recovered when $N=1$, where $G_f\simeq SO(2)$ is the $U(1)_f$ flavor symmetry that acts with opposite charges on $\psi_1$ and $\psi_2$, with conserved current $j_f^\mu =i \bar\psi \, \gamma^\mu \sigma_z\psi$.

\subsection{Lagrangian of the Large-$N$ CFT}
A well-defined large~$N$ limit is taken with $\Lambda\equiv e^2N$, $\lambda_4$, and $y$ fixed. In this way, the large-$N$ (massless) propagator of the scalar $\phi$ can be computed exactly and receives its leading contribution from resumming fermionic bubble diagrams,
\begin{equation}
\braket{\phi(p)\phi^*(-p)} = \frac{1}{p^2} \sum_{k=0}^{\infty} \left(-\frac{c\,\mathrm{tr}(\mathscr{Y}^\dagger \mathscr{Y})|p|}{2N} \frac{1}{p^2}\right)^k = \frac{1}{p^2+c|y|^2|p|} \simeq \frac{1}{c|y|^2|p|} \,,
\label{phi propagator}
\end{equation}
where $c$ is an $N$-independent positive constant and in the last step we have taken the low-energy limit $|p|\ll|y|^2$. We thus see that the field which has a well-defined IR propagator at large $N$ is $\sigma \equiv y^* \phi$, whose propagator scales as $1/|p|$. In particular, this implies that the scaling dimension of $\phi$, which is $1/2$ in the UV free theory, becomes $1+\mathcal{O}(1/N)$ in the IR CFT.\footnote{~This is a slight abuse of terminology, as the scalar field is not a gauge invariant operator. However, both in the UV (where $e^2,|y|^2\rightarrow 0$) and in the IR at leading order at large $N$, the scalar sector is the theory of a (generalized) free field.} Thus, both the quartic interaction $(\phi^* \phi)^2$ and the covariant kinetic term for $\phi$ are irrelevant operators. As expected from standard large-$N$ counting, the scalar field behaves like a complex Hubbard-Stratonovich field $\sigma$ of dimension $1+\mathcal{O}(1/N)$ and electric charge~$2$.\footnote{~At large $N$, the interaction with the photon is generated via loop effects, consistently with the Ward-Takahashi identity, see appendix \ref{appendix:rules}.}
Analogously, the large-$N$ propagator of the photon scales like $1/|p|$ in the low-energy limit $|p| \ll \Lambda$ (see appendix \ref{appendix:rules} for the explicit computation), so that the Maxwell term for the gauge field is irrelevant and can be neglected in the IR. Moreover, the $1/|p|$ behavior of the photon propagator (and, consequently, the photon scaling dimension) is exact to all orders in the large-$N$ expansion. This is consistent with the fact that $j_\mathcal{M}^\mu \sim \varepsilon^{\mu\nu\rho}f_{\nu\rho}$ is a conserved current that is a primary operator in a unitary 3d CFT, and hence it has a protected dimension equal to 2.

The discussion above implies that the tricritical CFT at large $N$ and low energies (namely $E\ll\Lambda,\lambda_4,|y|^2$) is described by the Lagrangian
\begin{equation}\label{largeNtricrit}
\mathscr{L}_{\text{CFT}} =  \bar\psi_i \left(\slashed{\partial}-i\slashed{a}\right)\psi_i
+ \frac{i}{2\sqrt{N}}Y_{ij} \, \sigma^* \psi^{t}_i \gamma^0 \psi_j + \frac{i}{2\sqrt{N}}Y^*_{ij} \,\sigma~ \bar\psi_i \gamma^0 \bar\psi^t_j~,
\end{equation}
where $Y = \sigma_x \otimes \mathbbm{1}_N$ is dimensionless, so that the Yukawa interaction is exactly marginal.
Our task is to compute the anomalous dimension of the fermion and scalar quadratic operators, which correspond to the massive deformations that can be added to the Lagrangian in \eqref{largeNtricrit},
\begin{equation}\label{relevantlagrangian}
\mathscr{L} = \mathscr{L}_{\text{CFT}} + M_{ij} \bar\psi_i\psi_j + \lambda \, \sigma^*\sigma \,,   
\end{equation}
where $\lambda$ is a coupling of unit mass dimension that plays the role of the scalar mass at large $N$,\footnote{~We are using a scheme where the tricritical point is reached by tuning $\lambda=0$.} and $M_{ij}$ is the Hermitian fermion mass matrix, which as a spurion transforms as $M \rightarrow U^\dagger M U$ under an $SU(2N)$ transformation $U$.
The Feynman rules for this Lagrangian are summarized in appendix \ref{appendix:rules}.

\subsection{Beta Functions and Anomalous Dimensions}
In order to compute the anomalous dimensions of relevant operators $\mathcal{O}_a$ (that have the same leading-order scaling dimension), one can deform the Lagrangian with $\sum_ag_a \mathcal{O}_a$, and compute the $\beta$ function of the couplings $g_a$, denoted as $\beta_a$. We have that
\begin{equation}\label{anodimformula}
\gamma_{ab} = \frac{\partial\beta_a}{\partial g_b}{\bigg{|}}_{\{g\}=0} \,,    
\end{equation}
where $\{g\}=0$ is the undeformed conformal fixed point. The anomalous dimensions are found by diagonalizing the matrix in \eqref{anodimformula}.

Our task is then to compute the beta functions of $M$ and $\lambda$ at leading order in the large-$N$ expansion and then extract the anomalous dimensions at the tricritical fixed point, where both $M=0$ and $\lambda=0$. A few comments are in order:
\begin{enumerate}
    \item There cannot be any mixing between the operators $\sigma^*\sigma$ and $\bar\psi_i\psi_j$, despite the fact that they have the same leading-order scaling dimension $\Delta=2$, as they carry different quantum numbers. This implies that the $\beta$ function of $M$ cannot depend linearly on $\lambda$, and vice versa. In particular, $\sigma^*\sigma$ is a singlet under both $G_f \simeq SO(2N)$ and time reversal. If $N>1$, the only fermion bilinear that is invariant under $G_f$ is the $SU(2N)$ singlet $\bar\psi_i\psi_i$, but this is odd under time reversal. If $N=1$, there is another singlet under $G_f \simeq U(1)_f$, which is $\bar\psi_i(\sigma_z)_{ij}\psi_j$, and it is even under time reversal.\footnote{~Consider the fermion bilinear $\bar\psi_i(T^a)_{ij}\psi_j$, where $T^a$ are the Hermitian generators of $U(2N)$, which transforms as $\bar\psi T^a\psi \rightarrow \bar\psi U^\dagger T^a U \psi $ under the $G_f$ global symmetry. Writing $U=PVP^{-1}$ as in footnote~\ref{footnote:unbrokensubgroup}, the singlet condition $T^a=U^\dagger T^a U$ is equivalent to $[P^{-1 }T^a P, V]=0$ for all $V\in SO(2N)$. If $N>1$ this implies that $P^{-1 }T^a P$ is a multiple of the identity, by Schur's lemma, namely $T^a$ itself is a multiple of the identity, and thus the only $G_f$ singlet is $\bar\psi\psi$. If $N=1$, instead, it implies that $P^{-1 }T^a P$ is a linear combination of the identity and the Levi-Civita symbol. The latter gives $T^a$ as a multiple of $\sigma_z$, namely the additional $G_f$ singlet $\bar\psi\sigma_z\psi$ for $N=1$.} However, as we discussed around \eqref{action_of_D}, it is odd under the $\mathbb{Z}_2^\mathcal{D}$ duality symmetry, that instead leaves $\sigma^*\sigma$ invariant. Thus, the operators $\sigma^*\sigma$ and $\bar\psi_i\psi_j$ do not mix.
    \item The only symmetric Yukawa matrix that is allowed by gauge invariance and the $G_f \simeq SO(2N)$ global symmetry is $Y=\sigma_x \otimes \mathbbm{1}_N$ up to multiplication by a (dimensionless) constant (see footnote \ref{footnote:preservingY}), so no other marginal Yukawa interaction can be generated at any order in the large-$N$ expansion.
\end{enumerate}
The details of the computation of the $\beta$ functions are summarized in appendix \ref{appendix:beta}. The result is (note that the $\beta$ function of the fermion mass is itself a $2N\times 2N$ matrix\footnote{~As a consistency check, note that under an $SU(2N)$ flavor transformation $U$, acting on the spurions as $M\rightarrow U^\dagger M U$ and $Y\rightarrow U^T Y U$, the $\beta$ function for the fermion mass transforms covariantly as $\beta_M \rightarrow U^\dagger \beta_M U$.})
\begin{equation}\label{betafunctions}
\begin{split}
\beta_{\lambda} &= - \frac{16}{3 \pi^2 N} \lambda \,, \\
\beta_M &= \frac{4}{3\pi^2 N}
\left(-8 M + \frac{12}{N} \, \mathrm{Tr}(M) \mathbbm{1}_{2N} \right) + \frac{4}{3\pi^2 N}
\left( M + 3 \, Y^\dagger M^T Y \right) \\
&= \frac{4}{3\pi^2 N} \left(-7 M + \frac{12}{N} \, \mathrm{Tr}(M) \mathbbm{1}_{2N} + 3 \, Y^\dagger M^T Y \right)  \,,
\end{split}
\end{equation}
where in the expression for $\beta_M$ the first contribution comes from the gauge sector whereas the second one comes from the Yukawa sector.

The scaling dimension of $\sigma^*\sigma$ can be extracted from the $\beta$ function of $\lambda$ and reads
\begin{equation}
\Delta[\sigma^* \sigma] = 2 - \frac{16}{3 \pi^2 N} \,.
\end{equation}
Notice that this result is always smaller than 2 and greater than the naive engineering dimension of the scalar bilinear in the UV, $\Delta[\phi^*\phi]=1$.

Concerning the scaling dimension of $\bar\psi_i\psi_j$, as a consistency check let us first consider the case without Yukawa interaction, so that the theory \eqref{largeNtricrit} reduces to ordinary QED$_3$ with $2N$ Dirac fermions. For the $SU(2N)$ singlet and adjoint fermion masses $m_s$ and $m_a$, we have
\begin{equation}
\begin{split}
\beta_{m_s} &= + \frac{64}{3\pi^2 N}m_s \equiv \gamma_s m_s \,,\\
\beta_{m_a} &=  -\frac{32}{3\pi^2 N}m_a \equiv \gamma_a m_a \,,
\end{split}
\end{equation}
which match known results for the anomalous dimensions $\gamma_s$ and $\gamma_a$ of the singlet and the adjoint fermion bilinears, respectively, in fermionic QED$_3$ (see e.g.~\cite{Chester:2016ref} and references therein).

We now consider the case of interest here, where $Y=\sigma_x \otimes \mathbbm{1}_N$. The $\beta$ function for the $SU(2N)$ singlet $\bar\psi_i\delta_{ij}\psi_j$ corresponds to taking $M=m_{\text s}\mathbbm{1}_{2N}$, and it is given by
\begin{equation}
\beta_{m_{\text s}} = \frac{80}{3\pi^2 N} m_{\text s} \,,   
\end{equation}
which implies
\begin{equation}
\Delta[\bar\psi \psi_{\text{singlet}}] = 2 + \frac{80}{3 \pi^2 N} \,.    
\end{equation}
Instead, for the $SU(2N)$ adjoint $\bar\psi_i \, T^a_{ij}\,\psi_j$ we need to distinguish between the unbroken and the broken generators, corresponding to the decomposition of the adjoint representation of $SU(2N)$ into the direct sum of the adjoint and the rank-$2$ symmetric traceless representations of $SO(2N)$.

The unbroken generators, transforming in the adjoint representation of $G_f \simeq SO(2N)$, satisfy $Y(T_{\text{u}}^a)^TY=-T_{\text{u}}^a$ and they are given by
\begin{equation}\label{unbrokengenerators}
    T^a_{\text{u}} = \begin{pmatrix}
        A && B \\
        B^\dagger && -A^T
    \end{pmatrix} \,, \qquad A^\dagger=A \quad \text{and} \quad B^T=-B \,.
\end{equation}
These are $2N^2-N$ independent generators, matching the dimension of $SO(2N)$. For $N=1$, we have one unbroken generator $T_u \propto \sigma_z$.

The broken generators, transforming in the rank-$2$ symmetric traceless representation of $G_f \simeq SO(2N)$, satisfy $Y(T_{\text{b}}^A)^TY=T_{\text{b}}^A$ and they are given by
\begin{equation}
    T^a_{\text{b}} = \begin{pmatrix}
        A && B \\
        B^\dagger && A^T
    \end{pmatrix} \,, \qquad A^\dagger=A \,,\quad \text{Tr}A=0 \,,\quad \text{and} \quad B^T=B \,.
\end{equation}
These are $2N^2+N-1$ independent generators, matching the difference between the dimensions of $SU(2N)$ and $SO(2N)$. For $N=1$, we have two broken generators $T^a_{\text b} \propto (\sigma_x,\sigma_y)$.

We then compute the beta function for the mass matrices $M_{\text u} = m^a_{\text u}T^a_{\text u}$ and $M_{\text b} = m^a_{\text b}T^a_{\text b}$ along the unbroken and broken directions, respectively. Using that both sets of generators are traceless and that they satisfy
\begin{equation}
   Y^\dagger M_{\text u}^T Y = - M_{\text u} \,, \qquad 
   Y^\dagger M_{\text b}^T Y = M_{\text b} \,,
\end{equation}
we get
\begin{equation}
\begin{split}
\beta_{m_\text{u}} &= -\frac{40}{3\pi^2 N} m_\text{u} \,, \\
\beta_{m_\text{b}} &= -\frac{16}{3\pi^2 N} m_\text{b} \,,
\end{split}
\end{equation}
which imply
\begin{equation}
\begin{split}
\Delta[\bar\psi \psi_{\text{unbroken}}] &= 2 - \frac{40}{3 \pi^2 N} \,, \\
\Delta[\bar\psi \psi_{\text{broken}}] &= 2 - \frac{16}{3 \pi^2 N} \,.
\end{split}
\end{equation}
Note that the scalar unitarity bound $\Delta\geq 1/2$ is satisfied for all $N$.

Finally, let us discuss the scaling dimensions of monopole operators $\mathcal{M}_{(q_\mathcal{M})}$, which have charge $q_\mathcal{M}$ under~$U(1)_\mathcal{M}$. For the theory \eqref{largeNtricrit} these were computed in \cite{Dupuis:2021flq}, and the ones for monopoles of lowest magnetic charges read\footnote{~At the next-to-leading order in the large-$N$ expansion we are working with, these are the scaling dimensions of all monopole operators with a given magnetic charge, irrespectively of their flavor/spin quantum numbers.}
\begin{equation}
\begin{split}
\Delta[\mathcal{M}_{(1)}] &= 2N(0.26510) + 0.10285 \,, \\  
\Delta[\mathcal{M}_{(2)}] &= 2N(0.67315) + 0.18663 \,, \\ 
\Delta[\mathcal{M}_{(3)}] &= 2N(1.18643) + 0.26528 \,, \\ 
\Delta[\mathcal{M}_{(4)}] &= 2N(1.78690) + 0.34262 \,.
\end{split}
\end{equation}
The leading-order result is the same as for ordinary QED$_3$ with $2N$ Dirac fermions, and shows that monopole operators are irrelevant at large $N$.

\subsection{Current-Current Two-Point Function}
In this section, we compute the current-current two-point function, both for the $SO(2N)$ flavor symmetry and the $U(1)_\mathcal{M}$ magnetic symmetry. The spacetime dependence of these correlators is fixed by conformal symmetry, up to a coefficient that we compute at leading order in the large-$N$ expansion.

The current of the $SO(2N)$ flavor symmetry is
\begin{equation}
    J^a_\mu = i \bar\psi_i \gamma_\mu (T^a_{\mathrm u})_{ij}\psi_j \,,
\end{equation}
where $T^a_{\mathrm{u}}$, with $a=1,\dots,2N^2-N$, are the $SO(2N)$ generators in \eqref{unbrokengenerators}. To match our normalization of the flavor current in the $N=1$ case (where $T_{\rm u}=\sigma_z$) we normalize them as $\tr(T^a_{\mathrm{u}}T^b_{\mathrm{u}})=2\delta^{ab}$. The two-point function can be obtained at leading order by computing a fermionic bubble diagram with two insertions. This is the same as the bubble diagram $\Pi_{\mu\nu}(p)$ contributing to the photon propagator, up to the replacement $\tr(\mathbbm{1}_{2N}) \rightarrow \tr(T^a_{\mathrm{u}}T^b_{\mathrm{u}})$, which we explicitly compute in \eqref{bubblephoton}. We thus get
\begin{equation}\label{2pointflavor}
\braket{J^a_\mu (p)J_\nu^b(-p)} = \frac{\tr(T^a_{\mathrm{u}}T^b_{\mathrm{u}})}{\tr(\mathbbm{1}_{2N})} \Pi_{\mu\nu}(p) = -\frac{1}{8} \frac{ p^2\delta_{\mu\nu} - p_\mu p_\nu }{|p|} \delta^{ab} \,.
\end{equation}

The current of the $U(1)_\mathcal{M}$ flavor symmetry is
\begin{equation}
j^\mathcal{M}_\mu  = \frac{i}{2\pi} \varepsilon_{\mu\nu\rho} \partial^\nu a^\rho \,.  
\end{equation}
Its two-point function is given by
\begin{equation}\label{2pointmagnetic}
\braket{j^\mathcal{M}_\mu (p) j^\mathcal{M}_\nu(-p)} = -\frac{\varepsilon_{\mu\rho\sigma}\varepsilon_{\nu\alpha\beta}}{4\pi^2} p^\rho p^\alpha \braket{a^\sigma(p)a^\beta(-p)} = -\frac{2}{\pi^2N} \frac{ p^2\delta_{\mu\nu} - p_\mu p_\nu }{|p|} \,,
\end{equation}
where we used the expression of the large-$N$ photon propagator in \eqref{photonpropagator}.

\subsection{Extrapolation to $N=1$}\label{subseclargeNextrap}

Let us now extrapolate the previous results to the case of our interest, where $N=1$.
\begin{itemize}
\item The scaling dimension of the time-reversal odd and duality even $U(1)_f$ singlet $\bar\psi\psi$ is
\begin{equation}
\Delta[\bar\psi \psi_{\text{singlet}}] = 2 + \frac{80}{3 \pi^2 N} \simeq 4.70 \,, 
\end{equation}
which suggests that this operator is highly irrelevant at the fixed point.

\item The scaling dimension of the time-reversal even and duality even $U(1)_f$ singlet $\sigma^*\sigma$ is
\begin{equation}
\Delta[\sigma^* \sigma] = 2 - \frac{16}{3 \pi^2 N} \simeq 1.46 \,,
\end{equation}
which is relevant at the fixed point.

\item The scaling dimension of the time-reversal even and duality odd $U(1)_f$ singlet $\bar\psi\sigma_z\psi$ is
\begin{equation}
\Delta[\bar\psi \psi_{\text{unbroken}}] = 2 - \frac{40}{3 \pi^2 N} \simeq 0.65 \,,    
\end{equation}
which is also relevant, and only slightly above the unitarity bound. 

\item The scaling dimension of the $U(1)_f$ doublet $(\bar\psi\sigma_x\psi,\bar\psi\sigma_y\psi)$ is
\begin{equation}
\Delta[\bar\psi \psi_{\text{broken}}] = 2 - \frac{16}{3 \pi^2 N} \simeq 1.46 \,,    
\end{equation}
which is relevant. 

\item The scaling dimensions of the monopole operators with lowest $U(1)_\mathcal{M}$ charge are
\begin{equation}
\begin{split}
\Delta[\mathcal{M}_{(1)}] &= 2N(0.26510) + 0.10285 \simeq 0.63 \,, \\  
\Delta[\mathcal{M}_{(2)}] &= 2N(0.67315) + 0.18663 \simeq 1.53 \,, \\ 
\Delta[\mathcal{M}_{(3)}] &= 2N(1.18643) + 0.26528 \simeq 2.64 \,, \\ 
\Delta[\mathcal{M}_{(4)}] &= 2N(1.78690) + 0.34262 \simeq 3.92 \,,
\end{split}
\end{equation}
which suggests that monopole operators with charges $q_\mathcal{M}=1,2,3$ are relevant at the fixed point. By contrast, all monopole operators with charges $q_\mathcal{M} \geq 4$ appear to be irrelevant.\footnote{~In \cite{Dupuis:2021flq} scaling dimensions of monopoles operators up to charge $q_\mathcal{M}=26$ were computed. They are monotonically increasing as the magnetic charge increases.} 
\end{itemize}
These results show evidence that there are two relevant operators for the $U(1)_f \times U(1)_\mathcal{M}$-preserving tricritical point, one being duality even ($\sigma^*\sigma$) and the other being duality odd ($\bar\psi \sigma_z \psi$), and both preserve time reversal~$\CT$ and charge conjugation~$\CC$. Notice that the results for their scaling dimensions are compatible with the available bootstrap bounds for theories with an $(O(2) \times O(2)) \rtimes S_2$ global symmetry \cite{Stergiou:2019dcv, Henriksson:2021lwn, Kousvos:2021rar}.
Moreover, the unit-charge minimal monopole is also relevant and triggers an RG flow towards the Fradkin-Shenker tricritical point.

As discussed in section \ref{sec:emergemirr}, recall that the $\mathbb{Z}_2$ mirror symmetry maps the $U(1)_f$-breaking fermion bilinear $\bar\psi(\sigma_x-i\sigma_y)\psi$ to the charge-2 monopole ${\mathcal{M}}_{(2)}$, which then must have the same scaling dimensions. In our estimate we see evidence for this, as their scaling dimensions ($1.46$ and $1.53$, respectively) agree within less than five percent.
Mirror symmetry exchanges the flavor and the magnetic current, so they should have the same two-point function. Extrapolating the leading-order results \eqref{2pointflavor} and \eqref{2pointmagnetic} to $N=1$, we find that the coefficients of the flavor and magnetic current two-point functions are $C_f=1/8 = 0.1250$ and $C_\mathcal{M} = 2/\pi^2 \simeq 0.2026$, respectively. As commonly expected in extrapolation of large-$N$ results to singlet operators, the mismatch between these coefficients is more pronounced with respect to the previous case. Notice also that these leading-order results are insensitive to the addition of the charge-2 scalar field, namely they are the same as in ordinary QED$_3$ with $2N$ Dirac fermions, where the emergence of mirror symmetry at $N=1$ is not expected in the symmetry-breaking scenario. We leave to the future the task of computing subleading corrections in $1/N$ to $C_{f,\mathcal{M}}$, which are sensitive to the presence of the charge-2 scalar.

Finally, all monopole operators with magnetic charge $q_\mathcal{M} \geq 4$ are irrelevant. This supports the scenario where the $U(1)_f \times U(1)_\mathcal{M}$-preserving CFT appears as a deconfined quantum multicritical point in the N\'eel-VBS phase diagram.

\section{Discussion}\label{sec:discussion}

In this work, we have explored the Fradkin-Shenker (FS) phase diagram from the point of view of $N_f=2$ QED$_3$ with a charge-2 Higgs field coupled to the fermions via a Yukawa interaction (HYQED). Although this theory has more symmetries than the FS model, we have proposed the SFS lattice model, which enjoys all of the symmetries of HYQED and may host a multicritical point described by it. The SFS model has in fact more symmetries than HYQED, implying that a $\mathbb{Z}_2$ mirror symmetry must be emergent at the multicritical point. We have shown that this is consistent with large-$N$ calculations (extrapolated to HYQED, with~$N = 1$); it is also a consequence of a multicritical duality we proposed between HYQED and the easy-plane $\mathbb{CP}^1$ (EP$\mathbb{CP}^1$) model. Finally, breaking the extra symmetries of the SFS model, we can deform to the FS model via operators that map to unit monopole operators in HYQED. These operators deform the phase diagram of HYQED into the phase diagram of the FS model, supporting the scenario that HYQED flows to the SFS CFT, with a further flow to the FS CFT triggered by relevant monopole operators.

Our work draws an unexpected connection between the Fradkin-Shenker model and N\'eel-VBS deconfined quantum critical points. In particular, our multicritical scenario for HYQED/EP$\mathbb{CP}^1$ is likely also realizable in the easy-plane N\'eel-VBS transition. We estimated that $O(2)$ spin rotations along with the rotational symmetry of a square lattice are sufficient to forbid all relevant monopole deformations at the multicritical point. It would be very interesting to find such a multicritical point in magnetic systems between two $U(1)$ breaking states and a $\bZ_2$ spin-liquid.

A numerical study of the SFS model would be very helpful to support our multicritical scenario for HYQED. The model in \eqref{eqnU1FSmodel} has a sign problem in the $Z$ basis, but a sign-problem-free formulation might be possible along the lines of \cite{MotrunichPaper}. It would also be interesting to see if the unbroken $\mathsf{C}_{e,m}$ symmetries act nontrivially in the FS CFT (cf.~figure \ref{fig:overview}), by testing if operators charged under $\pi/2$ lattice rotations have different scaling dimensions.

Recently, the authors of~\cite{Shi:2024pem} studied a field theory of two complex scalars $\varphi_e$, $\varphi_m$ carrying charge 1 under two $U(1)$ gauge fields $a_e$ and $a_m$, respectively, with a mutual Chern-Simons term $\frac{k}{2\pi} a_e \wedge da_m$, and they analyzed this theory as possibly describing the analogue of the FS multicritical point for the $\bZ_k$ toric code. This theory also has $U(1)$ monopole symmetries, and for $k \ge 4$ the monopole operators appear to be irrelevant~\cite{Shi:2024pem}. One can entertain the interesting possibility that for $k = 2$ their theory may describe the SFS CFT, i.e.~their model may be dual to HYQED and EP$\mathbb{CP}^1$. The symmetry of the theory studied in~\cite{Shi:2024pem} is
\[(U(1)_e \times U(1)_m) \rtimes (\bZ_2^D \times \bZ_2^{C} \times \bZ_2^{T_e})~, \label{cssym}\]
where $D$ is a duality-like symmetry exchanging $U(1)_e$ and $U(1)_m$, $C$ is charge-conjugation symmetry, $\varphi_{e,m} \mapsto \varphi_{e,m}^*$, $a_{e,m} \mapsto -a_{e,m}$, and $T_e$ is a mirror-like time-reversal symmetry, $\varphi_e \mapsto \varphi_e^*$, $\varphi_m \mapsto \varphi_m$, $a_e \mapsto - a_e$, $a_m \mapsto a_m$ (where we indicate the action of $T$ on the time component of the gauge fields, while the space components transform with the opposite sign). These are all part of our proposed $G_\text{SFS CFT}$, and the missing symmetries may emerge in this theory, as they do for both HYQED and EP$\C\P^1$. Note that unlike the latter two theories, these symmetries in~\eqref{cssym} are anomaly free, and the model indeed has a symmetric deformation to a trivial Higgs phase, where both scalars $\varphi_{e,m}$ condense. It would be interesting to revisit this possible duality in the future.

\section*{Acknowledgments}

\noindent We are grateful to Max Metlitski, Silviu Pufu, Zhengyan Darius Shi, and Ruben Verresen for useful discussions. TD thanks Juan Maldacena for collaboration on related topics. PN thanks Lorenzo Di Pietro for fruitful discussions and collaboration on related topics. The work of TD was supported in part by U.S. Department of Energy award DE-SC0009937, as well as the Simons Collaboration on Global Categorical Symmetries. PN is supported by the ERC-COG grant NP-QFT No.~864583 ``Non-perturbative dynamics of quantum fields: from new deconfined phases of matter to quantum black holes'' and by the MUR-FARE2020 grant No.~R20E8NR3HX ``The Emergence of Quantum Gravity from Strong Coupling Dynamics.'' RT acknowledges support from the Mani L.~Bhaumik Presidential Term Chair.

\appendix

\section{Simon's Diamagnetic Inequality for Lattice Bosons}\label{appdiamagneticinequality}

In this appendix, we give a proof of a lattice version of Simon's diamagnetic inequality~\cite{simon1976universal}, which was explained (though not explicitly stated) in~\cite{nie2013ground}. In particular, we consider the Hamiltonian
\[H(\theta) = - \sum_{j,k} \left( t_{jk} e^{i \theta_{jk}} c_j^\dagger c_k + \text{h.c.} \right) + \hat V \,,\]
where $c_j$ and $c_j^\dagger$ are bosonic creation and annihilation operators, and $\hat V$ is an arbitrary real function of the number operators $c_j^\dagger c_j$ (meaning it is diagonal in the occupation number basis), which may include chemical potential terms, Hubbard interactions, etc. We assume $t_{jk} \ge 0$. The parameter $\theta$ of the Hamiltonian refers to the set of phases $\theta_{jk}$.

Let $E_0(\theta)$ be the ground state energy of $H(\theta)$. The diamagnetic inequality is
\[E_0(\theta) \ge E_0(0) \qquad \text{for all } \theta = \{\theta_{jk}\}_{j,k} \,.\]
To prove the inequality, we follow the argument in \cite{simon1976universal}. Let
\[|\psi\rangle = \sum_{\vec n} \psi(\vec n) |\vec n\rangle \]
be a state in the occupation-number basis. We define
\[|{\rm abs}(\psi)\rangle = \sum_{\vec n} |\psi(\vec n)| |\vec n\rangle \,, \]
which is a normalized state if $|\psi\rangle$ is. Consider $\langle \psi| H(\theta) |\psi\rangle$, which contains terms
\[- t_{jk} e^{i\theta_{jk}} \psi(\vec n)^* \psi(\vec m) \langle \vec n| c_j^\dagger c_k | \vec m\rangle + \text{c.c.} \\
= -2 t_{jk} {\rm Re}(e^{i\theta_{jk}} \psi(\vec n)^* \psi(\vec m) \langle \vec n| c_j^\dagger c_k | \vec m\rangle) \\
\ge -2 t_{jk} |\psi(\vec n)| |\psi(\vec m)| \langle \vec n| c_j^\dagger c_k | \vec m\rangle,\label{simons1}\]
where we used $t_{jk} > 0$ and $\langle \vec n| c_j^\dagger c_k | \vec m\rangle \ge 0$. The last expression is the corresponding term that appears in $\langle {\rm abs}(\psi)|H(0)|{\rm abs}(\psi)\rangle$. The other terms from $\hat V$ are diagonal in the occupation number basis and so
\[\langle \psi|\hat V|\psi\rangle = \langle {\rm abs}(\psi)|\hat V|{\rm abs}(\psi)\rangle \,.\label{simons2}\]
Combining \eqref{simons1} and \eqref{simons2}, we get
\[\langle \psi| H(\theta)| \psi \rangle \ge \langle {\rm abs}(\psi)| H(0)|{\rm abs}(\psi)\rangle.\]
By the variational principle,
\[E_0(\theta) = \inf_{|\psi\rangle \text{ normalized}} \langle \psi|H(\theta)|\psi\rangle \,, \]
the diamagnetic inequality holds.

\section{Feynman Rules at Large $N$}
\label{appendix:rules}

In this appendix, we collect the Feynman rules for the large-$N$ Euclidean Lagrangian
\begin{equation}
\mathscr{L} =  \bar\psi_i \left(\slashed{\partial}-i\slashed{a}\right)\psi_i
+ \frac{i}{2\sqrt{N}}Y_{ij} \, \sigma^* \psi^{t}_i \gamma^0 \psi_j + \frac{i}{2\sqrt{N}}Y^*_{ij} \,\sigma~ \bar\psi_i \gamma^0 \bar\psi^t_j 
+ M_{ij} \bar\psi_i\psi_j + \lambda \, \sigma^*\sigma \,,   
\end{equation}
where $Y=\sigma_x \otimes \mathbbm{1}_N$, so that $\mathrm{Tr}(Y^\dagger Y)=2N$. 
The massless fermion propagator is
\begin{equation}
\braket{\psi_i(p)\bar\psi_j(-p)} = \frac{\delta_{ij}}{i\slashed{p}} = -\frac{i\slashed{p}}{p^2} \delta_{ij} \,,
\end{equation}
which also implies (note the relative minus sign due to the anticommuting nature of fermions)
\begin{equation}
\braket{\bar\psi^t_j(-p)\psi_i^t(p)} = +\frac{i\slashed{p}^t}{p^2} \delta_{ij} \,.
\end{equation}
In the following, when writing vertex rules, it is understood that external legs are amputated in the corresponding correlators and that momentum flow is defined to enter the vertex.
Since we are interested in computing $\beta$ functions at linear order in the relevant couplings $M_{ij}$ and $\lambda$, we treat the fermion mass as an interaction,
\begin{equation}
\braket{\bar\psi_i(p)\psi_j(-p)} = -M_{ij} \,, \qquad \braket{\psi^t_j(-p)\bar\psi^t_i(p)} = +M_{ij} \,,
\end{equation}
as well as the scalar mass,
\begin{equation}
\braket{\sigma^*(p)\sigma(-p)} = -\lambda \,.    
\end{equation}
The vertex with the photon is
\begin{equation}
\braket{\bar\psi_i(-p-q)a^\mu(p)\psi_j(q)} = +i\gamma^\mu \delta_{ij} \,, \qquad \braket{\psi_j^t(q)a^\mu(p)\bar\psi^t_i(-p-q)} = -i(\gamma^\mu)^t \delta_{ij}  \,,   
\end{equation}
whereas the vertices with the Hubbard-Stratonovich (HS) field are
\begin{equation}
\braket{\psi^t_i(-p-q)\sigma^*(p)\psi_j(q)} = -\frac{i}{\sqrt{N}} Y_{ij} \gamma^0 \,, \qquad
\braket{\bar\psi_i(-p-q)\sigma(p)\bar\psi^t_j(q)} = -\frac{i}{\sqrt{N}} Y^*_{ij} \gamma^0 \,.
\end{equation}

\subsection{Exact Scalar and Photon Propagators at Large $N$}
The tree-level propagator of the HS field is\footnote{~Here, we use the quadratic term as a regulator to have a finite tree-level propagator. We are interested in the massless propagator at large $N$, which requires to remove the regulator by putting $\lambda\rightarrow 0$. Notice that we could have equivalently regulated the tree-level propagator with the irrelevant kinetic term $\frac{1}{|y|^2}\partial_\mu\sigma^* \partial^\mu\sigma$, giving $\braket{\sigma(p)\sigma^*(-p)}^{\mathrm{tree}}=|y|^2/p^2$, and then take the IR limit $|p| \ll |y|^2$.}
\begin{equation}
\braket{\sigma(p)\sigma^*(-p)}^{\mathrm{tree}} = \frac{1}{\lambda} \,,    
\end{equation}
whereas the tree-level propagator of the photon is\footnote{~Here, we use as a regulator for the tree-level photon propagator the standard irrelevant Maxwell term $\frac{1}{4e^2}f_{\mu\nu}f^{\mu\nu}$, supplemented by a (non-local) 3d gauge-fixing term. Then, we will take the IR limit $|p|\ll \Lambda$, with $\Lambda\equiv e^2N$.}
\begin{equation}\label{propagatorphotontree}
\braket{a_\mu(p)a_\nu(-p)}^{\mathrm{tree}} = \frac{\Lambda}{Np^2}\left(\delta_{\mu\nu}-\frac{p_\mu p_\nu}{p^2}\right) + \frac{\xi}{N} \frac{p_\mu p_\nu}{|p|^3} \,,    
\end{equation}
where $\xi$ is an adimensional gauge-fixing constant, and $|p|\equiv\sqrt{p^2}$.
At large $N$ we can exactly resum the fermionic bubble diagrams to get the full propagators, and then reliably take the massless limit for the HS field, $\lambda\rightarrow 0$, and the IR limit for the photon, $\Lambda\rightarrow \infty$. We will repeatedly make use of the following integrals (removing polynomial divergences in the UV cutoff, namely in dimensional regularization),
\begin{equation}\label{usefulintegrals}
\begin{split}
\int\frac{d^3q}{(2\pi)^3}\frac{1}{(p+q)^2q^2} &= \frac{1}{8|p|} \,, \\
\int\frac{d^3q}{(2\pi)^3}\frac{p\cdot q + q^2}{(p+q)^2q^2} &= - \frac{|p|}{16} \,, \\
\int\frac{d^3q}{(2\pi)^3}\frac{q^\mu}{(p+q)^2q^2} &= -\frac{p^\mu}{16|p|} \,, \\
\int\frac{d^3q}{(2\pi)^3}\frac{q^\mu q^\nu}{(p+q)^2q^2} &= \frac{|p|}{64}\left(-\delta^{\mu\nu}+\frac{3p^\mu p^\nu}{p^2} \right) \,,
\end{split}
\end{equation}
and of the following properties of the Euclidean gamma matrices $\gamma^\mu=(-\sigma_y,\sigma_z,-\sigma_x)$,
\begin{equation}
    \mathrm{tr}(\gamma^\mu \gamma^\nu)=2\delta^{\mu\nu} \,, \quad \mathrm{tr}(\gamma^\mu\gamma^\nu\gamma^\rho)=2i\epsilon^{\mu\nu\rho} \,, \quad \mathrm{tr}(\gamma^\nu\gamma^\alpha\gamma^\mu\gamma^\beta)=2(\delta^{\mu\alpha}\delta^{\nu\beta}+\delta^{\mu\beta}\delta^{\nu\alpha}-\delta^{\mu\nu}\delta^{\alpha\beta}) \,.
\end{equation}

\begin{figure}[t]
	\centering
	\includegraphics[width=1\textwidth]{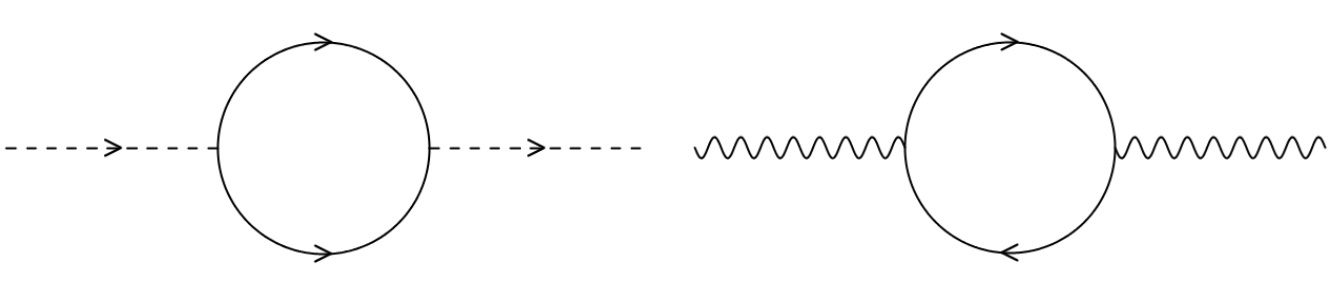}
	\caption{Fermion bubble diagrams contributing to the propagator of the HS field (left) and of the photon (right). Continuous, dashed, and wiggled lines represent the fermions, the HS field, and the photon, respectively. Here and in all subsequent diagrams, arrows denote the flow of electric charge and external legs are amputated.}
	\label{fig:Bubble}
\end{figure}

The bubble diagram contributing to the propagator of the HS field -- depicted in figure \ref{fig:Bubble} on the left -- is given by  (note the symmetry factor of $1/2$ due to identical fermions running in the loop)
\begin{equation}
\begin{split}
\Pi &=\frac{1}{2} \int \frac{d^3q}{(2\pi)^3}
(-1) \mathrm{tr}\left(
\frac{-i}{\sqrt{N}}Y^*_{kl}\gamma^0 \frac{+i(-\slashed{p}-\slashed{q})^t}{(p+q)^2}\delta_{lj} \frac{-i}{\sqrt{N}}Y_{ji}\gamma^0 \frac{-i\slashed{q}}{q^2}\delta_{ik}\right) \\
&= \frac{\mathrm{Tr}(Y^\dagger Y)}{2N} \int \frac{d^3q}{(2\pi)^3} \frac{\mathrm{tr}((\slashed{p}+\slashed{q})\slashed{q})}{(p+q)^2q^2}
= \int \frac{d^3q}{(2\pi)^3} \frac{2(p\cdot q + q^2)}{(p+q)^2q^2} = - \frac{|p|}{8} \,,
\end{split}
\end{equation}
where we have used that $\gamma^0(\gamma^\mu)^t\gamma^0=-\gamma^\mu$, $\mathrm{Tr}(Y^\dagger Y)=2N$, and \eqref{usefulintegrals}.
The massless propagator of the HS field is thus given by
\begin{equation}\label{sigmapropagator}
\braket{\sigma(p)\sigma^*(-p)}=\lim_{\lambda\rightarrow 0} \frac{1}{\lambda} \sum_{k=0}^\infty \left(-\frac{|p|}{8}\frac{1}{\lambda}\right)^k = \lim_{\lambda\rightarrow 0} \frac{1}{\lambda} \frac{1}{1+\frac{|p|}{8\lambda}} = \frac{8}{|p|} \,.    
\end{equation}

The bubble diagram contributing to the propagator of the photon -- depicted in figure \ref{fig:Bubble} on the right -- is given by
\begin{equation}
\begin{split}
&\Pi^{\mu\nu}=\int \frac{d^3q}{(2\pi)^3}
(-1) \mathrm{tr}\left( i\gamma^\nu\delta_{ij} \frac{-i(\slashed{p}+\slashed{q})}{(p+q)^2}\delta_{jk} \,i\gamma^\mu \delta_{kl} \frac{-i\slashed{q}}{q^2}\delta_{li}\right) \\
&= -2N \int \frac{d^3q}{(2\pi)^3} \frac{\mathrm{tr}(\gamma^\nu(\slashed{p}+\slashed{q})\gamma^\mu\slashed{q})}{(p+q)^2q^2} = -4N \int \frac{d^3q}{(2\pi)^3} \frac{p^\mu q^\nu + q^\nu p^\mu + 2 q^\mu q^\nu - \delta^{\mu\nu}(p\cdot q + q^2)}{(p+q)^2q^2} \,,
\end{split}
\end{equation}
where we have used that $\mathrm{tr}(\gamma^\nu\gamma^\alpha\gamma^\mu\gamma^\beta)=2(\delta^{\mu\alpha}\delta^{\nu\beta}+\delta^{\mu\beta}\delta^{\nu\alpha}-\delta^{\mu\nu}\delta^{\alpha\beta})$. Using \eqref{usefulintegrals}, we get
\begin{equation}\label{bubblephoton}
\Pi^{\mu\nu} = -4N \left(-\frac{2}{16}\frac{p^\mu p^\nu}{|p|} + \frac{|p|}{32}\left(-\delta^{\mu\nu}+3\frac{p^\mu p^\nu}{p^2} \right) + \delta^{\mu\nu} \frac{|p|}{16}\right) = -\frac{N|p|}{8} \left( \delta^{\mu\nu} - \frac{p^\mu p^\nu}{p^2} \right) \,.    
\end{equation}
As expected, $\Pi_{\mu\nu}$ is transverse, and it vanishes when contracted with the $\xi$-dependent part of the tree-level propagator \eqref{propagatorphotontree}. The photon propagator at finite $\Lambda$ is thus given by
\begin{equation}
\begin{split}
\braket{a_\mu(p)a_\nu(-p)}_\Lambda &= \frac{\xi}{N} \frac{p_\mu p_\nu}{|p|^3} + \frac{\Lambda}{Np^2}\left(\delta_{\mu\rho}-\frac{p_\mu p_\rho}{p^2}\right) \sum_{k=0}^{\infty} \left(-\frac{N|p|}{8} \frac{\Lambda}{Np^2} \right)^k \left( \delta_{\rho\nu} - \frac{p_\rho p_\nu}{p^2} \right) \\
&= \frac{\xi}{N} \frac{p_\mu p_\nu}{|p|^3} + \frac{\Lambda}{Np^2} \frac{1}{1 + \frac{\Lambda}{8 |p|}} \left( \delta_{\mu\nu} - \frac{p_\mu p_\nu}{p^2} \right)\,,    
\end{split}    
\end{equation}
and we can now reliably take the IR limit $\Lambda\rightarrow\infty$ to get the photon propagator
\begin{equation}\label{photonpropagator}
\braket{a_\mu(p)a_\nu(-p)} =\lim_{\Lambda\rightarrow\infty}\braket{a_\mu(p)a_\nu(-p)}_\Lambda = \frac{8}{N|p|} \left( \delta_{\mu\nu} - a \frac{p_\mu p_\nu}{p^2} \right) \,,
\end{equation}
where $a\equiv 1-\xi/8$ encodes the gauge dependence. Notice that this result matches the photon propagator of large-$N$ QED$_3$ with $2N$ Dirac fermions, as it is expected since the contribution of the single scalar field with charge 2 is negligible at leading order in the large-$N$ expansion. For this reason, the propagator above is the correct large-$N$ propagator also for the theory in \eqref{largeNLagrangian}.
In the following, we will perform computations with generic $a$ and verify that the gauge dependence cancels in physical observables.

\subsection{Effective Vertex between the Scalar and the Photon}

\begin{figure}[t]
	\centering
	\includegraphics[width=0.7\textwidth]{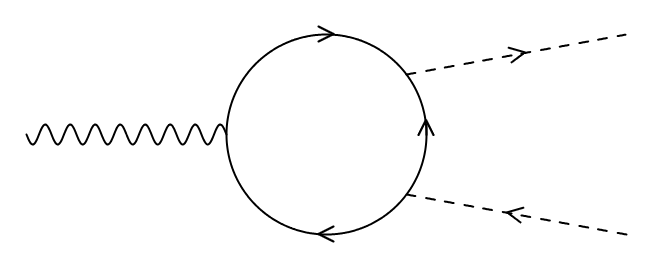}
	\caption{One-loop diagram generating the trilinear interaction between the scalar and the photon.}
	\label{fig:Induced vertex}
\end{figure}

Given that the HS field has electric charge 2, there is also an interaction with the photon. Since the Lagrangian does not have a covariant kinetic term for $\sigma$ (as this is irrelevant), such interaction actually arises as a loop effect at large $N$. As we will show below, at leading order in the large-$N$ expansion we only need the trilinear interaction $a^\mu\sigma^*\sigma$, which arises from a one-loop diagram with internal fermions attached to the three external legs, depicted in figure \ref{fig:Induced vertex}. It is given by
\begin{equation}
\begin{split}
&\braket{a^\mu(-p_1-p_2)\sigma(p_1)\sigma^*(p_2)} \\
&= \int\frac{d^3q}{(2\pi)^3} (-1) \mathrm{tr} \left( \frac{-i}{\sqrt{N}}Y_{kj}\gamma^0\frac{-i\slashed{q}}{q^2}\delta_{ji} i\gamma^\mu\delta_{in}\frac{-i(\slashed{p}_1+\slashed{p}_2+\slashed{q})}{(p_1+p_2+q)^2}\delta_{nm}\frac{-i}{\sqrt{N}}Y^*_{ml}\gamma^0\frac{+i(-\slashed{p}_2-\slashed{q})^t}{(p_2+q)^2}\delta_{lk}\right) \\
&= \frac{\mathrm{Tr}(Y^\dagger Y)}{N} \int\frac{d^3q}{(2\pi)^3} \frac{\mathrm{tr}(\slashed{q}\gamma^\mu(\slashed{p}_1+\slashed{p}_2+\slashed{q})(\slashed{p}_2+\slashed{q}))}{q^2(p_2+q)^2(p_1+p_2+q)^2} \\
&= 4 \int\frac{d^3q}{(2\pi)^3} \frac{q^\mu(p_2+q)^2+p_1^\mu q^2 + q^\alpha(\delta^{\mu\alpha}p_1\cdot p_2 + p_1^\mu p_2^\alpha - p_1^\alpha p_2^\mu)}{q^2(p_2+q)^2(p_1+p_2+q)^2} \,.
\end{split}
\end{equation}
It is easy to prove that this effective vertex is antisymmetric in the exchange $p_1 \leftrightarrow p_2$, by performing the change of variable $q'=-q-p_1-p_2$ in the last line. This property is required by the Ward-Takahashi identity associated to the $U(1)$ gauge invariance, which reads
\begin{equation}\label{wardidentity}
(p_1+p_2)_\mu \braket{a^\mu(-p_1-p_2)\sigma(p_1)\sigma^*(p_2)} =
2 \left( \braket{\sigma(p_1)\sigma^\star(-p_1)}^{-1} - \braket{\sigma(-p_2)\sigma^\star(p_2)}^{-1} \right) \,,   
\end{equation}
where on the right-hand side we have the difference between the inverse propagators of the HS field computed at momenta $+p_1$ and $-p_2$, respectively, and the factor of 2 is the gauge charge of $\sigma$.

In fact, we are actually only interested in this vertex when one of the two external HS legs carries zero momentum, so that either $p_1=0$ or $p_2=0$. Using \eqref{usefulintegrals}, we get
\begin{equation}\label{sigma photon vertex}
\begin{split}
\braket{a^\mu(-p)\sigma(p)\sigma^*(0)} &= 4 \int\frac{d^3q}{(2\pi)^3} \frac{q^\mu+p^\mu }{q^2(p+q)^2} =  + \frac{p^\mu}{4|p|} \,, \\
\braket{a^\mu(-p)\sigma(0)\sigma^*(p)} &=  4 \int\frac{d^3q}{(2\pi)^3} \frac{q^\mu}{q^2(p+q)^2} = - \frac{p^\mu}{4|p|} \,.
\end{split}
\end{equation}
Using the propagator \eqref{sigmapropagator}, one can easily verify that the vertex functions above satisfy the Ward-Takahashi identity \eqref{wardidentity}.

\section{Computation of the $\beta$ Functions at Large $N$}
\label{appendix:beta}

In this appendix, we first review the Wilsonian approach to extract $\beta$ functions from renormalization group constants. Then, we compute the latter at leading order in the large-$N$ expansion, and deduce the result in \eqref{betafunctions}.

\subsection{Renormalization Group Constants and $\beta$ Functions}
Our goal is now to compute the $\beta$ functions of $M_{ij}$ and $\lambda$ at leading order in the large-$N$ expansion. We adopt a Wilsonian renormalization approach, in which we declare our theory to be defined at the scale $\Lambda$.\footnote{~We use the same symbol as for the scale $e^2N$, as confusion should not arise.} After an RG step where we integrate out momenta between $\Lambda$ and $\Lambda'<\Lambda$, we will show below that the original Lagrangian, where all the fields are defined at the scale $\Lambda$,
\begin{equation}
\mathscr{L} =  \bar\psi_i \left(\slashed{D}\,\delta^{ij}+M^{ij}\right) \psi_j + \frac{i}{2\sqrt{N}}Y_{ij} \, \sigma^* \psi^{t}_i \gamma^0 \psi_j + \text{c.c.} + \lambda\, \sigma^*\sigma \,,
\end{equation}
where $\slashed{D}=\gamma^\mu(\partial_\mu-ia_\mu)$, becomes
\begin{equation}
\mathscr{L} = \bar\psi_i \left(Z_\psi\slashed{D}\,\delta^{ij}+Z_m^{ij}\right) \psi_j + \frac{i}{2\sqrt{N}}Z_y Y_{ij} \,\sigma^* \psi^{t}_i \gamma^0 \psi_j + \text{c.c.} + Z_\lambda\lambda\, \sigma^*\sigma \,,
\end{equation}
where $Z_i=Z_i(\Lambda/\Lambda')$ are the renormalization group constants.\footnote{~Gauge invariance guarantees that the renormalization constant of the electromagnetic interaction is the same $Z_\psi$ as for the fermion kinetic term.} We can rewrite this Lagrangian in terms of the renormalized fields, which are defined at the scale $\Lambda'$ and denoted with a primed superscript,
\begin{equation}
\psi'_i \equiv \sqrt{Z_\psi} \psi_i \,, \qquad \sigma' \equiv \sqrt{Z_\sigma}\sigma \,,    
\end{equation}
as
\begin{equation}
\mathscr{L} =  \bar\psi'_i \left(\slashed{D}\,\delta^{ij}+\frac{Z_m^{ij}}{Z_\psi}\right) \psi'_j + \frac{i}{2\sqrt{N}}\frac{Z_y}{Z_\psi \sqrt{Z_\sigma}} Y_{ij} \, \sigma'^* \psi'^{t}_i \gamma^0 \psi'_j + \text{c.c.} + \frac{Z_\lambda}{Z_\sigma} \lambda \, \sigma'^*\sigma' \,.
\end{equation}
Conformal invariance of the large-$N$ tricritical theory requires
\begin{equation}
\frac{Z_y}{Z_\psi\sqrt{Z}_\sigma} = 1 \,,  
\end{equation}
which we solve for $Z_\sigma$ as a function of $Z_y$ and $Z_\psi$.\footnote{~Similarly, we also have that the photon renormalization constant is trivial, namely $a_\mu'=a_\mu$. This implies that, as a consequence of gauge and conformal invariance, the $1/|p|$ scaling of the photon propagator is exact to all orders in $1/N$.}
The Lagrangian at the scale $\Lambda'$ can be recast in the same form as the Lagrangian at the scale $\Lambda$ by defining the couplings at the scale $\Lambda'$ as
\begin{equation}
M'_{ij} = \frac{Z_m^{ij}}{Z_\psi} \,, \qquad \lambda'= \frac{Z_\lambda}{Z_\sigma}\lambda = \frac{Z_\lambda Z_\psi^2}{Z_y^2} \lambda \,.
\label{renormalization constants}
\end{equation}
At leading order in the large-$N$ expansion we will show that
\begin{equation}
Z_m^{ij} = M^{ij} + \delta M^{ij} \,, \qquad Z_{\psi,y,\lambda} = 1 + \delta_{\psi, y, \lambda} \,,    
\end{equation}
where $\delta_i(\Lambda'/\Lambda)$ are $1/N$ constants. We can thus expand \eqref{renormalization constants} at leading order in $1/N$ to get
\begin{equation}
M'_{ij} = M_{ij} -\delta_\psi M_{ij} + \delta M_{ij}  \,, \qquad \lambda' = (1+\delta_\lambda+2\delta_\psi-2\delta_y)\lambda \,.
\end{equation}
We can now compute the $\beta$ functions of the couplings as
\begin{equation}
\beta_\lambda = \frac{\partial \lambda'}{\partial \log \Lambda'}  \,, \qquad \beta_M = \frac{\partial M'}{\partial \log \Lambda'}  \,.
\end{equation}
Since the original coupling constants do not depend on the scale $\Lambda'$, and the renormalization constants $\delta_i$ depend only on the combination $\log(\Lambda'/\Lambda)$ such that $\frac{\partial}{\partial\log\Lambda'}=-\frac{\partial}{\partial\log\Lambda}$,\footnote{~As usual, we only consider logarithmic divergences, which are the ones that contribute to $\beta$ functions. Any polynomial divergence can be reabsorbed by shifting the original couplings.}  we finally get
\begin{equation}\label{betafunctionsformulas}
\beta_\lambda = \frac{\partial}{\partial\log\Lambda} \left( - \delta_\lambda - 2 \delta_\psi + 2 \delta_y  \right) \lambda \,, \qquad (\beta_M)_{ij} = \frac{\partial}{\partial\log\Lambda} \left( \delta_\psi M_{ij} - \delta M_{ij} \right) \,.
\end{equation}
Our goal is now to compute the log-divergent part of the renormalization constants $\delta_i$ at leading order in the large-$N$ expansion. In our case, the logarithmic divergence arises from
\begin{equation}
\int \frac{d^3q}{(2\pi)^3} \frac{1}{|q|^3} = \frac{1}{2\pi^2} \log\left(\frac{\Lambda}{\Lambda'}\right) \,.   
\end{equation}
Moreover, since we want to extract the anomalous dimensions of the corresponding operators using \eqref{anodimformula}, we only need the $\beta$ functions at linear order in $\lambda$ and $M$, which will thus be treated as vertex insertions, keeping all propagators massless.\footnote{~As mentioned around \eqref{relevantlagrangian}, recall that that there cannot be any mixing between the operators $\sigma^*\sigma$ and $\bar\psi_i\psi_j$, so that at linear order the $\beta$ function of $M$ does not depend on $\lambda$, and vice versa.}

\subsection{Computation of $\delta_\psi$}

The value of $\delta_\psi$ can be extracted from the correction to the fermion propagator. Indeed, using the Lagrangian at the scale $\Lambda'$, the corrected fermion propagator is
\begin{equation}
\braket{\psi_i(p) \bar\psi_j(-p)} = -\frac{i\slashed{p}}{Z_\psi p^2} \delta_{ij} =  -\frac{i\slashed{p}}{p^2} \delta_{ij} + \left(-\frac{i\slashed{p}}{p^2} \delta_{ik}\right)(-i\delta_\psi \slashed{p} \, \delta_{kl}) \left(-\frac{i\slashed{p}}{p^2} \delta_{lj}\right) \,.   
\end{equation}
The second term is the contribution coming from the loop correction, so that we can extract $\delta_\psi$ from the amputed correction to the fermion propagator as
\begin{equation}
-i\delta_\psi \slashed{p} \, \delta_{kl} = \braket{\psi_k(p)\bar\psi_l(-p)}_{\mathrm{1-loop}} \,.    
\end{equation}
There are two such diagrams that contribute to the propagator correction, corresponding to sunset diagrams with a photon or with an HS propagator.

\begin{figure}[t]
	\centering
	\includegraphics[width=1\textwidth]{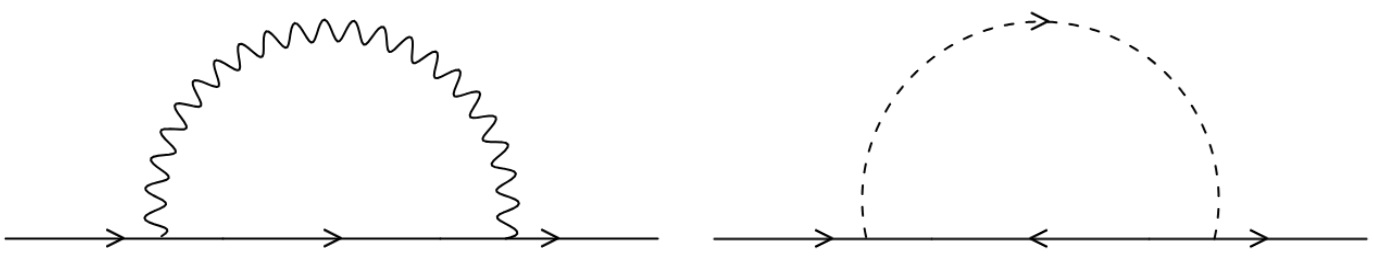}
	\caption{Diagrams contributing to the leading-order correction to the renormalization constant $\delta_\psi$ of the fermion propagator. Here, and in all subsequent diagrams, the internal scalar and photon lines represent the exact large-$N$ propagators \eqref{sigmapropagator} and \eqref{photonpropagator}, respectively.}
	\label{fig:Delta_psi}
\end{figure}

The photon correction -- depicted in figure \ref{fig:Delta_psi} on the left -- is given by
\begin{equation}
\begin{split}
-i\delta^\gamma_\psi \slashed{p} \, \delta_{kl} &=  \int \frac{d^3q}{(2\pi)^3} i\gamma^\mu \delta_{li} \frac{-i(\slashed{p}+\slashed{q})}{(p+q)^2}\delta_{ij} i\gamma^\nu \delta_{jk} \frac{8}{N|q|}\left(\delta_{\mu\nu}-a\frac{q_\mu q_\nu}{q^2}\right) \\
&= \frac{8i\delta_{kl}}{N} \int \frac{d^3q}{(2\pi)^3} \frac{(a-1)\slashed{p}-(a+1)\slashed{q}-2a\slashed{q} (p\cdot q)/q^2}{|q|(p+q)^2}\,,
\end{split}
\end{equation}
where we have used that $\gamma^\mu \gamma^\alpha \gamma^\nu = - \gamma^\alpha \gamma^\mu \gamma^\nu + 2 \delta^{\mu \alpha} \gamma^\nu$ and $\gamma^\mu \gamma_\mu = 3$.
To extract the linear term in the momentum we can expand the denominator as
\begin{equation}
\frac{1}{(p+q)^2}=\frac{1}{q^2}\left(1-\frac{2p\cdot q}{q^2}+\mathcal{O}(p^2)\right) \,,    
\end{equation}
so that
\begin{equation}
-i\delta^\gamma_\psi \slashed{p} \, \delta_{kl} = \frac{8i\delta_{kl}}{N} \int \frac{d^3q}{(2\pi)^3} \frac{(a-1)\slashed{p}+2\slashed{q} (p\cdot q)/q^2}{|q|^3} = \frac{8i\delta_{kl}}{N} \int \frac{d^3q}{(2\pi)^3} \frac{(a-1/3)\slashed{p}}{|q|^3} \,,
\end{equation}
where we have used rotational invariance to substitute
$\slashed{q}(p\cdot q)$ with $\slashed{p} \,q^2/3$. We then get
\begin{equation}
\delta^\gamma_\psi = \frac{4(1-3a)}{3\pi^2 N} \log\left(\frac{\Lambda}{\Lambda'}\right) \,.    
\end{equation}

The HS correction -- depicted in figure \ref{fig:Delta_psi} on the right -- is given by
\begin{equation}
-i\delta^\sigma_\psi \slashed{p} \, \delta_{kl} =  \int \frac{d^3q}{(2\pi)^3} \frac{-i}{\sqrt{N}} Y^*_{li} \gamma^0 \frac{+i(-\slashed{p}-\slashed{q})^t}{(p+q)^2} \delta_{ij} \frac{-i}{\sqrt{N}} Y_{jk} \gamma^0 \frac{8}{|q|} = -\frac{8i(YY^\dagger)_{lk}}{N} \int \frac{d^3q}{(2\pi)^3} \frac{\slashed{p}+\slashed{q}}{|q|(p+q)^2}\,,  
\end{equation}
and we can again expand the denominator and use rotational invariance to get
\begin{equation}
-i\delta^\sigma_\psi \slashed{p} \, \delta_{kl} = -\frac{8i(YY^\dagger)_{lk}}{N} \int \frac{d^3q}{(2\pi)^3} \frac{\slashed{p}-2\slashed{q}(p\cdot q)/q^2}{|q|^3} = -\frac{8i(YY^\dagger)_{lk}}{N} \int \frac{d^3q}{(2\pi)^3} \frac{\slashed{p}}{3|q|^3} \,.
\end{equation}
Using that $(YY^\dagger)_{lk}=\delta_{lk}$, we get
\begin{equation}
\delta^\sigma_\psi = \frac{4}{3\pi^2 N} \log\left(\frac{\Lambda}{\Lambda'}\right) \,,
\end{equation}
which is positive, as required by unitarity. All in all we have
\begin{equation}
\delta_\psi = \delta^\gamma_\psi + \delta^\sigma_\psi =\frac{4(1_y+1-3a)}{3\pi^2 N} \log\left(\frac{\Lambda}{\Lambda'}\right) \,,    
\end{equation}
where we denoted with a $y$ subscript the Yukawa contribution, for future convenience.

\subsection{Computation of $\delta_M$}

Since we treat the mass as a vertex insertion, we can compute the mass as the correction to the bilinear vertex with zero momentum for the external legs. Using the Lagrangian at the scale $\Lambda'$, the corrected mass vertex is
\begin{equation}
\braket{\bar\psi_i(0)\psi_j(0)} = -Z^{ij}_m = -M^{ij} - \delta M^{ij} \,.    
\end{equation}
The second term is the contribution coming from the loop correction, so that we can extract $\delta M_{ij}$ from the loop correction to the amputed mass vertex as
\begin{equation}
-\delta M_{ij} = \braket{\bar\psi_i(0)\psi_j(0)} _{\mathrm{1-loop}} \,.  
\end{equation}
There are three diagrams that contribute to this correction. Two are sunset diagrams with a photon or an HS field, and a mass insertion in the internal fermion propagator. The other is a sunset diagram with a photon, where the mass insertion is now done in an internal fermion loop correcting the photon propagator. (The analogous diagram that would correct the HS propagator vanishes, since it is proportional to $\mathrm{tr}\gamma^\mu=0$.)

\begin{figure}[t]
	\centering
	\includegraphics[width=1\textwidth]{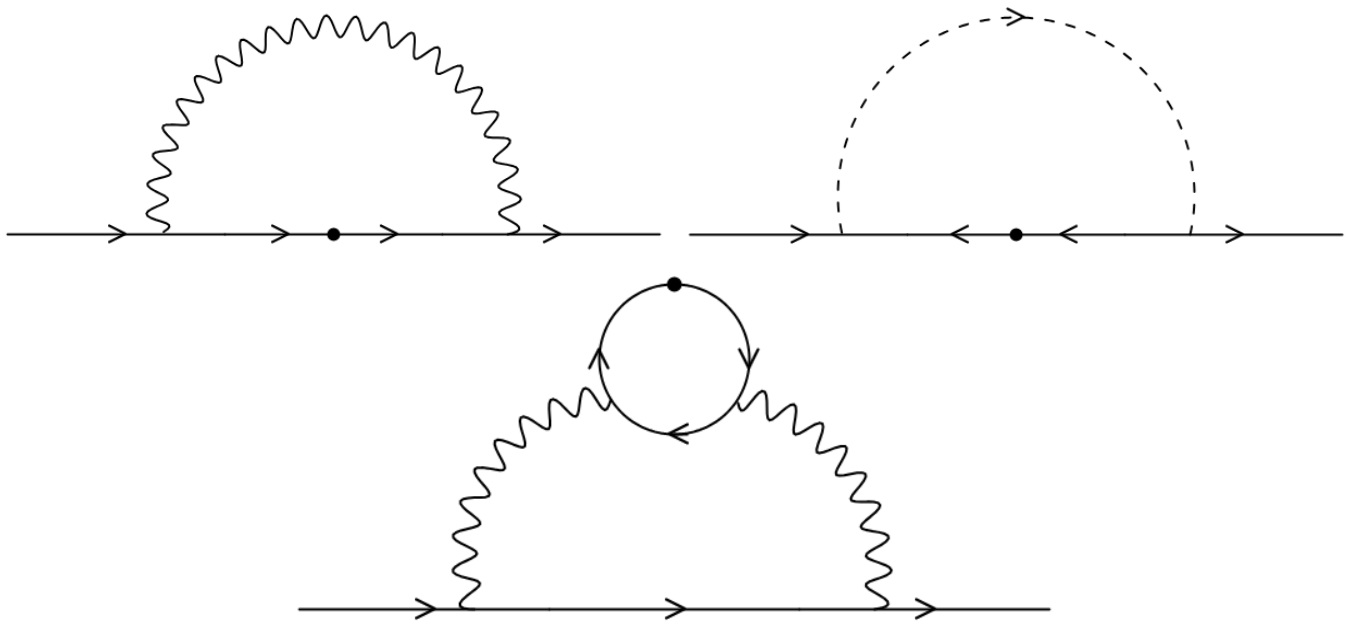}
	\caption{Diagrams contributing to the leading-order correction to the renormalization constant $\delta M_{ij}$ of the fermion mass. The black dot represents the insertion of the fermion mass vertex. On the bottom, we omit the analogous diagram where the mass insertion is done on the other fermionic line in the internal loop.}
	\label{fig:Delta_m}
\end{figure}

Let us consider the two diagrams of the first type. The photon correction -- depicted in figure \ref{fig:Delta_m} on the top left -- is given by
\begin{equation}
\begin{split}
-\delta M^{\gamma,1}_{ij} &= \int \frac{d^3q}{(2\pi)^3} i\gamma^\mu \delta_{ik} \frac{-i\slashed{q}}{q^2} \delta_{kl} ( - M_{lm} ) \frac{-i\slashed{q}}{q^2} \delta_{mn} i\gamma^\nu \delta_{nj} \frac{8}{N|q|}\left(\delta_{\mu\nu}-a\frac{q_\mu q_\nu}{q^2}\right) \\
&= - \frac{8 M_{ij}}{N} \int \frac{d^3q}{(2\pi)^3} \frac{3-a}{|q|^3} \,,  
\end{split}
\end{equation}
from which we get
\begin{equation}
\delta M^{\gamma,1}_{ij} = \frac{4(3-a)M_{ij}}{\pi^2 N} \log\left(\frac{\Lambda}{\Lambda'}\right) \,.   
\end{equation}
The HS correction -- depicted in figure \ref{fig:Delta_m} on the top right -- is given by
\begin{equation}
\begin{split}
-\delta M^{\sigma,1}_{ij} &= \int \frac{d^3q}{(2\pi)^3} \frac{-i}{\sqrt{N}}Y^*_{ik}\gamma^0 \frac{+i(-\slashed{q})^t}{q^2}\delta_{kl}(+M_{ml})\frac{+i(-\slashed{q})^t}{q^2} \delta_{mn}\frac{-i}{\sqrt{N}}Y_{nj} \gamma^0 \frac{8}{|q|} \\
&= \frac{8(Y^\dagger M^T Y)_{ij}}{N} \int \frac{d^3q}{(2\pi)^3} \frac{1}{|q|^3} \,,
\end{split}
\end{equation}
from which we get
\begin{equation}
\delta M^{\sigma,1}_{ij} = -\frac{4(Y^\dagger M^T Y)_{ij}}{\pi^2 N}\log\left(\frac{\Lambda}{\Lambda'}\right) \,.    
\end{equation}

Let us consider now the diagram of the second type, depicted in figure \ref{fig:Delta_m} on the bottom. We first compute the subdiagram given by a mass insertion in the fermion bubble with momentum $p$. For the fermion bubble correcting the photon propagator we have (taking into account the factor of 2 due to the fact that the mass insertion can be equivalently done in either of the two fermion propagators)
\begin{equation}
\begin{split}
&\Pi_M^{\mu\nu}(p) = 2\int \frac{d^3q}{(2\pi)^3} (-1) \mathrm{tr} \left( i\gamma^\nu \delta_{ni} \frac{-i(\slashed{p}+\slashed{q})}{(p+q)^2}\delta_{ij} (-M_{jk}) \frac{-i(\slashed{p}+\slashed{q})}{(p+q)^2}\delta_{kl}
i\gamma^\mu \delta_{lm} \frac{-i\slashed{q}}{q^2} \delta_{mn} \right) \\
&= - 2i \mathrm{Tr}M \int \frac{d^3q}{(2\pi)^3} \frac{\mathrm{tr}(\gamma^\nu \gamma^\mu \slashed{q})}{q^2(p+q)^2} 
= 4 \mathrm{Tr}M \epsilon^{\nu\mu\rho} \int \frac{d^3q}{(2\pi)^3} \frac{q_\rho}{q^2(p+q)^2}
= -\frac{\mathrm{Tr}M \epsilon^{\nu\mu\rho} p_\rho}{4|p|}  \,.
\end{split}
\end{equation}
Notice that the numerator is $\propto \mathrm{tr}(\gamma^\nu \gamma^\mu \slashed{q})$, so that the analogous diagram for the fermion bubble correcting the HS propagator vanishes, since it is $\propto \mathrm{tr}\slashed{q}=0$. This is clear from the fact that $\mathrm{Tr}M \neq 0$ breaks time-reversal invariance and indeed generates a Lorentz structure that is proportional to the Levi-Civita symbol (this amplitude is indeed responsible for generating a one-loop Chern-Simons term for the gauge field). We thus have another photon contribution to the mass renormalization constant given by
\begin{equation}
-\delta M^{\gamma,2}_{ij} = \int \frac{d^3q}{(2\pi)^3} i\gamma^\beta \delta_{ik} \frac{-i\slashed{q}}{q^2}\delta_{kl} i\gamma^\alpha \delta_{lj} \frac{8}{N|q|}\left(\delta_{\alpha\mu}-a\frac{q_\alpha q_\mu}{q^2}\right) \Pi_M^{\mu\nu}(-q) \frac{8}{N|q|} \left(\delta_{\nu\beta}-a\frac{q_\nu q_\beta}{q^2}\right) \,. 
\end{equation}
Using the antisymmetry of $\epsilon^{\nu\mu\rho}$, we have $\Pi_M^{\mu\nu}q_\mu=\Pi_M^{\mu\nu}q_\nu=0$ so that the gauge-dependent part of the integral vanishes, and we are left with
\begin{equation}
\begin{split}
-\delta M^{\gamma,2}_{ij} &= \frac{64i\delta_{ij}}{N^2} \gamma^\beta \gamma^\sigma \gamma^\alpha \int \frac{d^3q}{(2\pi)^3} \frac{q_\sigma}{q^4}\left(-\frac{\mathrm{Tr}M \epsilon^{\beta\alpha\rho} (-q_\rho)}{4|q|} \right) \\
&= \frac{16i\mathrm{Tr}M\delta_{ij}}{N^2} \epsilon_{\beta\alpha\rho} \gamma^\beta \gamma^\sigma \gamma^\alpha \int \frac{d^3q}{(2\pi)^3} \frac{q^2 \delta_{\rho\sigma}}{3|q|^5} \,.
\end{split}
\end{equation}
It is convenient to take the trace over spin indices over both sides to get
\begin{equation}
-2\,\delta M^{\gamma,2}_{ij} = \frac{16i\mathrm{Tr}M\delta_{ij}}{3N^2} \epsilon_{\beta\alpha\rho} (2i\epsilon^{\beta\rho\alpha})\int\frac{d^3q}{(2\pi)^3} \frac{1}{|q|^3} \,,  
\end{equation}
from which we get
\begin{equation}
\delta M^{\gamma,2}_{ij} = -\frac{16\mathrm{Tr}M\delta_{ij}}{\pi^2N^2} \log\left(\frac{\Lambda}{\Lambda'}\right) \,.   
\end{equation}
Since $\mathrm{Tr}(M)$ is $\mathcal{O}(N)$ we correctly have a $1/N$ renormalization constant. All in all we have
\begin{equation}
\delta M_{ij} = \delta M^{\gamma,1}_{ij} + \delta M^{\gamma,2}_{ij} + \delta M^{\sigma,1}_{ij} = \frac{4}{\pi^2N} \left( (3-a)M_{ij} - \frac{4}{N} \mathrm{Tr}M \delta_{ij} - (Y^\dagger M^T Y)_{ij}\right) \log\left(\frac{\Lambda}{\Lambda'}\right) \,.   
\end{equation}

\subsection{Computation of $\delta_\lambda$}
Analogously to the fermion mass, we treat the scalar mass as a vertex insertion, so that we can extract $\delta_\lambda$ from the loop correction to the amputed mass vertex as
\begin{equation}
-\delta_\lambda \lambda = \braket{\sigma^*(0)\sigma(0)}_{\mathrm{1-loop}} \,.    
\end{equation}
There are two contributions to such correction.

\begin{figure}[t]
	\centering
	\includegraphics[width=1\textwidth]{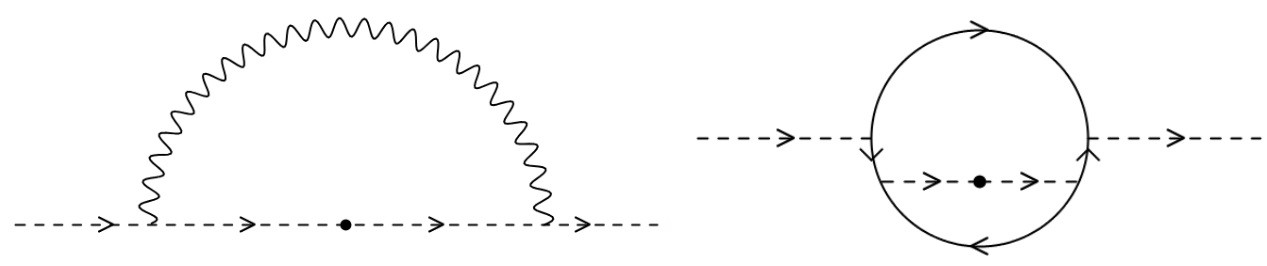}
	\caption{Diagrams contributing to the leading-order correction to the renormalization constant $\delta_\lambda$ of the scalar mass. The black dot represents the insertion of the scalar mass vertex. The effective trilinear vertices on the left correspond to the one-loop diagram in figure \ref{fig:Induced vertex}.}
	\label{fig:Delta_lambda}
\end{figure}

The first is a sunset diagram constructed with the effective vertex \eqref{sigma photon vertex} and with a mass insertion on the HS propagator -- depicted in figure \ref{fig:Delta_lambda} on the left -- which is given by
\begin{equation}
\begin{split}
-\delta^\gamma_\lambda \lambda &= \int \frac{d^3q}{(2\pi)^3} \left(+\frac{q^\mu}{4|q|}\right) \frac{8}{|q|} (-\lambda) \frac{8}{|q|} \left(-\frac{-q^\nu}{4|q|}\right) \frac{8}{N|q|} \left(\delta_{\mu\nu}-a\frac{q_\mu q_\nu}{q^2}\right) \\
&= -\frac{32\lambda}{N} \int \frac{d^3q}{(2\pi)^3} \frac{1-a}{|q|^3} \,,
\end{split}
\end{equation}
from which we get
\begin{equation}
\delta^\gamma_\lambda = \frac{16(1-a)}{\pi^2 N} \log\left(\frac{\Lambda}{\Lambda'}\right) \,.    
\end{equation}

The second is a two-loop diagram given by a mass insertion on the HS propagator internal to the fermion loop correcting the mass vertex -- depicted in figure \ref{fig:Delta_lambda} on the right -- which is given by\footnote{~The HS propagator must connect two points of the same fermion propagator running in the loop. These gives two possibilities, which however are topologically equivalent diagrams, and hence there is no factor of 2. Instead, the HS propagator cannot connect two points lying on different propagators, due to charge conservation.}
\begin{equation}
\begin{split}
-&\delta^\sigma_\lambda \lambda = \int \frac{d^3q}{(2\pi)^3} \int \frac{d^3 k}{(2\pi)^3} (-1) \frac{8}{|k|}(-\lambda) \frac{8}{|k|} \times \\
&\times \mathrm{tr}\left(
\frac{-i}{\sqrt{N}}Y_{qi} \gamma^0 \frac{-i\slashed{q}}{q^2} \delta_{ij} \frac{-i}{\sqrt{N}}Y^*_{jk} \gamma^0 \frac{+i(-\slashed{q})^t}{q^2} \delta_{kl} \frac{-i}{\sqrt{N}}Y_{lm} \gamma^0 \frac{-i(\slashed{q}+\slashed{k})}{(q+k)^2} \delta_{mn} \frac{-i}{\sqrt{N}}Y^*_{np} \gamma^0  \frac{+i(-\slashed{q})^t}{q^2} \delta_{pq} \right) \\
&= \frac{64 \lambda \mathrm{Tr}(Y^\dagger Y Y^\dagger Y)}{N^2} \int \frac{d^3q}{(2\pi)^3} \frac{1}{q^4} \int \frac{d^3 k}{(2\pi)^3} \frac{\mathrm{tr}((\slashed{q}+\slashed{k})\slashed{q})}{k^2(q+k)^2}
= \frac{128 \lambda}{N} \int \frac{d^3q}{(2\pi)^3} \frac{1}{q^4} \frac{|q|}{8} \,,
\end{split}
\end{equation}
where we have used that $Y^\dagger Y = \mathbbm{1}_{2N}$. We then get
\begin{equation}
\delta_\lambda^\sigma = - \frac{8}{\pi^2 N} \log\left(\frac{\Lambda}{\Lambda'}\right) \,.   
\end{equation}
All in all we have
\begin{equation}
\delta_\lambda = \delta_\lambda^\gamma + \delta_\lambda^\sigma =  \frac{8(1-2a)}{\pi^2 N} \log\left(\frac{\Lambda}{\Lambda'}\right) \,.  
\end{equation}

\subsection{Computation of $\delta_y$}
The value of $\delta_y$ can be extracted from the correction to the Yukawa vertex. Indeed, using the Lagrangian at the scale $\Lambda'$, the corrected vertex is
\begin{equation}
\braket{\psi_i^t(0)\psi_j(0)\sigma^*(0)} = - \frac{iZ_y}{\sqrt{N}}Y_{ij}\gamma^0 = - \frac{i}{\sqrt{N}}Y_{ij}\gamma^0 - \frac{i\delta_y}{\sqrt{N}}Y_{ij}\gamma^0 \,,
\end{equation}
so that we can extract $\delta_y$ from the correction to the vertex as
\begin{equation}
- \frac{i\delta_y}{\sqrt{N}}Y_{ij}\gamma^0 = \braket{\psi_i^t(0)\psi_j(0)\sigma^*(0)}_{\mathrm{1-loop}} \,.  
\end{equation}
There are three diagrams that contribute to the vertex correction.

\begin{figure}[t]
	\centering
	\includegraphics[width=1\textwidth]{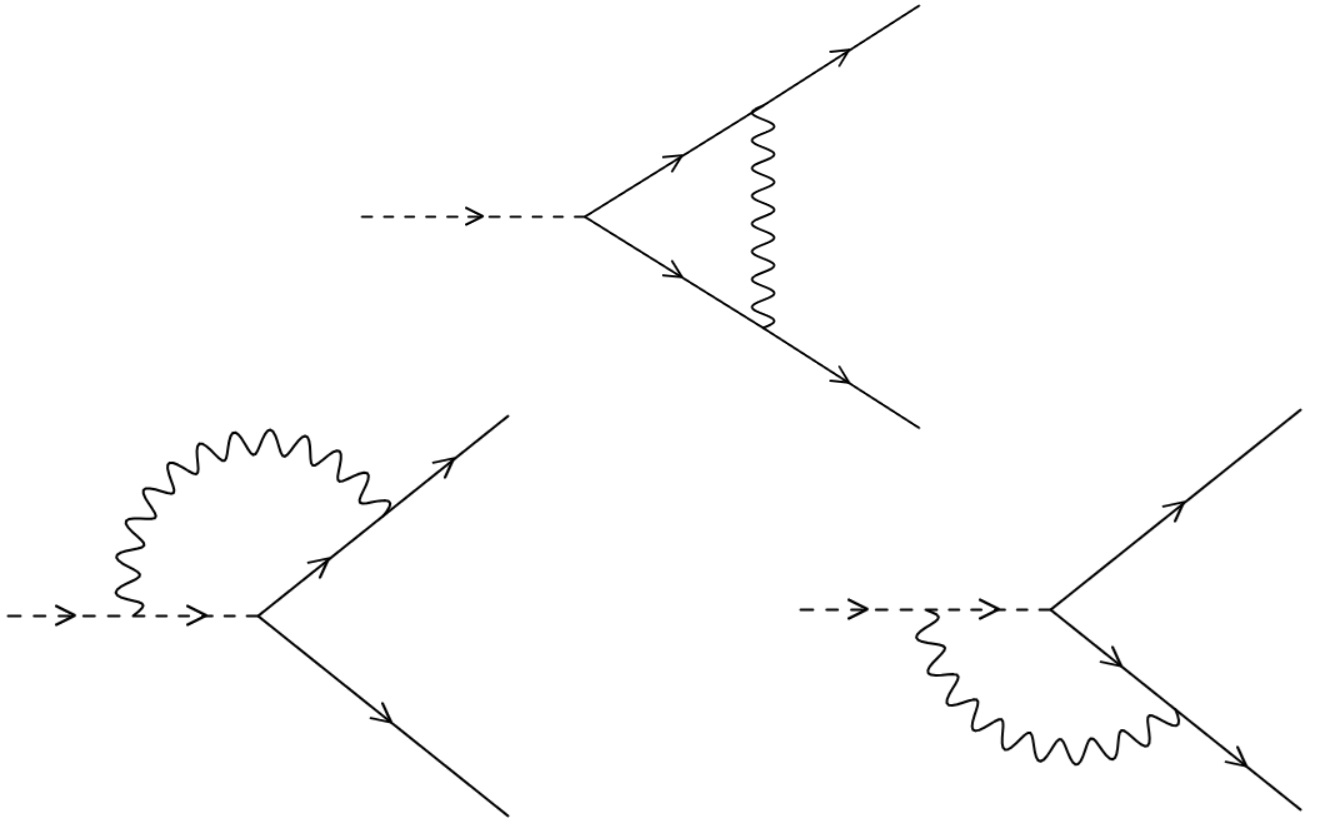}
	\caption{Diagrams contributing to the leading-order correction to the renormalization constant $\delta_y$ of the Yukawa vertex. The effective trilinear vertices on the bottom correspond to the one-loop diagram in figure \ref{fig:Induced vertex}.}
	\label{fig:Delta_y}
\end{figure}

The first is a one-loop diagram where a photon propagator connects the two fermion lines -- depicted in figure \ref{fig:Delta_y} on the top -- and it is given by\footnote{~Note that we cannot construct a diagram where the two fermion lines are connected by an HS propagator, due to charge conservation.} 
\begin{equation}
\begin{split}
- \frac{i\delta^1_y}{\sqrt{N}}Y_{ij}\gamma^0 &= \int \frac{d^3q}{(2\pi)^3} (-i(\gamma^\nu)^t \delta_{ik}) \frac{+i(-\slashed{q}^t)}{q^2}\delta_{kl} \left(-\frac{i}{\sqrt{N}}Y_{lm}\gamma^0 \right) \frac{-i\slashed{q}}{q^2}\delta_{mn}i\gamma^\mu \delta_{nj} \times \\
&\times \frac{8}{N|q|}\left(\delta_{\mu\nu}-a\frac{q_\mu q_\nu}{q^2}\right) = \frac{8i}{N\sqrt{N}}Y_{ij} \gamma^0 \int \frac{d^3q}{(2\pi)^3} \frac{\gamma^\nu \gamma^\mu}{|q|^3} \left(\delta_{\mu\nu}-a\frac{q_\mu q_\nu}{q^2}\right) \,,
\end{split}
\end{equation}
from which we get
\begin{equation}
\delta_y^1 = \frac{4(a-3)}{\pi^2 N} \log\left(\frac{\Lambda}{\Lambda'}\right) \,.  
\end{equation}

The other two diagrams are one-loop diagrams where a photon propagator connects one of the two fermion lines and the HS field (with the effective vertex \eqref{sigma photon vertex}) -- depicted in figure \ref{fig:Delta_y} on the bottom -- and they are given by
\begin{equation}
\begin{split}
- \frac{i\delta^2_y}{\sqrt{N}}&Y_{ij}\gamma^0 =  \int \frac{d^3q}{(2\pi)^3} \frac{-i}{\sqrt{N}}Y_{il}\gamma^0 \frac{-i\slashed{q}}{q^2} \delta_{lk} i\gamma^\nu \delta_{kj} \frac{8}{|q|} \left(+\frac{q^\mu}{4|q|}\right) \frac{8}{N|q|} \left(\delta_{\mu\nu}-a\frac{q_\mu q_\nu}{q^2} \right) \\
&+ \int \frac{d^3q}{(2\pi)^3} (-i(\gamma^\nu)^t \delta_{ik}) \frac{+i\slashed{q}^t}{q^2} \delta_{kl} \frac{-i}{\sqrt{N}}Y_{lj}\gamma^0 \frac{8}{|q|} \left(+\frac{q^\mu}{4|q|}\right) \frac{8}{N|q|} \left(\delta_{\mu\nu}-a\frac{q_\mu q_\nu}{q^2} \right) \,.
\end{split}
\end{equation}
The two contributions are equal and sum up to
\begin{equation}
- \frac{i\delta^2_y}{\sqrt{N}}Y_{ij}\gamma^0 = -\frac{32i}{N\sqrt{N}}Y_{ij}\gamma^0 \int \frac{d^3q}{(2\pi)^3} \frac{\slashed{q}\gamma^\nu (1-a)q_\nu}{|q|^5} \,,
\end{equation}
from which we get
\begin{equation}
\delta_y^2 = \frac{16(1-a)}{\pi^2 N} \log\left(\frac{\Lambda}{\Lambda'}\right) \,.   
\end{equation}
All in all we have
\begin{equation}
\delta_y = \delta_y^1 + \delta_y^2 = \frac{4(1-3a)}{\pi^2 N} \log\left(\frac{\Lambda}{\Lambda'}\right) \,.    
\end{equation}

\subsection{Beta Functions}
Let us collect here the results we got in a generic gauge for the various renormalization constants. They are (recall that the $y$ subscript in $\delta_\psi$ refers to the Yukawa contribution)
\begin{equation}
\begin{split}
\delta_\psi &= \frac{4(1_y+1-3a)}{3\pi^2 N} \log\left(\frac{\Lambda}{\Lambda'}\right) \,, \\
\delta M_{ij} &= \frac{4}{3\pi^2N} \left( (9-3a)M_{ij} - \frac{12}{N} \mathrm{Tr}M \delta_{ij} - 3(Y^\dagger M^T Y)_{ij} \right) \log\left(\frac{\Lambda}{\Lambda'}\right) \,, \\
\delta_\lambda &= \frac{8(1-2a)}{\pi^2 N} \log\left(\frac{\Lambda}{\Lambda'}\right) \,, \\
\delta_y &= \frac{4(1-3a)}{\pi^2 N} \log\left(\frac{\Lambda}{\Lambda'}\right) \,.
\end{split}
\end{equation}
Using the formulas \eqref{betafunctionsformulas} for the $\beta$ functions,
\begin{equation}
\beta_\lambda = \frac{\partial}{\partial\log\Lambda} \left( - \delta_\lambda - 2 \delta_\psi + 2 \delta_y  \right) \lambda \,, \qquad (\beta_M)_{ij} = \frac{\partial}{\partial\log\Lambda} \left( \delta_\psi M_{ij} - \delta M_{ij} \right) \,,
\end{equation}
we get that the gauge dependence cancels and the final result is
\begin{equation}
\begin{split}
\beta_\lambda &=  -\frac{16}{3\pi^2N}\lambda \,, \\  
\beta_M &= \frac{4}{3\pi^2N} \left( (-8+1_y)M + \frac{12}{N} \, \mathrm{Tr}M \mathbbm{1} + 3 \, Y^\dagger M^T Y \right) \,. 
\end{split}
\end{equation}
This proves equation \eqref{betafunctions}.

\input{triple-point-draft.bbl}

\end{document}

%% file: triple-point-draft.bbl
\providecommand{\href}[2]{#2}\begingroup\raggedright\endgroup